\documentclass[prb,aps,superscriptaddress,twocolumn,preprintnumbers,footinbib,longbibliography]{revtex4-2}

\usepackage{extarrows} 
\usepackage{graphicx}
\usepackage{dcolumn}
\usepackage{bm}
\usepackage{amsmath}
\usepackage{amsfonts}
\usepackage{amssymb}
\usepackage{hyperref}
\usepackage{MnSymbol}
\usepackage{upgreek}
\usepackage{subcaption}
\usepackage{verbatim}

\newcommand{\abscirc}{|\mkern-1mu\mathord{\circ}\mkern-1mu|}

\begin{document}

\title{Hyperdeterminant wavefunctions}

\author{Guan-Lin Lin}
\affiliation{Department of Physics, Boston College, Chestnut Hill, Massachusetts 02467, USA}

\author{Di Xiao}
\affiliation{Department of Materials Science and Engineering, University of Washington, Seattle, WA 98195, USA}

\author{Ying Ran}
\affiliation{Department of Physics, Boston College, Chestnut Hill, Massachusetts 02467, USA}

\date{\today}

\begin{abstract}
We systematically introduce hyperdeterminant wavefunctions as a variational-wavefunction-based theoretical framework for strongly correlated quantum states of matter, together with practical numerical simulation algorithms. This framework generalizes previously known fermionic parton constructions, yields reliable microscopics with intuitive physical pictures, and allows direct access to the fractionalized degrees of freedom together with associated microscopic effective field theories. We demonstrate the applications of this framework to fractional Chern insulators and quantum spin liquids. We comment that the hyperdeterminant states belong to a more general class of variational wavefunctions: the fused Gaussian states. 
\end{abstract}

\maketitle

\tableofcontents
\section{Introduction}
This long paper aims at developing a new theoretical framework to \emph{microscopically} investigate strongly correlated quantum states of matter, particularly when fractionalized degrees of freedom (d.o.f.) are relevant, e.g., in the context of fractional Chern insulators\cite{Tang2011, Sun2011, Neupert2011, Sheng2011, Regnault2011, Xiao2011, BergholtzLiu2013, ParameswaranRoySondhi2013} or quantum spin liquids\cite{Anderson1973, Lee2008, Balents2010, Zhou2017}. In generic quantum systems, microscopic investigations usually require numerical simulations, often extensive ones. In the past, various numerical techniques have been developed in this field, including exact diagonalization (ED)\cite{LauchliSudanMoessner2019, WietekLauchli2018, PrelovsekBonca2013}, density-matrix renormalization group (DMRG)\cite{White1992, Schollwoeck2011}, tensor-network (TN) methods\cite{Verstraete2008, Orus2014, GuLevinWen2009TERGTopological, GuVerstraeteWen2010GrassmannTN}, quantum Monte Carlo(QMC)\cite{Ceperley1995, Sandvik2010,AssaadEvertz2008QMC,XuQiZhangAssaadXuMeng2019CompactQEDQMC}, Gutzwiller-projected wavefunctions\cite{Gutzwiller1963, YokoyamaShiba1988, Gros1989}, and recently developed methods\cite{CarleoTroyer2017, Pfau2020FermiNet, HermannSchaetzleNoe2020PauliNet, LuoDaiFu2024MoireNN, TengDaiFu2025FQHNN} based on machine learning.

In these microscopic techniques, one often faces a trade-off: we hope to capture microscopic energetics accurately while also having intuitive physical pictures for fractionalized d.o.f. Unfortunately, these two goals, in many cases, cannot be achieved together within one framework. This issue becomes somewhat urgent, particularly given the experimental discovery of the fractional quantum anomalous Hall states\cite{Cai2023, Park2023, Zeng2023, Xu2023, Lu2024}, where intuitive physical pictures would be very helpful for interactions between theoretical and experimental efforts. 

The goal of an intuitive picture is to obtain a mean-field-like description of the quantum states under investigation. When fractionalized d.o.f. are present, we want to be able to access them directly. For example, when anyons\cite{LeinaasMyrheim1977, Wilczek1982,Halperin1984QuasiparticleStatistics,ArovasSchriefferWilczek1984FractionalStatistics}, or composite particles like composite fermions\cite{Jain1989, Jain2007Book, HalperinLeeRead1993, Son2015} are present in the quantum state, we hope to directly compute their band structures. However, since these are fractionalized d.o.f. carrying gauge charges, they cannot be directly seen in the many-body spectra of the electrons. On the other hand, most modern, accurate microscopic methods (e.g., ED, DMRG, TN, QMC) only have direct access to the electrons. Reading out the microscopics of the fractionalized d.o.f., although still possible\cite{Geraedts2016, Ippoliti2017}, becomes highly nontrivial.

This paper may be viewed as an attempt to identify a sweet spot where both goals can be achieved within the same framework. We introduce the concept of hyperdeterminant (Hdet) wavefunctions with the locality structure, which generalizes the previously known fermionic parton constructions\cite{Baskaran1987, AffleckMarston1988,   Jain1989IncompressibleQHStates, Wen1991, LeeNagaosaWen2006} for exotic states of matter. In some sense, Hdet wavefunctions are the natural generalization of Slater determinants, but for the situations when fractionalized d.o.f. are present. Namely, the Hdet wavefunctions are constructed from the fractionalized d.o.f., thereby allowing direct access to them.

As a class of variational wavefunctions, a crucial conceptual issue is the associated effective theories. Without knowing the associated effective theories, one may not directly know the quantum phase described by the variational wavefunctions. For this purpose, we develop a framework named the variational manifold path integral (VMPI). Applied to the Hdet wavefunctions, VMPI provides microscopic effective theories that capture physics from the lattice scale to the long distance, allowing one to sharply identify the quantum state of matter described by the Hdet wavefunction, as well as the collective mode dispersions. In simple terms, VMPI generalizes the time-dependent Hartree-Fock method of Slater determinants to Hdet wavefunctions, in a path-integral fashion.

Numerically simulating Hdet wavefunctions reliably is another challenge. Drastically different from the Slater determinants, simulating generic Hdet wavefunctions exactly is NP-hard. In the simplest limit, Hdet wavefunctions become products of determinants (e.g., a Laughlin wavefunction), which can be simulated by Monte Carlo. However, as we will see, deviating from this simplest limit is generally unavoidable if one wants to capture many important quantum phases at the microscopic level. To tackle this challenge, we develop a general method to approximately simulate Hdet wavefunctions: the projective expansion (PE), which allows one to systematically improve the approximation, order by order. The zeroth-order PE corresponds to a version of the mean-field approximation. Together with VMPI, PE allows one to compute the ground-state properties as well as the dynamical properties.

Hdet wavefunctions, the central topic of this paper, belong to an even larger class of variational wavefunctions, which we term the fused Gaussian states (FGS). FGS can capture more correlated quantum phases beyond the Hdet states, e.g., the FCI version of the $\nu=\frac{5}{2}$ nonabelian state\cite{MooreRead1991, GreiterWenWilczek1991}, as well as correlated superconducting states. We will comment on FGS at the end of this paper. Crucially, as Hdet states, the FGS-based framework can also achieve both goals: reliable microscopics together with intuitive physical pictures. Many of the methods developed here, including VMPI and PE, can be applied to FGS. 

This paper is organized as follows. In Sec.\ref{sec:def_math}, we introduce the definition of Hdet wavefunctions and their basic mathematical properties. However, not all Hdet wavefunctions are reasonable for low energy physics. In Sec.\ref{sec:locality}, we introduce the concept of the locality-structure of Hdet wavefunctions. Hdet wavefunctions admitting a locality structure become physically reasonable. To motivate the locality-structure, we first present explicit calculation results for the Hdet states in traditional FQH systems in Sec.\ref{sec:FQH_fusion}. Readers may skip Sec.\ref{sec:FQH_fusion} if one wants to directly see the definition of the locality-structure in Sec.\ref{sec:locality_def}. In Sec.\ref{sec:two_model_types}, we outline two types of model systems relevant for Hdet states: Type-(A) models featuring the full real-space d.o.f., and Type-(B) models confined within a topological band. Different model types lead to technical differences in the effective theory studies and simulation techniques presented afterward. 

In Sec.\ref{sec:VMPI} we present the effective theories for Hdet wavefunctions, by first introducing the general VMPI technique in Sec.\ref{sec:VMPI_setup} as a path integral version of the time-dependent variational principle. Before applying to Hdet states, we apply VMPI to Slater determinants in Sec.\ref{sec:VMPI_TDHF}. The motivation is two-fold. First, since the time-dependent Hartree-Fock(TDHF) is well known, this subsection may help readers quickly follow the VMPI formalism. Second, in order to have a general discussion of effective theories for Hdet states, we have to introduce a collection of somewhat abstract mathematical objects. It turns out that essentially the same mathematical objects will be present when VMPI is applied to Slater determinants, and it is much easier to introduce them there. In the only appendix of this paper, Appendix \ref{sec:AFM_example}, we explicitly demonstrate these mathematical objects with a simple, concrete TDHF calculation. In Sec.\ref{sec:Hdet_eff}, we apply VMPI to Hdet states, identifying the Ward identities for the gauge fluctuations. The benefit of keeping the discussion general is that, the calculation of the VMPI actions and the collective mode dispersions outlined in Sec.\ref{sec:Hdet_eff} can be viewed as well-defined algorithms. Together with PE, the microscopic effective theories for Hdet states can be computed directly.

In Sec.\ref{sec:proj_expansion}, we first describe the projective-expansion technique as a general method to approximately simulate Hdet states in Sec.\ref{sec:PE_setup}. In Sec.\ref{sec:PE_technical} we present the mathematical structure of PE with closed-form formulas, together with algorithms. For readers not interested in technical details/algorithms, this subsection can be skipped. To demonstrate its capabilities, we present benchmark results in Sec.\ref{sec:benchmark} for various quantum states and model systems. For instance, as advertised, the present framework allows direct access to the fractionalized d.o.f., and one may look into the parton band structures computed via the zeroth-order PE in FCI models in Fig.\ref{fig:parton_band}. Finally, in Sec.\ref{sec:discussion}, we comment on the generalization of Hdet states: FGS, as well as  future research directions. 

\section{Definition of Hdet wavefunctions and their mathematical properties}\label{sec:def_math}
\subsection{Definition of Hdet wavefunctions \label{sec:def}} 
Mathematically, the hyperdeterminant (Hdet) of a cubic tensor\cite{Cayley1845, Gelfand1994} is a natural generalization of the determinant for a square matrix. Here, by a cubic tensor, we mean a rank-$r$ tensor $T_{i_1,i_2,...i_r}$ with identical dimensions for all indices: $i_k=1,2,...,N$, $\forall k=1,2,...r$. Similar to the determinant of a rank-2 tensor (square matrix) $A_{ij}$, the (combinatorial) hyperdeterminant of a general cubic tensor $T$ is defined as:
\begin{widetext}
\begin{align}
\text{rank-2: }&\text{det}(A_{ij})\equiv  \sum_{P\in S_N}(-1)^P A_{1P(1)}A_{2P(2)}...A_{NP(N)}\notag\\
\text{rank-3: }&\text{Hdet}(T_{ijk})\equiv  \sum_{P,Q\in S_N}(-1)^P (-1)^Q T_{1P(1)Q(1)}T_{2P(2)Q(2)}...T_{NP(N)Q(N)}\notag\\
\text{rank-4: }&\text{Hdet}(T_{ijkl})\equiv  \sum_{P,Q,R\in S_N}(-1)^P (-1)^Q (-1)^R T_{1P(1)Q(1)R(1)}T_{2P(2)Q(2)R(2)}...T_{NP(N)Q(N)R(N)}\notag\\
\text{rank-$r$: }&\text{Hdet}(T)\equiv  \sum_{\sigma_2,...\sigma_r\in S_N}\prod_{\alpha=2}^{r} (-1)^{\sigma_\alpha} \prod_{i=1}^N T_{i \sigma_2(i)...\sigma_r(i)}
\label{eq:Hdet_def}
\end{align}
\end{widetext}

We know that Slater determinants play a fundamental role in investigating conventional quantum states of matter of electrons. Before defining an Hdet many-body wavefunction, we would like to bring the readers' attention to the following specific viewpoint of a Slater determinant many-body wavefunction. 

For a free-electron quantum many-body state, we define the overlap matrix $A$ as:
\begin{align}
A_{ij}\equiv \langle \psi_i^{(e)}|\phi_j^{(e)}\rangle.\label{eq:Slater_overlap_matrix}
\end{align}
Here, $|\psi_i^{(e)}\rangle$ labels a complete basis in the electron's single-particle Hilbert space $\mathcal H^{(e)}$: $i=1,2,...,\text{dim}\mathcal H^{(e)}$. $|\phi_j^{(e)}\rangle$, however, labels the $N_e$ (total electron number) \emph{filled } single-particle states in $\mathcal H^{(e)}$: $j=1,2,...,N_e$. Note that $A$ is a rectangular matrix, whose row-dimension could be much larger than the column-dimension. The quantum many-body state $|\Psi^{(e)}\rangle$ is well-defined as long as its overlap with an arbitrary many-body basis state $|\psi_{i_1}^{(e)}\psi_{i_2}^{(e)}...\psi_{i_{N_e}}^{(e)}\rangle$ is defined. Here, we define 
\begin{align}
|\psi_{i_1}^{(e)}\psi_{i_2}^{(e)}...\psi_{i_{N_e}}^{(e)}\rangle\equiv |\psi_{i_1}^{(e)}\rangle|\psi_{i_2}^{(e)}\rangle...|\psi_{i_{N_e}}^{(e)}\rangle
\label{eq:many_body_basis}
\end{align}
to be the simple direct product of single-particle states, which is \emph{not} antisymmetrized.

\begin{figure}
\begin{subfigure}{0.9\linewidth}
  \centering
\includegraphics[width=0.7\linewidth]{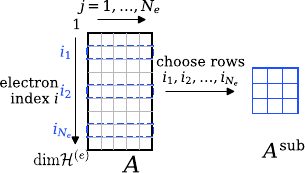}
  \caption{Slater-det in Eq.(\ref{eq:Slater_wfc_def})}
  \label{fig:Slater_submatrix}
\end{subfigure}%
\vspace{0.5cm}
\begin{subfigure}{.8\linewidth}
  \centering
\includegraphics[width=\linewidth]{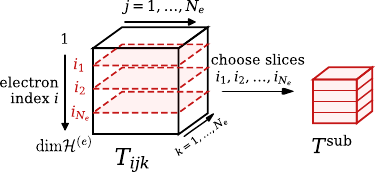}
  \caption{Hdet in Eq.(\ref{eq:Hdet_wfc_def})}
  \label{fig:Hdet_subtensor}
\end{subfigure}
\caption{Illustration of selecting submatrix/subtensor to define many-body wavefunctions.}
\label{fig:Slater_submatrix_Hdet_subtensor}
\end{figure}

The Slater-determinant wavefunction $|\Psi^{(e)}\rangle$ associated with $A$ is defined as:
\begin{align}\text{Slater-det: }
\langle \psi_{i_1}^{(e)}\psi_{i_2}^{(e)}...\psi_{i_{N_e}}^{(e)}|\Psi^{(e)}\rangle\equiv \text{det}(A^{\text{sub}}),\label{eq:Slater_wfc_def}
\end{align}
where $A^{\text{sub}}$ is the square submatrix obtained by stacking $i_1,i_2,...,i_{N_e}$ rows of the rectangular matrix $A$ (see Fig.\ref{fig:Slater_submatrix} for an illustration). Crucially, the exponentially-sized many-body state $|\Psi^{(e)}\rangle$ is fully characterized by a polynomial-sized matrix $A$.

Next, consider a general fermionic parton construction, in which an electron is split into a few species of fermionic partons. For instance, one may have $e\sim f^{(1)}\cdot f^{(2)}$ for a bosonic electron, and $e\sim f^{(1)}\cdot f^{(2)}\cdot f^{(3)}$ for a fermionic electron. In the past, different versions of fermionic parton constructions have been introduced in the context of fractionalized quantum states of matter (e.g., quantum spin liquids, fractional quantum Hall states), from either the real-space viewpoints\cite{Baskaran1987, AffleckMarston1988,   Jain1989IncompressibleQHStates, Wen1991, Wen-Symmetric-QSL, LeeNagaosaWen2006}, or from the tensor-product Hilbert space viewpoints\cite{PasquierHaldane1998,MurthyShankar2003HamiltonianFQHE,HuXiaoRan2024}. Regardless of which version one uses, the basic idea remains the same: by transitioning to an enlarged Hilbert space of partons, one can construct exotic, fractionalized states with intuitive, mean-field-like physical pictures. In a previous work involving some of us\cite{HuXiaoRan2024}, the core object for a general parton construction is identified: the \emph{tensor of fusion amplitudes}. For instance,
\begin{align}
e\sim f^{(1)}\cdot f^{(2)}:\;\; T_{ijk}\equiv& \langle \psi_i^{(e)}\abscirc
\phi^{(1)}_{j}\rangle|\phi^{(2)}_k\rangle,\notag\\
e\sim f^{(1)}\cdot f^{(2)}\cdot f^{(3)}:\;\; T_{ijkl}\equiv& \langle \psi_i^{(e)}\abscirc\phi^{(1)}_{j}\rangle|\phi^{(2)}_k\rangle|\phi^{(3)}_l\rangle.\label{eq:fusion_T_def}
\end{align}
Here, similar to the Slater-determinant case, $|\psi_i^{(e)}\rangle$ labels a complete basis in the electron's single-particle Hilbert space $\mathcal H^{(e)}$: $i=1,2,...,\text{dim}\mathcal H^{(e)}$. $|\phi_j^{(1)}\rangle, |\phi_k^{(2)}\rangle, |\phi_l^{(3)}\rangle$, on the other hand, labels the $N_e$ (total electron number) \emph{filled } single-particle states for each parton species: $j,k,l=1,2,...,N_e$. We also introduced an operator $"\circ"$ representing the fusion, which acts between $\mathcal H^{(e)}$ and $\otimes_p\mathcal H^{(p)}$ ($p=1,2,3..$). Note that despite one electron being split into a few species of partons, the total number of partons in each species is still equal to the total number of electrons.

The fusion tensor $T$ defines how to "glue" the partons back into electrons in a general sense, and should be viewed as a rectangular-shaped tensor because the first index $i$'s dimension, i.e., $\text{dim}\mathcal H^{(e)}$, is generically larger than $N_e$. The key observation of Ref\cite{HuXiaoRan2024} is that the (either bosonic or fermionic) electron many-body wavefunction in such parton constructions is really an Hdet wavefunction:
\begin{align}
\text{Hdet: }
\langle \psi_{i_1}^{(e)}\psi_{i_2}^{(e)}...\psi_{i_{N_e}}^{(e)}|\Psi^{(e)}\rangle\equiv \text{Hdet}(T^{\text{sub}}),\label{eq:Hdet_wfc_def}
\end{align}
where the cubic subtensor $T^{\text{sub}}$ is obtained by stacking $i_1,i_2,...,i_{N_e}$ slices of the rectangular tensor $T$ (see Fig.\ref{fig:Hdet_subtensor} for an illustration). Again, we define the many-body basis as in Eq.(\ref{eq:many_body_basis}), and the exponentially-sized many-body state $|\Psi^{(e)}\rangle$ is fully characterized by a polynomial-sized tensor $T$. 

\subsection{Mathematical properties of Hdet wavefunctions}
Before proceeding to physics, we list a few essential mathematical properties of the Hdet wavefunctions.

\subsubsection{Statistics theorem} For an even-rank (odd-rank) tensor $T$, the Hdet electronic wavefunction defined as Eq.(\ref{eq:Hdet_wfc_def}) is automatically a fermionic (bosonic) many-body state.

\emph{Proof:} This is a direct consequence of the definition of the combinatorial hyperdeterminant in Eq.(\ref{eq:Hdet_def}). Consider a permutation $\pi\in S_N$. Comparing with Eq.(\ref{eq:Hdet_wfc_def}), the overlap of $|\Psi^{(e)}\rangle$ with the permuted many-body basis $|\psi_{i_\pi(1)}^{(e)}\psi_{i_\pi(2)}^{(e)}...\psi_{i_{\pi(N_e)}}^{(e)}\rangle$ is the hyperdeterminant of the cubic subtensor $\tilde T^{\text{sub}}$ that can be obtained from $T^{\text{sub}}$ via permuting the first index by $\pi$. 
\begin{align}
&\text{Hdet}(\tilde T)=\sum_{\sigma_2,...\sigma_r\in S_N}\prod_{\alpha=2}^{r} (-1)^{\sigma_\alpha} \prod_{i=1}^N \tilde T_{i \sigma_2(i)...\sigma_r(i)}\notag\\
=&\sum_{\sigma_2,...\sigma_r\in S_N}\prod_{\alpha=2}^{r} (-1)^{\sigma_\alpha} \prod_{i=1}^N T_{\pi(i) \sigma_2(i)...\sigma_r(i)}\notag\\
=&((-1)^\pi)^{r-1}\sum_{\tilde \sigma_2,...\tilde \sigma_r\in S_N}\prod_{\alpha=2}^{r} (-1)^{\tilde \sigma_\alpha} \prod_{k=1}^N T_{k \tilde \sigma_2(k)...\tilde \sigma_r(k)}\notag\\
\end{align}
where we have defined $k\equiv\pi(i)$ (so $i=\pi^{-1}(k)$) $\tilde\sigma_\alpha=\sigma_\alpha\cdot\pi^{-1}$, and consequently $(-1)^{\tilde\sigma_\alpha}=(-1)^\pi\cdot(-1)^{\sigma_\alpha}$. Therefore, if $r$ is odd(even), the Hdet wavefunction is permutation symmetric(antisymmetric).

\subsubsection{Basis-independence theorem}  The fusion tensor $T$ defined in Eq.(\ref{eq:fusion_T_def}) is dependent on the basis-choice for both the electron and the partons. The natural question is whether the many-body Hdet wavefunction in Eq.(\ref{eq:Hdet_wfc_def}) depends on these basis choices or not. Here we show that \emph{the Hdet wavefunction is basis choice independent}.

\begin{figure}
\includegraphics[width=0.4\textwidth]{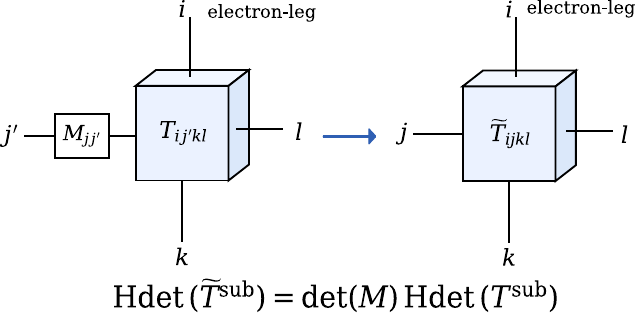}
\caption{Illustration of the tensor contraction with a matrix Eq.(\ref{eq:tensor_leg_contraction}), leading to a factor $\det(M)$ for the Hdet.}
\label{fig:basis_independence}
\end{figure}

First, the filled-parton single-particle states $\{|\phi_k^{(p)}\rangle\}$ (which do \emph{not} need to be orthonormal) span a sub-Hilbert space for the parton species-$p$. One may choose a different basis $\{|\tilde\phi_k^{(p)}\rangle\}$ that spans the \emph{same} sub-Hilbert space. Namely, we may perform a transformation of basis for the parton species-$p$:
\begin{align}
|\tilde\phi_k^{(p)}\rangle=\sum_{l=1}^{N_e} M_{kl}|\phi_l^{(p)}\rangle,
\end{align}
with an invertible $N_e\times N_e$ matrix $M$. For instance, if there are three species of partons, and $p=1$, the corresponding rank-4 fusion tensor is transformed as:
\begin{align}
T_{ijkl}\rightarrow \tilde T_{ijkl}\equiv \sum_{j'=1}^{N_e} M_{j j'}T_{ij'kl}\label{eq:tensor_leg_contraction}
\end{align}
Graphically, $\tilde T$ is obtained from $T$ by contracting the matrix $M$ with the  parton-leg-$p$ (see Fig.\ref{fig:basis_independence} for illustration). In this case, one can straightforwardly show that the Hdet wavefunction defined by $T$ or $\tilde T$ only differs by an overall constant factor $\text{det}(M)$. Namely, \emph{the many-body wavefunction is independent of the parton-basis choice.}

Second, one may choose an \emph{arbitrary} set of $N_e$ electronic orbitals $\{|\tilde\psi_{\alpha}^{(e)}\rangle\}$ ($\alpha=1,2,...,N_e$, which again do \emph{not} need to be an orthonormal set), that are linear superpositions of the basis $\{|\psi_i^{(e)}\rangle\}$:
\begin{align}
|\tilde\psi_\alpha^{(e)}\rangle=\sum_{i=1}^{\text{dim}\mathcal H^{(e)}} S_{\alpha i}|\psi_i^{(e)}\rangle,
\end{align}
characterized by a rectangular $N_e\times \text{dim}\mathcal H^{(e)}$ matrix $S$.
According to Eq.(\ref{eq:Hdet_wfc_def}), to compute the overlap between $|\tilde\psi_{1}^{(e)}\tilde\psi_{2}^{(e)}...\tilde\psi_{N_e}^{(e)}\rangle$ with the many-body state $|\Psi^{(e)}\rangle$, we only need to compute the hyperdeterminant of the cubic tensor $\mathbf T$ obtained from $T$ by contracting the electron-leg with the matrix $S^*$: 
\begin{align}
\langle \tilde\psi_{1}^{(e)}\tilde\psi_{2}^{(e)}...\tilde\psi_{N_e}^{(e)}|\Psi^{(e)}\rangle=\text{Hdet}(\mathbf T),\label{eq:Hdet_wfc_transformed}
\end{align}
where
\begin{align}
\mathbf T_{\alpha j k...}\equiv \sum_{i=1}^{\text{dim}\mathcal H^{(e)}}S^*_{\alpha i} T_{ijk...}.
\end{align}
The question is whether Eq.(\ref{eq:Hdet_wfc_transformed}) is consistent with the definition Eq.(\ref{eq:Hdet_wfc_def}). The answer is positive. 

The simplest way to see this is to consider the definition Eq.(\ref{eq:Hdet_def}), where the first index (i.e., electron-leg) is \emph{not} permuted. Consequently, one can straightforwardly show that:
\begin{align}
&\text{Hdet}(\mathbf T)\notag\\
=&\sum_{i_1,i_2,...i_{N_e}=1}^{\text{dim}\mathcal H^{(e)}} S^*_{1,i_1}S^*_{2,i_2}...S^*_{N,i_{N_e}}\langle\psi_{i_1}^{(e)}\psi_{i_2}^{(e)}...\psi_{i_{N_e}}^{(e)}|\Psi^{(e)}\rangle.
\end{align}
Namely, based on the definition Eq.(\ref{eq:many_body_basis}), Eq.(\ref{eq:Hdet_wfc_transformed}) is consistent with Eq.(\ref{eq:Hdet_wfc_def}). Therefore, \emph{the Hdet wavefunction is also independent of the basis-choice for the electron}.

\subsubsection{Second-quantized formulation} So far we have been using the first-quantized formulation to discuss Hdet wavefunctions. It may be helpful to translate the construction into the second-quantized formulation as well. For this purpose, one may define the fusion operator $\hat {\mathbf F}$ as:
\begin{align}
e\sim f^{(1)}f^{(2)}:\;\hat {\mathbf F}=&\sum_{ijk} T_{ijk} b_i^\dagger f_j^{(1)}f_k^{(2)}\notag\\
e\sim f^{(1)}f^{(2)}f^{(3)}:\;\hat {\mathbf F}=&\sum_{ijkl} T_{ijkl} c_i^\dagger f_j^{(1)}f_k^{(2)}f_{l}^{(3)}\notag\\
...\label{eq:full_fusion_orig}
\end{align}
\emph{Only in this subsection, for simplicity, we assume that the single particle states $|\psi_i^{(e)}\rangle$, $|\phi_j^{(1)}\rangle$, $|\phi_k^{(2)}\rangle$, etc, are orthonormal in their own Hilbert spaces, respectively.} The second-quantized operator $f_k^{(p)}$ is the fermionic annihilation operator associated with $|\phi_k^{(p)}\rangle$. The electron creation operator, $b_i^\dagger$ for bosonic electron (rank-odd $T$) and $c_i^\dagger$ for fermionic electron (rank-even $T$), is associated with $|\psi_i^{(e)}\rangle$. The fusion operator $"\circ"$ in Eq.(\ref{eq:fusion_T_def}) can be exactly interpreted as $\hat{\mathbf F}$:
\begin{align}
\circ=\hat{\mathbf F}.
\end{align}

One can check that the unsymmetrized overlap with an identical-particle state in the Fock space is equal to a second-quantized overlap:
\begin{align}
\langle \psi_{i_1}^{(e)}\psi_{i_2}^{(e)}...\psi_{i_{N_e}}^{(e)}|\Psi^{(e)}\rangle=\frac{1}{\sqrt{N_e!}}\langle0| e_{i_{N_e}}...e_{i_2}e_{i_1}|\Psi^{(e)}\rangle,
\end{align}
where $e_i\equiv c_i$ ($e_i\equiv b_i$) for fermionic(bosonic) electrons (even if more than one bosons fill the same single-particle level in the latter case). From this, it is easy to show that the electron many-body wavefunction in Eq.(\ref{eq:Hdet_wfc_def}) can be obtained by:
\begin{align}
|\Psi^{(e)}\rangle=\frac{1}{\sqrt{N_e!}}\hat{\mathbf F}^{N_e}|\Psi^{(1)}\rangle|\Psi^{(2)}\rangle...,\label{eq:fusion_Ne_power}
\end{align}
where the many-body state $|\Psi^{(p)}\rangle$ for parton species-$p$ is nothing but a single Slater-determinant state: 
\begin{align}
|\Psi^{(p)}\rangle=\prod_{k=1}^{N_e} f^{(p)\dagger}_{k}|0\rangle.
\end{align}

\subsubsection{Relation to products of determinants and CPD} Previously, it is known that a fractionalized quantum state can often be written as a product of determinants. One seminal example is the FQH Laughlin wavefunctions at filling $\nu=\frac{1}{m}$. In the disk geometry with symmetric gauge, $\Psi_{\text{Laughlin}}=\prod_{i<j}(z_i-z_j)^m e^{-\frac{\sum_i|z_i|^2}{4l_e^2}}$, which can be viewed as the $m$-th power of a Slater determinant (i.e., the $\nu=1$ integer quantum Hall state).

The product of determinants is a \emph{very special} limit of the general Hdet wavefunctions. It is straightforward to show that, based on the definition Eq.(\ref{eq:Hdet_def}), if a cubic tensor $T$ can be written as the product of square matrices in the following sense, then the hyperdeterminant of $T$ is the product of the determinants of the matrices:
\begin{align}
\text{If } T_{ijk}=A_{ij}\cdot B_{ik},\text{ then: } \text{Hdet}(T)=\text{det}(A)\text{det}(B),\label{eq:prod_det_two_parton}
\end{align}
and similarly:
\begin{align}
\text{If }& T_{ijkl}=A_{ij}\cdot B_{ik}\cdot C_{il},\notag\\
\text{ then: }& \text{Hdet}(T)=\text{det}(A)\text{det}(B)\text{det}(C).\label{eq:prod_det_three_parton}
\end{align}
In Sec.\ref{sec:FQH_fusion} we will show that the tensor $T$ for the Laughlin state can indeed be written as the product of matrices, but only after choosing a specific basis for the electron (the overcomplete real-space coherent-state basis $\{|z\rangle\}$ inside LLL). Clearly, whether the tensor can be written as a product of matrices crucially depends on the basis choice of the electron $\{|\psi_i^{(e)}\rangle\}$; but it does \emph{not} depend on the basis choice for partons, which only rotates the matrices $A,B...$. For example, if one wants to write down the same Laughlin state using the electron's momentum-space Bloch basis (e.g., in the Landau gauge), $T$ will not be a product of the matrices. Consequently, Laughlin states are not products of determinants in the momentum-space basis. But from the bases-independence theorem, they are still always hyperdeterminants.

A natural question is: under what condition can the tensor $T$ be written as a product of matrices? This is an important question because, in practice, one can easily simulate a product of determinants using the Monte Carlo algorithm, which is why Laughlin states have been accurately investigated numerically. To answer this question, it is instructive to consider the two-parton construction for bosonic electrons in Eq.(\ref{eq:prod_det_two_parton}). After fixing the index $i$, $T_{ijk}=A_{ij}B_{ik}$ can be interpreted as a singular value decomposition (SVD) of the matrix $T_{jk}(i)\equiv T_{ijk}$. Defining $A_j(i)\equiv A_{ij}$ and $B_{k}(i)\equiv B_{ik}$, one finds:
\begin{align}
T_{ijk}=A_{ij} B_{ik}\Longleftrightarrow T_{jk}(i)=A_j(i)B_{k}(i).
\end{align}
Namely, the square matrix $T_{jk}(i)$ is a product of two vectors, i.e., has only one nonzero singular value. Moving on to the three-parton construction for fermionic electrons in Eq.(\ref{eq:prod_det_three_parton}), the corresponding relation is:
\begin{align}
T_{ijkl}=A_{ij}B_{ik}C_{il}\Longleftrightarrow T_{jkl}(i)=A_j(i)B_{k}(i)C_{l}(i).\label{eq:CPD_one}
\end{align}
This is a generalization of SVD to the rank-3 tensor $T(i)$: the Canonical Polyadic Decomposition (CPD)\cite{CarrollChang1970, Harshman1970, KoldaBader2009} of the rank-3 tensor $T_{jkl}(i)$. For a general rank-$r$ tensor $\mathcal A$,  the CPD is defined as:
\begin{align}
\mathcal A_{i_1,i_2...,i_r}=\sum_{\alpha=1}^{R}\lambda_\alpha\cdot x^{(1)}_{\alpha,i_1}x^{(2)}_{\alpha,i_2}...x^{(r)}_{\alpha,i_r}\label{eq:CPD},
\end{align}
where the minimal integer $R$ for this decomposition, for historical reasons, is called the CPD-rank of the tensor $\mathcal A$ (to avoid confusion of terminology, we warn the reader that the CPD-rank and the rank (i.e., $r$) of a tensor have no direct relation). Mathematically, one may always absorb the CPD-values $\lambda_{\alpha}$ into the CPD-vectors $x$'s, which we will \emph{not} do due to physical reasons that will become clear in Sec.\ref{sec:locality}. 

Eq.(\ref{eq:CPD_one}) means the CPD-rank of the rank-3 tensor $T_{jkl}(i)$ is one -- a very special tensor. As already mentioned, the CPD-rank does \emph{not} depend on the parton-basis choice, which only rotates the CPD-vectors $x$'s. The CPD-rank only depends on the choice of the electron state $|\psi_i^{(e)}\rangle$ participating in the fusion tensor. The question now becomes: given a fusion tensor $T$, which defines $T_{jk..}(\tilde i)\equiv \langle \psi_{\tilde i}^{(e)} \abscirc\phi^{(1)}_j\rangle|\phi^{(2)}_k\rangle...$ ($\forall |\psi_{\tilde i}^{(e)}\rangle\in \mathcal H^{(e)}$), is it always possible to choose one electron state $|\psi_{\tilde i}^{(e)}\rangle$ such that $T(\tilde i)$'s CPD-rank is one?

The answer is negative. We will show in Sec.\ref{sec:FQH_fusion} that, even for the next FQH state in the Jain's composite-fermion sequence with $n=2$, the CPD-rank of $T(\tilde i)$ is larger than one for whatever choice of $|\psi_{\tilde i}^{(e)}\rangle$. For bosonic (fermionic) electrons, Jain's sequence is at filling fractions $\nu=\frac{n}{n(2s-1)+1}$ ($\nu=\frac{n}{2ns+1}$), where the integer $s=1,2,..$ characterizes the degrees of flux attachment. $n=1$ cases recover the Laughlin states. One can consider the state at $s=1$ and $n=2$ ($\nu=\frac{2}{3}$ for bosonic electrons, and $\nu=\frac{2}{5}$ for fermionic electrons). It turns out that, even after choosing the optimal $|\psi_{\tilde i}^{(e)}\rangle$ minimizing the CPD-rank, for $\nu=\frac{2}{3}$ bosonic case, the CPD-rank of $T(\tilde i)$ is three. And for $\nu=\frac{2}{5}$ fermionic case, the CPD-rank of $T(\tilde i)$ is four. Already in these traditional FQH systems, the many-body wavefunctions \emph{cannot} be written as a product of determinants. But, as pointed out in Ref.\cite{HuXiaoRan2024}, these are still hyperdeterminants.

\subsubsection{Exact Hdet is NP-hard}
Drastically different from computing the determinant of a matrix, exactly computing Hdet of a tensor is generally NP-hard\cite{HillarLim2013}. The Gaussian Elimination Algorithm, allowing a polynomial complexity for computing the determinant, is no longer applicable for (rank$>2$) tensors. One simple way to see the hardness of the Hdet is to consider another special limit of a rank-3 cubic tensor: $T_{ijk}=A_{ij}\delta_{jk}$. In this limit, the permutations in Eq.(\ref{eq:Hdet_def}) must satisfy $P=Q$ for a nonzero contribution, and consequently $\text{Hdet}(T)=\text{perm}(A)$. Here, we introduced the permanent of the square matrix $A$, which is an analog of the determinant but without the permutation sign. Permanent is related to the many-body wavefunctions of free bosons and is well-known to be NP-hard to compute\cite{Valiant1979, AaronsonArkhipov2011}.

From a different perspective, one may consider the CPD structure of a fusion tensor $T$. For instance, we will show in Sec.\ref{sec:FQH_fusion} that, after choosing the (overcomplete) coherent state basis for the electron, which is CPD-optimal, the fusion tensor $T$ for the bosonic $\nu=\frac{2}{3}$ Jain's composite-fermion state satisfies:
\begin{align}
T^{\nu=\frac{2}{3}}_{zjk}=\langle z^{(e)}\abscirc\phi_j^{(1)}\rangle|\phi^{(2)}_k\rangle=\sum_{\alpha=1}^3 \lambda_\alpha A_{\alpha,zj} B_{\alpha,zk},\;\; \forall z.
\end{align}
Here the complex real-space coordinate $z$ labels the electron's coherent states $|z^{(e)}\rangle$ that can be obtained after projecting the delta-function at $z$ into the LLL. Note that the equation above should be viewed as the CPD of the rank-2 tensor $T_{jk}(z)\equiv T^{\nu=\frac{2}{3}}_{zjk}$. A priori, both the CPD-rank $R=3$ and the CPD-values $\lambda_{\alpha}$ may depend on $z$, but the Galilean invariance in the system dictates that they will not.

After inserting into the definition Eq.(\ref{eq:Hdet_def}), one finds that the real-space wavefunctions for these systems are a summation of \emph{exponentially} many products of determinants:
\begin{widetext}
\begin{align}
&\langle z_1z_2...z_{N_e}|\Psi^{(e)}_{\nu=\frac{2}{3}}\rangle=\sum_{P,Q\in S_{N_e}}(-1)^P(-1)^Q\sum_{\alpha_1}\lambda_{\alpha_1}A_{\alpha_1,z_1P(1)}B_{\alpha_1,z_1 Q(1)}...\sum_{\alpha_{N_e}}\lambda_{\alpha_{N_e}}A_{\alpha_{N_e},z_{N_e}P(N_e)}B_{\alpha_{N_e},z_{N_e} Q(N_e)}\notag\\
=&\sum_{\{\alpha_i\}}\prod_{i=1}^{N_e}\lambda_{\alpha_i}\left[\sum_{P\in S_{N_e}}(-1)^PA_{\alpha_1,z_1P(1)}...A_{\alpha_{N_e},z_{N_e}P(N_e)}\right]\left[\sum_{Q\in S_{N_e}}(-1)^QB_{\alpha_1,z_1Q(1)}...B_{\alpha_{N_e},z_{N_e}Q(N_e)}\right]\notag\\
=&\sum_{\{\alpha_i=1,2,3\}}\prod_{i=1}^{N_e}\lambda_{\alpha_i}\cdot\text{det}(\mathbf A[\alpha_1,\alpha_2,...\alpha_{N_e}])\cdot \text{det}(\mathbf B[\alpha_1,\alpha_2,...\alpha_{N_e}]).
\end{align}
\end{widetext}
Here, if we consider $A_{\alpha,z_ij}$ as three matrices $A(\alpha)$ with $\alpha=1,2,3$ and $i$($j$) labeling rows(columns), the matrices $\mathbf A[\alpha_1,\alpha_2,...\alpha_{N_e}]$ (and similarly $\mathbf B[\alpha_1,\alpha_2,...\alpha_{N_e}]$) are formed by choosing rows among three $A(\alpha)$ matrices according to $\alpha_1,\alpha_2,...\alpha_{N_e}$. This is a summation of $3^{N_e}$ products of two determinants. 

Similarly, for the fermionic $\nu=\frac{2}{5}$ Jain's state, we will show:
\begin{align}
T^{\nu=\frac{2}{5}}_{zjkl}=\sum_{\alpha=1}^4 \lambda_\alpha A_{\alpha,zj} B_{\alpha,zk}C_{\alpha,zl},\;\; \forall z.
\end{align}
Its real-space many-body wavefunction is a summation of $4^{N_e}$ products of three determinants. 

From this analysis, the importance of the CPD structure of the fusion tensor becomes clear. At this point, we introduce the concept of \textbf{fusion channels}. \emph{If the fusion tensor $T(i)$ has CPD-rank $R$:
\begin{align}
T_{jk...}(i)\equiv \langle \psi_i^{(e)}\abscirc\phi^{(1)}_{j}\rangle|\phi^{(2)}_k\rangle...=\sum_{\alpha=1}^R\lambda_\alpha A^{(1)}_{\alpha,ij}A^{(2)}_{\alpha,ik}...\label{eq:CPD_structure}
\end{align}
we say there are $R$ fusion channels to the electron state $|\psi_i^{(e)}\rangle$}. We also say that the corresponding coefficient $\lambda_{\alpha}$ is the \textbf{fusion-amplitude} for the fusion-channel $\alpha$. The "..." indicates there may be more than two species of partons. Throughout this paper, the superscripts $^{(1)}$, $^{(2)}$ are reserved to denote the parton species. In the next section, we will explore the physical meanings of the fusion channels. 

\subsubsection{About uniqueness of the fusion-tensor $T$}
Due to the bases-independence theorem, WLOG, in this subsection, we assume the electronic single-particle basis $\{|\phi_i^{(e)}\rangle\}$ is orthonormal.

Suppose a many-body state $|\Psi^{(e)}_{\text{Slater}}\rangle$ is a Slater-determinant wavefunction associated with some overlap matrix $A_{ij}$ (see Eq.(\ref{eq:Slater_overlap_matrix})). Even if one does not know $A$, conceptually, it is straightforward to read it out from the many-body state $|\Psi^{(e)}_{\text{Slater}}\rangle$ itself, up to a unitary transformation among the filled states: $\tilde A_{ij}=\sum_{j'}A_{ij'}U_{j'j}$. For example, one can compute the single-body correlation matrix $\rho_{ii'}=\langle \Psi^{(e)}_{\text{Slater}}|c^\dagger_{i'}c_{i}|\Psi^{(e)}_{\text{Slater}}\rangle$. In terms of matrices $\rho=A\cdot A^\dagger$. After diagonalizing $\rho$, $A$ can be found. This means that the matrix $A$ is physical up to the expected unitary gauge redundancy $U$.

What happens for an Hdet wavefunction $|\Psi^{(e)}\rangle$ defined via a fusion tensor $T_{ijk...}$ as in Eq.(\ref{eq:Hdet_wfc_def})? Can one read out $T$ from $|\Psi^{(e)}\rangle$ up to expected gauge redundancies? As mentioned in the bases-independence theorem, any $GL(N_e,\mathcal C)$ (i.e., an invertible $N_e\times N_e$ complex matrix) transformation on a parton-index (i.e., $j,k,...$) does not change the electronic state $|\Psi^{(e)}\rangle$. In addition, obviously, any permutation between the parton indices, e.g., $j\leftrightarrow k$, does not change the electronic state as well. If we denote the number of parton species to be $n_{\text{p}}$, we have the expected $GL(N_e,\mathcal C)^{n_{\text{p}}}$ and $S_{n_{\text{p}}}$ ($S$ is the permutation group) gauge redundancies. 

This question is important because it tells us whether or not the fusion tensor $T$ in an Hdet wavefunction is as physical as the overlap matrix $A$ in a Slater-determinant wavefunction. At this point, we do not have rigorous mathematical results on this issue. However, it appears to us that generically $T$ should be fully fixed by $|\Psi^{(e)}\rangle$ up to the expected gauge redundancies above.

\section{The locality-structure of Hdet wavefunctions}\label{sec:locality}
So far, we have defined the Hdet wavefunctions mathematically. From the pure math viewpoint, one does not necessarily attach any physical meaning to the tensor $T$ as a fusion tensor, as defined in Eq.(\ref{eq:fusion_T_def}). One can even write down an \emph{arbitrary} tensor $T$, which nevertheless well defines a quantum many-body wavefunction according to Eq.(\ref{eq:Hdet_wfc_def}). However, not all tensors $T$'s are relevant to low-energy quantum many-body physics. A quantum Hamiltonian needs a notion of \emph{locality} even to define the spatial dimension in a physical sense, and the same is true for its low-energy states. 

For instance, for the ground state of a gapped local Hamiltonian, the connected correlators of local observables should have an exponential decay\cite{HastingsKoma2006}. Although a general sharp statement for the ground state of a gapless local Hamiltonian is not known to the authors' best knowledge, physical intuition indicates that the connected correlator of local observables should have a power-law decay, which cannot be too slow. In unitary conformal field theories, the bounds for the exponents of these power laws are known\cite{Mack1977, Rychkov2016}.

This raises an important question: \emph{what kind of fusion tensors $T$'s (and the associated Hdet wavefunctions) can be viewed as physically reasonable for low-energy physics?} For a Slater-determinant wavefunction, which appears in various mean-field analyses of electron systems, it is usually not a problem since one typically first writes down a physical parent free-fermion Hamiltonian. For Hdet wavefunctions describing strongly correlated, possibly fractionalized quantum phases of matter, we need a prescription to encode locality in the fusion tensor $T$, which we call the \emph{locality-structure} of $T$ and define in Sec.\ref{sec:locality_structure}. Namely, only an Hdet wavefunction whose fusion tensor $T$ admits the locality-structure is a physically reasonable low-energy wavefunction. This locality-structure immediately leads to an effective field theory description of the Hdet wavefunctions discussed in Sec.\ref{sec:Hdet_eff}.

\subsection{Fusion tensors for composite-fermion states in FQH}\label{sec:FQH_fusion}
To explore the locality structure of fusion tensors, we will start by introducing explicit analytical results from a concrete parton construction in the context of traditional FQH states as a demonstration. Already in the early development of the composite-fermion approach, Jain pointed out that a real-space parton construction can obtain his composite-fermion wavefunctions\cite{Jain1989IncompressibleQHStates, Jain2007Book}. This parton construction was later further developed to describe more general FQH states, both abelian and nonabelian\cite{Wen1991, BlokWen1990, WenZee1992, BarkeshliWen2010}. 

In this real-space parton construction, one first identifies the electron operator as a product of a few species of real-space parton operators. As a demonstration, for simplicity, we will focus on the $s=1$ composite fermion states (but $n$ can be a general positive integer, including the composite Fermi liquid state corresponding to $n\rightarrow \infty$.) For bosonic (fermionic) electrons, $s=1$ means that there are two (three) species of partons:
\begin{align}
b_{\mathbf r}^\dagger=& f_{\mathbf r}^{(1)\dagger}f_{\mathbf r}^{(2)\dagger}\notag\\
c_{\mathbf r}^\dagger=& f_{\mathbf r}^{(1)\dagger}f_{\mathbf r}^{(2)\dagger}f_{\mathbf r}^{(3)\dagger}\label{eq:parton_real_space}
\end{align}
If one assigns the electron's charge to be unity $q^{(e)}=1$, then each parton species-$p$ should carry a fractional charge $q^{(p)}$, so that the total charge is the electron's charge: $\sum_p q^{(p)}=1$. For instance, for the bosonic Jain's FQH sequence at filling $\nu=\frac{n}{n+1}$, $q^{(1)}=\frac{n}{n+1}$ and $q^{(2)}=\frac{1}{n+1}$. And for the fermionic Jain's FQH sequence at filling $\nu=\frac{n}{2n+1}$, $q^{(1)}=q^{(2)}=\frac{n}{2n+1}$ and $q^{(3)}=\frac{1}{2n+1}$.

Because each parton has a fractional charge, it sees a fractional strength of the magnetic field: the magnetic length of parton species-$p$ becomes $l_p\equiv \frac{l_e}{\sqrt{q^{(p)}}}$. But note that the total number of partons for a given species is the same as the number of electrons. Consequently, each parton species is at an integer filling fraction: for the bosonic (fermionic) Jain's FQH sequence: $\nu^{(1)}=1$ and $\nu^{(2)}=n$ ($\nu^{(1)}=\nu^{(2)}=1$ and $\nu^{(3)}=n$). At the mean-field level, these partons are treated as non-interacting and simply form gapped Slater-determinant states. This is the basic idea behind all parton constructions. By enlarging the Hilbert space via introducing partons, one can obtain an intuitive description of a fractionalized quantum state in terms of free-parton states. 

If one denotes a Slater-determinant parton-state by its integer filling fraction, the electron wavefunction is obtained simply by gluing all species of partons together in real space. For instance, for the bosonic Jain's sequence: $\tilde\Psi^{(e)}(\mathbf r_1,...\mathbf r_N)\equiv\Psi^{(1)}_{\nu=1}(\mathbf r_1,...\mathbf r_N)\cdot \Psi^{(2)}_{\nu=n}(\mathbf r_1,...\mathbf r_N)$, which we will abbreviate as 
\begin{align}
\text{bosonic Jain's states: }\tilde\Psi^{(e)}\equiv\Psi_{\nu=1}\cdot \Psi_{\nu=n}.
\end{align} 
And similarly,
\begin{align}
\text{fermionic Jain's states: }\tilde\Psi^{(e)}\equiv\Psi^2_{\nu=1}\cdot \Psi_{\nu=n}.
\end{align}
When $n\rightarrow \infty$, the last parton species carries zero electric charge and sees no magnetic field. A natural choice for $\Psi_{\nu=n\rightarrow\infty}=\Psi_{FL}$ is a Fermi liquid state with a circular fermi surface, leading to the composite Fermi liquid state for $\nu=1$ ($\nu=\frac{1}{2}$) for the bosonic (fermionic) case.

Note that $\tilde\Psi^{(e)}$ is a product of Slater determinants. When $n=1$, it recovers the Laughlin state successfully. A problem emerges when $n>1$: the wavefunction $\tilde \Psi^{(e)}$ is no longer inside the LLL. A wavefunction in the LLL should be holomorphic (up to the Gaussian factor), but the parton wavefunction $\Psi_{\nu=n>1}$ contains antiholomorphic coordinates $\bar z_i$. This problem originates from the construction Eq.(\ref{eq:parton_real_space}) at the very beginning, which explicitly relies on a real-space that goes outside the LLL. Physically, it is well known that a Chern band (e.g., LLL) cannot be Wannier-localized \cite{Thouless1982, Brouder2007, Read2017}. Inside the LLL, instead of the standard notation of real-space, the guiding center coordinates do not commute -- leading to a noncommutative space\cite{GirvinMacDonaldPlatzman1986, PasquierHaldane1998, Susskind2001}.

If we view the gluing procedure from Eq.(\ref{eq:parton_real_space}) as the \emph{first} projection from the parton Hilbert spaces to the electron Hilbert space, to obtain a wavefunction inside the LLL for $n>1$, it is crucial to perform a \emph{second} projection: $\mathcal P_{LLL}$, which simply eliminates all components of an electron wavefunction outside the LLL:
\begin{align}
\Psi^{(e)}\equiv \mathcal P_{LLL} \tilde\Psi^{(e)}.
\end{align}

Exactly what $\mathcal P_{LLL}$ does to the many-body wavefunction has been a puzzle. On the numerical side, once $\mathcal P_{LLL}$ is exactly implemented for $n>1$ states, it is practically known that $\Psi^{(e)}$ becomes exponentially hard to compute, and no general analytical results are known. This difficulty prompted the development of an approximation scheme by Jain and Kamilla\cite{JainKamilla1997} in which $\Psi^{(e)}$ is approximated as a single determinant of a matrix whose elements are many-body wavefunctions, though whether this approximation is under control remains unclear. 

On the conceptual side, it is not even clear whether $\mathcal P_{LLL}$ changes the long-distance universal physics of the many-body state. Without $\mathcal P_{LLL}$, the topological field theory (Chern-Simons theory) in the standard commutative real-space describing the long-distance physics of $\tilde \Psi^{(e)}$ can be obtained through a prescription developed by Wen and Zee\cite{WenZee1992}. However, in the presence of $\mathcal P_{LLL}$, it has been emphasized that noncommutative field theory needs to be used\cite{ Susskind2001QHNonCommutativeCS, Polychronakos2001MatrixCSQH, HellermanVanRaamsdonk2001QHNCFT,FradkinJejjalaLeigh2002NCCSDuality, DongSenthil2020NCFTCompositeFermiLiquids}. While significant progress has been made toward a noncommutative field-theoretic description of the bosonic Jain states\cite{GoldmanSenthil2022LLLBosonicJain}, a similar description for the fermionic Jain states remains an open problem.

Ref.\cite{HuXiaoRan2024} pointed out that it is exactly $\mathcal P_{LLL}$ that changes the product of determinant $\tilde\Psi^{(e)}$ into a hyperdeterminant $\Psi^{(e)}$. The fusion amplitude in Eq.(\ref{eq:fusion_T_def}) between partons and an electronic orbital $|\psi_i^{(e)}\rangle$ \emph{inside} the LLL, according to the real-space construction Eq.(\ref{eq:parton_real_space}), is nothing but:
\begin{align}
T_{ijk}\underset{\text{el.}}{\overset{\text{boson}}{=}} &\int d\mathbf r \langle \psi^{(e)}_i|\mathbf r^{(e)}\rangle\langle \mathbf r^{(1)}|\phi^{(1)}_j\rangle\langle\mathbf r^{(2)}|\phi^{(2)}_k\rangle,\notag\\
T_{ijkl}\underset{\text{el.}}{\overset{\text{fermi}}{=}} &\int d\mathbf r \langle \psi^{(e)}_i|\mathbf r^{(e)}\rangle\langle \mathbf r^{(1)}|\phi^{(1)}_j\rangle\langle\mathbf r^{(2)}|\phi^{(2)}_k\rangle\langle\mathbf r^{(3)}|\phi^{(3)}_l\rangle,
\end{align}
which consequently defines the Hdet many-body wavefunction $|\Psi^{(e)}\rangle$. In the present work, we will demonstrate the \emph{general} analytical form for this fusion tensor $T$, which will be shown useful even beyond the FQH context.

\subsubsection{Showcase: bosonic electron with two-parton fusion rule}
Here, we present the general fusion tensor for the real-space parton construction $e\sim f^{(1)}f^{(2)}$ Eq.(\ref{eq:parton_real_space}) in the following format. Assuming a filled single-particle state $|\phi^{(1)}_{aLL}\rangle$ for parton species-$1$ is in its $a$th-LL and a filled state $|\phi^{(2)}_{bLL}\rangle$ for parton species-$2$ is in its $b$th-LL ($a,b=0,1,2,...$) (both the parton fractional charges $q^{(1)},q^{(2)}$ are nonzero and $\sum_p q^{(p)}=1$), then the fusion amplitude to an electron's coherent state $|z^{(e)}\rangle$ turns out to be:
\begin{widetext}
\begin{align}
\langle z^{(e)}\abscirc\phi^{(1)}_{aLL}\rangle|\phi^{(2)}_{bLL}\rangle
=(-1)^a [q^{(1)}]^{\frac{b}{2}}[q^{(2)}]^{a+\frac{b}{2}}\sqrt{C^a_{a+b}}\sum_{l=0}^{a+b} (-1)^l \sqrt{C^l_{a+b}} \left(\frac{q^{(1)}}{q^{(2)}}\right)^{\frac{l}{2}}\langle z^{(1)}_{(a+b-l)}|\phi^{(1)}_{aLL}\rangle\langle z^{(2)}_{(l)}|\phi^{(2)}_{bLL}\rangle,\label{eq:boson_general_fusion}
\end{align}
\end{widetext}
where $C^l_{n}=\frac{n!}{(n-l)!l!}$ is the binomial coefficient. \emph{Note that, by using the electron's coherent state, the projection $\mathcal P_{LLL}$ is already implemented.} Also note that we allow both parton-$1$ and parton-$2$ to fill higher LLs. This is because, later, we will use these fusion amplitudes to describe states beyond the traditional FQH systems, e.g. those in FCI systems.

As a quick remark, there is a more symmetric and systematic way (e.g., easily generalizable to more parton species) to express the same fusion tensor in Eq.(\ref{eq:boson_general_fusion}):
\begin{align}
\langle z^{(e)}\abscirc\phi^{(1)}_{aLL}\rangle|\phi^{(2)}_{bLL}\rangle=\sum_{k,l}&\sqrt\frac{k!l!}{a!b!}(q^{(1)})^{\frac{a-k}{2}}(q^{(2)})^{\frac{b-l}{2}}X_{kl}\notag\\
&\cdot\langle z^{(1)}_{(k)}|\phi^{(1)}_{aLL}\rangle\langle z^{(2)}_{(l)}|\phi^{(2)}_{bLL}\rangle,
\end{align}
where $k,l$ are nonnegative integers satisfying $k+l=a+b$, and $X_{kl}$ is defined via the following binomial expansion:
\begin{align}
&\sum_{k,l} X_{kl} A^kB^l\notag\\
\equiv&[(q^{(1)}-1)A+q^{(2)}B]^a[q^{(1)}A+(q^{(2)}-1)B]^b.
\end{align}

On the RHS of Eq.(\ref{eq:boson_general_fusion}), we have introduced the generalized coherent states in the $n$th LL for partons: $|z_{(m)}\rangle_{nLL}$ ($m,n=0,1,2...$). To explain its definition, we need to have a brief discussion on the well-known guiding-center and cyclotron degrees of freedom for a single charged particle moving in a uniform magnetic field. With the canonical momentum $\boldsymbol\Pi=\mathbf p+q\mathbf A$, the cyclotron coordinate $\boldsymbol\eta$ and the guiding-center coordinate $\mathbf R$ are defined as:
\begin{align}
\boldsymbol\eta\equiv &\frac{l^2}{\hbar}\hat z\times \boldsymbol\Pi, &\mathbf R\equiv&\mathbf r-\boldsymbol\eta,
\end{align}
where $l$ is the magnetic length. $\boldsymbol\eta$ and $\mathbf R$ commute with each other, but
\begin{align}
[\hat\eta_x,\hat\eta_y]=&il^2,& [\hat R_x,\hat R_y]=-il^2. 
\end{align}
Namely, the single-particle Hilbert space can be decomposed as $\mathcal H=\mathcal H_{\mathbf R}\otimes \mathcal H_{\boldsymbol\eta}$. While $\boldsymbol\eta$ is the d.o.f. between different LLs, $\mathbf R$ is the d.o.f. within each LL. The commutation relation above allows one to construct the harmonic oscillator ladder operators:
\begin{align}
\hat a_{\mathbf R}\equiv &\frac{\hat R_x-i\hat R_y}{\sqrt 2 l},&\hat a^\dagger_{\mathbf R}=&\frac{\hat R_x+i\hat R_y}{\sqrt 2 l},\notag\\
\hat a_{\boldsymbol\eta}\equiv&\frac{\hat \eta_x+i\hat \eta_y}{\sqrt 2 l},&\hat a^\dagger_{\boldsymbol\eta}=&\frac{\hat \eta_x-i\hat\eta_y}{\sqrt 2 l}.
\end{align}

The magnetic translation along a vector $(x,y)$ only acts in the guiding-center subspace:
\begin{align}
\hat D_{\mathbf R}(z\equiv x+iy)\equiv \exp\Big[\frac{i}{l^2}(-y \hat R_x+x \hat R_y)\Big],
\end{align}
which satisfies the Girvin-MacDonald-Platzman (GMP) algebra\cite{GirvinMacDonaldPlatzman1986}
\begin{align}
\hat D_{\mathbf R}(\alpha ) \hat D_{\mathbf R}(\beta)=\exp\Big[\frac{\bar\alpha\beta-\alpha\bar\beta}{2l^2}\Big]\hat D_{\mathbf R}(\beta ) \hat D_{\mathbf R}(\alpha).
\end{align}

The generalized coherent state $|z_{(m)}\rangle_{nLL}$ is defined as:
\begin{align}
|z_{(m)}\rangle_{nLL}\equiv \big(\hat D_{\mathbf R}(z)|m\rangle_{\mathbf R}\big)\otimes|n\rangle_{\boldsymbol\eta}\label{eq:generalized_coherent_state_def}
\end{align}
where $|0\rangle_{\mathbf R}$ ($|0\rangle_{\boldsymbol\eta}$) is the vacuum of the ladder operator $\hat a_{\mathbf R}$ ($\hat a_{\boldsymbol\eta}$), and $|m\rangle_{\mathbf R}\equiv \frac{\hat a_{\mathbf R}^{\dagger m}}{\sqrt{m!}}|0\rangle_{\mathbf R}$ ($|n\rangle_{\boldsymbol\eta}\equiv \frac{\hat a_{\boldsymbol\eta}^{\dagger n}}{\sqrt{n!}}|0\rangle_{\boldsymbol\eta}$). 

The electron's coherent state $|z^{(e)}\rangle\in\mathcal H^{(e)}$ in the LLL on the LHS of Eq.(\ref{eq:boson_general_fusion}) is exactly the generalized coherent state when $n=m=0$, with the electron's magnetic length $l=l_e$. The parton generalized coherent state $|z^{(1)}_{(a+b-l)}\rangle_{aLL}\in \mathcal H^{(1)}$ ( $|z^{(2)}_{(l)}\rangle_{bLL}\in \mathcal H^{(2)}$) on the RHS of Eq.(\ref{eq:boson_general_fusion}), is for $m=(a+b-l), n=a$ ($m=l, n=b$) with the parton magnetic length $l=l^{(1)}$ ($l=l^{(2)}$). \textbf{Note that}, \emph{to save notation, we neglect the subscripts $_{aLL}$ and $_{bLL}$ for the generalized coherent states in Eq.(\ref{eq:boson_general_fusion}). And throughout this paper, the superscripts $^{(1)}$, $^{(2)}$ are reserved to denote the parton species.}

Below, we work within the symmetric gauge. One can show that
\begin{align}
|z_{(n)}\rangle_{nLL}=(-1)^n \cdot \mathcal P_{nLL}|\mathbf r_z\rangle, 
\end{align}
where $\mathbf r_z=(\text{Re}[z],\text{Im}[z])$, and $|\mathbf r_z\rangle$'s wavefunction is $\sqrt{2\pi}\;l\cdot\delta(\mathbf r-\mathbf r_z)$ (the normalization factor is chosen to simplify the expressions here). $\mathcal P_{nLL}=|n\rangle_{\boldsymbol\eta} \cdot_{\boldsymbol\eta}\langle n|$ in this single-particle case. 

The wavefunction of the generalized coherent state $|z_{(m)}\rangle_{nLL}$ can be written down, together with their overlaps within the same LL:
\begin{widetext}
\begin{align}
\langle \mathbf r_\omega|z_{(m)}\rangle_{nLL}=&(-1)^n\sqrt{\frac{m!}{n!}}\, e^{-\frac{|\omega|^2 + |z|^2 - 2\bar{z} \omega}{4l^2}}
\left( \frac{\bar{z} - \bar{\omega}}{\sqrt 2 l} \right)^{n - m}
L_m^{(n - m)}\left( \frac{|\omega - z|^2}{2l^2} \right),\text{ if } n\ge m,\notag\\
\langle\mathbf r_\omega|z_{(m)}\rangle_{nLL}=&(-1)^n\sqrt{\frac{n!}{m!}}\, e^{-\frac{|\omega|^2 + |z|^2 - 2\bar{z} \omega}{4l^2}}
\left( \frac{\omega-z}{\sqrt 2 l} \right)^{m - n}
L_n^{(m - n)}\left( \frac{|\omega - z|^2}{2l^2} \right),\text{ if } m\ge n,\notag\\
\langle \omega_{(n)}|z_{(m)}\rangle_{aLL}=&\sqrt{\frac{m!}{n!}}\, e^{-\frac{|\omega|^2 + |z|^2 - 2\bar{z} \omega}{4l^2}}
\left( \frac{\bar{z} - \bar{\omega}}{\sqrt 2 l} \right)^{n - m}
L_m^{(n - m)}\left( \frac{|\omega - z|^2}{2l^2} \right),\text{ if } n\ge m,\notag\\
\langle\omega_{(n)}|z_{(m)}\rangle_{aLL}=&\sqrt{\frac{n!}{m!}}\, e^{-\frac{|\omega|^2 + |z|^2 - 2\bar{z} \omega}{4l^2}}
\left( \frac{\omega-z}{\sqrt 2 l} \right)^{m - n}
L_n^{(m - n)}\left( \frac{|\omega - z|^2}{2l^2} \right),\text{ if } m\ge n,\notag\\\label{eq:generalized_coherent_state_wavefunction_overlap}
\end{align}
\end{widetext}
where $L^\alpha_n(x)=\frac{x^{-\alpha} e^x}{n!}\frac{d^n}{dx^n}(x^{n+\alpha}e^{-x})$ is the generalized Laguerre polynomials. Despite the similarity between the expressions of the wavefunctions and overlaps, conceptually they are very different objects: the $n$-index in the wavefunctions labels the LL, while the $n$-index in the overlaps labels the generalized coherent state in the bra. The crucial points are: (1) these wavefunctions are exponentially localized at $z$ by the magnetic length $l$. (2) $|z_{(m)}\rangle_{nLL}$ carries $z$-centered magnetic rotation angular momentum $L=m-n$.

With these results, we can have a physical interpretation for the fusion amplitude Eq.(\ref{eq:boson_general_fusion}). To be concrete, we list a few special cases: $a=b=0$, $a=0,b=1$ and $a=b=1$.
\begin{align}
&\langle z^{(e)}\abscirc\phi^{(1)}_{0LL}\rangle|\phi^{(2)}_{0LL}\rangle=\Xi^{00}_{00},\notag\\
&\langle z^{(e)}\abscirc\phi^{(1)}_{0LL}\rangle|\phi^{(2)}_{1LL}\rangle=[q^{(1)}q^{(2)}]^{\frac{1}{2}}\Xi^{01}_{10}-q^{(1)}\Xi^{01}_{01},\notag\\
&\langle z^{(e)}\abscirc\phi^{(1)}_{1LL}\rangle|\phi^{(2)}_{1LL}\rangle=-\sqrt{2}\,[q^{(1)}]^{\frac{1}{2}}[q^{(2)}]^{\frac{3}{2}}\Xi^{11}_{20}\notag\\
&+2q^{(1)}q^{(2)}\Xi^{11}_{11}-\sqrt{2}[q^{(1)}]^{\frac{3}{2}}[q^{(2)}]^{\frac{1}{2}}\Xi^{11}_{02},
\label{eq:boson_special_fusion}
\end{align}
where we defined a shorthand notation:
\begin{align}
\Xi^{ab}_{kl}\equiv\langle z^{(1)}_{(k)}|\phi^{(1)}_{aLL}\rangle\langle z^{(2)}_{(l)}|\phi^{(2)}_{bLL}\rangle .
\end{align}
Note that the parton states $|\phi^{(1)}\rangle$ and $|\phi^{(2)}\rangle$ can be \emph{any} orbitals in their LLs. If one replaces $|\phi^{(1)}\rangle\rightarrow |\phi_j^{(1)}\rangle$ and $|\phi^{(2)}\rangle\rightarrow |\phi_k^{(2)}\rangle$, \emph{these expressions in Eq.(\ref{eq:boson_special_fusion}) are exactly the CPD structure Eq.(\ref{eq:CPD_structure}) of the fusion tensor!} Note that, although the analytical result Eq.(\ref{eq:boson_general_fusion}) is obtained using the disk geometry, because the generalized coherent states are exponentially localized, on a finite-size system (e.g., finite-size torus), this fusion tensor is exponentially accurate when the system size is much larger than the involved parton magnetic lengths.

In the simplest case, $a=b=0$, there is only one fusion channel (i.e., CPD-rank $=1$). But when the parton fills higher LLs, the number of fusion channels increases. Generally, the number of fusion channels:
\begin{align}
R_{ab}=a+b+1.
\end{align}

For instance, for the $\nu=\frac{2}{3}$ Jain's state, parton-$1$ has $q^{(1)}=\frac{2}{3}$, filling the LLL. Parton-$2$ has $q^{(2)}=\frac{1}{3}$, filling both the LLL and the 1st LL. The results here show that, after choosing the electron's coherent state basis, there will be one fusion channel involving parton-$2$ in the LLL, and two fusion channels involving parton-$2$ in the 1st LL, so a total of three fusion channels (i.e., CPD-rank is three). 

One may ask the opposite question: which electron basis should one choose so that the number of fusion channels is minimal (i.e., CPD-rank optimal)? Although we do not have analytical proof at this point, our numerical simulation shows that the CPD-rank-optimal basis for electrons is indeed the coherent-state basis for FQH composite fermion states. For example, for $\nu=\frac{2}{3}$ Jain's state, the fusion tensor for an electron's complete basis (e.g., the Bloch basis in the LLL) $T_{ijk}$ can be easily computed on a torus, where $i=1,2,... \frac{3}{2}N_e$, $j,k=1,2,..., N_e$. Then one can numerically compute the fusion tensor (a square matrix) $T_{jk}(\tilde i)$ for an arbitrary state $| \psi_{\tilde i}^{(e)}\rangle\in H^{(e)}$ according to the bases-independence theorem. We find that numerically minimizing the CPD-rank (i.e., SVD-rank in the present case) of $T_{jk}(\tilde i)$ always leads to a coherent state $|\psi_{\tilde i}^{(e)}\rangle\propto|z^{(e)}\rangle$ at some position $z$.

Suppose we interpret the overlaps on the RHS as the parton propagation in Eq.(\ref{eq:boson_general_fusion}), clearly, \emph{a parton always needs to propagate to parton orbitals localized near the electron's position in order to fuse into an electron}, which we will call the \emph{locality structure} for the fusion tensor Eq.(\ref{eq:boson_general_fusion}). Motivated by this, we will soon define the locality structure for general fusion tensors.

Next, let's further understand the physical meaning of the locality structure here. If $a=b=0$, both partons propagate to the most localized state at the electron's position $z$: the coherent state $|z_{(0)}^{(p)}\rangle$ for parton species-$p$ in their LLL. If partons fill higher LLs, the generalized coherent state $|z_{(m)}^{(p)}\rangle_{nLL}$ appears in their fusion rule. Pictorially, the orbital $|z_{(m)}^{(p)}\rangle_{nLL}$ is most localized in $n$th LL if $m=0$, and more extended as $m$ increases. Physically, one may interpret it as a parton propagating to guiding-center orbitals in the \emph{vicinity} of $z$. The generalized coherent state $|z_{(m)}^{(p)}\rangle_{nLL}$ is nothing but the angular-momentum expansion of such guiding-center orbitals.

Before proceeding, to save notation, \emph{we introduce the abbreviation for the generalized coherent states} $|z^{(p)}_{(m)}\rangle_{nLL}$:
\begin{align}
|z;\mathbf m^{(p)}\rangle\equiv |z^{(p)}_{(m)}\rangle_{nLL}.
\end{align}
Namely, the bold-font $\bf m$ can be viewed as the combination of the indices $_{(m)}$ and $_{nLL}$.

In fact, one can obtain both the form and coefficients (up to an overall factor) in the CPD structure Eq.(\ref{eq:boson_general_fusion}) based on the locality structure and the Galilean invariance, without resorting to sophisticated calculations.

To see this, it is convenient to introduce another viewpoint on the locality structure -- the \emph{local-fusion operator}. The fusion tensor in Eq.(\ref{eq:boson_general_fusion}), (after adding subscripts for partons) $T_{zjk}$ has a locality structure, which can be written as:
\begin{align}
\langle z^{(e)}\abscirc\phi^{(1)}_j\rangle|\phi^{(2)}_k\rangle=\sum_{\alpha=1}^R \lambda_\alpha \langle z;\mathbf m_{\alpha}^{(1)}|\phi^{(1)}_j\rangle \langle z;\mathbf m_{\alpha}^{(2)}|\phi^{(2)}_k\rangle,\label{eq:boson_local_fusion}
\end{align}
where $|z;\mathbf m_{\alpha}^{(p)}\rangle$ are the generalized coherent states participating in the fusion tensor Eq.(\ref{eq:boson_general_fusion}) for the fusion-channel $\alpha$, and $\lambda_\alpha$ are the rather complicated coefficients involving $\sqrt{C_{a+b}^k}$. \emph{In principle, the fusion-amplitude $\lambda_\alpha=\lambda_\alpha(z)$ may depend on $z$, but at least in this showcase example, it is not after a proper gauge for the parton and electron states is chosen.} We then \emph{define a local-fusion operator}:
\begin{align}
\hat{ F}_{z}\equiv |z^{(e)}\rangle \cdot \Big[\sum_{\alpha=1}^R \lambda_{\alpha}\langle z;\mathbf m_{\alpha}^{(1)}|\langle z;\mathbf m_{\alpha}^{(2)}|\Big]\label{eq:boson_local_fusion_operator}
\end{align}
and \emph{define the integer $R$ as the number of local fusion channels}. Note that $\hat{ F}_z$ is an operator mapping a quantum state in the parton Hilbert space $\mathcal H^{(1)}\otimes \mathcal H^{(2)}$ to a quantum state in the electron's single-particle Hilbert space $\mathcal H^{(e)}$. (to avoid confusion with the bold-font many-body fusion operator $\hat{\mathbf F}$ in Eq.(\ref{eq:full_fusion_orig}) acting in the Fock spaces, we use regular font $\hat F_z$ here.) The fusion tensor can also be compactly expressed as:
\begin{align}
\langle z^{(e)}\abscirc\phi^{(1)}_j\rangle|\phi^{(2)}_k\rangle=\langle z^{(e)}|\hat{F}_z|\phi^{(1)}_j\rangle|\phi^{(2)}_k\rangle.\label{eq:parton_image_discussion}
\end{align}
Namely, $\hat{ F}_z$ fully specifies how the partons are glued into the electron coherent state $|z^{(e)}\rangle$. 

From this viewpoint, one may think of the state in the parton Hilbert space: $\Big[\sum_{\alpha=1}^R \lambda^*_{\alpha}|z;\mathbf m^{(1)}_\alpha\rangle| z;\mathbf m^{(2)}_\alpha\rangle\Big]$ as the "parton-image-state" of the electron local orbital $|z^{(e)}\rangle$. If there is only one fusion channel, then the parton-image-state is a \emph{product state} formed by different parton species. When the number of fusion channels increases, it indicates that the parton-image-state is an entangled quantum state between different parton species. The CPD-rank can be viewed as a measure for the entanglement between the parton species. Precisely, for the current two-parton fusion, the CPD-rank (i.e. SVD-rank) is directly related to the well-known 0th-order Renyi entanglement entropy: $S_0=\log (\text{SVD-rank})$. For more-parton fusion, the CPD-rank is associated with the so-called Schmidt measure in the context of quantum information theory\cite{EisertBriegel2001}.

Now we show that the local fusion operator $\hat{F}_z$ for the fusion tensor Eq.(\ref{eq:boson_general_fusion}) can be obtained via the Galilean invariance. First, the combination of $_{aLL}\langle z^{(1)}_{(a+b-l)}|$ and $_{bLL}\langle z^{(2)}_{(l)}|$ for the generalized coherent states on the RHS is simply a consequence of the $z$-centered angular momentum conservation: the first orbital carries $L^{(1)}=b-l$, and the second orbital carries $L^{(2)}=l-b$. Their summation must be zero because the electron coherent state carries $L^{(e)}=0$.

Second, the coefficients $\lambda_\alpha$ can also be obtained (up to an overall factor). The electron's wavefunction in the LLL must be a holomorphic function (up to Gaussian factors), and cannot depend on $\bar z$. Using the identity for the overlap between two generalized coherent states (special case of Eq.(\ref{eq:generalized_coherent_state_wavefunction_overlap})):
\begin{align}
\langle z_{(m)}|\omega_{(0)}\rangle_{nLL}=\frac{1}{\sqrt{m!}} \Big(\frac{\bar \omega-\bar z}{\sqrt2 l}\Big)^m e^{-\frac{|\omega|^2+|z|^2}{4l^2}+\frac{z\bar\omega}{2l^2}},\label{eq:z_w_overlap}
\end{align}
we know that, if one plugs in $|\phi^{(1)}_{aLL}\rangle\rightarrow |\omega_{(0)}^{(1)}\rangle_{aLL}$, $|\phi^{(2)}_{bLL}\rangle\rightarrow |\zeta_{(0)}^{(2)}\rangle_{bLL}$, the RHS of Eq.(\ref{eq:boson_general_fusion}) becomes a multinomial of $\bar z,\bar\omega,\bar\zeta$ together with Gaussian factors. Each term of the multinomial is $\propto (\bar\omega-\bar z)^{a+b-l}(\bar\zeta-\bar z)^l$. The binomial coefficients in $\lambda_\alpha$ are exactly such that the $\bar z$-dependence all cancels out so that the multinomial $\propto [(\bar\omega-\bar z)-(\bar\zeta-\bar z)]^{a+b}=(\bar\omega-\bar\zeta)^{a+b}$. In addition, it is easy to show that there is only one solution of $\lambda_\alpha$ (up to an overall factor) such that the $\bar z$-dependence in Eq.(\ref{eq:boson_general_fusion}) cancels out.

We have shown that the number of local fusion channels increases as the partons fill higher LLs. Precisely, in the $\nu=\frac{n}{n+1}$ bosonic Jain's sequence, the number of local fusion channels is $1+2+...+n=\frac{n(n+1)}{2}$ (even in the thermodynamic limit). What would happen for the bosonic $\nu=1$ composite Fermi liquid corresponding to $n\rightarrow \infty$? The number of local fusion channels would go to infinity in the thermodynamic limit for sure. On the other hand, for any finite-size system, the number of local fusion channels should be finite based on physical intuition. How does the number of local fusion channels depend on the system size for $\nu=1$?

Here, we list the two-parton fusion tensor for the case $q^{(1)}=1$ and $q^{(2)}=0$, as in the case of bosonic $\nu=1$ composite Fermi liquid:
\begin{widetext}
\begin{align}
&\langle z^{(e)}\abscirc\phi^{(1)}_{aLL}\rangle|\phi^{(2)}\rangle=\int \frac{d\alpha}{2\pi l_e^2}\int d\beta\; C^{aLL}_{\alpha\beta}(z)\cdot \langle (z+\alpha)^{(1)}_{(0)}|\phi^{(1)}_{aLL}\rangle\langle(z+\beta)^{(2)}|\phi^{(2)}\rangle,\notag\\
&\text{where } C^{aLL}_{\alpha\beta}(z)\equiv \frac{1}{2\pi l_e^2} \langle z^{(e)}|(z+\beta)^{(e)}\rangle\langle \mathbf r^{(1)}_{(z+\beta)}|(z+\alpha)^{(1)}_{(0)}\rangle_{aLL} =\frac{1}{2\pi l_e^2} e^{-\frac{|\beta|^2}{4l_e^2}}\cdot \langle \beta_{(0)}^{(1)} |\alpha^{(1)}_{(0)} \rangle_{aLL} \cdot e^{\frac{\bar\alpha z-\bar z \alpha}{4l_e^2}}\label{eq:boson_general_CFL_fusion}
\end{align}
\end{widetext}
A few remarks are in order. First, this result, obtained using the disk geometry, expresses the fusion rule as an integral over complex coordinates $\alpha$ and $ \beta$, describing local parton orbitals' spatial deviations from the electron's position. The crucial point for the local fusion tensor $C^{aLL}_{\alpha\beta}(z)$ is that it \emph{exponentially decays as either $\alpha$ or $\beta$ is much larger than the electron's magnetic length}. Therefore, this fusion tensor still has the locality structure. Second, for parton-$1$, we use the coherent state $|(z+\alpha)_{(0)}\rangle_{aLL}$ to describe the local parton orbital. For parton-$2$, however, since it does not see a magnetic field, we use the real-space $\delta$-wavefunction: $\delta(\mathbf r-(z+\beta))$, i.e., $|(z+\beta)^{(2)}\rangle$ to describe its local parton orbital. Finally, we allow parton-$1$ to fill its higher LL. This is because, later, we will use this fusion tensor to describe states beyond traditional FQH systems, e.g., in FCI systems.  

Although it appears that one needs to perform a continuous integral over $\alpha,\beta$, it turns out that, on a finite-size torus, the integral can be replaced by a double summation over discrete lattices (see Sec.\ref{sec:discrete_coherent_state_basis} for $\alpha$-lattice. For $\beta$-lattice of the charge neutral parton-$2$, the simplest way to see this is to consider a related construction in which $\beta$ d.o.f. is obtained by the tensor-product of the electron's and parton-$1$'s cyclotron d.o.f., as shown in Ref.\cite{HuXiaoRan2024}). Up to the exponential tail, one needs $\sim N_{s}^{(e)}$ (the number of flux quanta through the sample) lattice points for $\alpha$ and also $\sim N_{s}^{(e)}$ lattice points for $\beta$. As a matrix, the local fusion tensor $C^{aLL}_{\alpha\beta}(z)$'s size is $\sim N_s^{(e)}\times N_s^{(e)}$. After bringing this matrix to the diagonal form as in the local fusion operator Eq.(\ref{eq:boson_local_fusion_operator}), one finds that, \emph{on a finite size system, up to the exponential tail, the local fusion tensor $C$ in Eq.(\ref{eq:boson_general_CFL_fusion}) has $\sim N_s^{(e)}$ local fusion channels}.

The fact that the number of fusion channels $R$ scales as a power law with the system size indicates that partons are highly entangled in the CFL state \emph{within the LLL}. (On the other hand, if one constructs a CFL state in a lattice model outside a Chern band, $R$ could be a finite value even in the thermodynamic limit.)

\subsubsection{Showcase: fermionic electron with three-parton fusion rule}
Here, we briefly present the general fusion tensor for the real-space parton construction $e\sim f^{(1)}f^{(2)}f^{(3)}$ Eq.(\ref{eq:parton_real_space}) in the same format as the previous section. 

Assuming we have filled single-particle states $|\phi^{(1)}_{aLL}\rangle,|\phi^{(2)}_{bLL}\rangle,|\phi^{(3)}_{cLL}\rangle$ for parton species-$1,2,3$ respectively (the parton fractional charges $q^{(1)},q^{(2)},q^{(3)}$ are all nonzero and $\sum_p q^{(p)}=1$), from direct calculation, the fusion amplitude to an electron's coherent state $|z^{(e)}\rangle$ turns out to be:
\begin{align}
&\langle z^{(e)}\abscirc\phi^{(1)}_{aLL}\rangle|\phi^{(2)}_{bLL}\rangle|\phi^{(3)}_{cLL}\rangle\notag\\
=&\sum_{k,l,m}\sqrt{\frac{k!l!m!}{a!b!c!}} (q^{(1)})^{\frac{a-k}{2}}(q^{(2)})^{\frac{b-l}{2}}(q^{(3)})^{\frac{c-m}{2}}\cdot X_{klm}\notag\\
&\cdot \langle z^{(1)}_{(k)}|\phi^{(1)}_{aLL}\rangle\langle z^{(2)}_{(l)}|\phi^{(2)}_{bLL}\rangle\langle z^{(3)}_{(m)}|\phi^{(3)}_{cLL}\rangle,\label{eq:fermion_general_fusion}
\end{align}
where $k,l,m$ are non-negative integers satisfying $k+l+m=a+b+c$, and $X_{klm}$ is the following multinomial coefficient:
\begin{widetext}
    \begin{align}
\sum_{klm}X_{klm}A^kB^lC^m\equiv[(q^{(1)}-1)A+ q^{(2)}B+q^{(3)}C]^a\cdot [q^{(1)}A+(q^{(2)}-1)B+q^{(3)}C]^b\cdot [q^{(1)}A+q^{(2)}B+(q^{(3)}-1)C]^c.
\end{align}
\end{widetext}

As concrete examples:
\begin{align}
&\quad\langle z^{(e)}\abscirc \phi^{(1)}_{0LL}\rangle|\phi^{(2)}_{0LL}\rangle|\phi^{(3)}_{0LL}\rangle=\Xi^{000}_{000},\notag
\end{align}
\begin{align}
&\quad\langle z^{(e)}\abscirc \phi^{(1)}_{0LL}\rangle|\phi^{(2)}_{0LL}\rangle|\phi^{(3)}_{1LL}\rangle\notag\\
&=[q^{(1)}q^{(3)}]^{\frac{1}{2}}\Xi^{001}_{100}+[q^{(2)}q^{(3)}]^{\frac{1}{2}}\,\Xi^{001}_{010}-
(q^{(1)}+q^{(2)})\Xi^{001}_{001},\notag
\end{align}
\begin{align}
&\quad\langle z^{(e)}\abscirc \phi^{(1)}_{0LL}\rangle|\phi^{(2)}_{1LL}\rangle|\phi^{(3)}_{1LL}\rangle\notag\\
&=\sqrt{2}q^{(1)}[q^{(2)}q^{(3)}]^{\frac{1}{2}}\Xi^{011}_{200}-\sqrt{2}(q^{(1)}+q^{(3)})[q^{(2)}q^{(3)}]^{\frac{1}{2}}\Xi^{011}_{020}
\notag\\
&+[q^{(1)}q^{(3)}]^{\frac{1}{2}}(2q^{(2)}-1)\Xi^{011}_{110}+[q^{(1)}q^{(2)}]^{\frac{1}{2}}(2q^{(3)}-1)\Xi^{011}_{101}\notag\\
&+(q^{(1)}+2q^{(2)}q^{(3)})\Xi^{011}_{011}-\sqrt{2}(q^{(1)}+q^{(2)})[q^{(2)}q^{(3)}]^{\frac{1}{2}}\Xi^{011}_{002},
\label{eq:fermion_special_fusion}
\end{align}
where we defined:
\begin{align}
\Xi^{abc}_{klm}\equiv\langle z^{(1)}_{(k)}|\phi^{(1)}_{aLL}\rangle\langle z^{(2)}_{(l)}|\phi^{(2)}_{bLL}\rangle \langle z^{(3)}_{(m)}|\phi^{(3)}_{cLL}\rangle .
\end{align}
For general $a,b,c$ LL indices, the number of fusion channels
\begin{align}
R_{abc}=C^2_{a+b+c+2}=\frac{(a+b+c+1)(a+b+c+2)}{2}.\label{eq:fermion_general_R_abc}
\end{align}
For instance, consider the $\nu=\frac{2}{5}$ state in Jain's sequence, where $q^{(1)}=q^{(2)}=\frac{2}{5}$, and $q^{(3)}=\frac{1}{5}$. The parton-$1$ and parton-$2$ fill the LLL, while parton-$3$ fills both the LLL and 1st LL. Namely, we have $a=b=0$, while $c$ can be either $0$ or $1$. There are a total of 4 local fusion channels.

The locality structure for this fusion tensor can be expressed via the local fusion tensor and local fusion operator:
\begin{align}
&\langle z^{(e)}\abscirc\phi^{(1)}_j\rangle|\phi^{(2)}_k\rangle|\phi^{(3)}_l\rangle=\notag\\
=&\sum_{\alpha=1}^R \lambda_\alpha \langle z;\mathbf m^{(1)}_\alpha|\phi^{(1)}_j\rangle \langle z;\mathbf m^{(2)}_\alpha|\phi^{(2)}_k\rangle\langle z;\mathbf m_{\alpha}^{(3)}|\phi^{(3)}_l\rangle,\label{eq:fermion_local_fusion}
\end{align}
where $|z;\mathbf m_{\alpha}^{(p)}\rangle$ are the generalized coherent states participating in the fusion tensor Eq.(\ref{eq:fermion_general_fusion}). The corresponding local-fusion operator is:
\begin{align}
\hat{F}_{z}\equiv |z^{(e)}\rangle \cdot \Big[\sum_{\alpha=1}^R \lambda_{\alpha}\langle z;\mathbf m^{(1)}_\alpha|\langle z;\mathbf m^{(2)}_\alpha|\langle z;\mathbf m^{(3)}_{\alpha}|\Big]\label{eq:fermion_local_fusion_operator}
\end{align}

Finally, for the case that $q^{(1)},q^{(2)}\neq 0$ while $q^{(3)}=0$, relevant for the case of $\nu=\frac{1}{2}$ CFL, we have:
\begin{widetext}
\begin{align}
&\langle z^{(e)}\abscirc\phi^{(1)}_{aLL}\rangle|\phi^{(2)}_{bLL}\rangle|\phi^{(3)}\rangle=\int \frac{d\alpha}{2\pi (l^{(1)})^2}\int \frac{d\beta}{2\pi (l^{(2)})^2 }\int d\gamma\; C^{a,b}_{\alpha\beta\gamma}(z)\cdot \langle (z+\alpha)^{(1)}_{(0)}|\phi^{(1)}_{aLL}\rangle\langle(z+\beta)_{(0)}^{(2)}|\phi^{(2)}_{bLL}\rangle\langle (z+\gamma)^{(3)}|\phi^{(3)}\rangle,\notag\\
&\text{where } C^{a,b}_{\alpha\beta\gamma}(z)\equiv \frac{1}{ 2\pi l_e^2} \langle z^{(e)}|(z+\gamma)^{(e)}\rangle\langle \mathbf r^{(1)}_{(z+\gamma)}|(z+\alpha)^{(1)}_{(0)}\rangle_{aLL} \langle \mathbf r^{(2)}_{(z+\gamma)}|(z+\beta)^{(2)}_{(0)}\rangle_{bLL}\label{eq:fermion_general_CFL_fusion}
\end{align}
\end{widetext}
Again, the fusion amplitude above admits the locality structure: local fusion tensor $C^{a,b}_{\alpha\beta\gamma}(z)$ is exponentially suppressed when either $\alpha,\beta,\gamma$ is larger than the magnetic length. The apparent integral of $\alpha,\beta,\gamma$ over the length scale $\sim l_e$ can be replaced by a finite summation over discrete lattices (each variable $\sim N_s^{(e)}$), up to exponentially small tails.

\subsection{The locality-structure of the fusion tensor}\label{sec:locality_structure}
Here, we will present the definition for the locality-structure for the fusion tensor of a general Hdet wavefunction. Before that, we need to examine further the locality structure in the FQH showcase examples presented in the previous section to motivate the definition.

\subsubsection{Overcomplete discrete-coherent-state basis within a Chern band}\label{sec:discrete_coherent_state_basis}
We first comment on the electronic coherent state $|z^{(e)}\rangle$ participating in the local fusion tensors Eq.(\ref{eq:boson_local_fusion},\ref{eq:fermion_local_fusion}). 

\begin{figure}
\includegraphics[width=0.45\textwidth]{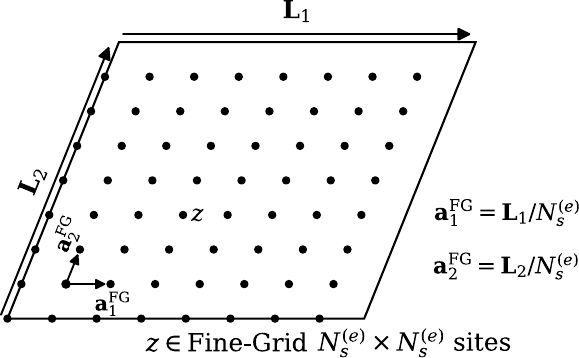}
\caption{The Fine-Grid on a finite torus with $N_s^{(e)}=8$.}
\label{fig:fine_grid}
\end{figure}

Very often, the electron coherent state basis $\{|z^{(e)}\rangle\}$ is discussed as if $z$ is treated as a \emph{continuous} variable. For instance, the usual many-body wavefunction of a FQH state indeed can be viewed as the overlap between the electron state with the continuous-coherent-state many-body basis: $\Psi^{(e)}(z_1,z_2,...,z_{N_e})=\langle z^{(e)}_1,z^{(e)}_2,...,z^{(e)}_{N_e}|\Psi^{(e)}\rangle$. However, in a finite-size system, it is physically unnecessary to introduce an infinite number of orbitals to capture the many-body state. Instead, one should be able to limit $z$ on a lattice, which we call a Fine-Grid, so that the corresponding discrete-coherent-states $\{|z^{(e)}\rangle,z\in \text{Fine-Grid}\}$ are still enough to form a good basis (with resolution of identity, etc). In fact, a recent work\cite{WangHaldane2019} has exactly demonstrated that one only needs the Fine-Grid on a finite torus whose size is $N_s^{(e)}\times N_s^{(e)}$. Namely, this Fine-Grid is generated by the minimally allowed magnetic translations $\mathbf a_1^{\rm FG},\mathbf a_2^{\rm FG}$ along the two periodic directions on the finite torus. (See Fig.\ref{fig:fine_grid} for an illustration of the Fine-Grid.) It is easy to show that, by the fact that the magnetic translations form a complete algebra, the discrete coherent states on the Fine-Grid feature the resolution of the identity:
\begin{align}
\sum_{z\in \text{Fine-Grid}}|z^{(e)}\rangle\langle z^{(e)}|=N_s^{(e)}\cdot \mathbf 1.\label{eq:dcs_resolution_of_identity}
\end{align}
We will denote $\{|z^{(e)}\rangle,z\in \text{Fine-Grid}\}$ as the \emph{discrete-coherent-state basis for the electrons in the LLL}. 

Note that this basis formed by $(N_s^{(e)})^2$ states is overcomplete, since the dimension of the electron single-particle Hilbert space is only $\text{dim}\mathcal H^{(e)}=N_s^{(e)}$. The overcompleteness is intrinsically related to the fact that we are working \emph{within} the electron's LLL in the showcase examples, which is a Chern band. On the one hand, a Chern band cannot be Wannier-localized\cite{Thouless1982, Brouder2007, Read2017}. Namely, there is no localized and orthonormal basis in the LLL. On the other hand, to expose the locality-structure of Hdet wavefunctions, we want to work with a set of localized electronic orbitals, even if they are overcomplete. 

In general, one may start with a lattice tight-binding model and identify an isolated Chern band whose Chern number $C$ may or may not be $\pm1$ (e.g., $C=\pm2$). Partially filling this Chern band may lead to FCI states that can be captured by Hdet wavefunctions. In this case, \emph{one can still work explicitly within the Chern band} to discuss the locality-structure of the fusion tensor. To this end, one also needs to introduce the corresponding electronic \emph{discrete-coherent-state basis in this Chern band}, which we require to be exponentially localized, overcomplete, and feature the standard resolution of the identity. Here, one straightforward prescription to generate the discrete-coherent-state basis is simply to project the original real-space orbitals in the tight-binding model into the Chern band. For instance, let's say there are $n_{orb}$ electronic orbitals per real-space unit cell, denoted by $\{|\mathbf r_{uc},i\rangle\}$, where $\mathbf r_{uc}$ is the discrete unit-cell real-space position and $i=1,2,...,n_{orb}$ labels the orbital within the unit-cell. One natural definition of a discrete-coherent-state basis is $\{\mathcal P_{CB}|\mathbf r_{uc},i\rangle \}$, where $\mathcal P_{CB}$ is the single-particle projection into the Chern band.

In summary, \emph{whenever we work \emph{within} a Chern band, to expose the locality-structure for the fusion tensor, we will give up the orthonormal basis and proceed with the overcomplete discrete-coherent-state basis. If one works within a topologically trivial band, or one works within a complete tight-binding model, however, one could simply use the orthonormal Wannier basis to discuss the locality-structure for the fusion tensor.}

\subsubsection{A change of viewpoint for $|z;\mathbf m^{(p)}\rangle$ and $|z^{(e)}\rangle$}\label{sec:change_of_viewpoint}

In the showcase examples, we already see that a parton-$p$ will always propagate to a parton orbital, some generalized coherent state $|z;\mathbf m^{(p)}\rangle$ at the electron's position $z$, to fuse into the electron. For fusion-channel $\alpha$, this propagation is captured by the overlap $\langle z;\mathbf m_{\alpha}^{(p)}|\phi^{(p)}\rangle$ in Eq.(\ref{eq:boson_local_fusion},\ref{eq:fermion_local_fusion}). 

These generalized coherent states are also \emph{not} orthonormal. If we follow the original interpretation of $\langle z;\mathbf m^{(p)}|\phi^{(p)}\rangle$, these non-orthogonal properties of the orbitals may lead to extra (unnecessary) complications in interpreting the Hdet wavefunctions. 

Here, we introduce a new viewpoint for the overlap $\langle z;\mathbf m^{(p)}|\phi^{(p)}\rangle$. From our Hdet construction, it is evident that the only important object is the value of the fusion tensor $T$. Therefore, one may re-interpret $\langle z;\mathbf m^{(p)}|\phi^{(p)}\rangle$ as long as its value is unchanged. 

In this new viewpoint, we interpret the collection of all the parton localized orbitals as an orthonormal real-space basis for parton-$p$: $\{|z;\mathbf m^{(p)}\rangle,z\in \text{Fine-Grid},\forall\mathbf m\}$. The state $|\phi^{(p)}\rangle$ is then considered as a linear superposition of this orthonormal basis:
\begin{align}
|\phi^{(p)}\rangle=\sum_{z\in \text{Fine-Grid},\mathbf m} \langle z;\mathbf m^{(p)}|\phi^{(p)}\rangle \cdot |z;\mathbf m^{(p)}\rangle.\label{eq:change_of_viewpoint}
\end{align}

For instance, one may choose $|\phi^{(p)}\rangle$ to be $|\phi^{(p)}_{\mathbf k}\rangle_{aLL}$, the Bloch state at a momentum point $\mathbf k$ in the $a$-th LL. Here, to define the Bloch basis, one needs to introduce two commuting magnetic translations $\hat T_1,\hat T_2$, whose corresponding real-space unit cell area encloses one parton flux quantum. There will be many parton real-space orbitals $\{|z;\mathbf m^{(p)}\rangle,z\in \text{Fine-Grid},\mathbf m\in aLL\}$ per unit cell (precisely, there are $\frac{N_s^{(e)}}{q^{(p)}}$ parton orbitals per $\mathbf m$ per unit cell). In this new viewpoint, as one tunes $\mathbf k$, $|\phi^{(p)}_{\mathbf k}\rangle_{aLL}$ forms a band via linear superpositions of these orthonormal parton orbitals. \emph{Crucially}, one can show that, in this new viewpoint, this band $|\phi^{(p)}_{\mathbf k}\rangle_{aLL}$ indeed carries Chern number $C=1$ (based on the parton-version of the resolution of identity Eq.(\ref{eq:dcs_resolution_of_identity}), see Eq.(\ref{eq:generalized_coherent_state_resolution_of_identity}) and discussions nearby).

The same change of viewpoint can be done for the electrons. Suppose we attempt to compute the Hdet many-body electron wavefunction in the discrete coherent-state basis on the Fine-Grid. Namely, given the full fusion tensor $T_{zjk..}$ where $z\in \text{Fine-Grid}$, we want to compute the \emph{value}:
\begin{align}
\langle z^{(e)}_1,z_2^{(e)},...,z^{(e)}_{N_e}|\Psi^{(e)}\rangle=\text{Hdet}(T^{\text{sub}}),
\end{align}
where $z^{(e)}_i\in \text{Fine-Grid}$, and the cubic tensor $T^{\text{sub}}$ is formed by stacking the $z_i$ slices of $T$. Because we only look for the value of the hyperdeterminant, it does not matter if one views $\{|z^{(e)}\rangle, z\in \text{Fine-Grid}\}$ as an orthonormal basis or not, as long as the fusion tensor $T$ is unchanged.

\subsubsection{A Grassmann tensor-network picture}
The main point from the previous discussion is that, for the purpose of computing the \emph{value} of Hdet wavefunctions in the showcase examples, it is perfectly fine to think of the discrete-coherent state basis $\{|z^{(e)}\rangle, z\in \text{Fine-Grid}\}$ and the parton generalized coherent states $\{|z;\mathbf m^{(p)}\rangle,z\in \text{Fine-Grid},\forall\mathbf m\}$ as orthonormal in their respective Hilbert spaces.

In this new viewpoint, the Hdet wavefunctions in the showcase examples have a simple picture. It is much easier to work with orthonormal bases since we can use the standard second-quantized formulation. Introducing the electron operator $b_z^\dagger$ ($c_z^\dagger$) for bosonic (fermionic) electron, and fermionic parton operator $f_{z,\mathbf m}^{(p)\dagger}$ for these orthonormal bases, the full fusion can be formulated as a single \emph{fusion operator} $\hat{\mathbf F}=\sum_z \hat {\mathbf F}_z$ :
\begin{align}
 e\sim & f^{(1)}f^{(2)}:\notag\\
\hat{\mathbf F} \equiv &\sum_{z}\hat {\mathbf F}_z,\;\; \hat {\mathbf F}_z\equiv \sum_{\alpha} \lambda_{\alpha} b_z^\dagger f_{z;\mathbf m^{(1)}_\alpha}^{(1)}f_{z;\mathbf m^{(2)}_\alpha}^{(2)};\notag\\
e\sim & f^{(1)}f^{(2)}f^{(3)}:\notag\\
\hat{\mathbf F}\equiv&\sum_{z}\hat {\mathbf F}_z,  \;\;\hat {\mathbf F}_z\equiv \sum_{\alpha} \lambda_{\alpha} c_z^\dagger f_{z;\mathbf m^{(1)}_\alpha}^{(1)}f_{z;\mathbf m^{(2)}_\alpha}^{(2)}f_{z;\mathbf m^{(3)}_\alpha}^{(3)};\label{eq:full_local_fusion}
\end{align}
where the summation over $z$ is for $z\in \text{Fine-Grid}$. 

Note that, by comparing with the fusion operator in Eq. (\ref{eq:full_fusion_orig}), we make two significant advances. First, $\hat{\mathbf F}$ is written as a summation of local terms -- each $\hat {\mathbf F}_z$ local-fusion operator is defined to be the term at position $z$ in Eq.(\ref{eq:full_local_fusion}). Second, in Eq.(\ref{eq:full_fusion_orig}), despite the coherent-state bases not being physically orthonormal, after the change of viewpoint, we can still formulate $\hat{\mathbf F}$ in second-quantized operators.

Up to an unimportant overall normalization factor, following Eq.(\ref{eq:fusion_Ne_power}), the Hdet wavefunction is obtained by applying $\hat{\mathbf F}^{N_e}$ to the tensor product of free parton states $|\Psi^{(p)}\rangle$:
\begin{align}
|\Psi^{(e)}\rangle=\hat{\mathbf F}^{N_e}|\Psi^{(1)}\rangle|\Psi^{(2)}\rangle...=\hat{\mathbf F}^{N_e}|\Psi^{MF}\rangle,
\end{align}
where we defined the full parton mean-field state $|\Psi^{MF}\rangle\equiv|\Psi^{(1)}\rangle|\Psi^{(2)}\rangle...$. 

Here, the exponent $N_e$ in $\hat{\mathbf F}^{N_e}$ makes sure that \emph{all} partons are fused into electrons, resulting in a fully electronic state (note that each $|\Psi^{(p)}\rangle$ is a Slater determinant state with fixed $N_e$ partons). On the other hand, it would be more elegant to revise our formulation so that $N_e$ is not present in the fusion. 

In fact, we can now take advantage of the new viewpoint, so that the result of the many-body fusion $\hat{\mathbf F}^{N_e}$ can be written as a product of (mutually commuting) local-fusion quantum gates $\{\hat{\mathbf C}_z\}$ not involving $N_e$:
\begin{align}
|\Psi^{(e)}\rangle=\hat{\mathbf F}^{N_e}|\Psi^{MF}\rangle=\prod_{z\in \text{Fine-Grid}}\hat {\mathbf C}_z|\Psi^{MF}\rangle.\label{eq:local_fusion_gate_feature}
\end{align}
Here, to avoid confusion with the local fusion operator $\hat{\mathbf F}_z$, we will call operators $\hat {\mathbf C}_z$ the \emph{local-fusion gates}.

The operator $\hat {\mathbf C}_z$ can be explicitly constructed. For this purpose, we consider the \emph{local} Fock spaces at the Fine-Grid site-$z$ for the electrons $\mathcal F_z^{(e)}$ and partons $\mathcal F_z^{(p)}$. Denoting $|0_z^{(e)}\rangle\in \mathcal F_z^{(e)}$ and  $|0_z^{(p)}\rangle\in\mathcal F_z^{(p)}$ as their vacuum states, also denoting $|0_z^{(\text{all } p)}\rangle\equiv \otimes_p|0_z^{(p)}\rangle\in \mathcal F_z^{(\text{all }p)}\equiv \otimes_p \mathcal F_z^{(p)}$ as the tensor product for all parton species, 
\begin{align}
&e\sim f^{(1)}\cdot f^{(2)}:\notag\\
&\hat {\mathbf C}_z\equiv|0_z^{(e)}\rangle\langle 0_z^{(\text{all } p)}|+\sum_{n=1}^{\infty}\frac{\hat{\mathbf F}_z^n}{n!}\bullet(|0_z^{(e)}\rangle\langle 0_z^{(\text{all } p)}|),\notag\\
&e\sim f^{(1)}\cdot f^{(2)}\cdot f^{(3)}:\notag\\
&\hat {\mathbf C}_z\equiv|0_z^{(e)}\rangle\langle 0_z^{(\text{all } p)}|+\hat{\mathbf F}_z\bullet(|0_z^{(e)}\rangle\langle 0_z^{(\text{all } p)}|).\label{eq:local_fusion_gate_def}
\end{align}
Here, we introduced an action $"\bullet"$ of $\hat{\mathbf F}_z^n$ to the operator $|0_z^{(e)}\rangle\langle 0_z^{(\text{all } p)}|$, which is defined by moving all the parton operators (maintaining the order) $f^{(p)}$ to the right of $|0_z^{(e)}\rangle\langle 0_z^{(\text{all } p)}|$ (so that they act on $\langle 0_z^{(\text{all } p)}|$ from the right), and moving all the electron operators ($b_z^\dagger$ or $c_z^\dagger$) to the left of $|0_z^{(e)}\rangle\langle 0_z^{(\text{all } p)}|$ (so they act on $|0_z^{(e)}\rangle$ from the left). For fermionic electrons, due to the Pauli exclusion principle, the power series truncates to $n=1$. 

For example:
\begin{align}
&e\sim f^{(1)}\cdot f^{(2)}\cdot f^{(3)}:\notag\\
&\hat {\mathbf C}_z=|0_z^{(e)}\rangle\langle 0_z^{(\text{all } p)}|\notag\\
&+\sum_\alpha \lambda_\alpha(c_z^\dagger|0_z^{(e)}\rangle)\cdot(\langle 0_z^{(\text{all } p)}| f^{(1)}_{z;\mathbf m^{(1)}_\alpha}f^{(2)}_{z;\mathbf m^{(2)}_\alpha}f^{(3)}_{z;\mathbf m^{(3)}_\alpha}).\label{eq:C_z_from_F_z_3_parton}
\end{align}
Namely, $\hat {\mathbf C}_z: \mathcal F^{(\text{all }p)}_z\mapsto  F^{(e)}_z$ is a linear map between the two Fock spaces. With $"\bullet"$ notation, we generally have:
\begin{align}
\hat{\mathbf C}_z=\exp(\hat{\mathbf F}_z)\bullet(|0_z^{(e)}\rangle\langle 0_z^{(\text{all } p)}|)
\end{align}

With the definition Eq.(\ref{eq:local_fusion_gate_def}), it is straightforward to show that the property Eq.(\ref{eq:local_fusion_gate_feature}) is satisfied. Here, the product of two gates $ \hat {\mathbf C}_{z}\hat {\mathbf C}_{z'}: \mathcal F^{(\text{all }p)}_z\otimes \mathcal F^{(\text{all }p)}_{z'} \mapsto  \mathcal F^{(e)}_z \otimes \mathcal F^{(e)}_{z'} $ should be understood as the tensor product. Explicitly, this map can be naturally obtained by moving all parton operators $f^{(p)}_z$ and $f^{(p)}_{z'}$ to the right (so that they act on the vacuum of partons from the right), and the electron operators to the left (so that they act on the vacuum of electrons from the left). It is important to note that when moving the fermion operators (either electron or parton) around, they satisfy anticommutation relations, and the fermion sign needs to be included.

Crucially, the local-fusion gate $\hat {\mathbf C}_z$'s total fermion parity is even. Consequently, \emph{the order of multiplication in $\prod_z \hat {\mathbf C}_z$ is immaterial}: $ \hat {\mathbf C}_{z}\hat {\mathbf C}_{z'}=\hat {\mathbf C}_{z'}\hat {\mathbf C}_{z}$. 

Such a local fusion structure is most conveniently illustrated using a Grassmann tensor-network language\cite{GuVerstraeteWen2010, Corboz2010}, in which a Grassmann variable replaces a second-quantized fermionic operator. It is well-known that, when $|\Psi^{(p)}\rangle$ is topological (e.g., with nonzero Chern number), a Gaussian tensor-network representation of $|\Psi^{(p)}\rangle$ has an obstruction when the virtual-bond dimension is finite \cite{DubailRead2015, Read2017} in the thermodynamic limit. However, as a conceptual language, it is fine to consider a finite-size system without specifying the virtual-bond dimension. 

\begin{figure}
\includegraphics[width=0.45\textwidth]{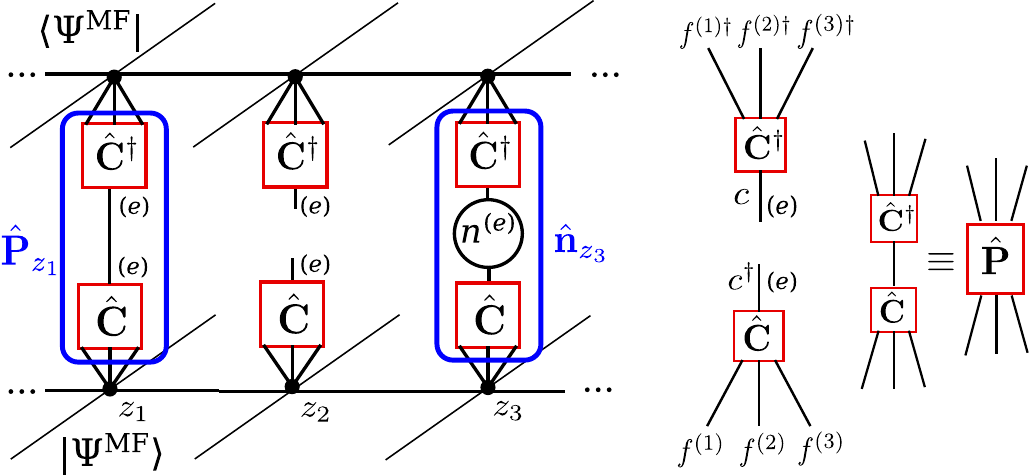}
\caption{Illustration of the Grassmann tensor-network picture with fusion gate $\hat{\mathbf C}_z$ and the gate $\hat{\mathbf P}_z$.}
\label{fig:fusion_gate_tn}
\end{figure}

In this picture, each Slater-determinant $|\Psi^{(p)}\rangle$ is represented as a layer of Gaussian Grassmann tensor-network\cite{Kraus2010, GuVerstraeteWen2010, DubailRead2015} on the Fine-Grid, with the local orbitals $|z;\mathbf m^{(p)}\rangle$ represented as the outgoing "physical-legs" of the parton (i.e., Grassmann variables). The local fusion $\hat {\mathbf C}_z$ can be viewed as a non-Gaussian Grassmann tensor (i.e., a quantum gate). The $\hat {\mathbf C}_z$ gate has incoming parton's physical-legs, and one outgoing electron's physical-leg. The electron's physical-leg will be a Grassmann (bosonic) variable if it is fermionic (bosonic). See Fig.\ref{fig:fusion_gate_tn} for an illustration.

In order to compute a physical observable $\hat O^{(e)}$ of the electron, one needs to compute $\langle\hat O^{(e)}\rangle\equiv\frac{\langle \Psi^{(e)}|\hat O^{(e)}|\Psi^{(e)}\rangle}{\langle \Psi^{(e)}|\Psi^{(e)}\rangle}$. As a concrete example, one may consider density-density correlator $\hat O^{(e)}=\hat n^{(e)}_\omega \hat n^{(e)}_\zeta$ (other observables can be easily generalized), where $\omega,\zeta\in \text{Fine-Grid}$, and $\hat n^{(e)}_z= b^\dagger_z b_z$ ($\hat n^{(e)}_z= c^\dagger_z c_z$) for the bosonic (fermionic) electron case. Ref.\cite{WangHaldane2019} points out that any four-fermion interaction energy of the electron can be computed based on $\langle\hat n^{(e)}_\omega \hat n^{(e)}_\zeta\rangle,\; \forall\omega,\zeta\in \text{Fine-Grid}$.

In this Grassmann tensor-network picture, 
\begin{align}
\langle\hat n^{(e)}_\omega \hat n^{(e)}_\zeta\rangle=\frac{\langle \Psi^{MF} |\prod_{z}\hat {\mathbf C}^\dagger_z \cdot\hat n^{(e)}_\omega\hat n^{(e)}_\zeta \cdot \prod_{z}\hat {\mathbf C}_z|\Psi^{MF}\rangle}{\langle \Psi^{MF} |\prod_{z}\hat {\mathbf C}^\dagger_z \cdot \prod_{z}\hat {\mathbf C}_z|\Psi^{MF}\rangle}.
\end{align}
We define
\begin{align}
\hat{\mathbf P}_z\equiv& \hat {\mathbf C}^\dagger_z \hat {\mathbf C}_z, & \hat{\mathbf n}_z\equiv& \hat {\mathbf C}^\dagger_z \hat n^{(e)}_z\hat {\mathbf C}_z.
\end{align}
These operators now directly act on the parton mean-field state, and have a simple geometric picture: the local gate $\hat{\mathbf P}_z$ is obtained by contracting the electron physical leg between the gates $\hat {\mathbf C}^\dagger_z$ and $\hat {\mathbf C}_z$, while the local gate $\hat{\mathbf n}_z$ is obtained by sandwiching $\hat n^{(e)}_z$ between them. See Fig.\ref{fig:fusion_gate_tn} for an illustration.

With these local gates, we have:
\begin{align}
\langle\hat n^{(e)}_\omega \hat n^{(e)}_\zeta\rangle=\frac{\langle \Psi^{MF} | \hat{\mathbf n}_\omega \hat{\mathbf n}_\zeta\prod_{z\neq \omega,\zeta}\hat {\mathbf P}_z|\Psi^{MF}\rangle}{\langle \Psi^{MF} |\prod_{z}\hat {\mathbf P}_z |\Psi^{MF}\rangle}.\label{eq:pair_correlation_gate}
\end{align}
For a general electronic observable $\hat O^{(e)}_D$, where $D\subset \text{Fine-Grid}$ is formed by some sites, one can similarly define the quantum gate 
\begin{align}
\hat{\mathbf O}_D\equiv \prod_{z\in D} \hat {\mathbf C}^\dagger_z \cdot \hat O^{(e)}_D \cdot\prod_{z\in D}\hat {\mathbf C}_z,
\end{align}
and
\begin{align}
\langle\hat O^{(e)}_D \rangle=\frac{\langle \Psi^{MF} |  \hat{\mathbf O}_D \prod_{z\notin D}\hat {\mathbf P}_z|\Psi^{MF}\rangle}{\langle \Psi^{MF} |\prod_{z}\hat {\mathbf P}_z |\Psi^{MF}\rangle}.\label{eq:general_observable_gate}
\end{align}

Eq.(\ref{eq:general_observable_gate}) captures the key consequence of the locality structure of Hdet wavefunctions. We will come back to it later because it implies systematic ways to numerically simulate Hdet wavefunctions. 

\subsubsection{Example: Jain's $\nu=\frac25$ state} \label{sec:locality_example_nu25}
As a concrete example, let's explicitly write down the parton operators $f^{(p)}_{z,\mathbf m^{(p)}}$ and the fusion operator $\hat{\mathbf F}_z$ for the fermionic Jain's $\nu=\frac{2}{5}$ state on the Fine-Grid, after the change of viewpoint. In this example, there are three species of partons: $n_{\rm p}=3$, carrying electric charge $q^{(1)}=q^{(2)}=\frac{2}{5}$ (filling LLL), $q^{(3)}=\frac15$ (filling both LLL and 1st LL). Based on Eq.(\ref{eq:fermion_special_fusion}), parton species $p=1,2$ each have two orbitals per site-$z$ on the Fine-Grid, corresponding to generalized coherent states $|z^{(p)}_{(0)}\rangle_{0LL}$ and $|z^{(p)}_{(1)}\rangle_{0LL}$, denoted as $\mathbf m^{(p)}=1,2$ respectively. For parton species $p=3$, we need three orbitals per site-$z$, corresponding to $|z^{(3)}_{(0)}\rangle_{0LL}$, $|z^{(3)}_{(0)}\rangle_{1LL}$,$|z^{(3)}_{(1)}\rangle_{1LL}$, denoted as $\mathbf m^{(3)}=1,2,3$ respectively. The fusion operator $\hat{\mathbf F}_z$, according to Eq.(\ref{eq:fermion_special_fusion}), has the form in Eq.(\ref{eq:full_local_fusion}) with $R=4$, and:
\begin{align}
\alpha=&1:&\lambda_1=&1,&\mathbf m^{(1)}_1=&1,&\mathbf m^{(2)}_1=&1,&\mathbf m^{(3)}_1=&1,\notag\\
\alpha=&2:&\lambda_2=&\frac{\sqrt 2}{5},& \mathbf m^{(1)}_2=&2,&\mathbf m^{(2)}_2=&1,&\mathbf m^{(3)}_2=&2,\notag\\
\alpha=&3:&\lambda_3=&\frac{\sqrt 2}{5},& \mathbf m^{(1)}_3=&1,&\mathbf m^{(2)}_3=&2,&\mathbf m^{(3)}_3=&2,\notag\\
\alpha=&4:&\lambda_4=&-\frac{4}{5},& \mathbf m^{(1)}_4=&1,&\mathbf m^{(2)}_4=&1,&\mathbf m^{(3)}_4=&3.\label{eq:nu25_example_local_fusion}
\end{align}
Namely:
\begin{align}
\hat{\mathbf F}_z=&c_z^\dagger\Big[f^{(1)}_{z,1}f^{(2)}_{z,1}f^{(3)}_{z,1}+\frac{\sqrt 2}{5}f^{(1)}_{z,2}f^{(2)}_{z,1}f^{(3)}_{z,2}\notag\\
&+\frac{\sqrt 2}{5}f^{(1)}_{z,1}f^{(2)}_{z,2}f^{(3)}_{z,2}-\frac{4}{5}f^{(1)}_{z,1}f^{(2)}_{z,1}f^{(3)}_{z,3}\Big]\label{eq:nu25_example_local_F}.
\end{align}
The corresponding local fusion gate $\hat {\mathbf C}_z$ as Eq.(\ref{eq:C_z_from_F_z_3_parton}) can be explicitly written down.

In order to compute physical quantities as in Eq.(\ref{eq:general_observable_gate}), we need to fully specify $|\Psi^{MF}\rangle$. Since it is a direct product of Slater determinants of partons, $|\Psi^{MF}\rangle$ is fully specified by the single-body reduced density matrix (RDM): $\langle f^{(p)\dagger}_{z_1,\mathbf m^{(p)}_1} f^{(p)}_{z_2,\mathbf m^{(p)}_2}\rangle$. Below we explicitly write down these RDM elements. 

According to the change of viewpoint Eq.(\ref{eq:change_of_viewpoint}), we should interpret the filled parton states as the linear superposition of the orthonormal local parton orbitals. Let's denote the filled parton states to be $|\phi^{(p)}_k\rangle_{aLL}$. For the parton species $p=1,2$, $a=0$ and $k=1,2,..,N_e$. For the parton species $p=3$, $a=0,1$ and $k=1,2,...,\frac{N_e}{2}$. The only technical detail is that the $|\phi^{(p)}_k\rangle$ state given in Eq.(\ref{eq:change_of_viewpoint}) is not normalized after the change of viewpoint. To normalize it, we introduce the resolution of identity of the generalized coherent state before the change of viewpoint. It is easy to show that:
\begin{align}
\sum_{z\in {\text{Fine-Grid}}} |z_{(m)}^{(p)}\rangle\langle z_{(n)}^{(p)}|_{aLL}= \delta_{mn}\frac{N_s^{(e)}}{q^{(p)}}\mathbf 1_{aLL},\label{eq:generalized_coherent_state_resolution_of_identity}
\end{align}
where $\mathbf 1_{aLL}$ is the identity in parton-$p$'s $aLL$. This equation follows from the fact that the operator on the LHS commutes with all magnetic translations, so it must be proportional to the identity. The proportional coefficient can be found by taking the trace.

Using this resolution of identity, it is straightforward to show that if $\{|\phi^{(p)}_k\rangle_{aLL}\}$ is an orthonormal basis before the change of viewpoint, it remains an orthogonal basis after the change of viewpoint, but with a norm:
\begin{align}
\sum_{m=0}^{N^{(p)}_{{\rm coh},aLL}-1}\sum_{z\in \text{Fine-Grid}} |\langle z_{(m)}^{(p)}|\phi^{(p)}\rangle_{aLL}|^2=N^{(p)}_{{\rm coh},aLL}\frac{N_s^{(e)}}{q^{(p)}},
\end{align}
where $N^{(p)}_{{\rm coh},aLL}$ is the number of generalized coherent state orbitals per site in the $aLL$ for parton-$p$. In the present $\nu=\frac{2}{5}$ example, $N^{(1)}_{{\rm coh},0LL}=N^{(2)}_{{\rm coh},0LL}=2$, while $N^{(3)}_{{\rm coh},0LL}=1,N^{(3)}_{{\rm coh},1LL}=2$. 

With this normalization factor after the change of viewpoint, one can write $f^{(p)\dagger}_{z,\mathbf m^{(p)}}$ operator as a linear superposition of the normalized filled state operator $f^{(p)\dagger}_{k,aLL}$. Since we know $\langle f^{(p)\dagger}_{k,aLL}f^{(p)}_{l,bLL}\rangle=\delta_{kl}\delta_{ab}$, the RDM $\langle f^{(p)\dagger}_{z_1,\mathbf m^{(p)}_1} f^{(p)}_{z_2,\mathbf m^{(p)}_2}\rangle$ can be computed (recall $\mathbf m$ is a combination of generalized coherent state index-$(m)$ and the LL index-$aLL$):
\begin{align}
&\langle f^{(p)\dagger}_{z_1,(m_1),aLL} f^{(p)}_{z_2,(m_2),bLL}\rangle\notag\\
=&\frac{q^{(p)}}{N^{(p)}_{{\rm coh},aLL}N_s^{(e)}}\delta_{ab}\left\langle (z_2)^{(p)}_{(m_2)}|(z_1)^{(p)}_{(m_1)} \right\rangle_{aLL},\label{eq:fdagf_FG}
\end{align}
where, on the RHS, the overlap between two generalized coherent states \emph{before the change of viewpoint} is given in Eq.(\ref{eq:generalized_coherent_state_wavefunction_overlap}).

\subsubsection{The definition of the locality-structure}\label{sec:locality_def}
The showcase examples motivate us to define the following general locality-structure of an Hdet wavefunction.

\emph{Definition: } Consider a Hdet construction involving $n_{\text{p}}$ parton species on a certain Real-Space-Lattice. Each parton many-body state is a Slater-determinant $|\Psi^{(p)}\rangle$ ($p=1,2,...,n_{\text{p}}.$), corresponding to filling $N_e$ single-particle states $|\phi^{(p)}_{j_p}\rangle$ ($j_p=1,2,...,N_e$). These single-particle states are linear superpositions of localized orthonormal basis $\{|\mathbf r;\mathbf m^{(p)}\rangle\}$:
\begin{align}
|\phi^{(p)}_{j_p}\rangle=\sum_{\mathbf r,\mathbf m^{(p)}}\langle\mathbf r;\mathbf m^{(p)}|\phi^{(p)}_{j_p}\rangle \cdot|\mathbf r;\mathbf m^{(p)}\rangle.
\end{align}
In addition, we choose a localized single-particle basis for electrons $\{|\mathbf r;\mathbf m^{(e)}\rangle\}$, indexed by $i$, that could be overcomplete. Here, for either electron's or parton's orbital, $\mathbf r\in $ Real-Space-Lattice labels its position, and $\mathbf m$ labels other quantum numbers. \emph{We say that the fusion tensor $T$ admits a locality structure if there exist local parton states $|\boldsymbol\varphi^{(p)}_{\alpha,i}\rangle$ such that}:
\begin{align}
&T_{i,j_1,j_2,...,j_{n_{\text{p}}}}\equiv\langle \mathbf r_i;\mathbf m_i^{(e)}\abscirc \prod_{p=1}^{n_{\text{p}}}\phi^{(p)}_{j_p}\rangle\notag\\
=&\sum_{\alpha=1}^{R_i}\lambda_{\alpha,i}\prod_{p=1}^{n_{\text{p}}}\langle\boldsymbol\varphi^{(p)}_{\alpha,i}|\phi^{(p)}_{j_p}\rangle,
\end{align}
where $|\boldsymbol\varphi^{(p)}_{\alpha,i}\rangle$, up to an exponentially small tail, has nonzero overlap with the real-space basis only in the vicinity of the electron's position $\mathbf r_i$: 
\begin{align}
|\langle\mathbf r;\mathbf m^{(p)}|\boldsymbol\varphi^{(p)}_{\alpha,i}\rangle | \lesssim e^{-\frac{|\mathbf r-\mathbf r_i|}{l}} \text{ when } |\mathbf r-\mathbf r_i|\gg l,
\end{align}
where $l$ is a certain finite length-scale. 

Practically, one may truncate the exponential tail so that $|\langle\mathbf r;\mathbf m^{(p)}|\boldsymbol\varphi^{(p)}_{\alpha,i}\rangle\neq 0$ only over a finite support $|\mathbf r-\mathbf r_i|<\tilde l$ with some finite length $\tilde l$. When the number of fusion channels $R_i$ is \emph{finite} (which is always physically true for a finite-size sample), it is convenient to introduce \emph{ancilla} parton orbitals at site-$\mathbf r_i$ by increasing the index $\mathbf m^{(p)}$, representing copies of the nearby parton orbitals. With these ancilla introduced, the fusion can be chosen to be on-site: $|\boldsymbol\varphi^{(p)}_{\alpha,i}\rangle = |\mathbf r_i;\mathbf m_{\alpha,i}^{(p)}\rangle$. Namely:
\begin{align}
&T_{i,j_1,j_2,...,j_{n_{\text{p}}}}\equiv\langle \mathbf r_i;\mathbf m_i^{(e)}\abscirc \prod_{p=1}^{n_{\text{p}}}\phi^{(p)}_{j_p}\rangle\notag\\
=&\sum_{\alpha=1}^{R_i}\lambda_{\alpha,i}\prod_{p=1}^{n_{\text{p}}}\langle\mathbf r_{i};\mathbf m_{\alpha,i}^{(p)}|\phi^{(p)}_{j_p}\rangle,\label{eq:onsite_fusion_parton_orbitals}
\end{align}

This \emph{on-site fusion} allows one to translate the above definition of the locality-structure into the formulation of a fused Gaussian Grassmann tensor-network as in Fig.\ref{fig:fusion_gate_tn}. The Hdet state $|\Psi^{(e)}\rangle$ can be obtained via $|\Psi^{(e)}\rangle=\hat{\mathbf F}^{N_e}|\Psi^{MF}\rangle\equiv \hat{\mathbf F}^{N_e}\prod_{p=1}^{n_{\text{p}}}|\Psi^{(p)}\rangle$ using the second-quantized fusion operator $\hat{\mathbf F}=\sum_{i} \hat{\mathbf F}_{i}$, where the on-site fusion operator
\begin{align}
\hat{\mathbf F}_{i}\equiv\sum_{\alpha=1}^{R_i} \lambda_{\alpha,i}\cdot c^{\dagger}_{i}\cdot \prod_{p=1}^{n_{\text{p}}}f^{(p)}_{i,\mathbf m^{(p)}_{\alpha,i}}.\label{eq:onsite_fusion_operator}
\end{align}
Here we introduced the second-quantized operator $c^{\dagger}_{i}$ for electrons. ($c_i^\dagger$ can be either bosonic or fermionic depending on whether $n_{\text{p}}$ is even or odd. Again, even if the electronic orbitals $\{|\mathbf r_i,\mathbf m_i^{(e)}\rangle\}$ may be overcomplete, one still can use the standard second-quantization formulation as if they are orthonormal, i.e., via the change of viewpoint.) The orthonormal second-quantized operators $f^{(p)}_{i,\mathbf m_i^{(p)}}$'s are located at site $\mathbf r_i$: $\mathbf m_i^{(p)}$ is \emph{defined} to be the collection of all parton-$p$ orbital labels in the local Fock space assigned to electronic orbital-$i$, and $f^{(p)}_{i,\mathbf m^{(p)}_{\alpha,i}}$ is participating in the fusion channel $\alpha$. 

Finally, by introducing additional ancilla parton orbitals, one can always make sure that if two different electronic orbitals $i,i'$ are located on the same site: $\mathbf r_i=\mathbf r_{i'}$ and $\mathbf m_{i}^{(e)}\neq \mathbf m_{i'}^{(e)}$, then $\mathbf m_{i}^{(p)}\cap\mathbf m_{i'}^{(p)}$ is empty. Namely, parton Fock spaces associated with $i$ and $i'$ are distinct. This property allows the convenient \emph{mutually commuting} fusion gate formalism:
\begin{align}
|\Psi^{(e)}\rangle=\hat{\mathbf F}^{N_e}|\Psi^{MF}\rangle=\prod_{i}\hat {\mathbf C}_{i}|\Psi^{MF}\rangle,\label{eq:general_Hdet_locality_state}
\end{align}
where:
\begin{align}
&\hat {\mathbf C}_i\equiv |0_i^{(e)}\rangle\langle 0_i^{(\text{all } p)}|+\sum_{n=1}^{\infty}\frac{\hat{\mathbf F}_i^n}{n!}\bullet(|0_i^{(e)}\rangle\langle 0_i^{(\text{all } p)}|)\notag\\
&\quad\;=\exp(\hat{\mathbf F}_i)\bullet(|0_i^{(e)}\rangle\langle 0_i^{(\text{all } p)}|),\label{eq:onsite_fusion_gate}\notag\\
\end{align}
where $|0_i^{(\text{all } p)}\rangle$ is the vacuum of all partons, whether they participate in this fusion gate or not. For fermionic electrons, the summation of $n$ truncates to $n=1$. For bosonic electrons, the summation truncates to a finite $n_{\rm max}$ determined by the fusion channels (since the partons are fermionic). After introducing gates $\hat{\mathbf P}_i\equiv \hat {\mathbf C}^\dagger_i\hat {\mathbf C}_i $, and observable  $\hat{\mathbf O}_D\equiv \prod_{i\in D} \hat {\mathbf C}^\dagger_i \cdot \hat O^{(e)}_D \cdot\prod_{i\in D}\hat {\mathbf C}_i$, Eq.(\ref{eq:general_observable_gate}) is directly generalized.

\section{Two types of model systems}\label{sec:two_model_types}
There are two common types of model systems giving rise to fractionalized states of matter: \textbf{(A) A model defined with full real-space degrees of freedom (e.g., a tight-binding model with interactions)}, and \textbf{(B) a model already projected into a topological band (e.g., a Chern band)}. As mentioned in Sec.\ref{sec:discrete_coherent_state_basis}, the treatments of the Hdet wavefunctions in these two model types are different. For type-(A) models, one can construct the locality-structure of the fusion tensor based on the real-space orthonormal Wannier orbitals of electrons. For type-(B) models, one needs to work with overcomplete discrete coherent states of electrons. In this section, we will present physical examples of how to construct Hdet wavefunctions for both types of model systems.

\subsection{Type-(A) example: QSL}\label{sec:QSL}
Traditionally, many quantum spin liquid (QSL) states are represented using the so-called Abrikosov-fermion formulation\cite{Abrikosov1965Pseudofermion,Wen-Symmetric-QSL}:
\begin{align}
\mathbf S^{\text{Abr.}}_{\mathbf r}=\frac{1}{2}\sum_{ab}\mathsf f^{(a)\dagger}_{\mathbf r}\boldsymbol{\sigma}_{ab} \mathsf f^{(b)}_{\mathbf r},\label{eq:Abrikosov_fermion}
\end{align}
where $\mathbf r$ labels the real-space site, and $a,b=\uparrow,\downarrow$ label the spin. By introducing spin-$\frac{1}{2}$ fermion $\mathsf f^{(a)}_{\mathbf r}$, the local Hilbert space is enlarged to $4$-dimensional: $\{|0\rangle,\mathsf f^{(\uparrow)\dagger}|0\rangle,\mathsf f^{(\downarrow)\dagger}|0\rangle,\mathsf f^{(\uparrow)\dagger}\mathsf f^{(\downarrow)\dagger}|0\rangle\}$. The physical spin-$\frac{1}{2}$ physical Hilbert space is recovered after enforcing the one-$\mathsf f$-fermion-per-site constraint: $\sum_a \mathsf f^{(a)\dagger}_{\mathbf r}\mathsf f^{(a)}_{\mathbf r}=1,\forall \mathbf r$.

In order to construct a QSL wavefunction, one first writes down a free fermion mean-field state of $\mathsf f$-fermions $|\Psi^{MF}\rangle$. If there is no pairing between $\mathsf f^{(\uparrow)}$ and $\mathsf f^{(\downarrow)}$, $|\Psi^{MF}\rangle$ would be a tensor product of two Slater determinants, one for each spin:  $|\Psi^{MF}\rangle=|\Psi^{(\uparrow)}\rangle\otimes |\Psi^{(\downarrow)}\rangle$. This is the typical case of the so-called $U(1)$ QSL\cite{Wen-Symmetric-QSL, LeeNagaosaWen2006, Ran2007}. 

To make a connection with our formulation of the Hdet wavefunction, it is convenient to map the $S=\frac{1}{2}$ model into a hard-core boson model, via the identification: 
\begin{align}
|\uparrow\rangle=&|1\rangle=b^\dagger|0\rangle,&|\downarrow\rangle=&|0\rangle\label{eq:spin_hardcore_boson_map}
\end{align} 
We may call these hard-core bosons the bosonic electrons. At the same time, we perform a particle-hole transformation for $\mathsf f^{(\downarrow)}$ only: $\mathsf f^{(\uparrow)}\equiv f, \mathsf f^{(\downarrow)}\equiv h^{\dagger}$. The mean-field state is still a product of two Slater determinants: $|\Psi^{MF}\rangle=|\Psi^{(\uparrow-f)}\rangle\otimes |\Psi^{(\downarrow-h)}\rangle$. In the language of the hard-core boson $b$ and $f,h$ partons, the projection into the physical local Hilbert space can be exactly represented using the local fusion operator $\hat {\mathbf F}_{\mathbf r}$:
\begin{align}
\hat {\mathbf F}^{\text{Abr.}}_{\mathbf r}=b_{\mathbf r}^\dagger  h_{\mathbf r}f_{\mathbf r}.
\end{align}
The corresponding local fusion gate $\hat{\mathbf C}_{\mathbf r}$ is:
\begin{align}
&\hat{\mathbf C}^{\text{Abr.}}_{\mathbf r}=|0_{\mathbf r}^{(e)}\rangle\langle 0_{\mathbf r}^{(\text{all }p)}|+|1_{\mathbf r}^{(e)}\rangle\langle 0_{\mathbf r}^{(\text{all }p)}|h_{\mathbf r}f_{\mathbf r}\notag\\
=&|\downarrow_{\mathbf r}\rangle\langle 0_{\mathbf r}^{(\text{all }p)}|+|\uparrow_{\mathbf r}\rangle\langle 0_{\mathbf r}^{(\text{all }p)}|h_{\mathbf r}f_{\mathbf r}.
\end{align}
Namely, \emph{the traditional Abrikosov-fermion construction of the U(1) QSL is nothing but a very special Hdet wavefunction with a locality structure: there is exactly one on-site fusion channel} ($R=1$). It is well-known that the spin wavefunction of the $U(1)$ QSL $|\Psi^{(\text{Abr.})}\rangle=\prod_{\mathbf r} \hat{\mathbf C}^{\text{Abr.}}_{\mathbf r}|\Psi^{MF}\rangle$ is a product of two determinants. In the past few decades, such Abrikosov-fermion QSL wavefunctions have been widely studied via variational Monte Carlo techniques\cite{Gros1989,Motrunich2005,Ran2007,Iqbal2011,Iqbal2013,Hu2015Chiral}.

The key point raised by the present work is that fusion need not be limited to a single channel. Namely, the parton images of the physical spin states $\{|\uparrow_{\mathbf r}\rangle,|\downarrow_{\mathbf r}\rangle\}$ do not need to be simple product states. In the presence of only $S_z$ spin rotation symmetry, it is straightforward to introduce multiple fusion channels in the previous hard-core picture. For example, one may introduce additional parton orbitals, labeled by $\mathbf m^{(a)}$, so that:
\begin{align}
\hat{\mathbf C}_{\mathbf r}
=|\downarrow_{\mathbf r}\rangle\langle 0_{\mathbf r}^{(\text{all }p)}|+\sum_{\alpha=1}^R \lambda_\alpha |\uparrow_{\mathbf r}\rangle\langle 0_{\mathbf r}^{(\text{all }p)}|h_{\mathbf r,\mathbf m^{(\downarrow)}_\alpha}f_{\mathbf r,\mathbf m^{(\uparrow)}_\alpha},
\end{align}
leading to a hyperdeterminant of a rank-3 fusion tensor $T$ for the spin (i.e., hard-core boson) wavefunction. 

In the above $S_z$-symmetric case, the parton image of $|\uparrow\rangle$ spin is an entangled state. But the parton image of $|\downarrow\rangle$ spin is a product state: the parton vacuum. However, in the presence of the full $SU(2)$ spin rotation symmetry, both parton images of $|\uparrow\rangle$ and $|\downarrow\rangle$ spins need to be entangled states for an Hdet structure. How to perform such a construction?

Interestingly, motivated by high-temperature superconductivity, Zhang and Sachdev\cite{ZhangSachdev2020} recently introduced an ancilla construction for correlated electronic states. In the special case of an undoped system, their construction exactly yields an Hdet QSL wavefunction respecting $SU(2)$ symmetry, whose fusion tensor has CPD-rank $R=2$. Below we introduce their construction for the undoped case and reveal the connection with the Hdet structure.

In the Zhang-Sachdev (ZS) construction for a spin-$1/2$ system, for each spin on site-$\mathbf r$, three spin-$\frac{1}{2}$ fermionic parton orbitals are introduced: $\mathsf f^{(a)}_{\mathbf r,u},\mathsf f^{(a)}_{\mathbf r,v},\mathsf f^{(a)}_{\mathbf r,w}$, where $a=\uparrow,\downarrow$ labels the spin, and $u,v,w$ labels the orbital. 

The physical spin is represented as the spin-singlet formed by the $u/v$ fermions together with the spin-carrying $w$-fermion. In terms of the fusion gate of $\mathsf f$:
\begin{align}
\hat{\mathbf C}_{\mathbf r}^{\text{ZS}}&=|\downarrow_{\mathbf r}\rangle\langle \big(\uparrow_{\mathbf r,u}\downarrow_{\mathbf r,v}-\downarrow_{\mathbf r,u}\uparrow_{\mathbf r,v}\big)\downarrow_{\mathbf r,w}|\notag\\
&+|\uparrow_{\mathbf r}\rangle\langle \big(\uparrow_{\mathbf r,u}\downarrow_{\mathbf r,v}-\downarrow_{\mathbf r,u}\uparrow_{\mathbf r,v}\big)\uparrow_{\mathbf r,w}|
\label{eq:ZS_undoped_fusion}
\end{align}
Crucially, both parton images of $|\uparrow\rangle$ and $|\downarrow\rangle$ are entangled states in an $SU(2)$ symmetric fashion.

To translate the ZS construction to the Hdet construction, it is convenient to consider a mapping between a spin-$\frac12$ system and a hard-core boson system different from Eq.(\ref{eq:spin_hardcore_boson_map}). Instead of one hard-core boson orbital per site, introduce two hard-core boson orbitals per site $b_{\mathbf r,\mathbf m^{(e)}}$ labeled by $\mathbf m^{(e)}=1,2$:
\begin{align}
|\uparrow\rangle=&|1_1,1_2\rangle=b_1^\dagger b_2^\dagger|0\rangle,& |\downarrow\rangle=&|1_1,0_2\rangle=b_1^\dagger |0\rangle.
\end{align}
Both boson orbitals are considered as physical d.o.f., but the boson-$1$ orbital is always filled on every site. 

Now consider a parton mean-field state $|\Psi^{MF}\rangle$ involving $SU(2)$ symmetric mixing between the $u/v/w$ orbitals, with a total of $3N_{site}$ filled single-particle parton states (half of which are spin-$\uparrow$). (In the simple limit when $w$-fermion does not mix with $u,v$-fermions, it is easy to see that Zhang-Sachdev construction goes back to the traditional Abrikosov fermion construction.) Next, similar to the construction before, we perform the particle-hole transformation for the spin-$\downarrow$ partons only, for all three orbitals. The mean-field state is a product of two Slater determinants, one for $\uparrow$-particles, one for $\downarrow$-holes: $|\Psi^{MF}\rangle=|\Psi^{(\uparrow-f)}\rangle\otimes |\Psi^{(\downarrow-h)}\rangle$. Now the local fusion operator, in the $b_{\mathbf r,\mathbf m^{(e)}}$ formalism, can be written as:
\begin{align}
\hat{\mathbf F}^{\rm ZS}_{\mathbf r,1}=&b_{\mathbf r,1}^\dagger(h_{\mathbf r,u}f_{\mathbf r,u}-h_{\mathbf r,v}f_{\mathbf r,v}),&\hat{\mathbf F}^{\rm ZS}_{\mathbf r,2}=&b_{\mathbf r,2}^\dagger h_{\mathbf r,w}f_{\mathbf r,w},
\end{align}
whose corresponding $\hat{\mathbf C}^{\rm ZS}_{\mathbf r,\mathbf m^{(e)}}$ fusion gate are defined according to Eq.(\ref{eq:onsite_fusion_gate}).
The physical spin wavefunction is obtained via:
\begin{align}
|\Psi^{\text{ZS}}_{QSL}\rangle\equiv\prod_{\mathbf r,\mathbf m^{(e)}=1,2} \hat{\mathbf C}_{\mathbf r,\mathbf m^{(e)}}^{\text{ZS}}|\Psi^{(\uparrow-f)}\rangle\otimes |\Psi^{(\downarrow-h)}\rangle.
\end{align}
It is easy to see that $|\Psi^{\text{ZS}}_{QSL}\rangle$ is exactly an Hdet wavefunction of a rank-3 tensor $T_{ijk}^{\text{ZS}}$ defined below, where $i=1,2,...,2N_{site};\; j,k=1,2,...\frac{3N_{site}}{2}$. 

Introducing the filled single-particle parton states $|\phi_j^{(\uparrow-f)}\rangle$ and  $|\phi_k^{(\downarrow-h)}\rangle$, and an indexing of sites $\mathbf r_l$ ($l=1,2,... N_{site}$), we define:
\begin{align}
\bullet\text{ If } &i=1,2,...,N_{site},\text{ (first-half)}:\notag\\
T_{ijk}^{\text{ZS}}&\equiv \langle 0| f_{\mathbf r_i,u} |\phi_j^{(\uparrow-f)}\rangle \langle 0 |h_{\mathbf r_i,u}|\phi_k^{(\downarrow-h)}\rangle \notag\\
&-\langle 0| f_{\mathbf r_i,v}|\phi_j^{(\uparrow-f)}\rangle \langle 0|h_{\mathbf r_i,v}|\phi_k^{(\downarrow-h)}\rangle.\notag\\
\bullet\text{ If } &i'\equiv (i-N_{site})=1,2,...,N_{site},\text{ (second-half)}:\notag\\
T_{ijk}^{\text{ZS}}&\equiv \langle 0| f_{\mathbf r_{i'},w} |\phi_j^{(\uparrow-f)}\rangle \langle 0 |h_{\mathbf r_{i'},w}|\phi_k^{(\downarrow-h)}\rangle .
\end{align}

For a given spin configuration $|\{s_\mathbf r\}\rangle$ ($s_\mathbf r=\uparrow,\downarrow$), one builds a cubic sub-tensor $T_{ijk}^{\text{sub}}$ of $T^{\text{ZS}}$ as follows. The first $i=1,2,...,N_{site}$ slices of $T_{ijk}^{\text{sub}}$ are fixed and are equal to the corresponding slices in the first-half of $T^{\text{ZS}}$. The last $i'\equiv (i-N_{site})=1,2,...,\frac{N_{site}}{2}$ slices of $T_{ijk}^{\text{sub}}$ are specified by the $\frac{N_{site}}{2}$ spin-$\uparrow$ positions in the spin configuration -- one picks the corresponding slices in second-half of $T^{\text{ZS}}$. As illustrated in Fig.\ref{fig:Hdet_subtensor}, $|\Psi^{\text{ZS}}_{QSL}\rangle$ is defined via:
\begin{align}
\langle\{s_\mathbf r\}|\Psi^{\text{ZS}}_{QSL}\rangle=\text{Hdet}(T^{\text{sub}}).
\end{align}

It is straightforward to further generalize the Zhang-Sachdev construction. One simple approach is introducing even more parton orbitals, leading to an Hdet structure with a larger CPD-rank. But even with the same number of parton orbitals (three spin-1/2 fermionic partons per site), because of the Clebsch-Gordon(CG) series $\frac{1}{2}\otimes\frac{1}{2}\otimes\frac{1}{2}=\frac{1}{2}\oplus\frac{1}{2}\oplus\frac{3}{2}$, one can introduce yet another linearly independent $SU(2)$-symmetric fusion channel, leading to a Hdet structure with CPD-rank $R=3$. Denoting the fusion amplitudes in the two relevant CG channels as $\lambda_1,\lambda_2$, one can show the generalized fusion operators:
\begin{align}
\hat {\mathbf F}_{\mathbf r,1}=&b_{\mathbf r,1}^\dagger \big[\lambda_1(h_{\mathbf r,u}f_{\mathbf r,u}-h_{\mathbf r,v}f_{\mathbf r,v})+\lambda_2(h_{\mathbf r,u}f_{\mathbf r,u}-h_{\mathbf r,w}f_{\mathbf r,w})\big]\notag\\
=&b_{\mathbf r,1}^\dagger \big[(\lambda_1+\lambda_2)h_{\mathbf r,u}f_{\mathbf r,u}-\lambda_1 h_{\mathbf r,v}f_{\mathbf r,v}-\lambda_2 h_{\mathbf r,w}f_{\mathbf r,w}\big],\notag\\
\hat {\mathbf F}_{\mathbf r,2}=&b_{\mathbf r,2}^\dagger[h_{\mathbf r,u}f_{\mathbf r,u}+h_{\mathbf r,v}f_{\mathbf r,v}+h_{\mathbf r,w}f_{\mathbf r,w}].
\end{align}
The corresponding rank-3 fusion tensor $T$ and $|\Psi_{\rm QSL}\rangle$ follow straightforwardly.

At the minimal level, these Hdet QSL constructions are expected to capture microscopic details of the spin wavefunctions beyond the traditional Abrikosov fermion method.

\subsection{Type-(B) example: FCI}\label{sec:type_B_model_FCI}
Here, we consider a model Hamiltonian $H_{\text{CB}}$ on a torus, describing a partially filled Chern band with Chern number $C=1$ together with interactions. $H_{\text{CB}}$ is fully constrained within the Chern band. It is well-known that such a Chern band can be mapped to the lowest Landau Level (LLL), preserving all crystalline symmetries\cite{Qi2011,JianQi2013,HuXiaoRan2024}. Following this mapping, $H_{\text{CB}}$ can be exactly mapped to a model $H_{\text{LLL}}$ inside the LLL, but with the lattice-potential/interaction breaking the Galilean invariance\cite{Qi2011,HuXiaoRan2024}. The lattice translation and rotation symmetries are mapped into the discrete magnetic translation and rotation symmetries in the LLL\cite{JianQi2013,HuXiaoRan2024}. We only need to demonstrate how to construct the fusion tensor $T$ for the Hdet trial wavefunctions $|\Psi^{(e)}\rangle$ for $H_{\text{LLL}}$.

After choosing the $(N^{(e)}_{s})^2$ discrete coherent states forming an overcomplete basis $\{|z^{(e)}\rangle\}$, $z\in$Fine-Grid, the construction of the fusion tensor $T$ is conceptually straightforward. Here, \emph{the main technical challenge in treating the overcomplete electronic basis is ensuring the correct linear dependence among the tensor slices $T_{z,...}$} with fixed $z$. Namely, based on the definition of the fusion tensor Eq.(\ref{eq:fusion_T_def}), among the $(N^{(e)}_{s})^2$ slices, only $N^{(e)}_{s}$ are linearly independent. Thanks to the fusion amplitudes analytically computed in the showcase examples Eq.(\ref{eq:boson_general_fusion},\ref{eq:boson_general_CFL_fusion},\ref{eq:fermion_general_fusion},\ref{eq:fermion_general_CFL_fusion}), this challenge is already resolved -- these amplitudes automatically ensure the correct linear dependence. 

Importantly, since the partons' states are constructed from full real-space local orbitals \emph{after the change of viewpoint}, as shown in Sec.\ref{sec:special_type_B}, we can write down the effective theories of Hdet states in the usual field-theory language, without resorting to the noncommutative field theories.

Given these fusion amplitudes, to construct the fusion tensor $T$, we need only specify the filled parton single-particle states $|\phi^{(p)}_j\rangle$. Comparing with FQH systems, the effect of the lattice-potential/interaction breaking the Galilean invariance in FCI systems is simply the \emph{hybridization between the parton Landau levels} (as well as changing the shape/topology/Bloch-states of the Fermi sea in the case of CFL) (see Sec.\ref{sec:special_type_B} for details).

As in all parton constructions, to obtain a symmetric electronic state, the parton states need to respect the symmetry in a projective fashion\cite{Wen-Symmetric-QSL}. In general, one must perform a careful analysis of the projective symmetry group (PSG) for the filled parton states. Here, however, the situation is slightly simpler, since the PSG transformation is already encoded in single-particle states in each parton Landau level. 

For example, consider a parton species-$p$ carrying electric charge $q^{(p)}=\frac{1}{3}$, as in the case of $\nu=\frac{1}{3}$ FCI states. We may construct $|\phi^{(p)}_j\rangle$ using the Bloch basis. 

First, we select a finite number of Landau levels for the parton, for instance, the LL index $n_{LL}=0,1,2,...,n_{\text{max}}$. Next, if the original model $H_{\text{CB}}$ has lattice translation primitive vectors $\mathbf a_1,\mathbf a_2$, then the corresponding magnetic translations $T^{(p)}_{\mathbf a_1},T^{(p)}_{\mathbf a_2}$ for the parton do not commute, as dictated by the Girvin-MacDonald-Platzman (GMP) algebra:
\begin{align}
T^{(p)}_{\mathbf a_1}T^{(p)}_{\mathbf a_2}=e^{i\frac{2\pi }{3}}T^{(p)}_{\mathbf a_2}T^{(p)}_{\mathbf a_1}
\end{align}
In order to construct the Bloch basis in a given parton LL with index $n_{LL}$, one has to enlarge the unit cell by three times. One choice is to consider the pair of commuting magnetic translations: $T^{(p)3}_{\mathbf a_1},T^{(p)}_{\mathbf a_2}$, and label their mutual eigenstate by the momentum $\mathbf k$. As a result, we obtain the Bloch basis: $\{|\phi^{(p)}_{\mathbf k}\rangle_{nLL}\}$. (The details can be found in Ref.\cite{HuXiaoRan2024})

The filled parton states $\{|\phi^{(p)}_{\mathbf k}\rangle\}$ can then be represented as a linear superposition of the Bloch basis at each fixed $\mathbf k$:
\begin{align}
|\phi^{(p)}_{\mathbf k}\rangle=\sum_{n=0}^{n_{\text{max}}} A^{(p)}_{n,\mathbf k} \cdot|\phi^{(p)}_{\mathbf k}\rangle_{nLL}.\label{eq:FCI_LL_superposition}
\end{align}
Given the superposition coefficients $A^{(p)}_{n,\mathbf k}$ for all parton species, the fusion tensor $T$ can be found by the multi-linear superposition of the fusion amplitudes in Eq.(\ref{eq:boson_general_fusion},\ref{eq:boson_general_CFL_fusion},\ref{eq:fermion_general_fusion},\ref{eq:fermion_general_CFL_fusion}). Notice that these fusion amplitudes are computed in the infinite plane. Therefore, on a finite-size torus, the fusion amplitudes will have finite-size corrections, which are exponentially small when the system size  $\gg l$ ($l$ is the magnetic length, which is comparable with the lattice constant of $H_{\text{CB}}$). For the system sizes that we studied in Sec.\ref{sec:benchmark}, we find numerically that these finite-size effects are negligible.

It is important to notice that $A^{(p)}_{n,\mathbf k}$ at different $\mathbf k$ points are still related by the magnetic translation $T^{(p)}_{\mathbf a_1}$, sending $|\phi^{(p)}_{\mathbf k}\rangle\rightarrow |\phi^{(p)}_{\mathbf k+\frac{\mathbf G_2}{3}}\rangle $; as well as the magnetic rotation $R$, sending the triplet subspace $(|\phi^{(p)}_{\mathbf k}\rangle, |\phi^{(p)}_{\mathbf k+\frac{\mathbf G_2}{3}}\rangle,|\phi^{(p)}_{\mathbf k+\frac{2\mathbf G_2}{3}}\rangle)\rightarrow (|\phi^{(p)}_{R\mathbf k}\rangle, |\phi^{(p)}_{R\mathbf k+\frac{\mathbf G_2}{3}}\rangle,|\phi^{(p)}_{R\mathbf k+\frac{2\mathbf G_2}{3}}\rangle)$. ($\mathbf G_2$ is the reciprocal vector conjugate to $\mathbf a_2$.) The details of the symmetry analysis are a bit technical, which we will present elsewhere due to the scope of this paper. But conceptually it is straightforward: the collection of symmetry-related $|\phi^{(p)}_{\mathbf k}\rangle$ (i.e., the projective $\mathbf k$-star) must form an irreducible projective representation of the space group. (For interested readers, these projective irreps, particularly at the high symmetry $\mathbf k$ points, are already manifest in each parton LL. By changing the parton LL index $n_{\text{LL}}$, all inequivalent projective irreps can be found.)

In Sec.\ref{sec:benchmark}, we will follow this prescription exactly to construct Hdet wavefunctions for FCI systems and numerically simulate them.

\section{Effective theories: Variational manifold path integral}\label{sec:VMPI}
In the context of fermionic-parton descriptions of fractionalized states of matter, the effective field theory involves fermionic fields coupled to gauge fields. To this end, there are two common ways to introduce the gauge fields into the formulation. The first way introduces gauge fields to implement the constraints to project the parton degrees of freedom back to physical degrees of freedom\cite{Baskaran1987,  AffleckZouHsuAnderson1988SU2Gauge, IoffeLarkin1989GaplessFermionsGaugeFields, LeeNagaosaWen2006, Wen1999ProjectiveConstructionQH, ReadSachdev1991, SachdevRead1991LargeNFrustratedDoped, Sachdev1992, Wen-Symmetric-QSL}, which we call constraint gauge fields. The second way, on the other hand, involves writing the conserved parton currents as the curl of gauge fields $\mathbf j=\nabla\times a$, which we call hydrodynamic gauge fields\cite{WenZee1992, BlokWen1990}. The second way leads to the K-matrix descriptions of general abelian topologically ordered states, which can be shown to be the dual theory of the constraint-gauge field description\cite{Wen1999ProjectiveConstructionQH}. 

Both methods have advantages and disadvantages. The hydrodynamic approach directly captures the topological part of the effective field theory and allows a consistent construction of the edge effective theories. On the other hand, it is unclear how to treat partons filling bands with higher Chern numbers, especially when such bands lead to nonabelian topological orders. Importantly, the hydrodynamic approach cannot provide microscopic information of the system (e.g., parton band structure). 

The constraint-gauge-field approach is more general and can capture parton bands with high Chern numbers. It also reveals microscopic information, such as parton band structures at the mean-field level, although it is unclear how to systematically improve beyond the mean-field (e.g., to obtain reliable energetics of the collective modes). Such a constraint-gauge-field approach usually starts with introducing a gauge group $GG$ (e.g., for Abrikosov-fermion construction, $GG=SU(2)$\cite{AffleckZouHsuAnderson1988SU2Gauge,AffleckMarston1988, Wen-Symmetric-QSL}), whose generators are parton fermion bilinears. The projection (i.e., fusion in this paper) into the physical degrees of freedom is implemented by requiring the physical state to be a $GG$ gauge singlet. The problem is that, when the number of parton orbitals is large and multiple fusion channels are present, like in the Hdet case, the $GG$ singlet condition does not enforce the fusion gate.

To have a simple demonstration of this technical problem, consider a local fusion operator for a hard-core boson type-(A) model with two parton species:
\begin{align}
\hat{\mathbf F}_{\mathbf r}=&b_{\mathbf r}^\dagger P_{\mathbf r},&P_{\mathbf r}\equiv&\sum_{\alpha=1}^R f^{(1)}_{\mathbf r,\alpha}f^{(2)}_{\mathbf r,\alpha}.\label{eq:GG_UR_construction}
\end{align}
This fusion operator has a $U(R)$ gauge redundancy: $ f_{\mathbf r}^{(1)\dagger}\rightarrow  f_{\mathbf r}^{(1)\dagger} U$, $f_{\mathbf r}^{(2)\dagger}\rightarrow  f_{\mathbf r}^{(2)\dagger}U^T$ ($f_{\mathbf r}^{(p)\dagger}$ is the row vector formed by $f_{\mathbf r,\alpha}^{(p)\dagger}$), leaving the fusion operator invariant. However, the $U(R)$ singlet condition is not enough to enforce this fusion: for instance, the four-parton state $(P_{\mathbf r}^\dagger)^2|0^{\rm all-p}\rangle$ is also a $U(R)$ singlet, and nonzero for $R>1$. One cannot specify the single "parton-image state" of the hard-core boson: $P_{\mathbf r}^\dagger|0^{\rm all-p}\rangle$ solely by the $U(R)$ singlet condition, and the traditional way to construct effective theories cannot even apply.

In addition, it is  known that the details of $GG$ are not relevant to the low-energy physics. As pointed out by a seminal work by Wen\cite{Wen-Symmetric-QSL}, the low-energy gauge fluctuations are controlled by the so-called invariant gauge group $IGG$. For example, in the context of $U(1)$ QSL, $IGG=U(1)$. Physically, the parton state Higgs out the additional gauge modes in $GG$, leaving only the $IGG$ at low-energy.

Here, we attempt to construct the effective theory avoiding the technical problem mentioned above. By introducing the concept of the variational manifold path integral, which is a path-integral version of the time-dependent variational principle, we show that the \emph{ Hdet wavefunctions are associated with full microscopic effective field theory descriptions} involving fermions coupled with fluctuating $GG$ lattice gauge fields. After the Higgs mechanism, the fluctuating $IGG$ lattice gauge fields remain at low energy. The physical outcome is two-fold. First, by writing down the effective field theories, one immediately knows the universal long-range properties associated with a Hdet ground state trial wavefunction, i.e., the nature of the \emph{quantum phase} it belongs to. Second, given the efficient simulation methods for Hdet wavefunctions (e.g., see Sec.\ref{sec:proj_expansion}), in principle, one can obtain a \emph{microscopic} effective field theory with reliable energetics, from which, for instance, one should be able to extract accurate dispersions of collective modes (e.g.,  magnetorotons in FCI.) to compare with results from other numerical methods (e.g., ED, QMC, DMRG). 

We will first introduce the concept of the variational manifold path integral as an approximation technique, and demonstrate it in the well-known time-dependent Hartree-Fock treatment. Next, we apply this technique to construct the effective theories for Hdet wavefunctions. 

\subsection{Variational manifold path integral: General Setup}\label{sec:VMPI_setup}
\subsubsection{VMPI as the quantum version of the time-dependent variational principle}
The variational manifold path integral (VMPI) introduced here is nothing but the path integral version of the well-known time-dependent variational principle(TDVP) \cite{Dirac1930, Frenkel1934, McLachlan1964, KramerSaraceno1981}. The basic idea of TDVP is to constrain the equation of motion of the quantum state in a variational manifold. Consider a manifold of \emph{normalized} many-body variational wavefunctions $|\Phi(\boldsymbol \lambda)\rangle$, \emph{smoothly} parameterized by a real vector $\boldsymbol \lambda=(\lambda_1,\lambda_2,...)$, Dirac–Frenkel TDVP introduces a Lagrangian in the real-time formulation:
\begin{align}
L_{\text{TDVP}}
=&\langle \Phi(\boldsymbol \lambda (t))|\, i\partial_t -H \,|\Phi(\boldsymbol \lambda (t))\rangle\notag\\
=&-\sum_a A_a(\boldsymbol\lambda)\,\dot\lambda_a -E(\boldsymbol\lambda),\label{eq:TDVP_Lag}
\end{align}
where  $H$ is the many-body Hamiltonian, $E(\boldsymbol\lambda)$ is the variational energy, and $A_a(\boldsymbol\lambda)$ is the Berry connection:
\begin{align}
A_a(\boldsymbol\lambda)\equiv -i\langle \Phi(\boldsymbol \lambda)|\partial_a|\Phi(\boldsymbol \lambda)\rangle.
\end{align}
Crucially, the parameter $\boldsymbol\lambda$ is treated \emph{classically} in TDVP, whose Euler-Lagrange equation of motion (EOM) is:
\begin{align}
-\sum_b F_{ab}(\boldsymbol\lambda)\,\dot\lambda_b=\partial_a E(\boldsymbol \lambda),
\label{eq:TDVP_EOM_realtime}
\end{align}
where $F_{ab}$ is the Berry curvature:
\begin{align}
F_{ab}(\boldsymbol\lambda) \equiv \partial_a A_b-\partial_b A_a=2\text{Im}\langle\partial_a\Phi|\partial_b\Phi\rangle.\label{eq:general_Berry_matrix}
\end{align}

Although it cannot be rigorously justified in most practical situations, TDVP has been widely applied as an approximation method in quantum many-body systems\cite{Haegeman2011, KramerSaraceno1981}, with the time-dependent Hartree-Fock (TDHF) treatment as a specific example. 

The standard practice of TDVP is to first identify the variational ground state $|\Phi(\bar {\boldsymbol \lambda})\rangle$ via the static problem $\partial_a E(\bar{\boldsymbol \lambda})=0$. Then, for small perturbations $\boldsymbol \lambda=\bar {\boldsymbol \lambda}+\delta \boldsymbol \lambda$, the linearized EOM describes coupled harmonic oscillators. Introducing the Hessian matrix $V_{ab}\equiv \partial_a\partial_b E(\bar{\boldsymbol \lambda})$:
\begin{align}
-\sum_b F_{ab}(\bar{\boldsymbol\lambda}) \,\delta\dot\lambda_b\approx \sum_b V_{ab}\,\delta\lambda_b,\label{eq:classical_HO_EOM}
\end{align}
from which one can find collective modes (e.g., gapless Goldstone mode) dispersion\cite{Thouless1960, Rowe1968, RingSchuck1980}. Equivalently, Eq.(\ref{eq:classical_HO_EOM}) can be obtained from the Gaussian order Lagrangian $L^{\rm G}_{\rm TDVP}$:
\begin{align}
L_{\rm TDVP}=&L^{\rm G}_{\rm TDVP}+O(\delta\boldsymbol\lambda^3), \;\;\text{ where:}\notag\\
L^{\rm G}_{\rm TDVP}\equiv&-\frac{1}{2}\sum_{ab}F_{ab}(\bar{\boldsymbol\lambda})\delta\lambda_a\delta\dot\lambda_b-\frac{1}{2}\sum_{ab}V_{ab}\delta\lambda_a\delta\lambda_b.\label{eq:LG_TDVP}
\end{align}

For instance, consider the case that $\boldsymbol\lambda$ labels the spatial-dependent spin-density-wave order parameter; one can obtain the spin-wave dispersion from the EOM Eq.(\ref{eq:classical_HO_EOM}). However, due to the classical treatment of variational parameters, one cannot automatically obtain the quantum mechanics of the collective modes: e.g., magnons. In order to obtain magnons, one would have to quantize the spin-wave manually.

The natural improvement of the TDVP is to introduce quantum mechanics of the collective modes in the first place: instead of treating $\boldsymbol\lambda$ as classical variables, one treats it quantum mechanically via a path integral, which exactly leads to VMPI. This path integral treatment is not only elegant, but also necessary for sophisticated quantum many-body systems, which we will comment on shortly in Sec.\ref{sec:Hdet_eff}.

In order to define a path integral in the variational manifold, one needs to define an integration measure. A minimal requirement for a well-defined integration measure is reparameterization-invariance. Up to an unknown scalar factor $f(\boldsymbol\lambda)$, this is fixed by the Fubini–Study quantum metric:
\begin{align}
d\mu(\boldsymbol\lambda)\equiv f(\boldsymbol\lambda)\sqrt {\det\big[g(\boldsymbol\lambda)\big]} \prod_a d\lambda_a,\label{eq:PI_measure}
\end{align}
where the quantum metric
\begin{align}
g_{ab}(\boldsymbol\lambda)\equiv \text{Re}[\langle\partial_a \Phi|\partial_b\Phi\rangle-\langle\partial_a \Phi|\Phi\rangle\langle\Phi|\partial_b\Phi\rangle].\label{eq:general_quantum_metric}
\end{align}

With this integration measure, the TDVP action can be formulated as a path integral. We begin from the saddle-point state time-evolution amplitude
\begin{align}
\langle \Phi(\bar{\boldsymbol\lambda})|\,\mathcal T\exp\!\Big[-i\int_{-\frac{T}{2}}^{\frac{T}{2}}dt\,H\Big] |\Phi(\bar{\boldsymbol\lambda})\rangle.
\end{align}
As in standard QFT analysis, eventually $T\rightarrow \infty(1-i\epsilon)$ when correlators are computed. Discretizing time into $N$ slices (labeled by $s$) $T=N\cdot\Delta t$, by inserting an operator for every time slice-$s$:
\begin{align}
\Omega\equiv\int d\mu(\boldsymbol\lambda)|\Phi(\boldsymbol\lambda)\rangle\langle\Phi(\boldsymbol\lambda)|,\label{eq:Omega_insertion}
\end{align}
we define VMPI amplitude as an approximation for the amplitude above:
\begin{align}
& Z_{\text{VMPI}}\notag\\
\equiv&\int\prod_{s=1}^{N-1} d\mu(\boldsymbol\lambda_s) \langle \Phi(\boldsymbol\lambda_{s+1})|e^{-i\Delta t H}|\Phi(\boldsymbol\lambda_{s})\rangle,\label{eq:VMPI_time_slices}
\end{align}
where we identified $\boldsymbol\lambda_0=\boldsymbol\lambda_N=\bar{\boldsymbol\lambda}$. In the continuum limit $N\rightarrow\infty$, we obtain the path integral of the TDVP action:
\begin{align}
Z_{\text{VMPI}}=\int\mathcal D[\boldsymbol\lambda(t)] \exp\Big[i\int dt L_{\text{TDVP}}(\boldsymbol\lambda(t))\Big].\label{eq:VMPI_Z}
\end{align}

In some situations, the unknown function $f(\boldsymbol\lambda)$ can be further fixed by other requirements, such as requiring the operator $\Omega$ to contain a resolution of identity of a sub-Hilbert space, or the invariance under certain symmetries. Fortunately, for most situations in this paper, we do not need to fix $f(\boldsymbol\lambda)$: We will primarily only consider the path integral up to Gaussian order fluctuations around its saddle point $\boldsymbol\lambda(t)=\bar{\boldsymbol\lambda}+\delta\boldsymbol\lambda(t)$. At the Gaussian order, one may set $f(\boldsymbol\lambda(t))=f(\bar{\boldsymbol\lambda})$ to be the value at the saddle point, as shown below. 

In typical interacting-field-theory analyses, the Gaussian order approximation can only be rigorously justified by additional expansion parameters, such as in the large-$N$ expansion or the semiclassical $\hbar$-expansion. For instance, in the $\hbar$-expansion, we have the scaling $\bar{\boldsymbol\lambda}\sim O(\hbar^0)$ and $\delta{\boldsymbol\lambda}\sim \hbar^{1/2}$. It is easy to show that the effect of the deviation $f(\bar{\boldsymbol\lambda}+\delta\boldsymbol\lambda(t))-f(\bar{\boldsymbol\lambda})$ only contributes a subleading term, down by a factor of $\hbar$ compared to the Gaussian order path integral taken with $f(\bar{\boldsymbol\lambda})$.

We would like to point out that we do not intend to justify the Gaussian-order approximation via rigorous expansion schemes in this paper; instead, we take a physics-motivated viewpoint: Gaussian-order fluctuations must be taken into account for quantum mechanics. So, at least one needs to identify the correct effective field theory for variational wavefunctions up to this order. \emph{In principle, one could generalize the VMPI analysis presented here beyond the Gaussian order, at which point the measure $f(\boldsymbol\lambda)$ will become an important issue.}

If the operator $\Omega$ contains a resolution of identity for the low-energy subspace of the quantum system of interest, then the VMPI treatment would be exact for the low-energy physics. However, in many interesting situations, including the situation in this paper, this may not be the case. As TDVP, in many situations, one may not be able to justify VMPI rigorously. Nevertheless, VMPI is a natural improvement of the TDVP, where quantum fluctuations of the variational parameters are considered.

\subsubsection{Zero-point motion, K\"{a}hler condition, and VMPI Wavefunctions}\label{sec:VMPI_wavefunction}
VMPI alone is sufficient to define all time-ordered correlators of the variational parameters, such as $\langle\mathcal T [\delta\lambda_a(t)\delta\lambda_b(0)]\rangle$. In this sense, it is already a full quantum field theory when $\boldsymbol\lambda$ describes spatially dependent fields. However, an important missing piece is how to translate $\delta\boldsymbol \lambda$ as quantum operators acting on physical wavefunctions in the original Hilbert space where the variational manifold $\{|\Phi(\boldsymbol\lambda)\rangle\}$ lives in. Without this Hilbert-space wavefunction translation, the correlators of $\delta\boldsymbol\lambda$ lose their microscopic meaning. It turns out that this translation becomes natural and canonical when an additional condition for the variational manifold is satisfied, the K\"{a}hler condition (defined below).

For example, the Gaussian order classical EOM Eq.(\ref{eq:classical_HO_EOM}) leads to classical eigenmodes as decoupled harmonic oscillators. In the VMPI treatment, we should expect these harmonic oscillators to be quantum mechanical with zero-point motion of $\delta\boldsymbol\lambda$, where we assume the number of variational parameters $N_{\boldsymbol\lambda}$ to be even. One may view one half of $\delta\boldsymbol\lambda$ as $q$'s, and the other half as $p$'s. Namely, even the ground state of the VMPI treatment should \emph{not} be viewed as a single state in the variational manifold $|\Phi(\bar{\boldsymbol\lambda})\rangle$. Instead, we expect VMPI to describe linear superpositions of variational wavefunctions:
\begin{align}
|\Psi_{\rm VMPI}\rangle=\int d\mu(\boldsymbol\lambda) |\Phi(\boldsymbol\lambda)\rangle C_\Psi(\boldsymbol\lambda),\label{eq:VMPI_superposition}
\end{align}
where $C_\Psi(\boldsymbol\lambda)$ is \emph{superposition coefficient}. This is the crucial conceptual improvement of VMPI over TDVP, in which one considers only a single $|\Phi(\boldsymbol\lambda)\rangle$. Since the variational states $\{|\Psi(\boldsymbol\lambda)\rangle\}$ are not orthogonal, $C_\Psi(\boldsymbol\lambda)$ is not unique for a given state. It is better to instead label $|\Psi_{\rm VMPI}\rangle$ by the VMPI \emph{wavefunction}, defined as:
\begin{align}
\Psi_{\rm VMPI}(\boldsymbol\lambda)\equiv \langle \Phi(\boldsymbol\lambda)|\Psi_{\rm VMPI}\rangle.\label{eq:VMPI_wavefunction}
\end{align}

In defining VMPI, the key object is the operator $\Omega$ in Eq.(\ref{eq:Omega_insertion}), which we do \emph{not} assume to contain an exact resolution of identity even for the low energy subspace of the Hamiltonian. The \emph{minimal requirement} for a physically reasonable VMPI wavefunction $|\Psi_{\rm VMPI}\rangle$, which describes the zero-point motion as well as possible quantum excitations for those harmonic oscillators, is to satisfy the condition:
\begin{align}
\text{Minimal Requirement: }\Omega |\Psi_{\rm VMPI}\rangle=|\Psi_{\rm VMPI}\rangle,\label{eq:VMPI_wavefunction_reproducing_condition}
\end{align}
under a properly defined normalization factor in the integration measure $d\mu(\boldsymbol\lambda)$. 

The time evolution of $|\Psi_{\rm VMPI}\rangle$ under VMPI is expected to lead to a Schr\"{o}dinger-type equation for $\Psi_{\rm VMPI}(\boldsymbol\lambda,t)$. Intuitively, within Gaussian order approximation, we expect this quantum equation of motion to be identical to that obtained by canonically quantizing $L_{\rm TDVP}^G$ in Eq.(\ref{eq:LG_TDVP}). It turns out that this intuition is not automatically satisfied. 

At this point, we introduce \emph{the K\"{a}hler condition: the variational manifold near the saddle point $|\Phi(\boldsymbol\lambda)\rangle$ is K\"{a}hler if and only if there exists a local complex coordinate $\boldsymbol\lambda[\bar{\mathbf z},\mathbf z]$ with $\mathbf z=(z_1,z_2,...,z_{\frac{N_{\boldsymbol\lambda}}{2}})$ and an unnormalized section $|\tilde \Phi(\boldsymbol\lambda)\rangle$, such that the $|\tilde \Phi(\mathbf z)\rangle\equiv|\tilde \Phi(\boldsymbol\lambda[\bar{\mathbf z},\mathbf z])\rangle$ is holomorphic in $\mathbf z$ in the neighborhood of the saddle point}.

Note that the variational wavefunction $|\Phi(\boldsymbol\lambda)\rangle$ is normalized. An unnormalized section $|\tilde \Phi(\boldsymbol\lambda)\rangle$ means that $|\tilde \Phi(\boldsymbol\lambda)\rangle=c(\boldsymbol\lambda)|\Phi(\boldsymbol\lambda)\rangle$ with some nonzero complex function $c(\boldsymbol\lambda)$ near the saddle point. A many-body state $|\tilde \Phi(\mathbf z)\rangle$ is holomorphic in $\mathbf z$ is defined as follows: In a fixed basis $\{|n\rangle\}$ spanning the full many-body Hilbert space (\emph{not} the variational manifold), the complex overlap function $\langle n|\tilde\Phi(\mathbf z)\rangle$ is holomorphic in $\mathbf z$(i.e., independent of $\bar{\mathbf z}$), $\forall n$.  WLOG, we will assume the saddle point is labeled by $\mathbf z=0$: $\boldsymbol\lambda[\boldsymbol0,\boldsymbol 0]=\bar{\boldsymbol\lambda}$.

Under this condition, the Schrödinger-type equation for $\Psi_{\rm VMPI}(\boldsymbol\lambda,t)$ is elegant, and the previously mentioned intuition is nicely satisfied. Although it is a strong condition for variational manifold, as we will show, \emph{Hdet wavefunctions, the central topic of this paper, satisfy the K\"{a}hler condition}. 

Note that, as shown in later discussions, the convenient parameterization of Hdet wavefunctions contains gauge redundancies: pure gauge transformations of $\boldsymbol\lambda$ do not change the physical Hdet wavefunction. In this case, the dimension of the variational manifold is less than $N_{\boldsymbol\lambda}$, and the map $\mathbf z\rightarrow |\tilde\Phi(\mathbf z)\rangle$ is many-to-one. However, based on the Holomorphic-Constant-Rank Theorem\cite{GunningRossi1965}, as long as this map has a constant Jacobian rank $r$ (i.e., the map is regular and the variational manifold is $2r$-dimensional), one can always perform a holomorphic change of variable $\mathbf z=\mathbf z(\boldsymbol\omega)$ such that in the new complex variable $|\tilde\Phi(\boldsymbol \omega)\rangle\equiv |\tilde\Phi(\mathbf z(\boldsymbol \omega))\rangle$ only depends on $(\omega_1,\omega_2,..,\omega_r)$. The map $(\omega_1,\omega_2,..,\omega_r)\rightarrow |\tilde\Phi(\boldsymbol \omega)\rangle$ is then bijective. Consequently, WLOG, \emph{in this subsection, we will assume the dimension of the variational manifold is $N_{\boldsymbol\lambda}$ and the map $\mathbf z\rightarrow |\tilde\Phi(\mathbf z)\rangle$ is locally bijective}. Mathematically, $\mathbf z$ is a local coordinate chart of the variational manifold.

The K\"{a}hler condition gives the variational manifold an elegant local geometry. Let's consider a generic overlap of two \emph{unnormalized} variational wavefunctions near the saddle point $\mathbf z=\mathbf 0$, labeled by $i$ and $f$. Defining
\begin{align}
\mathcal M(\bar{\mathbf z}_f,\mathbf z_i)\equiv \log [\langle\tilde\Phi(\mathbf z_f)|\tilde\Phi(\mathbf z_i)\rangle],\label{eq:M_definition}
\end{align}
the holomorphic and Hermiticity condition of overlap dictates its most general Taylor expansion up to the quadratic order:
\begin{align}
\mathcal M(\bar{\mathbf z}_f,\mathbf z_i)=&c+\mathbf a^T \mathbf z_i+\bar{\mathbf a}^T \bar{\mathbf z}_f\notag\\
+&\frac{1}{2}\big[\mathbf z_i^T P \mathbf z_i+ \bar{\mathbf z}_f^T \bar P \bar{\mathbf z}_f\big]+\mathbf z_i^T G \bar{\mathbf z}_f+O(\mathbf z^3),
\end{align}
where we use matrix formulation: $P^T=P, G^\dagger=G$, and $\mathbf z$ is a column vector. 

Perform a holomorphic gauge transformation:
\begin{align}
|\tilde\Phi(\mathbf z)\rangle\rightarrow e^{-h(\mathbf z)}|\tilde\Phi(\mathbf z)\rangle, 
\end{align}
after which:
\begin{align}
\mathcal M(\bar{\mathbf z}_f,\mathbf z_i)\rightarrow \mathcal M(\bar{\mathbf z}_f,\mathbf z_i)-h(\mathbf z_i)-\overline{ h(\mathbf z_f)}.
\end{align}
If we choose $h(\mathbf z)=\frac{c}{2}+\mathbf a^T\mathbf z +\frac{1}{2}\mathbf z^T P \mathbf z$, the function $\mathcal M$ only has the $G$-term left, which is the gauge we will be working in for the discussion below:
\begin{align}
\mathcal M(\bar{\mathbf z}_f,\mathbf z_i)=\mathbf z_i^T G \bar{\mathbf z}_f+O(\mathbf z^3)
\end{align}

When $\mathbf z_i=\mathbf z_f=\mathbf z$, $\mathcal M(\bar{\mathbf z},\mathbf z)=\mathbf z^T G \bar{\mathbf z}+O(\mathbf z^3)$ is the log of the positive wavefunction norm square of $|\tilde\Phi(\mathbf z)\rangle$, and $G$ encodes the real Fubini-Study quantum metric $g(\bar{\boldsymbol\lambda})$ (see Eq.(\ref{eq:general_quantum_metric})). To see this, consider the \emph{normalized} state:
\begin{align}
|\Phi(\mathbf z)\rangle\equiv e^{-\mathcal M(\bar{\mathbf z},\mathbf z)/2}|\tilde \Phi(\mathbf z)\rangle.\label{eq:normalized_Phi_as_z}
\end{align}
We warn the readers that here, we should have written it as $|\Phi(\bar{\mathbf z},\mathbf z)\rangle$, since the normalization factor depends on $\bar{\mathbf z}$. However, to simplify notation, we still use a single $\mathbf z$ to label $|\Phi(\mathbf z)\rangle$, which is the \emph{only} notation in this section violating holomorphic convention. 

Introducing the real coordinates
\begin{align}
z_\alpha\equiv \frac{1}{\sqrt{2}}(q_\alpha+ip_\alpha),\;\;\alpha=1,2,..,\frac{N_{\boldsymbol\lambda}}{2},\label{eq:pq_coord}
\end{align}
and the real and imaginary parts of $G$:
\begin{align}
G=S+iT,\;\; S^T=S,\;\;T^T=-T,
\end{align}
after grouping $(q,p)$ into the real variational parameter vector $\boldsymbol \lambda[\bar{\mathbf z},\mathbf z]=(\boldsymbol q,\boldsymbol p)$ (with $\bar{\boldsymbol\lambda}=\boldsymbol0$), direct calculation shows that the matrix $g(\bar{\boldsymbol\lambda})$ is:
\begin{align}
g(\bar{\boldsymbol\lambda})=\frac{1}{2}\begin{pmatrix}S&T\\-T&S\end{pmatrix}_{qp}.
\end{align}
Since $\mathbf z$ is a local coordinate chart, $g(\bar{\boldsymbol\lambda})$ must be positive definite, so is $G$. After a linear transformation of $\mathbf z$, $G$ can be brought to identity. We arrive at the \emph{canonical coordinate}:
\begin{align}
\mathcal M(\bar{\mathbf z}_f,\mathbf z_i)=\mathbf z_i^T\bar{\mathbf z}_f+O(\mathbf z^3).\label{eq:Kahler_Gaussian_M}
\end{align}

Based on Eq.(\ref{eq:M_definition},\ref{eq:normalized_Phi_as_z}), the overlap of normalized states in this coordinate is:
\begin{align}
\langle \Phi(\mathbf z_f)|\Phi(\mathbf z_i)\rangle=\text{exp}\Big[-\frac{\mathbf z^T_f \bar {\mathbf z}_f+\mathbf z^T_i \bar{\mathbf  z}_i}{2}+\mathbf z^T_i\bar{\mathbf z}_f+O(\mathbf z^3)\Big].\label{eq:VMPI_overlap_formula}
\end{align}
From this, the Berry curvature matrix $F(\bar{\boldsymbol\lambda})$ in the real $(q,p)$-coordinate can be computed (see Eq.(\ref{eq:general_Berry_matrix})):
\begin{align}
g(\bar{\boldsymbol\lambda})=\frac{1}{2}\begin{pmatrix}\mathbf 1&\mathbf 0\\\mathbf 0&\mathbf 1\end{pmatrix}_{qp},\;\;F(\bar{\boldsymbol\lambda})=\begin{pmatrix}\mathbf 0&\mathbf 1\\\mathbf -1&\mathbf 0\end{pmatrix}_{qp}.\label{eq:F_g_qp_coord}
\end{align}
As a coordinate-independent statement, $g,F$ satisfy 
\begin{align}
Fg^{-1}F=-4g,
\end{align}
a consequence of the K\"{a}hler condition.

The VMPI wavefunction in Eq.(\ref{eq:VMPI_wavefunction}) can now be written using $\mathbf z,\bar{\mathbf z}$:
\begin{align}
\Psi_{\rm VMPI}(\bar{\mathbf z},\mathbf z)\equiv \langle \Phi(\mathbf z) |\Psi_{\rm VMPI}\rangle.
\end{align}
We also consider another version of VMPI wavefunction, defined using the unnormalized variational state:
\begin{align}
\tilde \Psi(\bar{\mathbf z})\equiv \langle \tilde \Phi(\mathbf z) |\Psi_{\rm VMPI}\rangle,
\end{align}
which is explicitly holomorphic in $\bar{\mathbf z}$. Eq.(\ref{eq:normalized_Phi_as_z}) gives the relation between these two versions:
\begin{align}
\Psi_{\rm VMPI}(\bar{\mathbf z},\mathbf z)=e^{-\frac{\mathcal M(\bar{\mathbf z},\mathbf z)}{2}}\tilde\Psi(\bar{\mathbf z}).
\end{align}

We can now examine the minimal requirement Eq.(\ref{eq:VMPI_wavefunction_reproducing_condition}). Writing $\Omega=\int d\mu(\mathbf z_i)|\Phi(\mathbf z_i)\rangle\langle\Phi(\mathbf z_i)|$, this requirement is equivalent to:
\begin{align}
\tilde\Psi(\bar{\mathbf z}_f)=\int d\mu(\mathbf z_i) \text{exp}\big[\mathcal M(\bar{\mathbf z}_f,\mathbf z_i)-\mathcal M(\bar{\mathbf z}_i,\mathbf z_i)\big] \tilde\Psi(\bar{\mathbf z}_i).\label{eq:Kahler_reproducing_condition}
\end{align}

At the Gaussian order, it is safe to set the integral measure as:
\begin{align}
d\mu^{\rm G}(\mathbf z)\equiv\prod_\alpha \frac{d^2z_\alpha}{\pi}=\prod_\alpha \frac{d{\rm Re}[z_\alpha]d{\rm Im}[z_\alpha]}{\pi}.
\end{align}
In addition, with Eq.(\ref{eq:Kahler_Gaussian_M}), the minimal requirement at the Gaussian order becomes
\begin{align}
\tilde\Psi(\bar{\mathbf z}_f)=\int d\mu^{\rm G}(\mathbf z_i) \text{exp}\big[-\mathbf z_i^T\bar{\mathbf z}_i+\mathbf z_i^T \bar{\mathbf z}_f\big] \tilde\Psi(\bar{\mathbf z}_i),\label{eq:VMPI_Gaussian_reproducing_condition}
\end{align}
which, by standard Bargmann calculus, is always satisfied. It means that, although $\Omega$ may not be a resolution of identity in the low-energy subspace of the original Hamiltonian, it is a resolution of identity for the VMPI wavefunctions, at the Gaussian order.

\subsubsection{VMPI Schr\"{o}dinger equation, the ground state and excited states}
Representing the quadratic variational energy using the Hessian matrix $V$:
\begin{align}
\delta E[\bar{\mathbf z},\mathbf z]\equiv& \frac{1}{2} \boldsymbol\lambda[\bar{\mathbf z},\mathbf z]^T V\boldsymbol\lambda[\bar{\mathbf z},\mathbf z]\notag\\
\equiv&\mathbf z^T A \bar{\mathbf z}+\frac{1}{2}\Big[\mathbf z^TB \mathbf z+ \bar{\mathbf z}^T \bar{B} \bar{\mathbf z}\Big],
\end{align}
where we defined the $\mathbf z,\bar{\mathbf z}$-basis $\frac{N_{\boldsymbol\lambda}}{2}\times \frac{N_{\boldsymbol\lambda}}{2}$ matrices $A,B$, satisfying $A^\dagger=A,B^T=B$. The Gaussian order TDVP Lagrangian (see Eq.(\ref{eq:LG_TDVP})) in the $\mathbf z$ coordinate is:
\begin{align} 
L_{\rm TDVP}^{\rm G} &= \frac{i}{2} \left( \dot{\mathbf z}^T\bar{\mathbf z} - \mathbf z^T\dot{\bar{\mathbf z}} \right) - \delta E[\bar{\mathbf z},\mathbf z] \notag\\ &= \frac{i}{2} \left( \dot{\mathbf z}^T\bar{\mathbf z} - \mathbf z^T\dot{\bar{\mathbf z}} \right) - \mathbf z^T A\bar{\mathbf z} -\frac{1}{2} \left[ \mathbf z^TB\mathbf z + \bar{\mathbf z}^T\bar B\bar{\mathbf z} \right]. \label{eq:LG_TDVP_z}
\end{align}

Under VMPI time-evolution, we can now write down the Schr\"{o}dinger-type equation for $\tilde\Psi(\bar{\mathbf z},t)$. The short-time evolution over a single VMPI step (see Eq.(\ref{eq:VMPI_time_slices})) leads to the linear-$\Delta t$ change of VMPI wavefunction:
\begin{align}
&\Psi_{\rm VMPI}(\mathbf z_f,\bar{\mathbf z}_f,t+\Delta t)\notag\\
=&\int d\mu(\mathbf z_i) \langle \Phi(\mathbf z_f)|e^{-i\Delta t H}|\Phi(\mathbf z_i)\rangle\Psi_{\rm VMPI}(\mathbf z_i,\bar{\mathbf z}_i,t)\notag\\
\doteq&\int d\mu(\mathbf z_i) \langle \Phi(\mathbf z_f)|\Phi(\mathbf z_i)\rangle\big(1-i\Delta t \mathcal H_{\rm VMPI}(\bar{\mathbf z}_f,\mathbf z_i)\big)\notag\\
&\cdot\Psi_{\rm VMPI}(\mathbf z_i,\bar{\mathbf z}_i,t),\label{eq:VMPI_short_time_kernel}
\end{align}
where the Hamiltonian matrix element $\mathcal H_{\rm VMPI}(\bar{\mathbf z}_f,\mathbf z_i)$ is defined as:
\begin{align}
\mathcal H_{\rm VMPI}(\bar{\mathbf z}_f,\mathbf z_i)\equiv \frac{\langle\tilde \Phi(\mathbf z_f)|H|\tilde\Phi(\mathbf z_i)\rangle}{\langle\tilde \Phi(\mathbf z_f)|\tilde\Phi(\mathbf z_i)\rangle}.
\end{align}

Note that we use the unnormalized variational states in $\mathcal H_{\rm VMPI}$ for the holomorphic property, which, together with the Hermiticity property, lead to the most general quadratic expansion of $\mathcal H_{\rm VMPI}$:
\begin{align}
\mathcal H_{\rm VMPI}(\bar{\mathbf z}_f,\mathbf z_i)=&E(\bar{\boldsymbol\lambda})+\boldsymbol r^T\mathbf z_i+\bar{\boldsymbol r}^T\bar{\mathbf z}_f+\mathbf z_i^T C\bar{\mathbf z}_f\notag\\
+&\frac{1}{2}\Big[\mathbf z_i^T D\mathbf z_i+\bar{\mathbf z}_f^T \bar D\bar{\mathbf z}_f\Big]+O(\mathbf z^3),
\end{align}
The saddle-point condition dictates no linear term is present and $\boldsymbol r=0$. The matching condition with $\delta E[\bar{\mathbf z},\mathbf z]$ when $\mathbf z_i=\mathbf z_f=\mathbf z$ dictates $C=A$ and $D=B$. Namely, $\mathcal H_{\rm VMPI}\equiv\mathcal H_{\rm VMPI}^{\rm G}+O(\mathbf z^3)$, and the Gaussian order
\begin{align}
\mathcal H^{\rm G}_{\rm VMPI}(\bar{\mathbf z}_f,\mathbf z_i)=&E(\bar{\boldsymbol\lambda})+\mathbf z_i^T A\bar{\mathbf z}_f\notag\\
+&\frac{1}{2}\Big[\mathbf z_i^T B\mathbf z_i+\bar{\mathbf z}_f^T \bar B\bar{\mathbf z}_f\Big].
\end{align}
This is the point where the K\"{a}hler condition plays a crucial role: without the holomorphic property provided by the K\"{a}hler condition, even at the Gaussian order, the form of the matrix element $\mathcal H^{\rm G}_{\rm VMPI}(\bar{\mathbf z}_f,\mathbf z_i)$ \emph{cannot} be fixed by the diagonal element $\delta E[\mathbf z,\bar{\mathbf z}]$. Note that the TDVP Lagrangian only depends on the diagonal elements. Therefore, without the K\"{a}hler condition, the VMPI quantization and canonical quantization of $L^{\rm G}_{\rm TDVP}$ are generally different.

Utilizing Gaussian order $d\mu^{\rm G}(\mathbf z)$, Eq.(\ref{eq:VMPI_overlap_formula}) and the Bargmann calculus, one can replace $\mathbf z_i$ in $\mathcal H_{\rm VMPI}$ by the partial derivative column vector $\partial_{\bar{\mathbf z}_f}$ after performing the integral in Eq.(\ref{eq:VMPI_short_time_kernel}). We arrive at the Gaussian-order VMPI Schr\"{o}dinger equation for VMPI wavefunction $\tilde\Psi$:
\begin{align}
i\partial_t \tilde\Psi(\bar {\mathbf z},t)=H^{\rm G}_{\rm VMPI} \tilde\Psi (\bar {\mathbf z},t),
\end{align}
with the VMPI Hamiltonian defined as 
\begin{align}
H^{\rm G}_{\rm VMPI} = E(\bar{\boldsymbol\lambda})+\bar{\mathbf z}^T A^T\partial_{\bar{\mathbf z}}+\frac{1}{2}\Big[\partial_{\bar{\mathbf z}}^TB\partial_{\bar{\mathbf z}}+\bar{\mathbf z}^T\bar B\bar{\mathbf z} \Big],\label{eq:VMPI_Ham}
\end{align}
where, according to Bargmann calculus, $\partial_{\bar{\mathbf z}}$ in $A$-term stands to the right of $\bar{\mathbf z}$. 

$H^{\rm G}_{\rm VMPI}$ is exactly equivalent to the canonically quantized $L^{\rm G}_{\rm TDVP}$ in Eq.(\ref{eq:LG_TDVP_z}). In the $(q,p)$-coordinate introduced in Eq.(\ref{eq:pq_coord}) , the Berry's phase term in $L^{\rm G}_{\rm TDVP}$ becomes:
\begin{align}
L^{\rm Berry}_{\rm TDVP}=\frac{1}{2}\sum_\alpha(p_\alpha \dot q_\alpha-\dot p_\alpha q_\alpha)\doteq \sum_\alpha p_\alpha \dot q_\alpha.
\end{align}
where $\doteq$ means up to a total time-derivative. Canonical quantization replaces $(q,p)$ by operators, satisfying:
\begin{align}
[\hat q_\alpha,\hat p_\beta]=i\delta_{\alpha\beta}, \;\;[\hat q_\alpha,\hat q_\beta]=0, \;\;[\hat p_\alpha,\hat p_\beta]=0.
\end{align}
After introducing canonical boson operators:
\begin{align}
\hat b_\alpha\equiv \frac{1}{\sqrt{2}}(\hat q_\alpha+i\hat p_\alpha),\;\;  [\hat b_\alpha, \hat b^\dagger_\beta]=\delta_{\alpha\beta},\;\; [\hat b_\alpha, \hat b_\beta]=0,\label{eq:b_using_q_p}
\end{align}
we obtain the normal-ordered canonical quantization Hamiltonian:
\begin{align}
H^{\rm G}_{\rm can.}=E(\bar{\boldsymbol\lambda})+\hat {\boldsymbol b}^{\dagger T} A^T  \hat {\boldsymbol b}+\frac{1}{2}\Big[\hat {\boldsymbol b}^{T} B \hat {\boldsymbol b}+\hat {\boldsymbol b}^{\dagger T} \bar B  \hat {\boldsymbol b}^\dagger\Big].\label{eq:can_Ham}
\end{align}

If we use the normalized overcomplete coherent state basis $\{|\mathbf z\rangle\}$ satisfying $\hat b_{\alpha}|\mathbf z\rangle=z_{\alpha}|\mathbf z\rangle$, a normalized quantum state $|\psi\rangle$ in this canonical quantization Hilbert space can be written as a phase-space wavefunction $\psi(\mathbf z,\bar {\mathbf z})\equiv\langle \mathbf z |\psi\rangle$. $\psi(\mathbf z,\bar {\mathbf z})$ can be shown to have the form $\psi(\mathbf z,\bar {\mathbf z})=e^{-\frac{\mathbf z^T\bar{\mathbf z}}{2}}\tilde\psi(\bar{\mathbf z})$, where  $\tilde\psi(\bar{\mathbf z})$ is holomorphic in $\bar{\mathbf z}$. It is well known that when acting on $\tilde\psi(\bar{\mathbf z})$, the boson operators can be identified as:
\begin{align}
\hat b_{\alpha}\leftrightarrow \frac{\partial}{\partial \bar z_\alpha},\;\;\hat b^\dagger_{\alpha}\leftrightarrow \bar z_\alpha.\label{eq:VMPI_operator_dictionary}
\end{align}
Therefore, \emph{the VMPI Hamiltonian $H^{\rm G}_{\rm VMPI}$ in Eq.(\ref{eq:VMPI_Ham}) and $H^{\rm G}_{\rm can.}$ in Eq.(\ref{eq:can_Ham}) are equivalent}.

Under the K\"{a}hler condition, we now know that the Gaussian-order VMPI has full microscopic meaning. The ground state and excited states of the quadratic Hamiltonian $H^{\rm G}_{\rm can.}$ are easily obtained by Bogoliubov transformation -- leading to the eigen boson modes. Equivalently, the ground state and excited states for the VMPI wavefunction -- the linear superpositions of the original variational wavefunction $|\Phi(\boldsymbol\lambda)\rangle$, are also fully solved. The VMPI correlators of variational parameters such as $\langle\mathcal T [\delta\lambda_a(t)\delta\lambda_b(0)]\rangle$ can now be microscopically translated as the correlators of concrete quantum operators acting on VMPI wavefunctions using the dictionary Eq.(\ref{eq:b_using_q_p}) and Eq.(\ref{eq:VMPI_operator_dictionary}). 

Practically, the Bogoliubov transformation boils down to finding positive-frequency $\omega_\nu$ eigen boson operators:
\begin{align}
\hat{\gamma}^\dagger_\nu=\mathbf x^T_\nu\hat{\boldsymbol b}^\dagger-\mathbf y^T_\nu\hat{\boldsymbol b}, \;\;\;\;\hat{\gamma}_\nu=\mathbf x^\dagger_\nu\hat{\boldsymbol b}-\mathbf y^\dagger_\nu\hat{\boldsymbol b}^\dagger,
\end{align}
where (column) eigenvector $(\mathbf x_\nu,\mathbf y_\nu)$ satisfies:
\begin{align}
&\begin{pmatrix} A^T&\bar B\\ -B&-A\end{pmatrix}\begin{pmatrix}\mathbf x_\nu\\\mathbf y_\nu\end{pmatrix}=\omega_\nu\begin{pmatrix}\mathbf x_\nu\\\mathbf y_\nu\end{pmatrix},\notag\\
&\mathbf x_\nu^\dagger\mathbf x_\mu-\mathbf y_\nu^\dagger\mathbf y_\mu=\delta_{\nu\mu},\;\;\mathbf x_\nu^T\mathbf y_\mu-\mathbf y_\nu^T\mathbf x_\mu=0.
\end{align}
Collecting these column eigenvectors into matrices:
\begin{align}
\mathbf X\equiv(\mathbf x_1,\mathbf x_2...),\;\;\;\mathbf Y\equiv(\mathbf y_1,\mathbf y_2...),
\end{align}
the eigen boson operator vector is:
\begin{align}
\hat {\boldsymbol \gamma}=\mathbf X^\dagger \hat{\boldsymbol b}-\mathbf Y^\dagger \hat{\boldsymbol b}^\dagger, \;\;\;\hat {\boldsymbol \gamma}^\dagger=\mathbf X^T \hat{\boldsymbol b}^\dagger-\mathbf Y^T \hat{\boldsymbol b}.
\end{align}
The boson Hamiltonian is diagonalized: $H^{\rm G}_{\rm can.}=E(\bar{\boldsymbol\lambda})+E_{\rm VMPI}^{\rm zero-point}+\sum_\nu\omega_\nu \hat\gamma^\dagger_\nu\hat\gamma_\nu$ with VMPI zero-point energy:
\begin{align}
E_{\rm VMPI}^{\rm zero-point}\equiv \frac{1}{2}\Big(\sum_\nu\omega_\nu-{\rm Tr}A\Big).
\end{align}
In particular, the VMPI ground state wavefunction is
\begin{align}
&\tilde\Psi_{\rm VMPI,GS}(\bar {\mathbf z})=\frac{1}{\sqrt{ |\det\mathbf X|}}\exp\Big[\frac{1}{2}\bar {\mathbf z}^T\mathbf S\bar {\mathbf z}\Big],\notag\\
&\text{where } \mathbf S\equiv (\mathbf X^\dagger)^{-1}\mathbf Y^\dagger.\label{eq:VMPI_GS_wavefunction}
\end{align}
The one-boson excited state VMPI wavefunction is:
\begin{align}
&\tilde\Psi_{\rm VMPI,\nu}(\bar {\mathbf z})=\hat\gamma_\nu^\dagger \tilde\Psi_{\rm VMPI,GS}(\bar {\mathbf z})\notag\\
=&[\mathbf x_\nu^T\bar{\mathbf z}-\mathbf y_\nu^T\partial_{\bar{\mathbf z}}]\cdot\tilde\Psi_{\rm VMPI,GS}(\bar {\mathbf z})\notag\\
=&[(\mathbf x_\nu^T-\mathbf y_\nu^T\mathbf S)\bar{\mathbf z}]\cdot\tilde\Psi_{\rm VMPI,GS}(\bar {\mathbf z}),\label{eq:VMPI_excited_wavefunction}
\end{align}
where we used $\partial_{\bar{\mathbf z}}\tilde\Psi_{\rm VMPI,GS}(\bar {\mathbf z})=\mathbf S\bar{\mathbf z}\cdot\tilde\Psi_{\rm VMPI,GS}(\bar {\mathbf z})$.

In principle, the VMPI quantization (i.e., VMPI wavefunctions and the VMPI Schr\"{o}dinger equation) may be generalized beyond the Gaussian order, where the integral measure $d\mu(\mathbf z)$ needs to be solved to satisfy the minimal requirement Eq.(\ref{eq:Kahler_reproducing_condition}).

\subsection{Application of VMPI in Time-dependent Hartree-Fock}\label{sec:VMPI_TDHF}
\subsubsection{Thouless parameterization, saddle point analysis}
Next, we demonstrate the application of VMPI as an improvement of TDHF, where the variational manifold is formed by Slater determinants (pairing is not considered) -- \emph{the simplest Hdet wavefunctions for rank-2 fusion tensors}, and we will see that VMPI leads to a microscopic effective field theory. In this simple situation, obviously, one may obtain the effective theory following the traditional route: applying the Hubbard-Stratonovich transformation to decouple the electron interactions, then performing a saddle-point analysis. However, as we will see shortly in the application to Hdet wavefunctions, VMPI is a much more general construction. 

Many results in the current Slater-determinant VMPI application are equivalent to those from the well-known Random-Phase-Approximation (RPA) treatment. Nevertheless, the general framework and notation established in this subsection will prove to be very useful when we move on to the Hdet application.

A Slater determinant is fully specified by the 1-body reduced density matrix (RDM, i.e., the correlation matrix) $\rho_{ij}\equiv\langle c_j^\dagger c_i\rangle$, satisfying $\rho^2=\rho$, where $i=1,2...\text{dim}\mathcal H^{(e)}$ labels the real-space orbital basis of electrons ($i$ may be interpreted as $\{\mathbf r_i,\mathbf m_i\}$ where $\mathbf m$ labels the additional quantum numbers, e.g. spin.). Therefore, one may use $\rho$ as the variational parameter: $|\Phi(\rho)\rangle$. For a many-body Hamiltonian $H$, the variational energy function $E(\rho)\equiv \langle\Phi(\rho)|H|\Phi(\rho)\rangle$ can be used to define the effective hopping amplitudes $t$:
\begin{align}
t_{ij}(\rho)\equiv \frac{\partial E(\rho)}{\partial\rho_{ji}}\label{eq:TDHF_eff_hopping}
\end{align}
Here, since $\rho_{ij}=\rho_{ji}^*$ are not independent, the derivative should be interpreted as $\frac{\partial}{\partial \rho_{ji}}\equiv \frac{1}{2} (\frac{\partial}{\partial \text{Re}\rho_{ji}} -i \frac{\partial}{\partial\text{Im}\rho_{ji}})$ in terms of real derivatives. Consequently, $t$ is Hermitian. Note that, although the physical $\rho$ is constrained by $\rho^2=\rho$, $E(\rho)$ is a well-defined symbolic expression via Wick's theorem and consequently all the derivatives are well-defined. 

One may further define an effective free-fermion Hamiltonian $\hat h(\rho)$:
\begin{align}
\hat h(\rho)\equiv \sum_{ij} t_{ij}(\rho) c_i^\dagger c_j\;.
\end{align}
For the static Hartree-Fock, one needs to find the variational ground state, labeled by $\bar\rho$. The linear-order variation of $E(\rho)$ must vanish, which is equivalent to  $[t(\bar\rho),\bar\rho]=0$, and can be achieved by the self-consistent condition: the ground state of $\hat h(\bar\rho)$ is also labeled by $\bar\rho$.

We further define the Hessian $\mathcal K$ of $E(\rho)$ at $\bar\rho$:
\begin{align}
\mathcal K_{ij,kl}\equiv& \frac{\partial^2 E(\bar\rho)}{\partial\rho_{ji}\partial\rho_{lk}},&
[\hat{\mathcal K} X]_{ij}\equiv& \sum_{kl}\mathcal K_{ij,kl}X_{lk}.\label{eq:real_space_Hessian}
\end{align}
\emph{Consider the real linear space $M_{\rm h}$} formed by Hermitian matrices with the same size as $\rho_{ij}$, it is natural to define a real inner product:
\begin{align}
\llangle X,Y\rrangle\equiv \sum_{ij} X_{ij} Y_{ji},\label{eq:M_h_inner_product}
\end{align}
and the superoperator $\hat {\mathcal K}$ introduced above should be viewed as an operator acting in $M_{\rm h}$. Since $\mathcal K_{ij,kl}=\mathcal K_{kl,ij}$ and $\mathcal K_{ij,kl}=\mathcal K^*_{ji,lk}$, $\hat{\mathcal K}$ is symmetric: $\llangle X,\hat{\mathcal K}Y\rrangle=\llangle \hat{\mathcal K}X,Y\rrangle$.

For a deformation $\delta\rho$ around $\bar\rho$, up to quadratic order:
\begin{align}
E(\bar\rho+\delta\rho)=&E(\bar\rho)+\llangle t(\bar\rho),\delta\rho\rrangle+\frac{1}{2}\llangle\delta\rho,\hat{\mathcal K}\delta\rho\rrangle+O(\delta\rho^3).\label{eq:E_rho_expansion}
\end{align}

At this point, we introduce the Thouless parameterization of $\delta\rho$ near $\bar\rho$: $\rho=\bar\rho+\delta\rho$\cite{Thouless1960}. Consider a general small unitary transformation $|\Phi(\bar\rho+\delta\rho)\rangle = e^{i\hat\Xi(\kappa)}|\Phi(\bar\rho)\rangle$, with small Hermitian operator $\hat\Xi(\kappa)\equiv\sum_{\mathbf{ph}} (-i)\kappa_{\mathbf{ph}} c_{\mathbf p}^\dagger c_{\mathbf h}+h.c.$ where $\mathbf{p},\mathbf{h}$ label the \textbf{empty} and \textbf{filled} single-particle eigen-basis of $\bar\rho$ (and $t(\bar\rho)$), respectively. In this basis, $\bar\rho$ is diagonal: $\bar\rho_{\mathbf{hh'}}=\delta_{\mathbf{hh'}}$, and other matrix elements vanish. Up to the quadratic order of $\kappa$, the induced $\delta\rho$ can be found:
\begin{align}
\delta\rho_{\mathbf{ph}}=&\kappa_{\mathbf{ph}},&\delta\rho_{\mathbf{hp}}=&\kappa_{\mathbf{ph}}^*,\notag\\
\delta\rho_{\mathbf{hh'}}=&-\sum_\mathbf{p} \kappa_{\mathbf{ph}}^*\kappa_{\mathbf{ph'}},&\delta\rho_{\mathbf{pp'}}=&\sum_\mathbf{h} \kappa_{\mathbf{ph}}\kappa_{\mathbf{p'h}}^*.\label{eq:delta_rho_from_kappa}
\end{align}
Precisely, we also introduce the matrix form of $\hat\Xi(\kappa)$:
\begin{align}
\Xi(\kappa)_{\mathbf{ph}}\equiv &-i\kappa_{\mathbf{ph}},&\Xi(\kappa)_{\mathbf{hp}}\equiv &i\kappa^*_{\mathbf{ph}}.
\end{align}
The RDM for the rotated state $e^{i\hat\Xi(\kappa)}|\Phi(\bar\rho)\rangle$ is then:
\begin{align}
e^{i\Xi(\kappa)}\circ\bar\rho \equiv e^{i\Xi(\kappa)}\bar\rho e^{-i\Xi(\kappa)}=\bar\rho+\delta\rho.
\end{align}
Eq.(\ref{eq:delta_rho_from_kappa}) is simply the consequence of the identity:
\begin{align}
\delta\rho=i[\Xi(\kappa),\bar\rho]-\frac{1}{2}[\Xi(\kappa),[\Xi(\kappa),\bar\rho]]+O(\kappa^3).
\end{align}

At this point, we mention that, in a closely related parameterization of the current variational manifold formed by Slater determinants, it is obvious that \emph{the K\"{a}hler condition is satisfied}. Consider an unnormalized state:
\begin{align}
|\tilde \Phi(\tilde\kappa)\rangle\equiv \exp\big[\sum_{\mathbf{ph}} \tilde\kappa_{\mathbf{ph}} c_{\mathbf p}^\dagger c_{\mathbf h}\big]|\Phi(\bar\rho)\rangle.\label{eq:Det_Kahler_coord}
\end{align}
Comparing with $|\Phi(\tilde\kappa)\rangle$, the only difference is that the Hermitian conjugate term in $\hat\Xi(\tilde \kappa)$: $-\sum_{\mathbf{ph}}\tilde\kappa^*_{\mathbf{ph}}c_\mathbf h^\dagger c_{\mathbf p}$ (which annihilates $|\Phi(\bar\rho)\rangle$) is absent. The advantage here is that $|\tilde \Phi(\tilde\kappa)\rangle$ is clearly holomorphic in $\tilde \kappa$, and it is easy to show that $|\tilde \Phi(\tilde\kappa)\rangle$ is a Slater determinant:
\begin{align}
|\tilde \Phi(\tilde\kappa)\rangle=\prod_{\mathbf h} (c^\dagger_\mathbf h+\sum_{\mathbf p}\tilde \kappa_{\mathbf{ph}}c_{\mathbf p}^\dagger)|0\rangle.
\end{align}
By construction, this is a K\"{a}hler coordinate for the variational manifold. 

If one normalizes $|\tilde \Phi(\tilde\kappa)\rangle$, one can compare the $\kappa$ and $\tilde\kappa$ coordinates for the \emph{same} Slater determinant, leading to
\begin{align}
\tilde\kappa_{\mathbf {ph}}=\kappa_{\mathbf {ph}}+O(\kappa^3).
\end{align}
Therefore, for the present Gaussian order analysis, it does not matter if one uses $\kappa$ or $\tilde\kappa$.

Using $\kappa$ as $\boldsymbol\lambda$ to parameterize the small fluctuations, plugging into the TDVP Lagrangian Eq.(\ref{eq:TDVP_Lag}), one obtains up to the Gaussian order:
\begin{align}
L_{\text{TDVP}}=&\frac{i}{2} \left( \dot{\boldsymbol\kappa}^T\bar{\boldsymbol\kappa} - \boldsymbol\kappa^T\dot{\bar{\boldsymbol\kappa}} \right) -\boldsymbol\kappa^T\mathbf A\bar{\boldsymbol\kappa}\notag\\
&-\frac{1}{2}\Big[\boldsymbol\kappa^T\mathbf B\boldsymbol\kappa+\bar{\boldsymbol\kappa}^T\bar{\mathbf B}\bar{\boldsymbol\kappa}\Big]+O(\boldsymbol\kappa^3),
\label{eq:kappa_Lag}
\end{align}
where $\boldsymbol\kappa$ is the column vector obtained by reshaping the matrix $\kappa_{\mathbf{ph}}$ using the composite index $\mathbf{ph}$. Using the composite index, $\mathbf A,\mathbf B$ matrices are defined as:
\begin{align}
\mathbf A_{\mathbf{ph},\mathbf{p'h'}}\equiv&(\epsilon_\mathbf p-\epsilon_\mathbf h)\delta_{\mathbf{pp'}}\delta_{\mathbf{hh'}}+\mathcal K_{\mathbf{hp},\mathbf{p'h'}},\notag\\
\mathbf B_{\mathbf{ph},\mathbf{p'h'}}\equiv& \mathcal K_{\mathbf{hp},\mathbf{h'p'}},
\end{align}
satisfying $\mathbf A^\dagger=\mathbf A$, $\mathbf B^T=\mathbf B$.

The first term in Eq.(\ref{eq:kappa_Lag}) is the Berry's phase: $\langle\Phi(\kappa(t))|i\partial_t|\Phi(\kappa (t))\rangle$ which coincides with that of canonical bosons. Namely, \emph{the coordinate $\boldsymbol\kappa$ is already canonical and can be literally identified as $\mathbf {z}$ in our general VMPI quantization discussion, and all results from Eq.(\ref{eq:VMPI_overlap_formula}) to Eq.(\ref{eq:VMPI_excited_wavefunction}) can be immediately applied.}

To further improve the notation, we define the tangent subspace of Hermitian matrices at the saddle point:
\begin{align}
T_{\bar\rho}\equiv \{X=X^\dagger, \bar\rho X\bar\rho=0,(1-\bar\rho)X(1-\bar\rho)=0\}\subset M_{\rm h},
\end{align}
i.e., those Hermitian matrices with only nonzero $\mathbf{ph}$ and $\mathbf{hp}$ blocks. The Thouless parameterization $\kappa$ is an explicit coordinate for $T_{\bar\rho}$: $\kappa_{\mathbf{hp}}=\kappa^*_{\mathbf{ph}}$. 

Next, the bare kinetic energy $(\epsilon_\mathbf p-\epsilon_\mathbf h)$-term and the Berry's phase term can be captured by defining superoperators $\hat {\mathcal G}$ and $\hat{\mathcal F}$ acting in $T_{\bar\rho}$: $\forall X\in T_{\bar\rho}$,
\begin{align}
(\hat {\mathcal G} X)_{\mathbf{ph}}\equiv &(\epsilon_\mathbf p-\epsilon_\mathbf h)X_{\mathbf{ph}},&(\hat {\mathcal G} X)_{\mathbf{hp}}\equiv& (\epsilon_\mathbf p-\epsilon_\mathbf h)X_{\mathbf{hp}},\notag\\
(\hat {\mathcal F} X)_{\mathbf{ph}}\equiv& -iX_{\mathbf{ph}},&(\hat {\mathcal F} X)_{\mathbf{hp}}\equiv& iX_{\mathbf{hp}}.\label{eq:T_rho_superoperators}
\end{align}
$\hat{\mathcal G}$ is symmetric and positive definite ($\llangle X,\hat{\mathcal G}Y\rrangle=\llangle \hat{\mathcal G}X,Y\rrangle$, $\llangle X,\hat{\mathcal G}X\rrangle>0$), while $\hat{\mathcal F}$ is antisymmetric ($\llangle X,\hat{\mathcal F}Y\rrangle=-\llangle \hat{\mathcal F}X,Y\rrangle$). 

Physically, the superoperator $\hat{\mathcal F}$ is nothing but the Berry's curvature in the tangent space: consider two small $\kappa_1,\kappa_2\in T_{\bar\rho}$, then
\begin{align}
&\llangle \kappa_1,\hat{\mathcal F}\kappa_2\rrangle\notag\\
=& 2\cdot\rm{Im}\big[\langle  \Phi(e^{i\Xi(\kappa_1)}\circ\bar\rho)| \Phi(e^{i\Xi(\kappa_2)}\circ\bar\rho)\rangle\big]_{\kappa_1\kappa_2\text{-part}}+O(\kappa^3)\notag\\
=& 2\cdot\rm{Im}\big[\langle e^{i\hat\Xi(\kappa_1)} \Phi(\bar\rho)| e^{i\hat\Xi(\kappa_2)}\Phi(\bar\rho)\rangle\big]_{\kappa_1\kappa_2\text{-part}}+O(\kappa^3)\notag\\
=&-i\langle\Phi(\bar\rho)|[\hat\Xi(\kappa_1),\hat\Xi(\kappa_2)]|\Phi(\bar\rho)\rangle+O(\kappa^3).\label{eq:F_many_body_Berry}
\end{align}
As we will see later, when we study Hdet wavefunctions, the corresponding $\hat{\mathcal F}$ superoperator can still be defined via the first equal sign, although it cannot generally be exactly computed, and the latter lines no longer hold.

The TDVP Lagrangian can now be compactly written as:
\begin{align}
L_{\rm{TDVP}}=-\frac{1}{2}\llangle \kappa,\hat{\mathcal F}\dot\kappa\rrangle-\frac{1}{2}\llangle \kappa,(\hat{\mathcal G}+\hat{\mathcal{K}})\kappa\rrangle+O(\kappa^3)
\label{eq:kappa_Lag_1}
\end{align}
Comparing with the notation in Eq.(\ref{eq:LG_TDVP}) in the general setup section, $F$ should be identified with $\hat{\mathcal F}$, $V$ should be identified with $\hat{\mathcal G}+\hat{\mathcal{K}}$.

One may add a source term involving $J(t)\in T_{\bar\rho}$ in the Lagrangian to facilitate the computation of correlation functions:
\begin{align}
&L_{J-\kappa}\equiv \llangle J,\kappa\rrangle =\sum_{\mathbf{ph}}\Big[J^*_{\mathbf{ph}}(t)\,\kappa_{\mathbf{ph}}(t)+J_{\mathbf{ph}}(t)\,\kappa^*_{\mathbf{ph}}(t)\Big]
\label{eq:kappa_source_definition}
\end{align}

Up to the Gaussian order of $\kappa$, the unknown function $f(\kappa)$ in the integration measure does not matter since it only gives higher-order terms. We have the generating functional in VMPI formulation:
\begin{align}
Z_{\kappa}[J]\equiv\int \mathcal D[\kappa^*(t),\kappa(t)] \exp\!\Big[i\int dt\,( L_{\text{TDVP}}+L_{J-\kappa})\Big].\label{eq:VMPI_def}
\end{align}
The time-ordered $\kappa$ correlator can be computed as:
\begin{align}
\langle \mathcal T \kappa_{\mathbf{ph}}(t)\kappa_{\mathbf{p'h'}}(t')\rangle=\left.\frac{\delta^2 [-i\ln Z_{\kappa}[J]]}{\delta J^*_{\mathbf{ph}}(t)\delta J^*_{\mathbf{p'h'}}(t')}\right|_{J=0}.
\end{align}
The poles of this $\kappa$-correlator lead to bosonic collective modes.

\subsubsection{Hubbard-Stratonovich transformation and the effective theory of TDHF}
But our goal is not the $\kappa$-theory, since it is fully bosonic -- the fermions (electrons here) were bosonized as $\kappa$. Next, we will refermionize the $\kappa$-theory, and show that it is equivalent to a fermion theory coupling with Hubbard-Stratonovich (HS) fields, up to the Gaussian order.

For this purpose, it is convenient to first perform the standard HS transformation for the residual interaction $\mathcal K$-term in Eq.(\ref{eq:E_rho_expansion}) before inserting the Thouless parameterization. Let the subspace 
\begin{align}
V_{\rm HS}\equiv \text{Im}(\hat{\mathcal K})\subset M_{\rm h}\label{eq:V_HS_def}
\end{align}
be the image of the superoperator in Eq.(\ref{eq:real_space_Hessian}). Precisely speaking, $V_{\rm HS}$ is spanned by eigenmatrices of $\hat{\mathcal K}$ with nonzero eigenvalues (note that, since $\hat{\mathcal K}$ is symmetric in the real linear space $M_{\rm h}$, its eigenvalues are all real.). Introducing the projector superoperator $\hat{\mathcal P}_{\rm HS}$ to the subspace $V_{\rm HS}$, we define $\hat{\mathcal K}_{\rm HS}^{-1}\equiv \hat{\mathcal P}_{\rm HS}\hat{\mathcal K}^{-1}\hat{\mathcal P}_{\rm HS}$. The HS transformation is:
\begin{align}
&e^{-\frac{i}{2}\llangle \delta\rho,\hat{\mathcal K}\delta\rho\rrangle}=e^{-\frac{i}{2}\sum_{ij}\delta\rho_{ij} [\hat{\mathcal K}\delta\rho]_{ji}}\notag\\
=&\mathcal N\int D[\chi] \exp\!\Big[-i\sum_{ij}\chi_{ij}\delta\rho_{ji}+\frac{i}{2}\sum_{ij}\chi_{ij}\big[(\hat{\mathcal K}^{-1}_{\rm HS}\chi)\big]_{ji}\Big]\notag\\
=&\mathcal N\int D[\chi]\exp\!\Big[-i\llangle \chi,\delta\rho\rrangle+\frac{i}{2}\llangle \chi,\hat{\mathcal K}^{-1}_{\rm HS}\chi\rrangle \Big],
\label{eq:realtime_HS_identity}
\end{align}
where $\mathcal N$ is an unimportant normalization constant, and $D[\chi]$ is the integration over Hermitian matrices $\chi\in V_{\rm HS}$.

We now can rewrite the VMPI as a field theory involving both fluctuating $\kappa$ and $\chi$:
\begin{align}
Z_{\kappa}[J]=\int \mathcal D[\kappa^*(t),\kappa(t),\chi(t)] \exp\!\Big[i\int dt\,  (L_{\kappa,\chi}+L_{J-\kappa})\Big],
\end{align}
where
\begin{align}
L_{\kappa,\chi}=&-\frac{1}{2}\llangle \kappa,\hat{\mathcal F}\dot\kappa\rrangle-\frac{1}{2}\llangle \kappa,\hat{\mathcal G}\kappa\rrangle\notag\\
&-\llangle \chi,\kappa\rrangle+\frac{1}{2}\llangle \chi,\hat{\mathcal K}^{-1}_{\rm HS}\chi\rrangle +O((\chi,\kappa)^3)
\label{eq:kappa_chi_Lag}
\end{align}

The $\kappa-\chi$ theory above is still fully bosonic. Now we show that the $\kappa-\chi$ theory is equivalent to an effective fermion-$\chi$ theory up to the Gaussian order. Introducing complex Grassmann variables $\psi_i(t)$, we define the effective Lagrangian
\begin{align}
L_{\text{eff}}\equiv& \sum_{ij}\Big[\bar\psi_i (i\partial_t\delta_{ij}-t_{ij}(\bar\rho))\psi_j-\chi_{ij}(\bar\psi_i\psi_j-\bar\rho_{ji})\Big]\notag\\
&+\frac{1}{2}\llangle \chi,\hat{\mathcal K}^{-1}_{\rm HS}\chi\rrangle, \label{eq:eff_Lag}
\end{align}
and the corresponding source term and generating functional:
\begin{align}
L_{\psi-J}=& \sum_{\mathbf{ph}}[J^*_{\mathbf{ph}}(t)\bar\psi_\mathbf h(t)\psi_\mathbf p(t)+J_{\mathbf{ph}}(t)\bar\psi_\mathbf p(t)\psi_\mathbf h(t)]\notag\\
Z_{\text{eff}}[J]=&\int \mathcal D[\bar\psi(t),\psi(t),\chi(t)] \exp\!\Big[i\int dt\, (L_{\text{eff}}+L_{\psi-J} )\Big]. \label{eq:eff_PI}
\end{align}
The point is that, up to Gaussian order, the two generating functionals are identical $Z_{\rm eff}[J]=Z_{\kappa}[J]$. In addition, after integrating out $\kappa$ in $\kappa-\chi$ theory, the resulting $\chi$-only action is identical to that obtained by integrating out fermions in the effective theory (up to Gaussian order). For example, the correlator $\langle\mathcal T[\rho_{ij}(t)\rho_{kl}(t')]\rangle_{\text{VMPI}}$ using the path integral Eq.(\ref{eq:VMPI_def}) will be identical (up to Gaussian order) to the correlator $\langle \mathcal T[\bar\psi_j(t)\psi_i(t) \bar\psi_l(t')\psi_k(t')]\rangle_{\text{eff}}$ using the path integral Eq.(\ref{eq:eff_PI}). And the collective modes in the original $\kappa$-theory are reproduced in the effective theory. Namely, we have refermionized the VMPI action.

Finally, we explain the identification between the fermionic theory $L_{\text{eff}}$ in Eq.(\ref{eq:eff_Lag}) and the bosonic theory $L_{\kappa,\chi}$ in Eq.(\ref{eq:kappa_chi_Lag}) in an intuitive fashion, since we will use similar technique for the effective theories of Hdet wavefunctions soon. The only difference between them lies in the choice of formulation in the path integral. Consider a fixed $\chi(t)$ path. The fermion part of the path integral of $L_{\text{eff}}$ can be obtained by the single-body (time-dependent) kernel:
\begin{align}
\hat h_{\chi}(t)\equiv \sum_{ij} [t_{ij}(\bar\rho)+\chi_{ij}(t)]c_i^\dagger c_j.\label{eq:TD_ham}
\end{align}
If one inserts the resolution of identity of usual fermion coherent states into the time slices, $L_{\text{eff}}$ is obtained in terms of Grassmann variables $\bar\psi,\psi$. On the other hand, it can be shown that all Slater determinants labeled by $\rho_{ij}$ also form a resolution of identity for the $N$-electron Hilbert space:
\begin{align}
\mathbf 1_N=\int d\mu(\rho) |\Phi(\rho)\rangle\langle\Phi(\rho)|, \label{eq:Grassmannian_ID}
\end{align}
where the measure $\mu(\rho)$ (up to an unimportant overall constant) is exactly the Fubini–Study quantum metric in Eq.(\ref{eq:PI_measure}) with $f(\rho)=1$. Precisely speaking, the integration here is over the complex Grassmannian manifold $\text{Gr}(N,M)\equiv \frac{U(M)}{U(N)\times U(M-N)}$ where $M$ is the dimension of the single-particle Hilbert space $\mathcal H^{(e)}$. The Thouless parameterization is a local coordinate for this manifold. It would be more convenient if $\text{Gr}(N,M)$ has a global coordinate, but unfortunately, it does not. If one inserts Eq.(\ref{eq:Grassmannian_ID}) into the time slices of the path integral of Eq.(\ref{eq:TD_ham}), then $L_{\kappa,\chi}$ is obtained.

The resolution of identity Eq.(\ref{eq:Grassmannian_ID}), although being an elegant mathematical fact, is \emph{not} crucial to establish the equivalence of the two theories at the Gaussian order. As noted earlier, the details of the factor $f(\rho)$ in the integration measure do not affect the path integral at Gaussian order.

Here, we are dealing with a particularly simple case of Slater determinants: the Berry's phase term in $L_{\kappa,\chi}$ (the first term) Eq.(\ref{eq:kappa_chi_Lag}) can be computed exactly, which is just a different formulation of the fermion Berry's phase: $\sum_i\bar\psi_i(i\partial_t)\psi_i$. We will see that the situation for Hdet wavefunctions is more complicated.

\subsubsection{Continuous Symmetry, Ward Identity and Goldstone zero modes}
TDHF is known to be a conserving approximation, leading to gapless Goldstone modes when a continuous symmetry is spontaneously broken. Here we show the algebraic structure behind this statement. Soon, in our discussion of the effective theories of the Hdet wavefunctions, the similar algebraic structure will be present.

Consider a $U(1)_Q$ symmetry of the Hamiltonian $H$, generated by $\hat Q$
\begin{align}
\hat Q=\sum_{ij} Q_{ij}c_i^\dagger c_j,
\end{align}
in both the operator and matrix formulation:
\begin{align}
\hat D(\theta)\equiv& e^{i\theta \hat Q},& D(\theta)\equiv& e^{i\theta Q}.
\end{align}
Performing the unitary rotation for the state $|\Phi(\rho_Q(\theta))\rangle\equiv \hat D(\theta)|\Phi(\rho)\rangle$, one finds the $\rho_{ij}$ matrix transforms as 
\begin{align}
\rho\mapsto \rho_Q(\theta)= D(\theta)\rho D^\dagger(\theta),
\end{align}
Since the variational energy is symmetry invariant $E\big(\rho_Q(\theta)\big)=E(\rho)$, Eq.(\ref{eq:TDHF_eff_hopping}) leads to the covariance of $t(\rho)$
\begin{align}
t\big(\rho_Q(\theta)\big)=D(\theta)\,t(\rho)\,D^\dagger(\theta).
\label{eq:t_covariance}
\end{align}
This is the origin of the Ward identity in the TDHF analysis.

Suppose $\bar\rho$ is a saddle point spontaneously breaking the $U(1)_Q$-symmetry:
\begin{align}
\bar\rho_Q(\theta)\equiv D(\theta)\bar\rho D^\dagger(\theta)\neq \bar\rho,
\end{align}
then $\bar\rho_Q(\theta)$ is also a saddle point: 
\begin{align}
[t(\bar\rho_Q(\theta)),\bar\rho_Q(\theta)]=0.
\end{align}
For a small $\theta$, the symmetry-induced linear-order change $\bar\rho_Q(\theta)-\bar\rho$ is:
\begin{align}
(\delta\rho_Q)_{ij}\equiv\theta\cdot\left.\frac{d\bar\rho_Q(\theta)_{ij}}{d\theta}\right|_{\theta=0}=i\theta[Q,\bar\rho]_{ij}.
\label{eq:delta_rho_Q_realspace}
\end{align}
Similarly, we define the symmetry-induced linear-order variation of the effective hopping:
\begin{align}
(\chi_Q)_{ij}\equiv\theta\cdot \left.\frac{d\,t(\bar\rho_Q(\theta))_{ij}}{d\theta}\right|_{\theta=0}=i\theta[Q,t(\bar\rho)]_{ij}.
\label{eq:chi_Q_realspace}
\end{align}
Both $\delta\rho_Q,\chi_Q$  matrices are Hermitian.

Since $t(\rho)$ is the first derivative of $E(\rho)$, the variation of $t$, i.e., $\chi_Q$ is related to the second derivative of $E(\rho)$ -- the Hessian $\mathcal K$. Precisely:
\begin{align}
i\theta[Q,t(\bar\rho)]_{ij}=(\chi_Q)_{ij}=\sum_{kl}\mathcal K_{ij,kl}(\delta\rho_Q)_{lk}=[\hat{\mathcal K}\,\delta\rho_Q]_{ij}.
\label{eq:realspace_Ward}
\end{align}
Eq.(\ref{eq:realspace_Ward}) can be identified as the Ward identity in the real-space basis. If the generator $Q$ were \emph{not} a symmetry, although $[\hat{\mathcal K}\,\delta\rho_Q]_{ij}$ is still the linear-order change of the effective hopping for the rotated state $\hat D(\theta)|\Phi(\rho)\rangle$, it would not be directly related to $i\theta[Q,t(\bar\rho)]_{ij}$.

In the particle--hole ($\mathbf{ph}$) basis which diagonalizes both $\bar\rho$ and $t(\bar\rho)$, together with the Thouless parameterization, Eq.(\ref{eq:delta_rho_Q_realspace},\ref{eq:chi_Q_realspace}) lead to:
\begin{align}
\kappa_{Q,\mathbf{ph}}\equiv(\delta\rho_Q)_{\mathbf{ph}}=i\theta Q_{\mathbf{ph}},\;(\chi_Q)_{\mathbf{ph}}=-(\epsilon_{\mathbf p}-\epsilon_{\mathbf h})\kappa_{Q,\mathbf{ph}}.
\label{eq:kappa_chi_Q_ph_basis}
\end{align}
Note that, $\kappa_Q$ only has off-diagonal $\mathbf{ph},\mathbf{hp}$ components, while $\chi_Q$ generally also has $\mathbf{pp'},\mathbf{hh'}$ components. In other words, we know $\delta\rho_Q=\kappa_Q\in T_{\bar\rho}$ and $\chi_Q\in V_{\rm HS}$ (from Eq.(\ref{eq:realspace_Ward})), but $\chi_Q$ may not be in $T_{\bar\rho}$. The relation $\kappa_{Q,\mathbf{ph}}=i\theta Q_{\mathbf{ph}}$ leads to a useful identity for later discussion:
\begin{align}
e^{i\theta \hat Q}|\Phi(\bar\rho)\rangle=e^{i\theta\sum_{\mathbf h}Q_{\mathbf h\mathbf h}}\cdot e^{i\hat\Xi(\kappa_Q)}|\Phi(\bar\rho)\rangle+O(\theta^2).\label{eq:Q_rotated_vs_kappa_rotated}
\end{align}
Namely $e^{i\theta \hat Q}|\Phi(\bar\rho)\rangle$ and $e^{i\hat\Xi(\kappa_Q)}|\Phi(\bar\rho)\rangle$ only differ by a gauge choice at $\theta$-linear order.

We now introduce the projection superoperator $\hat{\mathcal P}_T$ to the $T_{\bar\rho}$ subspace, and similar to $\hat{\mathcal K}^{-1}_{\rm HS}\equiv\hat{\mathcal P}_{\rm HS} \hat {\mathcal K}^{-1}\hat{\mathcal P}_{\rm HS}$, we define $\hat {\mathcal G}^{-1}_{T}\equiv \hat{\mathcal P}_T \hat {\mathcal G}^{-1}\hat{\mathcal P}_T$. The above relations can then be compactly written as:
\begin{align}
\hat{\mathcal P}_T\chi_Q=&-\hat{\mathcal G}\kappa_Q, &\chi_Q=&\hat{\mathcal K}\kappa_Q,\notag\\
-\hat{\mathcal G}_T^{-1}\chi_Q=&\kappa_Q,&\hat{\mathcal K}_{\rm HS}^{-1}\chi_Q=&\hat{\mathcal P}_{\rm HS}\kappa_Q,\label{eq:kappa_Q_chi_Q_relations}
\end{align}
leading to the \emph{Ward identity} in the $\kappa$-language:
\begin{align}
(\hat{\mathcal G}+\hat{\mathcal P}_T\hat{\mathcal K})\kappa_Q=&0,& \llangle \kappa_Q,(\hat{\mathcal G}+\hat{\mathcal K})\kappa_Q\rrangle=&0,\label{eq:Ward_kappa}
\end{align}
and the corresponding one in the $\chi$-language:
\begin{align}
(\hat{\mathcal P}_{\rm HS}\hat{\mathcal G}^{-1}_T+\hat{\mathcal K}_{\rm HS}^{-1})\chi_Q=&0,&\llangle\chi_Q,(\hat{\mathcal G}^{-1}_T+\hat{\mathcal K}_{\rm HS}^{-1})\chi_Q\rrangle=&0.\label{eq:Ward_chi}
\end{align}

The superoperator $\hat{\mathcal G}^{-1}_T$ acting on $T_{\bar\rho}$:
\begin{align}
(\hat {\mathcal G}^{-1}_T X)_{\mathbf{ph}}\equiv &\frac{1}{\epsilon_\mathbf p-\epsilon_\mathbf h}X_{\mathbf{ph}},&(\hat {\mathcal G}^{-1}_T X)_{\mathbf{hp}}\equiv& \frac{1}{\epsilon_\mathbf p-\epsilon_\mathbf h}X_{\mathbf{hp}},\label{eq:G_inv_superoperator}
\end{align}
has a familiar form of the first-order perturbation theory and a simple physical interpretation. Consider a single-body Hamiltonian $\hat h_\chi \equiv \sum_{ij}[t_{ij}(\bar\rho)+\chi_{ij}]c_i^\dagger c_j$ and its ground state RDM $\rho(\chi)$. We know $\rho(\chi=0)=\bar\rho$. $\hat{\mathcal G}^{-1}_T$ is exactly describing the linear-order change: $\rho(\chi)-\bar\rho\equiv \delta\rho(\chi) =-\hat{\mathcal G}^{-1}_T\chi+O(\chi^2)$. Since $\delta\rho=\kappa$ to the linear order, $\kappa=-\hat{\mathcal G}^{-1}_T\chi$ is a static relation between the Thouless parameter of $\rho(\chi)$ and $\chi$.

To see how the Ward identity leads to Goldstone zero modes, we first work out the linearized EOM from the $\kappa$-theory $L_{\rm TDVP}$ in Eq.(\ref{eq:kappa_Lag_1}). In frequency space $\kappa(t)=\int \frac{d\omega}{2\pi}\kappa(\omega) e^{-i\omega t}$, the $\kappa$-EOM is 
\begin{align}
\Big[-i\omega\hat{\mathcal F}+\hat{\mathcal G}+\hat {\mathcal P}_T \hat{\mathcal K}\Big]\kappa(\omega)=0.\label{eq:kappa_eigen_eqn}
\end{align}
Therefore Eq.(\ref{eq:Ward_kappa}) dictates $\kappa_Q$ being $\omega=0$ mode. 

To find the $\chi$-EOM, we first integrate out $\kappa$ in $L_{\kappa,\chi}$ in Eq.(\ref{eq:kappa_chi_Lag}), giving rise to the $\chi$-action in the frequency space (up to Gaussian order):
\begin{align}
S_\chi=\frac{1}{2}\int\frac{d\omega}{2\pi}\chi(-\omega)\Big[\hat{\mathcal P}_{\rm HS}(-i\omega\hat{\mathcal F}+\hat{\mathcal G})_T^{-1}+\hat{\mathcal K}_{\rm HS}^{-1}\Big]\chi(\omega),\label{eq:chi_action}
\end{align}
where, similar to $\hat{\mathcal G}_T^{-1}$, the superoperator inverse $(\hat {\mathcal O})^{-1}_T$ for $\hat {\mathcal O}$ acting within $T_{\bar\rho}$ should be interpreted as $\hat{\mathcal P}_T\hat {\mathcal O}^{-1} \hat{\mathcal P}_T$. The linearized $\chi$-EOM in terms of eigen equation is then:
\begin{align}
\Big[\hat{\mathcal P}_{\rm HS}(-i\omega\hat{\mathcal F}+\hat{\mathcal G})_T^{-1}+\hat{\mathcal K}_{\rm HS}^{-1}\Big]\chi(\omega)=0,\label{eq:chi_eigen_eqn}
\end{align}
and Eq.(\ref{eq:Ward_chi}) dictates $\chi_Q$ being $\omega=0$ mode. 

For the purpose of collective modes, whether one wants to work with $\kappa$ or $\chi$ languages is a personal preference -- they are dual degrees of freedom capturing the same physics.  $\chi$ describes the Weiss-field fluctuations while $\kappa$ describes the corresponding density-matrix fluctuations. For example, one can check that every $\chi$-eigenmode satisfying Eq.(\ref{eq:chi_eigen_eqn}) can be mapped to a $\kappa$-eigenmode satisfying Eq.(\ref{eq:kappa_eigen_eqn}) with the same frequency via: 
\begin{align}
\chi\rightarrow\kappa: \kappa(\omega)=-(-i\omega\hat{\mathcal F}+\hat{\mathcal G})^{-1}_T\chi(\omega). \label{eq:kappa_chi_eigenmode_map}
\end{align}
For the purpose of later Hdet discussions, we are particularly interested in the effective theory in the form of $L_{\rm eff}$ in Eq.(\ref{eq:eff_Lag}), and will now focus on the $\chi$-language. Eq.(\ref{eq:kappa_chi_eigenmode_map}) is convenient since the zero-point motion and quantum excitation of the $\chi$-eigenmode have a concrete meaning in the VMPI wavefunction $\Psi_{\rm VMPI}(\kappa)$, which is written in the $\kappa$-language.

In our subsequent discussion of effective theories of Hdet wavefunctions, we will investigate the emergent gauge dynamics and demonstrate the $U(1)$ gauge fluctuations as a concrete example. Although the continuous symmetry and the gauge redundancy are fundamentally different physically, the gauge dynamics (e.g., the Maxwell equation, the Chern-Simons(CS)-term) emerge due to the Ward identity for the gauge redundancy, quite similar to how the Goldstone mode dynamics emerge due to the Ward identity for spontaneous symmetry breaking. The lessons learned here will prove very helpful.

\subsubsection{Type-I and Type-II Goldstone dispersions}
In the later Hdet discussion, we will encounter two kinds of $U(1)$ gauge dynamics in the long-wavelength limit: the gapless photon phases governed by the Maxwell term, and the topological phases governed by the CS term. The crucial difference between them is that the Maxwell term contains second time derivatives, while the CS term contains first time derivatives. 

Interestingly, in parallel, even in the current TDHF demonstration, there are also two kinds of long-wavelength Goldstone mode dispersions: $\omega\sim |\mathbf q|$ (Type-I) as in an antiferromagnet, and $\omega\sim \mathbf q^2$ (Type-II) as in a ferromagnet. As we will see, the physical origin for these two Goldstone-mode types is quite similar to the physical origin for Maxwell and CS gauge dynamics: the effective Lagrangian for type-I Goldstone modes only contains second time derivatives, while for type-II Goldstone modes it contains first time derivatives. Again, the lessons we learn here will prove useful when we microscopically compute the Maxwell and CS terms in the Hdet gauge dynamics.

Note that we are now focusing on the $\chi$-language to describe Goldstone modes. When $\chi$ was first introduced via the HS transformation, it did not have dynamics (i.e., time derivatives). The $\chi$-dynamics is generated upon integrating out $\kappa$ as in Eq.(\ref{eq:chi_action}). We will first present a general analysis of \emph{translationally invariant} systems, which is fully microscopic and can be used to compute the full collective-mode dispersions. We then focus on long-wavelength behavior to understand the origin of type-I vs. type-II Goldstone dispersions.

Quite generally, we assume that the effective hopping $t_{ij}(\bar\rho)$ is short-ranged. Introducing a double index: $i=(\mathbf R,\boldsymbol\upalpha)$, where $\mathbf R$ labels the (possibly enlarged) unit cell position, while $\boldsymbol\upalpha$ labels orbitals within the unit cell, we have
\begin{align}
t(\bar\rho)_{(\mathbf R,\boldsymbol\upalpha),(\mathbf R+\boldsymbol\delta,\boldsymbol\upbeta)}=t_{\boldsymbol\upalpha\boldsymbol\upbeta;\boldsymbol\delta}.\label{eq:t_alpha_beta_delta}
\end{align}

It is convenient to perform the lattice Fourier transformation to transform all Hermitian matrices in $M_{\rm h}$ to the momentum space. Recall that $M_{\rm h}$ is a real linear space. For the purpose of the Fourier transformation, we complexify $M_{\rm h}$ into $M_{\rm h}^{\mathbb C}$. We define a $\mathbf q$-space vector for $M_{\rm h}^{\mathbb C}$: $X_{\boldsymbol\upalpha\boldsymbol\upbeta;\boldsymbol\delta}(\mathbf q)$ based on its tensor value:
\begin{align}
X_{(\mathbf R,\boldsymbol\upalpha),(\mathbf R+\boldsymbol\delta,\boldsymbol\upbeta)}=\frac{1}{\sqrt {N_{\rm uc}}}\sum_{\mathbf q}e^{i\mathbf q\cdot\mathbf R}X_{\boldsymbol\upalpha\boldsymbol\upbeta;\boldsymbol\delta}(\mathbf q),\label{eq:lattice_Fourier}
\end{align}
where $N_{\rm uc}$ is the number of unit-cells and $\mathbf q\in$ Brillouin Zone (BZ). For each fixed $\mathbf q\in \rm BZ$, we denote the complex linear space spanned by all $X_{\boldsymbol\upalpha\boldsymbol\upbeta;\boldsymbol\delta}(\mathbf q)$ as $M_{\rm h}(\mathbf q)$. Thus:
\begin{align}
M_{\rm h}^{\mathbb C}=\bigoplus_\mathbf q M_{\rm h}(\mathbf q).
\end{align}
In this formalism, $t(\bar\rho)= t_{\boldsymbol\upalpha\boldsymbol\upbeta;\boldsymbol\delta}(\mathbf 0)\in M_{\rm h}(\mathbf 0)$.

The original inner product $\llangle X,Y\rrangle=\sum_{ij} X_{ij}Y_{ji}$ on $M_{\rm h}$ in Eq.(\ref{eq:M_h_inner_product}) can be extended as a \emph{bilinear form} to $M_{\rm h}^{\mathbb C}$ maintaining its form. One finds that:
\begin{align}
\llangle X,Y \rrangle=\sum_\mathbf q \llangle X(-\mathbf q),Y(\mathbf q)\rrangle,
\end{align}
where the $\mathbf q$-space bilinear form is
\begin{align}
\llangle X(-\mathbf q),Y(\mathbf q)\rrangle\equiv \sum_{\boldsymbol\upalpha\boldsymbol\upbeta;\boldsymbol\delta} e^{i\mathbf q\cdot \boldsymbol\delta}X_{\boldsymbol\upalpha\boldsymbol\upbeta;\boldsymbol\delta}(-\mathbf q)Y_{\boldsymbol\upbeta\boldsymbol\upalpha;-\boldsymbol\delta}(\mathbf q).\label{eq:q_mq_bilinear}
\end{align}
By momentum conservation, this bilinear form can be nonzero only for such a $-\mathbf q,\mathbf q$ pair.

In addition, $\forall X(\mathbf q)\in M_{\rm h}(\mathbf q)$, one can show its Hermitian conjugate $X(\mathbf q)^\dagger\in M_{\rm h}(-\mathbf q)$ is:
\begin{align}
   [X(\mathbf q)^\dagger]_{\boldsymbol\upalpha\boldsymbol\upbeta;\boldsymbol\delta}(-\mathbf q)=e^{-i\mathbf q\cdot\boldsymbol\delta}X^*_{\boldsymbol\upbeta\boldsymbol\upalpha;-\boldsymbol\delta}(\mathbf q)
\end{align}
The original real linear space $M_{\rm h}$ is formed by those $X\in M_{\rm h}^{\mathbb C}$ satisfying $X(-\mathbf q)=X(\mathbf q)^\dagger$.

Finally, the Hermitian conjugate operation allows one to define an inner product within $M_{\rm h}(\mathbf q)$ for a fixed $\mathbf q$:
\begin{align}
\langle X(\mathbf q),Y(\mathbf q)\rangle_{\mathbf q}\equiv \llangle X(\mathbf q)^\dagger,Y(\mathbf q) \rrangle=\sum_{\boldsymbol\upalpha\boldsymbol\upbeta;\boldsymbol\delta} X^*_{\boldsymbol\upalpha\boldsymbol\upbeta;
\boldsymbol\delta}(\mathbf q)Y_{\boldsymbol\upalpha\boldsymbol\upbeta;
\boldsymbol\delta}(\mathbf q).\label{eq:same_q_inner_product}
\end{align}

With these understandings for the $\mathbf q$-space algebraic structure, we now consider the important subspaces $T_{\bar\rho},V_{\rm HS}\subset M_{\rm h}$. After complexification we have $T^{\mathbb C}_{\bar\rho},V^{\mathbb C}_{\rm HS}\subset M_{\rm h}^{\mathbb C}$, which can be further decomposed into fixed-$\mathbf q$ subspaces:
\begin{align}
T^{\mathbb C}_{\bar\rho}=&\bigoplus_{\mathbf q}T_{\bar\rho}(\mathbf q),&T_{\bar\rho}(\mathbf q)\equiv&T^{\mathbb C}_{\bar\rho}\cap M_{\rm h}(\mathbf q),\notag\\
V^{\mathbb C}_{\rm HS}=&\bigoplus_{\mathbf q}V_{\rm HS}(\mathbf q),&V_{\rm HS}(\mathbf q)\equiv&V^{\mathbb C}_{\rm HS}\cap M_{\rm h}(\mathbf q).\label{eq:T_VHS_q_def}
\end{align}

The general $\chi$-field in action Eq.(\ref{eq:chi_action}), in momentum space, becomes $\chi(\mathbf q)\in V_{\rm HS}(\mathbf q)$. Due to translation symmetry, all superoperators defined previously act within fixed-$\mathbf q$ subspaces. For instance:
\begin{align}
(-i\omega\hat{\mathcal F}+\hat{\mathcal G})^{-1}_T(\mathbf q):&\;\;T_{\bar\rho}(\mathbf q)\rightarrow T_{\bar\rho}(\mathbf q),\notag\\
\hat{\mathcal P}_{\rm HS}(\mathbf q),\hat{\mathcal K}_{\rm HS}^{-1}(\mathbf q):&\;\;V_{\rm HS}(\mathbf q)\rightarrow V_{\rm HS}(\mathbf q).\label{eq:superoperators_q_space}
\end{align}
In fact, it is easy to show that, w.r.t. the inner product defined in Eq.(\ref{eq:same_q_inner_product}), the superoperators $\hat{\mathcal K}(\mathbf q),\hat{\mathcal K}_{\rm HS}^{-1}(\mathbf q),\hat{\mathcal G}(\mathbf q),\hat{\mathcal G}^{-1}_T(\mathbf q),\hat {\mathcal P}_T(\mathbf q),\hat {\mathcal P}_{\rm HS}(\mathbf q)$ are \emph{all Hermitian} in respective subspaces that they act on, while $\hat{\mathcal F}(\mathbf q)$ is anti-Hermitian.

One may solve the linearized EOM in $\mathbf q $-space to find general collective mode dispersions:
\begin{align}
&\hat{\mathcal D}(\mathbf q,\omega)\chi(\mathbf q,\omega)=0,\;\;\text{where }\notag\\
\hat{\mathcal D}(\mathbf q,\omega)\equiv& \hat{\mathcal P}_{\rm HS}(\mathbf q)(-i\omega\hat{\mathcal F}+\hat{\mathcal G})_T^{-1}(\mathbf q)+\hat{\mathcal K}_{\rm HS}^{-1}(\mathbf q),\label{eq:chi_q_kernel}
\end{align}
and $\hat{\mathcal D}(\mathbf q,\omega): V_{\rm HS}(\mathbf q)\rightarrow V_{\rm HS}(\mathbf q)$ is Hermitian.

Next, we will prepare for the long-wavelength analysis of the Goldstone modes. At this point, we need to make further assumptions for the real-space form of superoperator $\hat {\mathcal K}$. Quite generally, we assume that the Hessian $\mathcal K$ in Eq.(\ref{eq:real_space_Hessian}) admits a translation invariant form:
\begin{align}
&\mathcal K_{(\mathbf R,\boldsymbol\upalpha),(\mathbf R+\boldsymbol\delta,\boldsymbol\upbeta),(\mathbf R',\boldsymbol\upgamma),(\mathbf R'+\boldsymbol\delta',\boldsymbol\uplambda)}\notag\\
=&\sum_{A,B=1}^{N_w} w^A_{\boldsymbol\upalpha\boldsymbol\upbeta;\boldsymbol\delta}M_{AB}(\mathbf R-\mathbf R')w^B_{\boldsymbol\upgamma\boldsymbol\uplambda;\boldsymbol\delta'},\label{eq:K_M_form}
\end{align}
where $\{w^A_{\boldsymbol\upalpha\boldsymbol\upbeta;\boldsymbol\delta}\}$ ($A=1,2,..N_w$) labels a linearly-independent set of fixed Hermitian \textbf{bond tensors} satisfying $w^A_{\boldsymbol\upalpha\boldsymbol\upbeta;\boldsymbol\delta}=w^{A*}_{\boldsymbol\upbeta\boldsymbol\upalpha;-\boldsymbol\delta}$. In addition, we assume that $M_{AB}(\boldsymbol\Delta)$ is not too long-ranged so that its Fourier transform is a continuous function near $\mathbf q=0$. It is well-known that, for sufficiently long-ranged interactions, the would-be Goldstone-mode dispersion can become gapped near $\mathbf q=0$. Note that, the exact Goldstone zero-mode analysis in the previous section is still valid even in that case, however, due to the singularity of the momentum space kernel near $\mathbf q=0$, those zero modes cannot be lifted into dispersive modes. If the underlying interaction is truly short-ranged, then $M_{AB}(\boldsymbol\Delta)$ will have a finite support of $\boldsymbol\Delta$ and its Fourier transform is a smooth function; in addition, the number of bond tensors $N_w$ participating in $\mathcal K$ will be finite, constrained by a finite support of $\boldsymbol\delta$.

The Fourier transform of the Hermitian bond tensors leads to  $N_w$ vectors $\{w^A_{\boldsymbol\upalpha\boldsymbol\upbeta;\boldsymbol\delta}(\mathbf q)\}\in M_{\rm h}(\mathbf q)$ for every $\mathbf q$, whose tensor values $w^A_{\boldsymbol\upalpha\boldsymbol\upbeta;\boldsymbol\delta}$ are \emph{independent} of $\mathbf q$. The crucial consequence of the real-space form of $\mathcal K$ in Eq.(\ref{eq:K_M_form}) is that, after acting on an arbitrary $\mathbf q$-space basis vector $X_{\boldsymbol\upalpha\boldsymbol\upbeta;\boldsymbol\delta}(\mathbf q)$ as in Eq.(\ref{eq:lattice_Fourier}), it is obvious that $\hat {\mathcal K} X_{\boldsymbol\upalpha\boldsymbol\upbeta;\boldsymbol\delta}(\mathbf q)$ must be in the $N_w$-dimensional linear subspace $V_w(\mathbf q)$ spanned by $\{w^A_{\boldsymbol\upalpha\boldsymbol\upbeta;\boldsymbol\delta}(\mathbf q)\}$. Since $\hat {\mathcal K}(\mathbf q)$ is Hermitian w.r.t. inner product Eq.(\ref{eq:same_q_inner_product}), one may view $\hat{\mathcal K}(\mathbf q)$ as an operator $\hat{\mathcal K}(\mathbf q): V_w(\mathbf q)\rightarrow V_w(\mathbf q)$.

We define:
\begin{align}
K_{AB}(\mathbf q)\equiv \langle w^A(\mathbf q),\hat{\mathcal K}(\mathbf q)w^B(\mathbf q)\rangle_\mathbf q,\label{eq:K_full_rank}
\end{align}
which is automatically a Hermitian matrix of $A,B$. Finally, we \emph{assume} that the matrix $K_{AB}(\mathbf q)$ is full-rank (at least for small $\mathbf q$'s). This allows us to immediately identify $V_{\rm HS}(\mathbf q)\equiv{\rm Im}(\hat{\mathcal K}(\mathbf q))=V_w(\mathbf q)$, and $N_w=\rm{dim}[V_{\rm HS}(\mathbf q)]$.

Now, we consider the general situation where there may be more than one $U(1)$ symmetries are spontaneously broken. Denoting the number of linearly independent broken symmetry generators as $n_{\rm G}$, we label these generators by $Q_a$ ($a=1,2..n_{\rm G}$). Because both $Q_a$ and $t(\bar\rho)$ are translational invariant, their commutator as in Eq.(\ref{eq:chi_Q_realspace}) -- the Goldstone zero mode $\chi_{Q_a}$ must be translational invariant and in $ V_{\rm HS}(\mathbf 0)$. We denote these vectors as $u^a_{\boldsymbol\upalpha\boldsymbol\upbeta;\boldsymbol\delta}(\mathbf 0)$:
\begin{align}
\chi_{Q_a}=i\theta_a[Q_a,t(\bar\rho)]\equiv \theta_a \cdot \mathsf u^a_{\boldsymbol\upalpha\boldsymbol\upbeta;\boldsymbol\delta}(\mathbf 0)\in V_{\rm HS}(\mathbf 0).\label{eq:chi_a_u_a}
\end{align}
Consequently, the Hermitian bond tensors $\{\mathsf u^a_{\boldsymbol\upalpha\boldsymbol\upbeta;\boldsymbol\delta}\}$ $(a=1,2,..,n_{\rm G})$ are inside the real linear space spanned by $\{w^A_{\boldsymbol\upalpha\boldsymbol\upbeta;\boldsymbol\delta}\}$ ($A=1,2,..,N_w$). WLOG, we can  choose the tensors $\{w^A_{\boldsymbol\upalpha\boldsymbol\upbeta;\boldsymbol\delta}\}$ ($A=1,2,..,N_w$) so that $w^a_{\boldsymbol\upalpha\boldsymbol\upbeta;\boldsymbol\delta}= \mathsf u^a_{\boldsymbol\upalpha\boldsymbol\upbeta;\boldsymbol\delta}$ ($a=1,2,...,n_{\rm G}$). 

We denote the subspace spanned by $\{w^A_{\boldsymbol\upalpha\boldsymbol\upbeta;\boldsymbol\delta}(\mathbf q)\}$ ($A=1,2,..,n_{\rm G}$) as $V_{\rm G}(\mathbf q)$ (the Goldstone subspace), and the subspace spanned by $\{w^A_{\boldsymbol\upalpha\boldsymbol\upbeta;\boldsymbol\delta}(\mathbf q)\}$ ($A=n_{\rm G}+1,..,N_w$) as $V_{\perp}(\mathbf q)$. Consequently, $V_{\rm HS}(\mathbf q)=V_{\rm G}(\mathbf q)\oplus V_{\perp}(\mathbf q)$.

We can write down the $\chi$-action in Eq.(\ref{eq:chi_action}) in the momentum-frequency space:
\begin{align}
S_\chi=\int \frac{d\omega}{4\pi}\sum_\mathbf q \llangle \chi(-\mathbf q,-\omega),\hat {\mathcal D}(\mathbf q,\omega)\chi(\mathbf q,\omega)\rrangle,
\end{align}
where we used the bilinear form defined in Eq.(\ref{eq:q_mq_bilinear}).

Expanding the general $\chi$-field using the $\{w^A(\mathbf q)\}$ basis:
\begin{align}
\chi(\mathbf q,\omega)&=\sum_{A=1}^{n_{\rm G}}\phi_A(\mathbf q,\omega) w^A(\mathbf q)\notag\\
&+\sum_{B=1}^{N_w-n_{\rm G}}\eta_{B}(\mathbf q,\omega) w^{B+n_{\rm G}}(\mathbf q),\label{eq:chi_Goldstone_decomp}
\end{align}
we have, in concrete matrix form:
\begin{align}
S_\chi=&\int \frac{d\omega}{4\pi}\sum_\mathbf q \big(\phi(-\mathbf q,-\omega),\eta(-\mathbf q,-\omega)\big)\notag\\
&\cdot \begin{pmatrix} D_{\rm{GG}}(\mathbf q,\omega)&D_{\rm{G}\perp}(\mathbf q,\omega)\\D_{\perp\rm{G}}(\mathbf q,\omega)&D_{\perp\perp}(\mathbf q,\omega)\end{pmatrix}_{N_w\times N_w}\cdot\begin{pmatrix}\phi(\mathbf q,\omega)\\ \eta(\mathbf q,\omega)\end{pmatrix},\label{eq:phi_eta_action}
\end{align}
where $\rm G,\perp$ label components in $V_{\rm G},V_\perp$, and
\begin{align}
D_{AB}(\mathbf q,\omega)&\equiv\llangle w^A(-\mathbf q),\hat{\mathcal D}(\mathbf q,\omega) w^B(\mathbf q)\rrangle\notag\\
&=\langle w^A(\mathbf q),\hat{\mathcal D}(\mathbf q,\omega) w^B(\mathbf q)\rangle_\mathbf q.\label{eq:kernel_D_TDHF}
\end{align}
The last equality follows from the hermiticity of the tensors $\{w^A\}$. For simplicity, we further assume that the fermion $\mathbf{ph}$ spectrum is fully gapped, since the general coupling between Goldstone bosons with gapless fermion matter requires sophisticated analysis\cite{WatanabeVishwanath2014}. Under this assumption, $D_{AB}$ is a regular function of $(\mathbf q,\omega)$: e.g. it cannot have $|\mathbf q|$-linear dependence.  Moreover, $\chi$ being a Hermitian matrix and $\hat{\mathcal D}$ being a Hermitian superoperator dictates that:
\begin{align}
\phi_A(-\mathbf q,-\omega)=&\phi_A^*(\mathbf q,\omega),\;\;\eta_A(-\mathbf q,-\omega)=\eta_A^*(\mathbf q,\omega),\notag\\
D_{AB}(-\mathbf q,-\omega)=&D^*_{AB}(\mathbf q,\omega)=D_{BA}(\mathbf q,\omega)\label{eq:D_hermiticity_constraint}
\end{align}
Namely the full $D_{AB}(\mathbf q,\omega)$ matrix is Hermitian, together with the EOM:
\begin{align}
\begin{pmatrix} D_{\rm{GG}}(\mathbf q,\omega)&D_{\rm{G}\perp}(\mathbf q,\omega)\\D_{\perp\rm{G}}(\mathbf q,\omega)&D_{\perp\perp}(\mathbf q,\omega)\end{pmatrix}\cdot\begin{pmatrix}\phi(\mathbf q,\omega)\\ \eta(\mathbf q,\omega)\end{pmatrix}=0.
\end{align}

Now we are ready to perform the \emph{long-wavelength} analysis. The Ward identity Eq.(\ref{eq:Ward_chi}) can be written as:
\begin{align}
\hat {\mathcal D}(\mathbf 0,0) w^A(\mathbf 0)=&0,&A=&1,2,.., n_{\rm G}.
\end{align}
Since the $D_{AB}$ matrix is Hermitian, we know that:
\begin{align}
D_{\rm{GG}}(\mathbf 0,0)=&0,&D_{\rm{G}\perp}(\mathbf 0,0)=&0,&D_{\perp\rm{G}}(\mathbf 0,0)=&0.\label{eq:D_G_identity}
\end{align}
Quite generally, we assume that the orthogonal block is nonsingular (i.e. gapped) and denote it as $\Delta\equiv D_{\perp\perp}(\mathbf 0,0)$. 

For a small $\mathbf q$, the solution of EOM leads to:
\begin{align}
\eta(\mathbf q,\omega)&=-D_{\perp\perp}(\mathbf q,\omega)^{-1}D_{\perp \rm G}(\mathbf q,\omega)\phi(\mathbf q,\omega)\notag\\
&\sim O(|\mathbf q|,\omega)\phi(\mathbf q,\omega),
\end{align}
since $\Delta^{-1}$ is finite. Physically, it means that, to the leading order of the gradient expansion, the small-$\mathbf q$ Goldstone mode only contains $\phi$ components, and can be generated by a spatial-dependent (with the wavevector $\mathbf q$) symmetry rotation.

Upon integrating out the gapped $\eta$-modes, we arrive at the effective action in the Goldstone subspace $V_{\rm G}(\mathbf q)$:
\begin{align}
S_{\rm G}=\int \frac{d\omega}{4\pi}\sum_{\mathbf q}\phi(-\mathbf q,-\omega)\Pi(\mathbf q,\omega )\phi(\mathbf q,\omega),
\end{align}
where the reduced kernel is given by the Schur complement:
\begin{align}
\Pi(\mathbf q,\omega )\equiv D_{\rm{GG}}(\mathbf q,\omega)-D_{\rm G\perp}(\mathbf q,\omega )D^{-1}_{\perp\perp}(\mathbf q,\omega )D_{\perp\rm G}(\mathbf q,\omega )\label{eq:D_red}
\end{align}
is also a regular function. Following Eq.(\ref{eq:D_G_identity}), the reduced Ward identity is: 
\begin{align}
\Pi(\mathbf 0,0)=0.
\end{align}
One may parameterize its small-$(\mathbf q,\omega)$ expansion using constants $\mathsf B_{ab},\mathsf C_{ab},\mathsf G^{IJ}_{ab}$ ($a,b=1,..,n_{\rm G}$.):
\begin{align}
 \Pi_{ab}(\mathbf q,\omega )=&i\omega\cdot \mathsf B_{ab}+\omega^2\cdot \mathsf C_{ab}+\sum_{u,v}\mathbf q_u\mathbf q_v\cdot \mathsf G^{uv}_{ab}\notag\\
&+O((|\mathbf q|,\omega)^3),
\end{align}
where the $\mathbf q$-linear term is forbidden by the stability of variational energy minimum reached by $\bar\rho$, while $|\mathbf q|$-linear term is forbidden by regularity. (If the state $|\Phi(\bar\rho)\rangle$ carries background flows, e.g., supercurrent in a Bose-Hubbard model, a Doppler term $\sim \omega\cdot\mathbf q$ may be present. Here we assume it is not such a state.) From Eq.(\ref{eq:D_hermiticity_constraint}), we know that $\mathsf B$ is real antisymmetric, $\mathsf C$ and $\mathsf G$ are real symmetric (w.r.t. $a,b$). Generically $\mathsf C, \mathsf G \neq 0$. 

Type-I (type-II) Goldstone modes correspond to those $a,b$ components with $\mathsf B_{ab}= 0$ ($\mathsf B_{ab}\neq 0$), leading to dispersion $\omega\sim |\mathbf q|$ ($\omega\sim |\mathbf q|^2$). The key point is that, as dictated by Eq.(\ref{eq:D_red}), \emph{$\mathsf B$ is completely determined by $D_{\rm{GG}}(\mathbf q=0,\omega)$}: due to Eq.(\ref{eq:D_G_identity}), $D_{\rm G\perp},D_{\perp\rm G}$ both at most host $\omega$-linear term, and consequently the correction $D_{\rm G\perp}(\mathbf 0,\omega )D^{-1}_{\perp\perp}(\mathbf 0,\omega )D_{\perp\rm G}(\mathbf 0,\omega )\sim O(\omega^2)$. Based on the definition of the $\chi$-kernel $\hat{\mathcal D}$ in Eq.(\ref{eq:chi_q_kernel}), because 
\begin{align}
(-i\omega\hat{\mathcal F}+\hat{\mathcal G})^{-1}_T=\hat{\mathcal G}^{-1}_T+i\omega\hat{\mathcal G}^{-1}_T \hat{\mathcal F}\hat{\mathcal G}^{-1}_T+O(\omega^2),\label{eq:chi_kernel_expansion}
\end{align}
using Eq.(\ref{eq:chi_a_u_a}): $w^a(\mathbf 0)=\frac{1}{\theta_a}\chi_{Q_a}$, one finds:
\begin{align}
\mathsf B_{ab}=&\frac{1}{\theta_a\theta_b}\cdot\llangle \chi_{Q_a},\hat{\mathcal G}^{-1}_T\hat{\mathcal F}\hat{\mathcal G}^{-1}_T\chi_{Q_b}\rrangle=\frac{1}{\theta_a\theta_b}\cdot \llangle \kappa_{Q_a},\hat{\mathcal F}\kappa_{Q_b}\rrangle\notag\\
\end{align}
where we used the hermiticity of $\hat{\mathcal G}^{-1}_T$ and Eq.(\ref{eq:kappa_Q_chi_Q_relations}). 

$\mathsf B_{ab}$ is real and antisymmetric, directly related to the many-body Berry's curvature as in Eq.(\ref{eq:general_Berry_matrix}): using Eq.(\ref{eq:F_many_body_Berry},\ref{eq:Q_rotated_vs_kappa_rotated}), and the gauge invariance of Berry's curvature, 
\begin{align}
\mathsf B_{ab}=&\frac{2}{\theta_a\theta_b}\cdot\text{Im}\big[\langle e^{i\hat\Xi(\kappa_{Q_a})}\Phi(\bar\rho)|e^{i\hat\Xi(\kappa_{Q_b})}\Phi(\bar\rho)\rangle\big]\notag\\
=&\frac{2}{\theta_a\theta_b}\cdot\text{Im}\big[\langle e^{i\theta_a\hat Q_a}\Phi(\bar\rho)|e^{i\theta_b\hat Q_b}\Phi(\bar\rho)\rangle\big]_{\theta_a\theta_b-\text{part}}\notag\\
=&(-i)\langle \Phi(\bar\rho)|[\hat Q_a,\hat Q_b]|\Phi(\bar\rho)\rangle\notag\\
=&(-i)\text{Tr}(\bar\rho[ Q_a, Q_b]).\label{eq:B_ab_calculation}
\end{align}
For example, in a ferromagnetic state with moment along the $z$-direction with $\hat Q_a=\hat S_x$, $\hat Q_b=\hat S_y$, the total spins along $x,y$-directions, one finds $\mathsf B_{ab}\propto\langle \hat S_z\rangle\neq 0$. However, for an antiferromagnetic state, $\mathsf B_{ab}=0$.

\subsubsection{Zero-point motion of Goldstone modes in VMPI wavefunctions}
The advantage of VMPI over TDVP is the built-in quantum fluctuation of variational parameters. In the context of Goldstone mode fluctuations, their zero-point motion could qualitatively change the long-range behavior. For instance, consider a 1+1D single-band $t-J$ model at half-filling. For simplicity of presentation, let's assume $J$-term is easy-plane and ferromagnetic. When $|J|\gg t$, the static variational ground state $|\Phi(\bar\rho)\rangle$ will be ferromagnetic with a fully gapped electronic bandstructure. If we choose the ferromagnetic moment in $|\Phi(\bar\rho)\rangle$ along $\mathbf S_x$-direction, the Goldstone mode will be described by the fluctuation of the moment in $\mathbf S_y$-direction. Namely, in terms of the effective theory $L_{\rm eff}$ in Eq.(\ref{eq:eff_Lag}), the gapped fermions are polarized along $\mathbf S_x$-direction, but their $\mathbf S_y$-direction moment is coupled with the fluctuating bosonic field ($\chi_Q$-field) down to the lowest energy. 

Standard Gaussian-order field theory calculation for $L_{\rm eff}$ shows a power-law at long distances for the equal-time correlator $\langle \mathbf S^+(\mathbf R) \mathbf S^-(\mathbf 0)\rangle\sim |\mathbf R|^{-\eta}$,  due to the zero-point motion of the Goldstone mode with dispersion $\omega\propto |\mathbf q|$ -- the well-known Kosterlitz-Thouless physics. 

The point is that this physics is built into the VMPI treatment, both in terms of a microscopic field theory and \emph{microscopic wavefunctions}. Precisely, as a \emph{linear-superposition} of Slater determinants, the VMPI ground state $|\Psi\rangle_{\rm VMPI,GS}$ (see Eq.(\ref{eq:VMPI_GS_wavefunction} and replacing $\mathbf z\rightarrow\boldsymbol\kappa$) already includes the zero-point motion of the Goldstone modes: if one measures the spin-spin correlation function in $|\Psi\rangle_{\rm VMPI,GS}$ (\emph{not} $|\Phi(\bar\rho)\rangle$), the power-law order will be reproduced.

\subsection{Application of VMPI in Hdet wavefunctions}\label{sec:Hdet_eff}
In this subsection, applying the technique of VMPI, we study the effective theories of Hdet wavefunctions mainly focusing on type-(A) models, while commenting on type-(B) models in subsection \ref{sec:special_type_B}. For simplicity of presentation, here we again focus on the situation where \emph{there is only one electron's orbital per real-space site}, labeled by $\mathbf r$. Each parton-$p$'s mean-field state is labeled by its RDM $\rho^{(p)}$. These mean-field states are fused into an Hdet wavefunction via the local fusion operators $\hat {\mathbf F}_\mathbf r$ as in Eq.(\ref{eq:onsite_fusion_operator}) (or, equivalently, the local fusion gates $\hat {\mathbf C}_\mathbf r$ as in Eq.\ref{eq:onsite_fusion_gate}):

\begin{align}
\hat{\mathbf F}_{\mathbf r}\equiv\sum_{\alpha=1}^{R} \lambda_{\alpha}\cdot c^{\dagger}_{\mathbf r}\cdot \prod_{p=1}^{n_{\text{p}}}f^{(p)}_{\mathbf r,\mathbf m^{(p)}_\alpha},\label{eq:r_fusion_operator}
\end{align}
where, for simplicity, we assume that $\lambda_{\alpha}$ is independent of $\mathbf r$.

Throughout this subsection, we assume that these fusion operators are \emph{fixed}. Namely, all the CPD values $\lambda_{\alpha}$ in Eq.(\ref{eq:onsite_fusion_operator}) are \emph{not} treated as variational parameters. This is motivated by the observations for both type-(A) and type-(B) models. For type-(A) models, the electronic local states are orthonormal. By introducing enough parton ancilla orbitals, one can always make different fusion channels not share parton orbitals. It is then possible to absorb $\lambda_{\alpha}$ into one parton's mean-field state without modifying the Hdet many-body wavefunction, after which all $\lambda_{\alpha}$ can be set to unity. For type-(B) models, the overcompleteness of the Fine-Grid coherent states fixes the fusion amplitudes $\lambda_{\alpha}$ in the FQH and FCI examples (see Eq.(\ref{eq:boson_general_fusion},\ref{eq:boson_general_CFL_fusion},\ref{eq:fermion_general_fusion},\ref{eq:fermion_general_CFL_fusion}). 

\emph{In our presentation, we often focus on the case of two parton species $n_{\text{p}}=2$, i.e., rank-3 Hdet wavefunctions}. The generalization to higher rank Hdet wavefunctions is straightforward.

An $n_{\rm p}=2$ \emph{Hdet wavefunction $|\Phi(\rho)\rangle$} is then parameterized by two parton RDMs:
\begin{align}
\rho\equiv\{\rho^{(1)},\rho^{(2)}\}.
\end{align}
We will consider translational-invariant systems, possibly with an enlarged unit cell (e.g., due to magnetic translation). Consistent with the TDHF demonstration in the previous section, for each parton species-$p$, we denote the parton orbital by: 
\begin{align}
    i=(\mathbf r,\mathbf m)=({\mathbf R},\boldsymbol\upalpha),
\end{align}
where $\mathbf R$ labels the unit cell, and $\boldsymbol\upalpha$ collects all parton orbitals within the unit cell (i.e., different real-space sites $\mathbf r$ in one unit cell \emph{and} the orbital label $\mathbf m$). The real-space position of the orbital is denoted as $\mathbf r_i$ or ${\mathbf r}(\mathbf R,\boldsymbol\upalpha)$, and the local orbital index is denoted as $\mathbf m_i$. The parton RDM is then:
\begin{align}
\rho^{(p)}_{ij}\equiv \langle f^{(p)\dagger}_j f^{(p)}_i\rangle.
\end{align}
We warn the reader that, in general, the range of the index-$i$ may depend on the parton species-$p$, because in general, different parton species may have different numbers of orbitals per site. So technically, index-$i$ should be interpreted as $i^{(p)}$. However, to save notation, we neglect the superscript $^{(p)}$, which should not cause confusion.

Defining the parton-$p$' \emph{normalized} mean-field state $|\Psi^{(p)}(\rho^{(p)})\rangle$ to be the Slater-determinant specified by $\rho^{(p)}$, the full parton mean-field state is
\begin{align}
|\Psi^{MF}(\rho)\rangle\equiv |\Psi^{(1)}(\rho^{(1)})\rangle\otimes|\Psi^{(2)}(\rho^{(2)})\rangle,\label{eq:Hdet_MF_state}
\end{align}
and via fusion:
\begin{align}
|\Psi^{(e)}\{\Psi^{MF}\}\rangle\equiv&\hat{\mathbf C}|\Psi^{MF}\rangle,\text{ where: }\hat{\mathbf C}\equiv\prod_\mathbf r\hat{\mathbf C}_{\mathbf r},\notag\\
|\Phi\{\Psi^{MF}\}\rangle\equiv &\frac{|\Psi^{(e)}\{\Psi^{MF}\}\rangle}{\sqrt{\langle\Psi^{(e)}\{\Psi^{MF}\}|\Psi^{(e)}\{\Psi^{MF}\}\rangle}}.
\end{align}
Throughout this paper, $\Phi$ denotes the normalized electronic state. $|\Phi\{\Psi^{MF}\}\rangle$ should be viewed as the map from the parton's  mean-field state $\Psi^{MF}$ to the electronic normalized state $\Phi$. As a shorthand notation:
\begin{align}
|\Phi(\rho)\rangle\equiv |\Phi\{\Psi^{MF}(\rho)\}\rangle.
\end{align}

\emph{Crucially}, the mean-field states $\{|\Psi^{MF}(\rho)\rangle\}$, as direct products of Slater determinants, form a manifold satisfying the K\"{a}hler condition -- the parton's Thouless parameterization, as mentioned near Eq.(\ref{eq:Det_Kahler_coord}), is an explicit construction for its K\"{a}hler coordinate. Since the fusion operator is \emph{fixed, even after fusion, the Hdet states $\{|\Phi(\rho)\rangle\}$ still form a K\"{a}hler manifold. }

\subsubsection{Gauge covariant VMPI and the temporal gauge field $a_0$}
Formally, one can immediately repeat the VMPI analysis using Hdet state $|\Phi(\rho)\rangle$, similar to the former TDHF demonstration. However, there are \emph{two} challenges: (1) Due to the NP-hardness of Hdet calculation, generally, both the \emph{Berry's phase} term $\langle\Phi(\rho)|i\partial_t|\Phi(\rho)\rangle$ and $E(\rho)$ cannot be computed exactly. (2) The \emph{gauge redundancy} in the $\{\rho^{(1)},\rho^{(2)}\}$ parameterization must be treated, leading to fluctuating gauge fields.

The challenge-(1) is a technical one, yet extremely important. Fortunately, as we will see in Sec.\ref{sec:proj_expansion}, Berry's phase and $E(\rho)$ can still be approximately computed, order by order. In this subsection, to avoid confusion, we adapt the following notation: \emph{a subscript "$_{\rm approx}$" means the quantity is computed in a certain approximate scheme for Hdet states; a subscript "$_{\rm exact}$" means the quantity is computed using exact Hdet states; the absence of these subscripts means the quantity can be either the approximate one or the exact one.} 

Challenge-(2) captures the crucial physics and needs a thorough discussion. It also impacts our approximate scheme for Hdet calculations in Sec.\ref{sec:proj_expansion}.

As in all parton constructions, the parton parameterization for the physical wavefunction is redundant. In the present case, many different $\{\rho^{(1)},\rho^{(2)}\}$ may label the same physical Hdet wavefunction -- the key ingredient of gauge theories. This gauge redundancy is described by the gauge group $GG$, defined as follows. Let the number of parton orbitals on a given site be $n^{(p)}_{\rm orb}$ (labeled by $\mathbf m^{(p)}$), the unitary rotation among these parton orbitals forms the $U(n^{(p)}_{\rm orb})$ group. \emph{Within the Hdet construction}, $GG$ is defined to be the maximal subgroup of $\prod_p U(n^{(p)}_{\rm orb})$ leaving the fusion gate $\hat{\mathbf C}_{\mathbf r}$ invariant:
\begin{align}
GG \equiv \{g\in \prod_p U(n^{(p)}_{\rm orb}) |\hat{\mathbf C}_{\mathbf r} \hat U_\mathbf r(g)=&\hat{\mathbf C}_{\mathbf r}\}.\label{eq:GG_def}
\end{align}
Here $\hat U_\mathbf r(g)$ is the second-quantized operator acting on the parton Fock space on site-$\mathbf r$. For example, in the Hdet construction Eq.(\ref{eq:GG_UR_construction}), $GG=U(R)$.

Let $N_G$ be the dimension of $GG$, and denote its generators as Hermitian matrices $T^A$, $A=0,...,N_G-1$.  $T^A$ should be viewed as the direct sum of matrices in each parton's local orbital space: Let $T^{A,(p)}$ be the corresponding $n^{(p)}_{\rm orb}\times n^{(p)}_{\rm orb}$ matrix for parton-$p$ with elements $T^{A,(p)}_{\mathbf m,\mathbf m'}$, then $T^A=\oplus T^{A,(p)}$. $\{T^A\}$ form the Lie algebra $gg$ of $GG$, whose structure constant is defined as:
\begin{align}
[T^A,T^B]=&if^{AB}_C T^C.\label{eq:gg_def}
\end{align}

The site-dependent $GG$ gauge transformation is labeled by $\theta\equiv \{\theta^A(\mathbf r)\}$. We define matrices on the full parton real-space basis, as well as second-quantized operators:
\begin{align}
&[T_{\mathbf r}^{A,(p)}]_{ij}\equiv \delta_{\mathbf r,\mathbf r_i}\delta_{\mathbf r,\mathbf r_j}T_{\mathbf m_i,\mathbf m_j}^{A,(p)},\notag\\
&T_{\mathbf r}^{A}\equiv \oplus_p T_{\mathbf r}^{A,(p)},\;\;\hat T^A_{\mathbf r}\equiv \sum_{p,ij}[T_{\mathbf r}^{A,(p)}]_{ij}f_i^{(p)\dagger}f_j^{(p)}.
\end{align}
With this notation, the onsite gauge transformation in Eq.(\ref{eq:GG_def}) is parameterized as $\hat U_{\mathbf r}(g)=\exp \Big[i\sum_{A}\theta_{\mathbf r}^A\hat T^A_{\mathbf r}\Big]$. The site-dependent gauge transformation $\theta$ has both a matrix form and a second-quantized operator form:
\begin{align}
&\Lambda^{(p)}[\theta]\equiv \sum_{\mathbf r,A}\theta^A(\mathbf r)T^{A,(p)}_{\mathbf r},\; \Lambda[\theta]\equiv\oplus_p\Lambda^{(p)}[\theta],\;\Theta[\theta]\equiv e^{i\Lambda[\theta]}\notag\\
&\hat\Lambda[\theta]\equiv\sum_{p,ij}\Lambda^{(p)}_{ij}[\theta]f_i^{(p)\dagger}f_j^{(p)}, \;\hat\Theta[\theta]\equiv e^{i\hat\Lambda[\theta]}.\label{eq:Lambda_Theta_def}
\end{align}
The map $\Lambda:\{\theta^A(\mathbf r)\}\rightarrow M_{\rm h}$ is injective, so $\theta$ and $\Lambda[\theta]$ may be viewed as equivalent objects.

The gauge transformation on $\rho$ is denoted as:
\begin{align}
\rho^{(p)}\rightarrow \tilde\rho^{(p)}= \Theta^{(p)}[\theta]\rho^{(p)} \Theta^{(p)\dagger}[\theta]\equiv G(\theta)\circ\rho.
\label{eq:rho_gauge_covariance}
\end{align}

For the $n_{\rm p}=2$ case, one \emph{always} has a $U(1)_I\subset GG$, and it is convenient to denote its generator as $A=0$:
\begin{align}
T^{A=0,(p)}\equiv\text{q}^{(p)}\mathbf 1, \;\;&\text{where gauge charge: }\notag\\
&\text{q}^{(1)}=1,\text{q}^{(2)}=-1.
\end{align}
The special property for this $U(1)_I$ subgroup is that: the spatial independent $e^{i\theta T^{0}}$ always leaves $\rho$ invariant as in Eq.(\ref{eq:rho_gauge_covariance}), since $\Theta^{(p)}$ is an overall phase factor. 

In an \emph{exact} treatment of Hdet wavefunction, we have:
\begin{align}
|\Phi(G(\theta)\circ\rho)\rangle_{\rm exact}=e^{i\phi(\theta,\rho)}|\Phi(\rho)\rangle_{\rm exact}.\label{eq:exact_gauge_inv}
\end{align} 
Here, there is generally a nonzero phase $\phi(\theta,\rho)$. Its origin can be traced back to Eq.(\ref{eq:Hdet_MF_state}), where we choose a map from $\rho$ to the parton mean-field state $|\Psi^{MF}(\rho)\rangle$. However, this map is not unique: for example, it is perfectly fine to redefine $|\Psi^{MF}(\rho)\rangle\rightarrow e^{i\chi(\rho)}|\Psi^{MF}(\rho)\rangle$, and consequently the exact physical state transforms as $|\Phi(\rho)\rangle_{\rm exact}\rightarrow e^{i\chi(\rho)}|\Phi(\rho)\rangle_{\rm exact}$, and $\phi(\theta,\rho)$ transforms as: 
\begin{align}
\phi(\theta,\rho)\rightarrow \phi(\theta,\rho)+\chi(G(\theta)\circ\rho)-\chi(\rho).
\end{align}

For the purpose of explicit computation of the VMPI action shortly, it is convenient to introduce a related state:
\begin{align}
|\Phi(G(\theta);\rho)\rangle\equiv& |\Phi\big\{e^{i\hat\Lambda[\theta]}\Psi^{MF}(\rho)\big\}\rangle.\label{eq:Phi_G_rho_comma}
\end{align}
Clearly, the parton gauge transformation operator $e^{i\hat\Lambda[\theta]}$ transforms the parton RDM as Eq.(\ref{eq:rho_gauge_covariance}). Therefore, $|\Psi^{MF}(G(\theta)\circ\rho)\rangle$ and $e^{i\hat\Lambda[\theta]}|\Psi^{MF}(\rho)\rangle$ can only differ by an overall phase factor. In addition, due to the definition of $GG$, we have: $\prod_{\mathbf r}\hat{\mathbf C}_{\mathbf r}e^{i\hat\Lambda[\theta]}=\prod_{\mathbf r}\hat{\mathbf C}_{\mathbf r}$, and consequently:
\begin{align}
|\Phi(G(\theta);\rho)\rangle_{\rm exact}=|\Phi(\rho)\rangle_{\rm exact},\label{eq:exact_gauge_inv_1}
\end{align}
which is similar to Eq.(\ref{eq:exact_gauge_inv}), but without the phase factor $e^{i\phi(\theta,\rho)}$, dictating:
\begin{align}
|\Psi^{MF}(G(\theta)\circ\rho)\rangle=e^{i\phi(\theta,\rho)}e^{i\hat\Lambda[\theta]}|\Psi^{MF}(\rho)\rangle.\label{eq:phi_theta_rho}
\end{align}

Gauge invariance will play a crucial role in our discussion. In the exact treatment, gauge invariance is already captured by Eqs. (\ref{eq:exact_gauge_inv}) and (\ref{eq:exact_gauge_inv_1}). However, because exactly simulating Hdet states is NP-hard, practical approximation schemes are necessary for large system sizes. For the purpose of VMPI, two types of quantities are needed: (1) The static properties like \begin{align}
E(\rho)\equiv\langle \Phi(\rho)| H|\Phi(\rho)\rangle,
\end{align}
and (2) the dynamical properties like the Berry's phase from the overlap $\langle \Phi(\rho+\delta\rho)|\Phi(\rho)\rangle$. An approximation scheme must satisfy some fundamental conditions for the approximate VMPI to be physically reasonable, which we define below.

We define an approximation scheme for \emph{computing the electronic observables in Hdet states} to be \emph{statically conserving} iff, $\forall$ electronic observable $O^{(e)}$, parton mean-field state $|\Psi^{MF}\rangle$, phase factor $e^{i\alpha}$, and gauge transformation of $\theta$, it satisfies:
\begin{align}
&\langle\Phi\{e^{i\alpha}e^{i\hat\Lambda[\theta]}\Psi^{MF}\}|O^{(e)}|\Phi\{e^{i\alpha}e^{i\hat\Lambda[\theta]}\Psi^{MF}\}\rangle_{\rm approx}\notag\\
=&\langle\Phi\{\Psi^{MF}\}|O^{(e)}|\Phi\{\Psi^{MF}\}\rangle_{\rm approx}.\label{eq:stat_conserving_def}
\end{align}
We also define an approximation scheme for \emph{computing overlaps of electronic Hdet states} to be \emph{dynamically conserving} iff, $\forall$ parton mean-field state $|\Psi^{MF}_i\rangle,|\Psi^{MF}_f\rangle$, phase factors $e^{i\alpha_i},e^{i\alpha_f}$ and gauge transformation of $\theta$, it satisfies a gauge-invariance condition:
\begin{align}
&\langle\Phi\{e^{i\alpha_f}e^{i\hat\Lambda[\theta]}\Psi^{MF}_f\}|\Phi\{e^{i\alpha_i}e^{i\hat\Lambda[\theta]}\Psi_i^{MF}\}\rangle_{\rm approx}\notag\\
=&e^{i(\alpha_i-\alpha_f)}\langle\Phi\{\Psi_f^{MF}\}|\Phi\{\Psi_i^{MF}\}\rangle_{\rm approx},\label{eq:dyna_conserving_def}
\end{align}
as well as the elementary Hermiticity property:
\begin{align}
\langle\Phi\{\Psi^{MF}_f\}|\Phi\{\Psi_i^{MF}\}\rangle_{\rm approx}=\langle\Phi\{\Psi^{MF}_i\}|\Phi\{\Psi_f^{MF}\}\rangle^*_{\rm approx}\label{eq:Hermiticity_conserving_def}
\end{align}
\emph{In this paper, we ONLY consider approximation schemes that are both statically-conserving and dynamically-conserving}. For example, the projective-expansion(PE) introduced in Sec.\ref{sec:proj_expansion} satisfies both conditions at any order. 

The static-conserving condition is very natural, leading to the gauge invariance of the energy function: $E(\rho)=E(G(\theta)\circ\rho)$. On the other hand, compared to Eq.(\ref{eq:exact_gauge_inv_1}), the dynamical-conserving condition Eq.(\ref{eq:dyna_conserving_def}) is a much weaker requirement, in which the gauge transformation appears in both the ket and the bra -- it is a double-sided gauge-invariance condition. In an exact treatment, we would have a single-sided gauge-invariance: $\langle \Phi(\rho_f)|\Phi(G(\theta);\rho_i)\rangle_{\rm exact}=\langle \Phi(\rho_f)|\Phi(\rho_i)\rangle_{\rm exact}$ -- a consequence of the exact implementation of Gauss's law. It is unrealistic to expect the approximate overlap to also be single-sided gauge invariant, which would have required the approximate Hdet state to have Gauss' law completely implemented.

More generally, there may be even larger gauge redundancies not captured by $\prod_p U(n_{\rm orb}^{(p)})$ when constructing states \emph{beyond the present Hdet formulation}. For instance, consider the \emph{special case} that the parton orbitals for different fusion channels are \emph{distinct}, i.e., $f^{(p)}_{\mathbf r,\mathbf m_\alpha}\neq f^{(p)}_{\mathbf r,\mathbf m_\beta}$ if $\alpha\neq \beta$. In this case, the orbital label $\mathbf m$ is not necessary and one may write $f^{(p)}_{\mathbf r,\alpha}\equiv f^{(p)}_{\mathbf r,\mathbf m_\alpha}$. The fusion operator Eq.(\ref{eq:r_fusion_operator}) becomes $\hat{\mathbf F}_{\mathbf r}\equiv\sum_{\alpha=1}^{R} \lambda_{\alpha}\cdot c^{\dagger}_{\mathbf r}\cdot \prod_{p=1}^{n_{\text{p}}}f^{(p)}_{\mathbf r,\alpha}$. If one allows different parton species to hybridize yet using this $\hat{\mathbf F}_{\mathbf r}$ to construct the physical wavefunction, one would have at least $SU(n_{\text{p}})$ gauge redundancy, where the $W\in SU(n_{\text{p}})$ gauge transformation is $f^{(p)}_{\mathbf r,\alpha}\rightarrow \sum_{p'}W(\mathbf r)_{pp'}f^{(p')}_{\mathbf r,\alpha}$. Obviously, the fusion operator $\hat {\mathbf F}_\mathbf r$ in this special case is an $SU(n_{\text{p}})$ singlet. The previous $U(1)_I\subset GG$ redundancy is nothing but the diagonal subgroup of $SU(n_{\text{p}}=2)$.  Therefore, if we use $\rho^{(\text{all }p)}$ to denote the single RDM of the hybridized parton mean-field state, then different $\rho^{(\text{all }p)}$'s related by $SU(n_{\text{p}})$ gauge transformations are labeling the identical physical wavefunction. Here, the hybridization of parton species leads to the hyperpfaffian-type mathematical structure\cite{Barvinok1995, LuqueThibon2002} for the physical wavefunction after fusion, beyond the current Hdet discussion, and we leave its detailed investigation as a future project. For the moment, we simply call such physical states constructed from species-hybridized mean-field states $\rho^{(\text{all } p)}$ via local fusion $\hat {\mathbf F}_\mathbf r$ Hpfaffian states. \emph{Generally, as long as the fusion operator $\hat{\mathbf F}_{\mathbf r}$ (or equivalently the fusion gate $\hat{\mathbf C}_{\mathbf r}$) is $SU(n_{\text{p}})$ invariant, the corresponding Hpfaffian construction has the $SU(n_{\text{p}})$ gauge redundancy.}

No matter what the gauge group of the fusion is, the gauge group for the low-energy fluctuations, however, is really determined by the saddle point $\bar\rho$. This low-energy gauge group is denoted as $IGG$ by Wen and is the remaining gauge group not Higgs-ed out by $\bar\rho$. Precisely speaking, this $IGG$ group is identified as the gauge transformations that even leave the parton mean-field state invariant. In the present Hdet construction, for $n_{\rm p}=2$, we already know $U(1)_I\subset IGG$. \emph{In this subsection, we will assume that $IGG=U(1)_I$}. (Conceptually, the $IGG=U(1)^{n_{\text{p}}-1}$-gauge VMPI presented here can be straightforwardly generalized to the situations $IGG=SU(n_{\text{p}})$ in a Hpfaffian construction.)

The space-time-dependent $GG$ gauge transformations, from the VMPI viewpoint, lead to fluctuations $\delta\rho_{\text{pg}
}$ from $\bar\rho$ that do not cost any energy-- the pure gauge modes. At long wavelengths, the $IGG$ part of the pure gauge modes will qualitatively affect the low-energy physics. One must consistently track these gauge modes in VMPI.

To keep track of such gauge modes, for every time slice-$s$ ($s=1,...,N-1.$) in the VMPI (see Eq.(\ref{eq:VMPI_time_slices})), we explicitly insert the pure gauge redundancies in the integral measure: 
\begin{align}
&\int d\mu(\rho_s)\equiv\int d\mu(\rho_{ij;s}^{(1)},\rho_{ij;s}^{(2)}) \rightarrow \notag\\
&\int \prod_\mathbf r d \eta_s(\mathbf r)\int d\mu(G(\eta_s)\circ\rho_s),\label{eq:a_0_insertion}
\end{align}
where $d \eta_s(\mathbf r)$ is the $GG$ invariant Haar measure, and modify the projector:
\begin{align}
|\Phi(\rho_s)\rangle\langle \Phi(\rho_s)|\rightarrow |\Phi(G(\eta_s)\circ\rho_s)\rangle\langle \Phi(G(\eta_s)\circ\rho_s)|.\label{eq:a_0_insertion_projector}
\end{align}
We \emph{require} that the measure is gauge invariant:
\begin{align}
d\mu(G(\eta)\circ\rho)=d\mu(\rho).
\end{align}

We now define the time-component of the $GG$ gauge field on the time-links.
\begin{align}
\Theta^{-1}[\eta_s]\Theta[\eta_{s+1}]=\exp \Big[i\Delta t\Lambda[a^{\text{full}}_{0;s}]+O(\Delta t^2)\Big],
\end{align}
where, to save notation later, we introduced the superscript "$^{\text{full}}$" since we will soon perform the saddle-point analysis $a^{\text{full}}_0=\bar a_0+a_0$. In the continuum-$t$ limit:
\begin{align}
\Lambda[a^{\text{full}}_{0}(t)]\equiv -i\Theta^{-1}[\eta(t)]\partial_t\Theta[\eta(t)]=\sum_{\mathbf r,A}a_0^{\text{full},A}(\mathbf r,t)T^A_{\mathbf r},\label{eq:GG_a0_def}
\end{align}
where $a_0^{\text{full},A}(\mathbf r,t)$ are real fields since $\Lambda[a^{\text{full}}_{0}(t)]$ is Hermitian.

The path integral $\int \mathcal D[\eta]$ can be replaced by $\int \mathcal D[a^{\text{full}}_0]$ after setting boundary condition $\eta(\mathbf r,-\frac{T}{2})=0$. We have
\begin{align}
Z_{\text{VMPI}}=\int \mathcal D[\rho(t),a^{\text{full}}_0(t)] e^{i\int dt L_{\text{TDVP}}},\label{eq:Hdet_Z_TDVP}
\end{align}
with
\begin{align}
L_{\text{TDVP}}=&\langle\Phi(\rho(t),a^{\text{full}}_0(t))|i\partial_t-H|\Phi(\rho(t),a^{\text{full}}_0(t))\rangle\notag\\
=&L_{\text{Berry}}-E(\rho(t)),\label{eq:Hdet_L_TDVP_def}
\end{align}
where the gauge-covariant state is defined by
\begin{align}
|\Phi(\rho(t),a^{\text{full}}_0(t))\rangle\equiv |\Phi\Big[G\big(\eta(t)\big)\circ\rho(t)\Big]\rangle,\label{eq:Hdet_Phi_rho_a_0}
\end{align}
where $\eta(t)$ and $a^{\text{full}}_0(t)$ are related via Eq.(\ref{eq:GG_a0_def}), and the Berry's phase term is:
\begin{align}
L_{\text{Berry}}\equiv\langle\Phi(\rho(t),a^{\text{full}}_0(t))|i\partial_t|\Phi(\rho(t),a^{\text{full}}_0(t))\rangle.\label{eq:gauge_Berry}
\end{align}

Under a space-time dependent gauge transformation $\theta(t)$ satisfying the boundary condition $\theta(\mathbf r,-\frac{T}{2})=0$, we have:
\begin{align}
\Theta[\eta(t)]\rightarrow& \Theta[\eta(t)]\Theta^{-1}[\theta(t)],&\rho(t)\rightarrow& G(\theta(t))\circ\rho(t),
\end{align}
and the states on the RHS of Eq.(\ref{eq:a_0_insertion_projector}) are explicitly gauge invariant: $|\Phi([G(\eta_s)\circ G^{-1}(\theta_s)]\circ G(\theta_s)\circ \rho_s)\rangle=|\Phi(G(\eta_s)\circ\rho_s)\rangle$. Consequently, $L_{\rm Berry}$ is \emph{gauge invariant} by construction. Since $E(\rho(t))$ is already gauge invariant due to the static-conserving condition, $L_{\rm TDVP}$ is gauge invariant. In the continuum-time limit, we have the gauge transformation labeled by $\{\theta(\mathbf r,t)\}$:
\begin{align}
\Lambda[a^{\text{full}}_{0}]&\rightarrow \Theta[\theta]\Lambda[a^{\text{full}}_{0}]\Theta^{-1}[\theta]+i(\partial_t\Theta[\theta])\Theta^{-1}[\theta],\notag\\
 \rho(t)&\rightarrow G(\theta(t))\circ\rho(t).\label{eq:gauge_transform_a_0_rho}
\end{align}
If one would like to explicitly write down the transformation of each real field $a_0^{\text{full},A}$, we can define the adjoint-representation matrix $R^B_A[\theta]$:
\begin{align}
\Theta[\theta]T^A_{\mathbf r}\Theta^{-1}[\theta]\equiv \sum_B R^B_A[\theta(\mathbf r)] T_{\mathbf r}^B,
\end{align}
and expand:
\begin{align}
i(\partial_t\Theta[\theta])\Theta^{-1}[\theta]=\sum_{\mathbf r,B} \zeta^B(\theta(\mathbf r,t),\dot\theta(\mathbf r,t)) T_{\mathbf r}^B.
\end{align}
Then
\begin{align}
a_0^{\text{full},B}(\mathbf r,t)\rightarrow &\sum_A R_A^B[\theta(\mathbf r,t)]a_0^{\text{full},A}(\mathbf r,t)\notag\\
&+\zeta^B(\theta(\mathbf r,t),\dot\theta(\mathbf r,t)).\label{eq:a_0_gauge_transform}
\end{align}
For an infinitesimal gauge transformation $\theta\ll 1$:
\begin{align}
\Theta[\theta]=1+i\Lambda[\theta]+O(\theta^2), 
\end{align}
Eq.(\ref{eq:gauge_transform_a_0_rho}) gives:
\begin{align}
\Lambda[a_0^{\text{full}}]\rightarrow \Lambda[a_0^{\text{full}}]-\partial_t\Lambda[\theta]+i[\Lambda[\theta],\Lambda[a_0^{\text{full}}]]+O(\theta^2).
\end{align}
Using Eq.(\ref{eq:gg_def}), the individual components transform as:
\begin{align}
a_0^{\text{full},C}(\mathbf r,t)&\rightarrow a_0^{\text{full},C}(\mathbf r,t)-\partial_t\theta^C(\mathbf r,t)\notag\\
&-\sum_{AB}f^{AB}_C\theta^A(\mathbf r,t)a_0^{\text{full},B}(\mathbf r,t)+O(\theta^2).\label{eq:a_0_gauge_transform_small_theta}
\end{align}

Although the projector on the RHS of Eq.(\ref{eq:a_0_insertion_projector}) is conceptually elegant for demonstrating the gauge invariance of $L_{\rm Berry}$, it is not convenient for its practical computation due to the unspecified phase $\phi(\theta,\rho)$. To practically compute $L_{\rm Berry}$, we replace the projector as:
\begin{align}
&|\Phi(G(\eta_s)\circ\rho_s)\rangle\langle \Phi(G(\eta_s)\circ\rho_s)|\notag\\
\rightarrow& |\Phi(G(\eta_s);\rho_s)\rangle\langle \Phi(G(\eta_s);\rho_s)|,\label{eq:projector_replace}
\end{align}
where the state introduced in Eq.(\ref{eq:Phi_G_rho_comma}) is used. Based on the dynamical-conserving condition Eq.(\ref{eq:dyna_conserving_def}),  $S_{\rm Berry}\equiv \int dt L_{\rm Berry}$ remains unchanged under the replacement. 

Precisely, because $|\Psi^{MF}(G(\eta_s)\circ\rho_s)\rangle$ and $e^{i\hat\Lambda[\eta_s]}|\Psi^{MF}(\rho_s)\rangle$ only differ by an overall phase factor $e^{i\alpha_s}$, setting $\theta=0$ in Eq.(\ref{eq:dyna_conserving_def}):
\begin{align}
&\langle \Phi(G(\eta_{s+1});\rho_{s+1})|\Phi(G(\eta_s);\rho_s) \rangle\notag\\
=&e^{i(\alpha_{s+1}-\alpha_s)}\langle \Phi(G(\eta_{s+1})\circ\rho_{s+1})|\Phi(G(\eta_s)\circ\rho_s) \rangle.
\end{align}
Taking the continuum-$t$ limit, $L_{\rm Berry}$ before and after the replacement Eq.(\ref{eq:projector_replace}) only differ by a total derivative $\frac{d\alpha}{dt}$, which does not change physics. However, after the replacement, $L_{\rm Berry}$ is gauge-invariant up to a total derivative (see below).

From now on, \emph{we will only use the projector after the replacement Eq.(\ref{eq:projector_replace}) to define and compute $L_{\rm Berry}$. Instead of introducing new notations, we redefine:}
\begin{align}
|\Phi(\rho(t),a^{\text{full}}_0(t))\rangle\equiv |\Phi\big[G(\eta(t));\rho(t)\big]\rangle,\label{eq:Hdet_Phi_rho_a_0_1}
\end{align}
which replaces Eq.(\ref{eq:Hdet_Phi_rho_a_0}), so that Eq.(\ref{eq:Hdet_Z_TDVP},\ref{eq:Hdet_L_TDVP_def},\ref{eq:gauge_Berry}) can be reused. Defining:
\begin{align}
Q^A_{\mathbf r}(\rho)\equiv i\left.\frac{d}{ds} \left\langle\Phi\Big\{e^{is\hat T^A_\mathbf r}\Psi^{MF}(\rho)\Big\}|\Phi\{\Psi^{MF}(\rho)\}\right\rangle\right|_{s=0},\label{eq:Gauss_Q_def}
\end{align}
$L_{\text{Berry}}$ can be separated into two terms using the dynamical-conserving condition Eq.(\ref{eq:dyna_conserving_def}):
\begin{align}
L_{\text{Berry}}=&\langle\Phi(\rho(t))|i\partial_t|\Phi(\rho(t))\rangle-\sum_{\mathbf r,A}a^{\text{full},A}_0(\mathbf r,t)Q_{\mathbf r}^A(\rho(t)).\label{eq:gauge_berry_two_term}
\end{align}
To appreciate the physical meaning of the $a^{\text{full}}_0$-term, it is instructive to consider the mean-field approximation, where we use the parton mean-field states to compute overlaps (which is equivalent to the zeroth-order PE in Sec.\ref{sec:proj_expansion}, denoted by the subscript "$_{[0]}$"):
\begin{align}
&\left.Q^A_{\mathbf r}(\rho)\right|_{[0]}=\langle \Psi^{MF}(\rho)|\hat T^A_\mathbf r|\Psi^{MF}(\rho)\rangle.\label{eq:mf_second_term}
\end{align}
This is just the gauge charge density on the mean-field level. Since there is only one term of $a^{\text{full}}_0$ in $L_{\rm TDVP}$, the $a^{\text{full}}_0$ saddle-condition is:
\begin{align}
\left.\frac{\delta L_{\rm Berry}}{\delta a^{\text{full}}_0}\right|_{\bar a_0,\bar\rho}=0\Rightarrow Q^A_{\mathbf r}(\bar\rho)=0,\forall \mathbf r,A,\label{eq:a_0_saddle}
\end{align}
which is the static Gauss's law constraint.

Generally, under a gauge transformation $\{\theta(\mathbf r,t)\}$, $a^{\text{full}}_0(t),\rho(t)$ transform as Eq.(\ref{eq:gauge_transform_a_0_rho}). Based on Eq.(\ref{eq:phi_theta_rho},\ref{eq:dyna_conserving_def}), $L_{\text{Berry}}$ transforms as:
\begin{align}
L_{\text{Berry}}\rightarrow L_{\text{Berry}}-\frac{d}{dt}\phi(\theta,\rho),\label{eq:L_berry_gauge_total_deriv}
\end{align}
i.e., $L_{\text{Berry}}$ is gauge invariant up to a total derivative.

Notice that throughout the discussion from Eq.(\ref{eq:a_0_insertion}) to Eq.(\ref{eq:gauge_berry_two_term}), no "$_{\rm exact}$" or "$_{\rm approx}$" subscript is present, indicating these generally hold for both \emph{exact and approximate} treatments of Hdet states. Careful readers may have already noted that, for an \emph{exact} treatment, $a^{\text{full}}_0$ actually does \emph{not} enter the Lagrangian! Indeed, if we substitute Eq.(\ref{eq:exact_gauge_inv_1}) in Eq.(\ref{eq:Gauss_Q_def}), $a^{\text{full}}_0$-contribution vanishes. It is actually unnecessary to introduce $a^{\text{full}}_0$ in the exact treatment since the path integrals $\int \mathcal D[\rho(t),a^{\text{full}}_0(t)]_{\rm exact}$ and $\int \mathcal D[\rho(t)]_{\rm exact}$ only differ by an overall constant.

In fact, there is no surprise here. In a gauge theory, the physical meaning of $a^{\text{full}}_0$ is to implement Gauss's law. Without $a^{\text{full}}_0$ the \emph{approximate} action will not be gauge invariant, and \emph{the sole purpose of introducing $a^{\text{full}}_0$ is to restore the gauge invariance}. After integrating out $a^{\text{full}}_0$, the states that participate in the path integral are all gauge-invariant. But in the exact treatment, Gauss's law is already implemented by the exact fusion $\hat{\mathbf C}$, which is qualitatively equivalent to the situation where $a^{\text{full}}_0$ is already integrated out in an approximate treatment. We will return to this point shortly when we perform the saddle point analysis for Hdet VMPI. As a \emph{preview}, it is helpful to note now that integrating out $a^{\text{full}}_0$ does not affect any physical gauge dynamics.

For example, we will show that the saddle-point analysis for Hdet VMPI leads to a general quadratic action for (2+1)-D $U(1)_I$ gauge dynamics:
\begin{align}
S^{\rm gauge}=\int\frac{d\omega}{4\pi}\sum_{\mathbf q,\mu,\nu} \mathbf a_\mu(-\mathbf q,-\omega) \Pi_{\mu\nu}(\mathbf q,\omega) \mathbf a_\nu(\mathbf q, \omega),\label{eq:general_gauge_action}
\end{align}
where $\mu,\nu=0,1,2$, $\mathbf a_0\equiv a_0^{{\text{full}},A=0}-\bar{a}_0^{A=0}$, and $S_{\rm gauge}$ is obtained after integrating out the matter fields characterized by $\rho(t)$. Under a gauge transformation labeled by $\{\theta^{A=0}(\mathbf r,t)\}$, $\mathbf a_0\rightarrow \mathbf a_0-\partial_t\theta^{A=0}$, and $\mathbf a_u\rightarrow \mathbf a_u-\partial_u\theta^{A=0}$, where we introduce $u,v=1,2$ to label the spatial components of the $U(1)_I$ gauge fields. In $(\mathbf q,\omega)$-space:
\begin{align}
\mathbf a_\mu(\mathbf q,\omega)\rightarrow \mathbf a_\mu(\mathbf q,\omega) - i q_\mu \theta^{A=0}(\mathbf q,\omega),\label{eq:standard_U1_gauge_transform}
\end{align}
where $q_\mu\equiv (-\omega,\mathbf q)$. The gauge invariance of $S_{\rm gauge}$ leads to the well-known gauge Ward identity:
\begin{align}
\sum_\mu q_\mu\Pi_{\mu\nu}(\mathbf q,\omega)=\sum_\nu\Pi_{\mu\nu}(\mathbf q,\omega)q_\nu=0. \label{eq:usual_gauge_ward}
\end{align}

The gauge Ward identity constrains the form of $\Pi$. Let's assume the matter fields that we integrated out are all gapped. In an \emph{approximate} Hdet treatment, such as the projective-expansion that we will introduce in Sec.\ref{sec:proj_expansion}, the gauge kernel $\Pi$ will consequently be a \emph{regular} function of $(\mathbf q,\omega)$. To the linear-order (quadratic-order) of $(\mathbf q,\omega)$, the allowed Chern-Simons term (Maxwell term) is given by $\Pi^{\rm CS}$ ($\Pi^{\rm Max}$):
\begin{align}
\Pi^{\rm CS}_{\mu\nu}=&\frac{-ik}{2\pi}\epsilon_{\mu\nu\lambda}q_\lambda,\notag\\
\Pi^{\rm Max}_{00}=&\sum_{uv}\chi_E^{uv}\mathbf q_u\mathbf q_v,\;\;\; \Pi^{\rm Max}_{0u}=\Pi^{\rm Max}_{u0}=\omega \sum_v\chi_E^{uv}\mathbf q_v,\notag\\
\Pi^{\rm Max}_{uv}=&\omega^2\chi_E^{uv}-\chi_B(\mathbf q^2\delta_{uv}-\mathbf q_u\mathbf q_v).\label{eq:CS_Max_Pi}
\end{align}
Below, \emph{for simplicity of presentation, we will assume the isotropic form} of $\chi_E^{uv}=\chi_E\cdot\delta_{uv}$. In real space, we recover the familiar gauge actions under the Einstein convention:
\begin{align}
S^{\rm CS}=&\frac{k}{4\pi}\int d\mathbf r dt \epsilon_{\mu\nu\lambda }\mathbf a_\mu\partial_\nu \mathbf a_\lambda , \notag\\
S^{\rm Max}=&\frac{1}{2}\int d\mathbf r dt\Big[\chi_E (\partial_t \mathbf a_u-\partial_u \mathbf a_0)^2-\chi_B(\epsilon_{uv}\partial_u \mathbf a_v)^2\Big].\label{eq:CS_Max_real_space}
\end{align}

Note that the gauge actions Eq.(\ref{eq:CS_Max_Pi},\ref{eq:CS_Max_real_space}) have explicit $\mathbf a_0$-dependence. Therefore, they should really come from an approximate Hdet treatment, and should be interpreted as $\Pi^{\rm CS}_{\rm approx},\Pi^{\rm Max}_{\rm approx}$ and $S^{\rm CS}_{\rm approx},S^{\rm Max}_{\rm approx}$.

In an \emph{exact} Hdet treatment, however, there should be no $\mathbf a_0$-dependence as previously mentioned, qualitatively corresponding to integrating out $\mathbf a_0$ in $S^{\rm gauge}$ in Eq.(\ref{eq:general_gauge_action}). We then obtain the spatial-component-only reduced gauge action:
\begin{align}
S^{\rm gauge}_{\rm reduced}=\int\frac{d\omega}{4\pi}\sum_{\mathbf q,u,v} \mathbf a_u(-\mathbf q,-\omega) \mathsf S_{uv}(\mathbf q,\omega) \mathbf a_v(\mathbf q, \omega),\label{eq:reduced_gauge_action}
\end{align}
where $u,v=1,2$ and
\begin{align}
\mathsf S_{uv}(\mathbf q\neq 0,\omega)\equiv \Pi_{uv}(\mathbf q,\omega)-\Pi_{u0}(\mathbf q,\omega)\Pi_{00}^{-1}(\mathbf q,\omega)\Pi_{0v}(\mathbf q,\omega).\label{eq:Pi_to_S}
\end{align}
It is easy to show that the usual gauge Ward identity Eq.(\ref{eq:usual_gauge_ward}) leads to the reduced gauge Ward identity:
\begin{align}
\sum_u \mathbf q_u\mathsf S_{uv}(\mathbf q,\omega)=\sum_v\mathsf S_{uv}(\mathbf q,\omega)\mathbf q_v=0,\label{eq:reduced_gauge_Ward}
\end{align}
which means that the longitudinal spatial component of $\mathbf a_u\propto \mathbf q_u$ does not enter the action. In fact, this longitudinal spatial component is a gauge redundancy, and only the transverse component $\mathbf a_u\perp \mathbf q_u$ is physical. 

For instance, for the pure Maxwell-kernel $\Pi^{\rm Max}$, we obtain the reduced kernel
\begin{align}
\mathsf S_{uv}^{\rm Max}(\mathbf q\neq 0,\omega)=(\chi_E\omega^2-\chi_B|\mathbf q|^2) \Big(\delta_{uv}-\frac{\mathbf q_u\mathbf q_v}{\mathbf q^2}\Big),\label{eq:reduced_Max_kernel}
\end{align}
leading to a linearly dispersive, transverse photon mode. On the other hand, if one starts with $S^{\rm CS+Max}=S^{\rm CS}+S^{\rm Max}$,  integrating out $\mathbf a_0$ leads to a reduced kernel
\begin{align}
\mathsf S_{uv}^{\rm CS+Max}(\mathbf q\neq 0,\omega)=&\Big(\chi_E\omega^2-\chi_B|\mathbf q|^2-\frac{1}{\chi_E}\big(\frac{k}{2\pi}\big)^2\Big)\notag\\
&\cdot \Big(\delta_{uv}-\frac{\mathbf q_u\mathbf q_v}{\mathbf q^2}\Big),\label{eq:reduced_CS_Max_kernel}
\end{align}
describing a massive transverse photon mode. 

The $\mathbf q=0$ gauge kernel must be treated separately: $\Pi^{\rm CS+Max}_{00}(\mathbf 0,\omega)=0$ vanishes and one should directly work with $\Pi$ without the Schur complement Eq.(\ref{eq:Pi_to_S}). In fact, $\mathbf a_0(\mathbf q=0,\omega)$ is not present in $S^{\rm CS+Max}$ due to gauge invariance. For spatial gauge field $\mathbf a_u$, the $\mathbf q=0$ modes correspond to $\mathbf r$-independent flat connections, i.e., fluxes $\varphi_x,\varphi_y$ through the torus holes:
\begin{align}
\mathbf a_x(\mathbf r,t)=&\frac{\varphi_x(t)}{L_x},&\mathbf a_y(\mathbf r,t)=&\frac{\varphi_y(t)}{L_y}. \label{eq:torus_flux_varphi}
\end{align}
One finds:
\begin{align}
&\mathsf S_{uv}^{\rm CS+Max}(\mathbf q=0,\omega)=\Pi_{uv}^{\rm CS+Max}(\mathbf q=0,\omega)\notag\\
=&i\omega\frac{k}{2\pi}\epsilon_{uv}+\chi_E\omega^2\delta_{uv},\label{eq:CS_Max_kernel_q_0}
\end{align}
and the corresponding Lagrangian:
\begin{align}
L(\mathbf q=0,t)=\frac{k}{4\pi}(\varphi_y\dot\varphi_x-\varphi_x\dot\varphi_y)+\frac{\chi_E}{2}(\frac{L_y}{L_x}\dot\varphi^2_x+\frac{L_x}{L_y}\dot\varphi^2_y).
\end{align}
When $k\neq 0$, we obtain $|k|$-fold ground state degeneracy on a torus from the VMPI of $L(\mathbf q=0)$\cite{Wen1989, WenNiu1990}. The pure CS limit $\mathsf S^{\rm CS}$ corresponds to $\chi_E=\chi_B=0$, leading to an infinite mass gap and only the ground state sector, i.e., a topological quantum field theory. Here, we see the convenience of VMPI: because of the path integral, the theory is intrinsically quantum-mechanical. 

In summary, in an \emph{exact} Hdet treatment, we expect to obtain the reduced gauge action $S^{\rm gauge}_{\rm reduced}$ in Eq.(\ref{eq:reduced_gauge_action}) and reduced, $\mathbf a_u$-only kernels consistent with $\mathsf S^{\rm Max}$ in Eq.(\ref{eq:reduced_Max_kernel}) or $\mathsf S^{\rm CS+Max}$ in Eq.(\ref{eq:reduced_CS_Max_kernel},\ref{eq:CS_Max_kernel_q_0}), which do not miss any physics compared with the $\Pi$-kernels in Eq.(\ref{eq:CS_Max_Pi}) with $\mathbf a_0$-dependence.

\subsubsection{Saddle point analysis and gauge Ward identity in the $\kappa$-language}
At this point, we introduce saddle-point analysis for general calculation schemes, either approximate or exact. Denoting $a_0^{\rm full}=\bar a_0+a_0$, one can always find the saddle point of $L_{\text{TDVP}}$ in the static-saddle gauge, where $\bar a_0$ and $\bar\rho$ are time-independent. Here, $\bar\rho$ also needs to satisfy the saddle equation of $a_0$ variation: $\left.\frac{\delta L}{\delta a_0}\right|_{\bar a_0,\bar\rho}=0$: Eq.(\ref{eq:a_0_saddle}). \emph{We will consider fluctuations around this saddle point $\{\bar a_0,\bar\rho\}$}. Due to the gauge invariance, the $\{\theta(\mathbf r,t)\}$ gauge transformed one describes the same physical saddle point: $\bar\rho\rightarrow \bar\rho_\theta(t)\equiv G(\theta(t))\circ\bar\rho$, and $\bar a_{0}(\mathbf r)\rightarrow \bar a_{0,\theta}(\mathbf r,t)$ which is obtained via Eq.(\ref{eq:a_0_gauge_transform}). These fluctuations are precisely the nonphysical pure gauge modes that should be singled out. Below, we will always work in the static-saddle gauge: $a_0^{\rm full}(\mathbf r,t)=\bar a_0(\mathbf r)+a_0(\mathbf r,t)$.

The $\bar a_0$-saddle equation is already given in Eq.(\ref{eq:a_0_saddle}). Next, we present the $\bar \rho$-saddle equation. Defining a static energy including the $\bar a_0$ contribution:
\begin{align}
E_{\rm st}(\rho;\bar a_0)\equiv E(\rho)+\sum_{\mathbf r,A}\bar a_0^A(\mathbf r)Q^A_{\mathbf r}(\rho),\label{eq:E_st_def}
\end{align}
the effective hopping and Hessian for each parton species $p=1,2$ can be defined:
\begin{align}
t^{(p)}_{ij}(\rho;\bar a_0)\equiv&\frac{\partial E_{\rm st}}{\partial \rho^{(p)}_{ji}},&\mathcal K^{(pq)}_{ij,kl}\equiv&\left.\frac{\partial^2 E_{\rm st}}{\partial\rho^{(p)}_{ji}\,\partial \rho^{(q)}_{lk}}\right|_{\bar\rho,\bar a_0}.\label{eq:Hdet_t_K}
\end{align}
Up to quadratic order:
\begin{align}
&E_{\rm st}(\bar\rho+\delta\rho;\bar a_0)=E_{\rm st}(\bar\rho;\bar a_0)+\sum_{p,ij}t^{(p)}_{ij}(\bar\rho;\bar a_0)\,\delta\rho^{(p)}_{ji}\notag\\
&+\frac{1}{2}\sum_{p,q}\sum_{ij,kl}\mathcal K^{(pq)}_{ij,kl}\,\delta\rho^{(p)}_{ji}\delta\rho^{(q)}_{lk}+O(\delta\rho^3).\label{eq:E_rho_component_wise}
\end{align}
Importantly, $\delta\rho^{(1)}$ and $\delta\rho^{(2)}$ are coupled via the inter-species Hessian components $\mathcal K^{(12)}$ and $\mathcal K^{(21)}$. We also define the effective parton single-body Hamiltonian:
\begin{align}
\hat h(\rho;\bar a_0)\equiv \sum_p\sum_{ij} t^{(p)}_{ij}(\rho;\bar a_0) f^{(p)\dagger}_i f^{(p)}_j\;.\label{eq:parton_eff_ham}
\end{align}
The $\bar\rho$ saddle condition is the vanishing condition for the linear $\delta\rho$-term in Eq.(\ref{eq:E_rho_component_wise}) for all tangent $\delta\rho$, which is equivalent to the self-consistency condition: $[t(\bar\rho;\bar a_0),\bar\rho]=0$, or the ground state RDM of $\hat h(\bar\rho;\bar a_0)$ is also $\bar\rho$.

Now it is a good moment to introduce the vector space of Hermitian matrices relevant for Hdet wavefunctions and the associated inner product. We first define the \emph{real} linear space $M^{(p)}_{\rm h}$ for each parton species-$p$ formed by Hermitian matrices of the same size as $\rho^{(p)}_{ij}$. Their direct sum defines the full \emph{real} linear space $M_{\rm h}$:
\begin{align}
M_{\rm h}\equiv M^{(1)}_{\rm h}\oplus M^{(2)}_{\rm h}.
\end{align}
An element $X\in M_{\rm h}$ can be represented as $X=\{X^{(1)},X^{(2)}\}$, where $X^{(p)}\in M^{(p)}_{\rm h}$. We define the inner product in $M_{\rm h}$:
\begin{align}
\llangle X,Y\rrangle\equiv \sum_{p=1}^2 \sum_{ij} X^{(p)}_{ij}Y^{(p)}_{ji}.\label{eq:Hdet_M_h_inner}
\end{align}

We also define the superoperator $\hat{\mathcal K}$ acting on $M_{\rm h}$:
\begin{align}
[\hat{\mathcal K} X]^{(p)}_{ij}\equiv \sum_{q=1}^2\sum_{kl}\mathcal K^{(pq)}_{ij,kl} X^{(q)}_{lk}.
\end{align}
Eq.(\ref{eq:E_rho_component_wise}) can be compactly written as:
\begin{align}
E_{\rm st}(\rho;\bar a_0)=&E_{\rm st}(\bar\rho;\bar a_0)+\llangle t(\bar\rho;\bar a_0),\delta\rho\rrangle\notag\\
&+\frac{1}{2}\llangle\delta\rho,\hat{\mathcal K}\delta\rho \rrangle+O(\delta\rho^3).
\end{align}
One can further show that $\hat{\mathcal K}$ is symmetric: $\forall X,Y\in M_{\rm h}$,
\begin{align}
    \llangle X, \hat{\mathcal K} Y\rrangle= \llangle \hat{\mathcal K} X, Y\rrangle,
\end{align}
which follows from identities $\mathcal K^{(pq)}_{ij,kl}=\mathcal K^{(qp)}_{kl,ij}$ and $\mathcal K^{(pq)}_{ij,kl}=\mathcal K^{(pq)*}_{ji,lk}$.

We now examine the gauge transformation of $E_{\rm st}$. Under a time-independent gauge transformation $\{\theta(\mathbf r)\}$: $\rho\rightarrow G(\theta)\circ\rho$, and static-conserving condition leads to $E(\rho)=E(G(\theta)\circ\rho)$. For the second term, from Eq.(\ref{eq:a_0_gauge_transform}), we know 
\begin{align}
\bar a_0^B(\mathbf r)\rightarrow \bar a^B_{0,\theta}(\mathbf r)\equiv\sum_A R_A^B(\theta(\mathbf r))\bar a_0^A(\mathbf r),\label{eq:bar_a_0_transform},
\end{align}
which is equivalent to $\Lambda[\bar a_0]\rightarrow \Theta[\theta]\Lambda[\bar a_0]\Theta^\dagger[\theta]$. One can show from the definition Eq.(\ref{eq:Gauss_Q_def}) and the dynamical-conserving condition:
\begin{align}
Q_{\mathbf r}^B(G(\theta)\circ \rho)=[R^{-1}]^A_B(\theta(\mathbf r)) Q^A_{\mathbf r}(\rho).\label{eq:Q_A_transform}
\end{align}
Therefore, $E_{\rm st}$ is gauge invariant:
\begin{align}
E_{\rm st}(\rho;\bar a_0)=E_{\rm st}(G(\theta)\circ\rho;\bar a_{0,\theta}),
\end{align}
and Eq.(\ref{eq:Hdet_t_K}) gives to the gauge covariance of the effective hopping,
\begin{align}
t^{(p)}\big(G(\theta)\circ\rho;\bar a_{0,\theta}\big)=\Theta^{(p)}\,t^{(p)}(\rho;\bar a_0)\,\Theta^{(p)\dagger}, 
\label{eq:Hdet_t_covariance}
\end{align}
which in turn leads to the gauge Ward identity. 

Next, we introduce the tangent space $T_{\bar\rho}\subset M_{\rm h}$ and the relevant superoperators, in parallel to the discussion near Eq.(\ref{eq:delta_rho_from_kappa}-\ref{eq:T_rho_superoperators}). Introducing the Thouless parameterization for each parton species:
\begin{align}
\hat\Xi^{(p)}(\kappa^{(p)})\equiv\sum_{\mathbf{ph}}\Big[-i\,\kappa^{(p)}_{\mathbf{ph}}f^{(p)\dagger}_{\mathbf p}f^{(p)}_{\mathbf h}+h.c.
\Big],\label{eq:Hdet_Thouless_param}
\end{align}
which transforms the parton mean-field state as $e^{i\hat\Xi^{(p)}(\kappa^{(p)})}|\Psi^{(p)}(\bar\rho^{(p)})\rangle=|\Psi^{(p)}(e^{i\hat\Xi^{(p)}(\kappa^{(p)})}\circ\bar\rho^{(p)})\rangle$.

Defining the tangent space $T^{(p)}_{\bar\rho}\subset M^{(p)}_{\rm h}$:
\begin{align}
T^{(p)}_{\bar\rho}\equiv &\{X^{(p)}|\;\bar\rho^{(p)} X^{(p)}\bar\rho^{(p)}=0,\notag\\
&(1-\bar\rho^{(p)})X^{(p)}(1-\bar\rho^{(p)})=0\}\subset M^{(p)}_{\rm h},
\end{align}
$\kappa^{(p)}$ is an explicit coordinate for $T^{(p)}_{\bar\rho}$. We define the full tangent space:
\begin{align}
T_{\bar\rho}\equiv T_{\bar\rho}^{(1)}\oplus T_{\bar\rho}^{(2)}\subset M_{\rm h},
\end{align}
and an element $\kappa\in T_{\bar\rho}$ can be written as:
\begin{align}
\kappa=\{\kappa^{(1)},\kappa^{(2)}\}.
\end{align}
We also define $\hat{\mathcal P}_T$ to be the projector into $T_{\bar\rho}$.

One may parameterize the Hdet state near the saddle point by $\kappa$ as:
\begin{align}
|\Phi(\kappa)\rangle\equiv |\Phi(\rho(\kappa))\rangle\equiv |\Phi(e^{i\Xi(\kappa)}\circ\bar \rho)\rangle,
\end{align}
where 
\begin{align}
&\Big(e^{i\Xi(\kappa)}\circ\bar \rho\Big)^{(p)}\equiv e^{i\Xi^{(p)}(\kappa^{(p)})}\circ\bar\rho^{(p)}\notag\\
=&e^{i\Xi^{(p)}(\kappa^{(p)})}\bar\rho^{(p)}e^{-i\Xi^{(p)}(\kappa^{(p)})}.
\end{align}

Using the parton version of Eq.(\ref{eq:delta_rho_from_kappa}), one can write the Gaussian order of $E_{\rm st}(\rho(\kappa);\bar a_0)-E_{\rm st}(\bar\rho;\bar a_0)\equiv \delta E_{\rm st}(\kappa)$ as
\begin{align}
\delta E_{\rm st}(\kappa)=\frac{1}{2}\llangle \kappa,(\hat{\mathcal G}+\hat{\mathcal K}) \kappa\rrangle+O(\kappa^3),
\end{align}
where the symmetric superoperator $\hat{\mathcal G}$ acting in $T_{\bar\rho}$ is defined as: $\forall X\in T_{\bar\rho}$,
\begin{align}
(\hat {\mathcal G} X)^{(p)}_{\mathbf{ph}}\equiv&(\epsilon^{(p)}_\mathbf p-\epsilon^{(p)}_\mathbf h)X^{(p)}_{\mathbf{ph}},&(\hat {\mathcal G} X)^{(p)}_{\mathbf{hp}}\equiv&(\epsilon^{(p)}_\mathbf p-\epsilon^{(p)}_\mathbf h)X^{(p)}_{\mathbf{hp}}.\notag\\
\end{align}

Now, we consider a small gauge transformation $\theta\ll 1$ and the $\theta$ linear-order change of the effective hopping $t^{(p)}_{ij}$:
\begin{align}
\big(\chi_\theta^{(p)}\big)_{ij}\equiv\left.\frac{d}{ds}\,t^{(p)}_{ij}\big(G(s\theta)\circ\bar\rho;\bar a_{0,s\theta}\big)\right|_{s=0},
\end{align}
using Eq.(\ref{eq:Hdet_t_covariance}), one finds
\begin{align}
\big(\chi_\theta^{(p)}\big)_{ij}&=i\big[\Lambda^{(p)}[\theta],t^{(p)}(\bar\rho;\bar a_0)\big]_{ij}
\label{eq:Hdet_chi_theta}
\end{align}
On the other hand, because $t$ is the first-order derivative of $E_{\rm st}$, $\chi_{\theta}$ can be computed via the second-order derivative of $E_{\rm st}$ involving the Hessian $\mathcal K$, similar to how we obtain Eq.(\ref{eq:realspace_Ward}) in the TDHF demo. From Eq.(\ref{eq:rho_gauge_covariance}), the $\theta$ linear-order change $G(\theta)\circ\bar\rho$ is:
\begin{align}
\big(\kappa_\theta^{(p)}\big)_{ij}&\equiv \left.\frac{d}{ds}\Big[G(s\theta)\circ\bar\rho\Big]^{(p)}\right|_{s=0}=i\big[\Lambda^{(p)}[\theta],\bar\rho^{(p)}\big]_{ij}.
\label{eq:Hdet_kappa_theta}
\end{align}
$\kappa_{\theta}\in T_{\bar\rho}$. But due to the second term of $E_{\rm st}$, the Hessian $\mathcal K$ is not enough:
\begin{align}
\big(\chi_\theta^{(p)}\big)_{ij}=&\left.\frac{d}{ds}\,\frac{\partial}{\partial\rho^{(p)}_{ji}}E_{\rm st}\big(G(s\theta)\circ\bar\rho;\bar a_{0,s\theta}\big)\right|_{s=0}\notag\\
=&[\hat {\mathcal K}\kappa_{\theta}]^{(p)}_{ij}+\sum_{\mathbf r,A} (\delta_\theta \bar a_0)^A(\mathbf r)\left.\frac{\partial Q^A_{\mathbf r}(\rho)}{\partial \rho^{(p)}_{ji}}\right|_{\bar\rho},
\end{align}
where $\delta_\theta \bar a_0$ is the $\theta$ linear-order of $\bar a_{0,\theta}$ in Eq.(\ref{eq:bar_a_0_transform}):
\begin{align}
\delta_\theta \bar a_0^C(\mathbf r)\equiv-\sum_{A,B}f^{AB}_C\theta^A(\mathbf r)\bar a^B_0(\mathbf r),
\end{align}
or equivalently:
\begin{align}
\Lambda[\delta_\theta \bar a_0]=i[\Lambda[\theta],\Lambda[\bar a_0]].
\end{align}

Define a superoperator: $\hat {\mathcal B}: \Lambda[\xi]\rightarrow M_{\rm h}$ in terms of components:
\begin{align}
\big[\hat {\mathcal B}\Lambda[\xi]\big]^{(p)}_{ij}\equiv\sum_{\mathbf r,A}\xi^A(\mathbf r)\left.\frac{\partial Q^A_{\mathbf r}(\rho)}{
\partial\rho_{ji}^{(p)}
}\right|_{\bar\rho}, \label{eq:Hdet_B_def}
\end{align}
and consequently:
\begin{align}
\sum_{\mathbf r,A}\xi^A(\mathbf r)Q^A_{\mathbf r}(\rho(\kappa))=\llangle\hat {\mathcal B}\Lambda[\xi],\kappa \rrangle+O(\kappa^2).
\end{align}
One arrives at 
\begin{align}
\chi_\theta=\hat{\mathcal K}\,\kappa_\theta+\hat {\mathcal B}\Lambda[\delta_\theta\bar a_0].
\label{eq:real_space_gauge_Ward}
\end{align}
This is the static gauge Ward identity in the real-space basis. In addition, $t(\bar\rho;\bar a_0)=t(\bar\rho)+\hat {\mathcal B}\Lambda[\bar a_0]$, where $t(\bar\rho)$ is the derivative of $E(\rho)$ alone.

As a concrete example, if one uses the mean-field approximation to compute overlaps (see Eq.(\ref{eq:mf_second_term})), one has:
\begin{align}
\hat {\mathcal B}_{[0]}\Lambda[a_0]=\Lambda[a_0],
\end{align}
i.e., the mean-field $\hat {\mathcal B}_{[0]}$ is equivalent to identity. 

Next, we move on to write down the Berry's phase term $L_{\rm Berry}$ in Eq.(\ref{eq:gauge_berry_two_term}) in terms of $\kappa$ and $a_0$. Let's start with the first term in Eq.(\ref{eq:gauge_berry_two_term}): $\langle\Phi(\kappa)|i\partial_t|\Phi(\kappa)\rangle$, and define the Berry curvature superoperator $\hat{\mathcal F}: T_{\bar\rho}\rightarrow T_{\bar\rho}$:
\begin{align}
\llangle \kappa_1,\hat{\mathcal F}\kappa_2\rrangle\equiv& 2\cdot\left.\rm{Im}\big[\langle  \Phi(\kappa_1)| \Phi(\kappa_2)\rangle\big]\right|_{\kappa_1\kappa_2-\text{part}},\label{eq:gauge_Berry_F}
\end{align}
which is antisymmetric following the Hermiticity property in Eq.(\ref{eq:dyna_conserving_def}): $\llangle X,\hat{\mathcal F} Y\rrangle=-\llangle \hat{\mathcal F}X,Y\rrangle$. Up to a total derivative, this term is determined by $\hat{\mathcal F}$ up to the Gaussian order:
\begin{align}
\langle\Phi(\kappa)|i\partial_t|\Phi(\kappa)\rangle=-\frac12\llangle \kappa, \hat{\mathcal F} \dot\kappa\rrangle+\text{total deriv.}+O(\kappa^3)
\end{align}

As an important general property, $\forall$ gauge transformations $\theta,\eta$, we have:
\begin{align}
\llangle \kappa_\eta,\hat{\mathcal F} \kappa_\theta\rrangle=\sum_{\mathbf r,A,B,C}\eta^A(\mathbf r)\theta^B(\mathbf r)f^{AB}_C Q_{\mathbf r}^C(\bar\rho) =0,\label{eq:Hdet_approx_F_identity}
\end{align}
where we use the saddle condition Eq.(\ref{eq:a_0_saddle}). This is a direct consequence of dynamically-conserving condition Eq.(\ref{eq:dyna_conserving_def}):
\begin{align}
&\llangle \kappa_\eta,\hat{\mathcal F} \kappa_\theta\rrangle=2{\rm Im}\frac{d^2}{dsdt}\left.\langle \Phi(G(s\eta);\bar\rho) |\Phi(G(t\theta);\bar\rho)\rangle \right|_{s=t=0}\notag\\
=&2{\rm Im}\frac{d^2}{dsdt}\left.\langle \Phi(\bar\rho) |\Phi\{e^{-is\hat\Lambda[\eta]}e^{it\hat\Lambda[\theta]}\Psi^{MF}(\bar\rho)\}\rangle \right|_{s=t=0},
\end{align}
Using $e^{-is\hat\Lambda[\eta]}e^{it\hat\Lambda[\theta]}=e^{-is\hat\Lambda[\eta]+it\hat\Lambda[\theta]+\frac{st}{2}[\hat\Lambda[\eta],\hat\Lambda[\theta]]}+O((s,t)^3)$ and the Lie algebra Eq.(\ref{eq:gg_def}), Eq.(\ref{eq:Hdet_approx_F_identity}) is established. Here, note that we replaced $|\Phi(G(\theta)\circ\bar\rho)\rangle$ by $|\Phi(G(\theta);\bar\rho)\rangle$ in this calculation, but only the former one is what we used to define $\kappa_{\theta}$ in Eq.(\ref{eq:Hdet_kappa_theta}). However, $e^{i\hat{\Lambda}[\theta]}|\Psi^{MF}(\bar\rho)\rangle$ and $|\Psi^{MF}(G(\theta)\circ\bar\rho)\rangle$ differ only by a phase factor (see Eq.(\ref{eq:phi_theta_rho})), based on the dynamical-conserving condition Eq.(\ref{eq:dyna_conserving_def}), this replacement can change the Berry connection but cannot change the Berry curvature. 

For an exact Hdet calculation, we have an even stronger identity: $\forall$ gauge transformation $\theta$:
\begin{align}
\hat{\mathcal F}_{\rm exact} \kappa_\theta=0,\label{eq:Hdet_exact_F_identity}
\end{align}
since the gauge transformation does not change the exact Hdet wavefunction $|\Phi(G(\theta);\bar\rho)\rangle_{\rm exact}=|\Phi(\bar\rho)\rangle_{\rm exact}$.

Next, we move on to the second term in Eq.(\ref{eq:gauge_berry_two_term}). The $\bar a_0$-part of this term is already included in $E_{\rm st}$. For the remaining $a_0$-part, Eq.(\ref{eq:a_0_saddle},\ref{eq:Hdet_B_def}) allows us to write it up to the Gaussian order:
\begin{align}
-\sum_{\mathbf r,A}a^{A}_0(\mathbf r,t)Q_{\mathbf r}^A(\rho(t))=-\llangle \hat {\mathcal B}\Lambda[a_0],\kappa\rrangle+O((a_0,\kappa)^3),
\end{align}

Generally, one can show that $\hat{\mathcal P}_T\hat {\mathcal B}\Lambda[a_0]$ is fully determined by $\hat {\mathcal F}$ via the following identity:
\begin{align}
\hat{\mathcal P}_T\hat {\mathcal B}\Lambda[a_0]=&\hat{\mathcal F}\kappa_{a_0},\notag\\
\Leftrightarrow\llangle\kappa,\hat {\mathcal B}\Lambda[a_0]\rrangle=&\llangle\kappa,\hat{\mathcal F}\kappa_{a_0}\rrangle,\forall \kappa\in T_{\bar\rho}.\label{eq:B_F_relation}
\end{align}
The simplest way to see this is to consider the parallel transport around a small loop involving four nearby electronic states $|\Phi\{\Psi^{MF}\}\rangle$, labeled by four parton mean-field states $|\Psi^{MF}\rangle$:
\begin{align}
&|\Psi^{MF}(\bar\rho)\rangle\rightarrow |\Psi^{MF}(\rho(\kappa))\rangle\rightarrow |\Psi^{MF}(G(a_0)\circ\rho(\kappa))\rangle\notag\\
&\rightarrow |\Psi^{MF}(G(a_0)\circ\bar\rho)\rangle\rightarrow|\Psi^{MF}(\bar\rho)\rangle.
\end{align}
The electronic (not parton's) Berry's phase for this small loop, to the order of $a_0\kappa$, gives the RHS. But if one replaces $|\Psi^{MF}(G(a_0)\circ\rho)\rangle$ by $e^{i\hat \Lambda[a_0]}|\Psi^{MF}(\rho)\rangle$, the electronic Berry's phase gives the LHS. Due to the dynamical-conserving condition Eq.(\ref{eq:dyna_conserving_def}), Berry's phase must be identical before and after the replacement.

Putting together, and neglecting total derivatives, we arrive at:
\begin{align}
L_{\rm TDVP}=&-\llangle \hat{\mathcal B}\Lambda[a_0],\kappa\rrangle-\frac{1}{2}\llangle\kappa,(\hat{\mathcal F}\partial_t+\hat{\mathcal G}+\hat{\mathcal K})\kappa\rrangle\notag\\
&+O((a_0,\kappa)^3)\label{eq:Hdet_L_TDVP}
\end{align}
Under a space-time dependent gauge transformation $\theta(\mathbf r,t)\ll 1$, from Eq.(\ref{eq:a_0_gauge_transform_small_theta},\ref{eq:Hdet_kappa_theta}), we have: 
\begin{align} 
a_0\rightarrow& a_0-\dot \theta+\delta_\theta \bar a_0,\;\;\;\kappa\rightarrow \kappa+\kappa_\theta,\label{eq:kappa_gauge_transformation} 
\end{align} 
where the first transformation can be equivalently written as: $\Lambda[a_0]\rightarrow \Lambda[a_0-\dot\theta]+i[\Lambda[\theta],\Lambda[\bar a_0]]$. 

To understand the gauge transformation of the static part of $L_{\rm TDVP}$, it is  easy to show that 
\begin{align}
\hat{\mathcal P}_T\chi_\theta=-\hat {\mathcal G}\kappa_\theta.\label{eq:gauge_G_kappa_to_chi}
\end{align}
For example, Eq.(\ref{eq:Hdet_kappa_theta}) leads to $\big(\kappa^{(p)}_\theta\big)_{\mathbf{ph}}=i\big(\Lambda^{(p)}[\theta]\big)_{\mathbf{ph}}$, and Eq.(\ref{eq:Hdet_chi_theta}) leads to $\big(\chi_\theta^{(p)}\big)_{\mathbf{ph}}=-(\epsilon^{(p)}_\mathbf p-\epsilon^{(p)}_\mathbf h)\big(\kappa^{(p)}_\theta\big)_\mathbf{ph}$. 

Therefore, together with Eq.(\ref{eq:real_space_gauge_Ward}):
\begin{align}
(\hat {\mathcal G}+\hat{\mathcal P}_T\hat {\mathcal K})\kappa_\theta+\hat{\mathcal P}_T\hat{\mathcal B}\Lambda[\delta_\theta\bar a_0]=&0,\label{eq:gauge_static_Ward_kappa}
\end{align}
which is the static gauge Ward identity in the $\kappa$-language. With this, one can show that $L_{\text{TDVP}}$ is gauge invariant up to a total time derivative: 
\begin{align} 
L_{\text{TDVP}}\rightarrow &L_{\text{TDVP}}-\frac{1}{2}\frac{\partial}{\partial t}\llangle \kappa_\theta,\hat{\mathcal F} \kappa\rrangle,
\end{align}
where Eq.(\ref{eq:Hdet_approx_F_identity},\ref{eq:B_F_relation}) are also used.

The equation of motion(eom) of $L_{\rm TDVP}$ in Eq.(\ref{eq:Hdet_L_TDVP}) should give the excitation spectrum in this Gaussian theory. In the frequency space, this eigen-equation is:
\begin{align}
[-i\omega \hat{\mathcal F}+\hat {\mathcal G}+\hat{\mathcal P}_T\hat{\mathcal K}]\kappa(\omega)+\hat{\mathcal P}_T\hat{\mathcal B}\Lambda[a_0(\omega)]=0,\label{eq:kappa_eigen_eqn_Hdet}
\end{align}
under the linearized Gauss's law constraint on $\kappa$ implemented by the $a_0$-eom: $\forall \delta a_0$:
\begin{align}
\llangle \hat{\mathcal B} \Lambda[\delta a_0(-\omega)],\kappa(\omega)\rrangle=0\Leftrightarrow \llangle\kappa,\hat{\mathcal F}\kappa_{\delta a_0}\rrangle=0\label{eq:kappa_a0_eom}
\end{align}
This is the Hdet analog of the eom Eq.(\ref{eq:kappa_eigen_eqn}), but with $a_0$ fluctuations. Note that this Gauss's law constraint Eq.(\ref{eq:kappa_a0_eom}) is not an additional condition in an \emph{exact} Hdet study, due to the identity Eq.(\ref{eq:Hdet_exact_F_identity}). Again, Gauss's law is already implemented in an \emph{exact} Hdet study, corresponding to integrating out $a_0$. On the other hand, Eq.(\ref{eq:kappa_a0_eom}) plays an important role in an approximate Hdet study.

Again, one way to understand the concrete meaning of Eq.(\ref{eq:kappa_a0_eom}) is to consider a mean-field approximation, corresponding to the zeroth-order projective expansion in Sec.\ref{sec:proj_expansion}: $\hat{\mathcal B}_{[0]}$ is equivalent to the identity. Since $\kappa=\delta\rho$ to the linear order, the fluctuation of $\kappa(t)$ should maintain the local gauge charge neutrality $\langle \Psi^{MF}(\kappa(t))|\hat T_{\mathbf r}^{A}| \Psi^{MF}(\kappa(t))\rangle=0$ at any $t$ up to $O(\kappa^2)$ -- this is the dynamical Gauss's law constraint at the mean-field level.

Since we are dealing with a gauge theory, there are solutions $[\kappa(\omega),a_0(\omega)]$ of Eq.(\ref{eq:kappa_eigen_eqn_Hdet},\ref{eq:kappa_a0_eom}) corresponding to the redundancy in the formalism: the pure gauge modes. These pure gauge modes are nonphysical and should be removed to isolate the physical modes. This is exactly the point of the gauge Ward identity, which allows one to sharply identify them. In the frequency space, Eq.(\ref{eq:kappa_gauge_transformation}) leads to pure gauge modes:
\begin{align}
\text{pure gauge: }\kappa^{\rm pg}(\omega)=\kappa_{\theta(\omega)},\;\;a^{\rm pg}_0(\omega)=i\omega\theta(\omega)+\delta_{\theta(\omega)}\bar a_0,
\end{align}
which automatically satisfy the \emph{gauge Ward identity} (i.e., Eq.(\ref{eq:kappa_eigen_eqn_Hdet},\ref{eq:kappa_a0_eom})):
\begin{align}
[-i\omega \hat{\mathcal F}+\hat {\mathcal G}+\hat{\mathcal P}_T\hat{\mathcal K}]\kappa_{\theta(\omega)}+\hat{\mathcal P}_T\hat{\mathcal B}\Lambda[i\omega\theta(\omega)+\delta_{\theta(\omega)} \bar a_0]=&0,\notag\\
\llangle\kappa_{\theta},\hat{\mathcal F}\kappa_{\delta a_0}\rrangle=&0,
\label{eq:kappa_gauge_Ward}
\end{align}
where we used Eq.(\ref{eq:Hdet_approx_F_identity},\ref{eq:B_F_relation},\ref{eq:gauge_static_Ward_kappa})

\subsubsection{HS transformation and spatial $IGG$ gauge fields}
As in our TDHF demo, the next step is to perform the HS transformation to decouple the Hessian $\hat{\mathcal K}$. We define the subspace $V_{\rm HS}=\text{Im}[\hat{\mathcal K}]\subset M_{\rm h}$ that the $\chi$-field lives in, the projector superoperator onto $V_{\rm HS}$: $\hat {\mathcal P}_{\rm HS}$, as well as the inverse of $\hat{\mathcal K}$ restricted to $V_{\rm HS}$: $\hat{\mathcal K}_{\rm HS}^{-1}\equiv \hat {\mathcal P}_{\rm HS} \hat{\mathcal K} ^{-1}\hat {\mathcal P}_{\rm HS}$ and the inverse of $\hat{\mathcal G}$ restricted to $T_{\bar\rho}$: $\hat{\mathcal G}_T^{-1}\equiv \hat{\mathcal P}_T\hat{\mathcal G}^{-1}\hat{\mathcal P}_T$. From Eq.(\ref{eq:Hdet_chi_theta},\ref{eq:real_space_gauge_Ward}), we know $\chi_\theta-\hat{\mathcal B}\Lambda[\delta_\theta\bar a_0]$ must be $\in V_{\rm HS}$. It is convenient to define:
\begin{align}
\boldsymbol\delta_\theta\chi\equiv \chi_\theta-\hat{\mathcal B}\Lambda[\delta_\theta\bar a_0],\label{eq:delta_theta_chi_def}
\end{align}
and we list the identities following Eq.(\ref{eq:real_space_gauge_Ward},\ref{eq:gauge_G_kappa_to_chi}):
\begin{align}
\hat{\mathcal P}_T\chi_\theta=&-\hat{\mathcal G}\kappa_\theta, &\boldsymbol \delta_\theta\chi=&\hat{\mathcal K}\kappa_\theta,\notag\\
-\hat{\mathcal G}_T^{-1}\chi_\theta=&\kappa_\theta,&\hat{\mathcal K}_{\rm HS}^{-1}\boldsymbol\delta_\theta\chi=&\hat{\mathcal P}_{\rm HS}\kappa_\theta.\label{eq:gauge_kappa_Q_chi_Q_relations}
\end{align}
These lead to the static Ward identity in the $\kappa$-language Eq.(\ref{eq:gauge_static_Ward_kappa}), and the corresponding one in the $\chi$-language:
\begin{align}
&(\hat{\mathcal P}_{\rm HS}\hat{\mathcal G}^{-1}_T+\hat{\mathcal K}_{\rm HS}^{-1})\boldsymbol\delta_\theta\chi+\hat{\mathcal P}_{\rm HS}\hat{\mathcal G}^{-1}_T\hat{\mathcal B}\Lambda[\delta_\theta \bar a_0]\notag\\
=&\hat{\mathcal P}_{\rm HS}\hat{\mathcal G}^{-1}_T\chi_\theta+\hat{\mathcal K}_{\rm HS}^{-1}\boldsymbol\delta_\theta\chi=0.\label{eq:gauge_static_Ward_chi}
\end{align}

It is worth mentioning that, generally, only $\boldsymbol\delta_\theta\chi\in V_{\rm HS}$, while $\chi_{\theta}$ may be outside. For the special gauge transformation $\theta^{A=0}(\mathbf r,t)$, since $U(1)_I$ is in the center of the $GG$, we have $\delta_{\theta^0}\bar a_0=0$, and consequently $\boldsymbol\delta_{\theta^0}\chi=\chi_{\theta^0}\in V_{\rm HS}$.

The $(\kappa,\chi,a_0)$-theory obtained after HS transformation can be written as:
\begin{align}
L_{\kappa,\chi,a_0}=&-\llangle \hat{\mathcal B}\Lambda[a_0],\kappa\rrangle-\frac{1}{2}\llangle \kappa,(\hat{\mathcal F}\partial_t+\hat {\mathcal G})\kappa\rrangle\notag\\
&-\llangle \chi,\kappa\rrangle+\frac{1}{2}\llangle \chi,\hat{\mathcal K}^{-1}_{\rm HS}\chi\rrangle +O((a_0,\chi,\kappa)^3).\notag\\
\label{eq:gauge_kappa_chi_Lag}
\end{align}
In this theory, the gauge transformation labeled by $\theta(\mathbf r,t)$ is given by:
\begin{align}
a_0(t)\rightarrow& a_0(t)-\partial_t\theta+\delta_\theta\bar a_0,&\kappa(t)\rightarrow& \kappa(t)+\kappa_\theta(t),\notag\\
\chi(t)\rightarrow& \chi(t)+\boldsymbol\delta_\theta \chi(t),&&\label{eq:kappa_chi_a0_gauge_transformation}
\end{align}
and the quadratic part of $L_{\kappa,\chi,a_0}$ is gauge invariant (up to a total derivative). 

Next, we will focus on translation-invariant systems and work in the $\mathbf q$-space. The notation/definition introduced in Eq.(\ref{eq:t_alpha_beta_delta}--\ref{eq:superoperators_q_space}) can be directly reused in the present Hdet discussion. In addition, we make the same assumption for the real-space form of $\mathcal K$ as in Eq.(\ref{eq:K_M_form}), following the similar discussion/assumptions near Eq.(\ref{eq:K_full_rank}), we know $V_{\rm HS}(\mathbf q)\equiv {\rm Im}\hat{\mathcal K}=V_w(\mathbf q)$, where $V_w$ is spanned by a fixed set of $\mathbf q$-space vectors (of \textbf{bond tensors}) $\{w^I(\mathbf q)\}, \;I=1,2,..,N_w$, whose tensor values are \emph{independent} of $\mathbf q$. The only new ingredient is that $w^I(\mathbf q) \in M_{\rm h}(\mathbf q)=M^{(1)}_{\rm h}(\mathbf q)\oplus M^{(2)}_{\rm h}(\mathbf q)$, and $w^I(\mathbf q)=w^{I,(1)}(\mathbf q)\oplus w^{I,(2)}(\mathbf q)$'s tensor values should be interpreted as the combination of $w^{I,(1)}_{\boldsymbol\upalpha \boldsymbol\upbeta\boldsymbol;\boldsymbol\delta}(\mathbf q)$ and $ w^{I,(2)}_{\boldsymbol\upalpha \boldsymbol\upbeta;\boldsymbol\delta}(\mathbf q)$ due to the presence of two parton species (for simplicity, we neglect the superscript $^{(p)}$ in $\boldsymbol\upalpha^{(p)}$, which should not cause confusion).

We now prepare for the long-wavelength analysis of the spatial $IGG=U(1)_I$ gauge fields, which are described by a part of the $\chi$-field ($\delta_{\theta^0}\chi$) living in $V_{\rm HS}$. Analogously, in the Goldstone-mode analysis in the previous TDHF demo, this is achieved by the Goldstone basis in Eq.(\ref{eq:chi_a_u_a}) spanning a subspace of $V_{\rm HS}(\mathbf q)$. 

We first need to characterize all the pure gauge modes $\delta_\theta\chi$. Due to the possible enlarged unit-cell, in the long-wavelength analysis, these modes can be separated into one acoustic branch and $N_{\rm subl.}-1$ optical branches, where $N_{\rm subl.}$ is the number of sublattices (i.e., real-space sites) inside one unit cell. Using $s=1,2...,N_{\rm subl.}$ to label the sublattice sites, we choose a set of $N_{\rm subl.}$ orthogonal $N_{\rm subl.}$-dimensional real vectors $\{e_m\}$, $m=0,1,...N_{\rm subl.}-1$, satisfying:
\begin{align}
\sum_{s}e_m(s) e_n(s)=&\delta_{mn}N_{\rm subl.}, & e_0(s)=1.
\end{align}
The uniform $e_0$ vector will describe the acoustic pure gauge mode, while $e_{\rm op}$ $({\rm op}=1,2,...,N_{\rm subl.}-1)$ will describe the optical pure gauge modes. The gauge transformation labeled by $\theta(\mathbf r,t)$ can be rewritten using the unit-cell coordinate $\mathbf R$: for every generator $T^A$ of $GG$, $A=0,...,N_G-1$, $\theta^{A,s}(\mathbf R,t)\equiv \theta^A(\mathbf r(\mathbf R,s),t)$ and expressed in the $(\mathbf q,\omega)$ space:
\begin{align}
\theta^{A,s}(\mathbf R,t)\equiv \frac{1}{\sqrt{N_{\rm uc}}} \sum_{\mathbf q}\int \frac{d\omega}{2\pi} e^{i\mathbf q\cdot[\mathbf R+\boldsymbol\tau(s)]-i\omega t}\theta^{A,s}(\mathbf q,\omega),\label{eq:theta_FT}
\end{align}
where $\boldsymbol\tau(s)$ is the sublattice position inside the unit cell: $\mathbf r(\mathbf R,s)=\mathbf R+\boldsymbol\tau(s)$. We can further express $\theta$ in terms of the acoustic-optical gauge parameters $\theta_m$: 
\begin{align}
\theta^{A,s}(\mathbf q,\omega)=\sum_{m=0}^{N_{\rm subl.}-1}\theta^A_m(\mathbf q,\omega) e_m(s),\label{eq:gauge_theta_acoustic_optical}
\end{align}
where $\theta^A_m(\mathbf q,\omega)\equiv\frac{1}{N_{\rm subl.}}\sum_{s=1}^{N_{\rm subl.}} e_m(s)\theta^{A,s}(\mathbf q,\omega)$.

The sublattice site of a parton orbital $\boldsymbol\upalpha$ will be denoted as $s_{\boldsymbol\upalpha}$. Based on Eq.(\ref{eq:Hdet_chi_theta},\ref{eq:delta_theta_chi_def}), in the $\mathbf q$-space, a general pure gauge mode in the $\chi$-language is:
\begin{align}
\boldsymbol\delta_\theta\chi(\mathbf q,\omega)=\sum_{A=0}^{N_G-1}\sum_{m=0}^{N_{\rm subl.}-1}\theta_m^A(\mathbf q,\omega) \mathsf u^A_m(\mathbf q).\label{eq:gauge_chi_theta_u}
\end{align}
Here, 
\begin{align}
\mathsf u^A_m(\mathbf q)\equiv\chi^A_{m}(\mathbf q)-\hat{\mathcal B}(\mathbf q)\Lambda[\delta_m^A\bar a_0](\mathbf q)=\hat{\mathcal K}(\mathbf q)\kappa^A_m(\mathbf q),\label{eq:u_A_m_def}
\end{align}
where
\begin{align}
\Lambda[\delta_m^A\bar a_0](\mathbf q)\equiv& i[\Lambda_m^A(\mathbf q),\Lambda[\bar a_0]],\notag\\
 \chi^A_{m}(\mathbf q)\equiv&i[\Lambda_m^A(\mathbf q),t(\bar\rho;\bar a_0)],\notag\\
 \kappa_m^A(\mathbf q)\equiv&i[\Lambda_m^A(\mathbf q),\bar\rho],\label{eq:many_bond_tensors}
\end{align}
and the bond tensor value of $\Lambda_m^A(\mathbf q)$ is
\begin{align}
[\Lambda_m^{A,(p)}(\mathbf q)]_{\boldsymbol\upalpha\boldsymbol\upbeta;\boldsymbol\delta}\equiv \delta_{\boldsymbol\delta,\mathbf 0}\delta_{s_{\boldsymbol\upalpha},s_{\boldsymbol\upbeta}}e^{i\mathbf q\cdot \boldsymbol\tau(s_{\boldsymbol\upalpha})} e_m(s_{\boldsymbol\upalpha})T^{A,(p)}_{\mathbf m_{\boldsymbol\upalpha},\mathbf m_{\boldsymbol\upbeta}},\label{eq:Lambda_A_m_def}
\end{align}
where we used the notation developed in Eq.(\ref{eq:t_alpha_beta_delta}). The saddle $\bar a_0$ can be expanded in the $\Lambda_m^{A}(\mathbf q=0)$ bond tensor basis:
\begin{align}
\Lambda[\bar a_0]=\sum_{A,m} \bar a^A_{0,m} \Lambda_m^{A}(\mathbf 0),
\end{align}
where $\bar a^A_{0,m}\equiv \frac{1}{N_{\rm subl.}}\sum_s e_m(s)\bar a_0^{A,s}$ and $\bar a_0^{A,s}$ is $\bar a^A_0$'s value on sublattice-$s$.
Using Lie algebra Eq.(\ref{eq:gg_def}), $\Lambda[\delta_m^A\bar a_0](\mathbf q)$ can be expanded in the $\Lambda_m^{A}(\mathbf q)$ bond tensor basis:
\begin{align}
\Lambda[\delta_m^A\bar a_0](\mathbf q)=-\sum_{B,C,n,l}\bar a^B_{0,n}f_C^{AB}\mathsf N_{lmn}\Lambda^C_l(\mathbf q),\label{eq:Lambda_delta_m_a_expansion}
\end{align}
where $\mathsf N_{lmn}\equiv \frac{1}{N_{\rm subl.}}\sum_s e_l(s)e_m(s)e_n(s)$.

The bond tensor value of $\chi^A_{m}(\mathbf q)$ is:
\begin{align}
&\big(\chi_m^{A,(p)}\big)_{\boldsymbol\upalpha\boldsymbol\upbeta;\boldsymbol\delta}(\mathbf q)\notag\\
\equiv&i\,e^{i\mathbf q\cdot\boldsymbol\tau(s_{\boldsymbol\upalpha})}e_m(s_{\boldsymbol\upalpha})\sum_{\mathbf m'}T^{A,(p)}_{\mathbf m_{\boldsymbol\upalpha},\mathbf m'}t^{(p)}_{(s_{\boldsymbol\upalpha},\mathbf m'),\boldsymbol\upbeta;\boldsymbol\delta}\notag\\
&-i\,e^{i\mathbf q\cdot[\boldsymbol\delta+\boldsymbol\tau(s_{\boldsymbol\upbeta})]}e_m(s_{\boldsymbol\upbeta})\sum_{\mathbf m'}t^{(p)}_{\boldsymbol\upalpha,(s_{\boldsymbol\upbeta},\mathbf m');\boldsymbol\delta}T^{A,(p)}_{\mathbf m',\mathbf m_{\boldsymbol\upbeta}}.
\label{eq:chi_Am_explicit}
\end{align}
As a special case, the $U(1)_I$ components $\mathsf u_m^{A=0}$:
\begin{align}
&(\mathsf u^{0,(p)}_m)_{\boldsymbol\upalpha\boldsymbol\upbeta;\boldsymbol\delta}(\mathbf q)\notag\\
\equiv& i {\rm q}^{(p)}\big[e^{i\mathbf q\cdot \boldsymbol\tau(s_{\boldsymbol\upalpha})}e_m(s_{\boldsymbol\upalpha})-e^{i\mathbf q\cdot [\boldsymbol\delta+\boldsymbol\tau(s_{\boldsymbol\upbeta})]}e_m(s_{\boldsymbol\upbeta})\big] t^{(p)}_{\boldsymbol\upalpha\boldsymbol\upbeta;\boldsymbol\delta},\label{eq:u_op_tensors}
\end{align}
Eq.(\ref{eq:gauge_chi_theta_u}) is exactly analogous to Eq.(\ref{eq:chi_a_u_a}) in the Goldstone mode analysis in the TDHF demo. But here, since there are $N_G$ gauge redundancy variables $\theta^A$ per site, we have $\mathsf u^A_m(\mathbf q)$, $\forall\mathbf q$. 

Let us consider the long-wavelength $\mathbf q \rightarrow \mathbf 0$ limit. All optical branches (${\rm op}=1,2..,N_{\rm subl.}-1$, $A=0,...,N_G-1$) have a nonzero limit: $\mathsf u^A_{\rm op}(\mathbf q)\rightarrow \mathsf u^A_{\rm op}(\mathbf 0)\neq 0$, as well as non-$IGG$ acoustic branches ($A=1,...,N_G-1$): $\mathsf u^A_{0}(\mathbf q)\rightarrow \mathsf u^A_{0}(\mathbf 0)\neq 0$. This is precisely the meaning of the Higgs mechanism: these gauge modes \emph{transform} the saddle nontrivially. On the other hand, the $IGG$ acoustic branch $\mathsf u^0_0$ has the following limiting behavior:
\begin{align}
\mathsf u^0_0(\mathbf q)\rightarrow (-i)[\mathbf q_x \mathsf w^x(\mathbf q)+\mathbf q_y \mathsf w^y(\mathbf q)]+O(\mathbf q^2),\label{eq:u_0_q_0_limit}
\end{align}
where $\mathsf w^u(\mathbf q)\in V_{\rm HS}(\mathbf q)$ ($u=x,y$) have the following $\mathbf q$-independent tensor value:
\begin{align}
&\mathsf w^{(p),u}_{\boldsymbol\upalpha\boldsymbol\upbeta;\boldsymbol\delta}(\mathbf q)\equiv i{\rm q}^{(p)}[\mathbf d(\boldsymbol\upalpha\boldsymbol\upbeta;\boldsymbol\delta)]_u t^{(p)}_{\boldsymbol\upalpha\boldsymbol\upbeta;\boldsymbol\delta},\;\text{ where }\notag\\
&\mathbf d(\boldsymbol\upalpha\boldsymbol\upbeta;\boldsymbol\delta) \equiv \boldsymbol\delta+\boldsymbol\tau(s_{\boldsymbol\upbeta})-\boldsymbol\tau(s_{\boldsymbol\upalpha}).\label{eq:w_xy_tensors}
\end{align}

Consider the $\chi$-fluctuation parameterized by real-fields $\{\mathbf a_x(\mathbf R,t),\mathbf a_y(\mathbf R,t)\}$, in the $(\mathbf q,\omega)$-space:
\begin{align}
\mathbf a_u(\mathbf R,t)=&\frac{1}{\sqrt{N_{\rm uc}}}\sum_{\mathbf q}\int \frac{d\omega}{2\pi}e^{i\mathbf q\cdot\mathbf R-i\omega t}\mathbf a_u(\mathbf q,\omega),\notag\\
\chi_{\mathbf a}(\mathbf q,\omega)\equiv& \mathbf a_x(\mathbf q,\omega) \mathsf w^x(\mathbf q)+\mathbf a_y(\mathbf q,\omega) \mathsf w^y(\mathbf q),
\end{align}
where $\chi_{\mathbf a}(\mathbf q,\omega)$ lies in a subspace $\subset V_{\rm HS}(\mathbf q)$ spanned by $\{\mathsf w^x(\mathbf q),\mathsf w^y(\mathbf q)\}$. Under a $\theta^{A=0}$ gauge transformation:
\begin{align}
\delta_{\theta^0}\chi(\mathbf q,\omega)=-i \sum_{u=x,y}\mathbf q_u \theta^0_0(\mathbf q,\omega)\mathsf w^u(\mathbf q)+O(\mathbf q^2).
\end{align}
Comparing with $\chi_{\mathbf a}$ gives:
\begin{align}
\mathbf a_u(\mathbf q,\omega)\rightarrow \mathbf a_u(\mathbf q,\omega)-i\mathbf q_u\theta^0_0(\mathbf q,\omega).
\end{align}
At long-wavelength, $\mathbf a_u$ $(u=x,y)$ are nothing but the spatial components of the $U(1)_I$ gauge fields, as one can see via the Peierls coupling: $t^{(p)}_{ij}\rightarrow t^{(p)}_{ij} e^{-i{\rm q}^{(p)}\mathbf a\cdot (\mathbf r_i-\mathbf r_j)}$. Importantly, among the two long-wavelength spatial gauge fields $\mathbf a_x,\mathbf a_y$, there is only one acoustic pure gauge longitudinal mode that is nonphysical, leading to a remaining transverse physical gauge mode.

\subsubsection{Effective theory of Hdet wavefunction}
Now it is clear that the gauge invariant theory $L_{\kappa,\chi,a_0}$ in Eq.(\ref{eq:gauge_kappa_chi_Lag}) can be viewed as the matter field $\kappa$ coupled with fluctuating $GG$ gauge fields: $a_0$ is the temporal component, while the hopping on bond $t(\bar\rho;\bar a_0)+\chi$ encodes the spatial components. $\kappa$ also couples with other, non-gauge bosonic fields described by the other parts of $\chi$. The only difference from the familiar effective theories is that the matter field is bosonic -- the Thouless parameter $\kappa$ for the underlying fermionic mean-field state. Here we refermionize the theory to obtain an effective theory of fermionic matter fields coupled to a dynamical $GG$ gauge field.

First of all, let's write the full action of $L_{\kappa,\chi,a_0}$ in the $(\mathbf q,\omega)$ space:
\begin{align}
&S_{\kappa,\chi,a_0}=\sum_{\mathbf q}\int \frac{d\omega}{2\pi}\Big[-\llangle \tilde\chi(-\mathbf q,-\omega),\kappa(\mathbf q,\omega)\rrangle\notag\\
&-\frac{1}{2}\llangle \kappa(-\mathbf q,-\omega),(-i\omega\hat{\mathcal F}+\hat{\mathcal G})(\mathbf q)\kappa(\mathbf q,\omega)\rrangle\notag\\
&+\frac{1}{2}\llangle \chi(-\mathbf q,-\omega),\hat{\mathcal K}^{-1}_{\rm HS}\chi(\mathbf q,\omega)\rrangle\Big]+O((\kappa,\chi,a_0)^3),\label{eq:S_kappa_chi_a0}
\end{align}
where we combined $a_0$ and $\chi$ into a field $\tilde \chi$:
\begin{align}
\tilde\chi\equiv \chi+\hat{\mathcal B}\Lambda[a_0].
\end{align}

Similar to the TDHF demo in Eq.(\ref{eq:eff_Lag}), we now write down a Gaussian-order candidate effective theory of fermions ($\bar\psi,\psi$), $\chi$ and $a_0$:
\begin{align}
&L_{\rm eff,[0]}\equiv \sum_{p=1}^{2}\sum_{ij}\Big[\Lambda^{(p)}[a_0]_{ij}+\chi^{(p)}_{ij}\Big]\bar\rho^{(p)}_{ji}+\frac{1}{2}\llangle\chi,\hat{\mathcal K}_{\rm HS}^{-1}\chi\rrangle\notag\\
&+\sum_{p,ij}\bar\psi_i^{(p)}\Big[i\partial_t\delta_{ij}-\Lambda^{(p)}[a_0]_{ij}-t^{(p)}_{ij}(\bar\rho;\bar a_0)-\chi^{(p)}_{ij}\Big]\psi^{(p)}_j.
\end{align}
Here, we define the gauge transformation labeled by $\theta(\mathbf r,t)$ as:
\begin{align}
\bar\rho^{(p)}\rightarrow& \Theta^{(p)}\bar\rho^{(p)}\Theta^{(p)\dagger},\notag\\
\Lambda^{(p)}[\bar a_0]\rightarrow& \Theta^{(p)}\Lambda^{(p)}[\bar a_0]\Theta^{(p)\dagger},\notag\\
t^{(p)}(\bar\rho;\bar a_0)+\chi\rightarrow& \Theta^{(p)}(t^{(p)}(\bar\rho;\bar a_0)+\chi)\Theta^{(p)\dagger},\notag\\
\Lambda^{(p)}[a_0]\rightarrow &\Theta^{(p)}\Lambda^{(p)}[a_0]\Theta^{(p)\dagger}+i(\partial_t\Theta^{(p)})\Theta^{(p)\dagger},\notag\\
\psi^{(p)}\rightarrow&\Theta^{(p)}\psi^{(p)},\;\; \bar\psi^{(p)}\rightarrow \bar\psi^{(p)}\Theta^{(p)\dagger},\label{eq:fermion_chi_a_0_gauge_transformation}
\end{align}
where $\Theta$ is a shorthand notation for $\Theta[\theta]$ in Eq.(\ref{eq:Lambda_Theta_def}). $\psi$ minimally couples with $GG$ gauge field and $L^{\rm{0th}}_{\rm eff}$ is gauge invariant (up to a total derivative from the $\bar\rho$ background term). 

Again, the reason for the subscript "$_{[0]}$" in $L_{\rm eff,[0]}$ is that, the 0th-order projective-expansion introduced in Sec.\ref{sec:proj_expansion} is a version of time-dependent mean-field approximation, in which one still uses the Slater-determinant mean-field state $|\Psi^{MF}(\rho)\rangle$ in Eq.(\ref{eq:Hdet_MF_state}) to compute wavefunction overlaps, and consequently the Berry's operator is known $\hat {\mathcal F}=\hat {\mathcal F}_{[0]}$ -- the canonical boson Berry's phase as in the TDHF demo (see Eq.(\ref{eq:T_rho_superoperators})) and $\hat{\mathcal B}_{[0]}=\mathbf 1$. If one uses the 0th-order projective-expansion to approximately compute the action $S_{\kappa,\chi,a_0}$, after using Thouless parameterization of $L_{\rm eff,[0]}$, it is obvious that $S_{\kappa,\chi,a_0}$ is exactly reproduced by $L_{\rm eff,[0]}$ up to the Gaussian order.

However, as long as one goes beyond the 0th-order projective-expansion to compute $S_{\kappa,\chi,a_0}$, $L_{\rm eff,[0]}$ is no longer an exact refermionized theory at the Gaussian order. Nevertheless, one may always introduce counterterms into $L_{\rm eff}$ so that it still reproduces $S_{\kappa,\chi,a_0}$ at the Gaussian order, in the sense that after integrating out the matter fields (i.e., $\kappa$ and $\bar\psi,\psi$), the same Gaussian order theory of $(\chi,a_0)$ is obtained. For example, the dispersion of collective modes will be exactly reproduced. Precisely,
\begin{align}
S_{\rm eff}\equiv S_{\rm eff,[0]}+S_{\rm ct}[\chi,a_0],
\end{align}
where the counterterm $S_{\rm ct}$ describes additional couplings between $\chi,a_0$-fields:
\begin{align}
&S_{\rm ct}[\chi,a_0]=\sum_{\mathbf q}\int \frac{d\omega}{4\pi}\Big[\llangle \tilde\chi(-\mathbf q,-\omega),(-i\omega \hat{\mathcal F}+\hat{\mathcal G})_T^{-1}\tilde \chi(\mathbf q,\omega)\rrangle\notag\\
&-\llangle \tilde\chi_{[0]}(-\mathbf q,-\omega),(-i\omega \hat{\mathcal F}_{[0]}+\hat{\mathcal G})_T^{-1}\tilde \chi_{[0]}(\mathbf q,\omega)\rrangle\Big].
\end{align}
Here $\tilde\chi_{[0]}\equiv \chi+\hat{\mathcal B}_{[0]}\Lambda[a_0]$.

$S_{\rm eff}$ is gauge invariant up to the Gaussian order (and up to a total derivative). At the 0th-order projective-expansion approximation, $S_{\rm ct}$ vanishes. Beyond the 0th-order projective-expansion(PE), the crucial question is whether the low-energy physics is modified qualitatively or not by $S_{\rm ct}$, e.g., whether there are additional topological CS terms contributed by $S_{\rm ct}$. One can argue that the answer will be negative, at least in the framework of projective expansion. As we will see in Sec.\ref{sec:PE_local}, as a local expansion, PE is not expected to modify global properties, such as the topological terms. Although we suspect the answer is generally negative even in an exact treatment of Hdet states with the locality structure, we do not have a sharp statement to make. 

In summary, the microscopic Gaussian theory $L_{\kappa,\chi,a_0}$ can be refermionized as $L_{\rm eff}$, in which Grassmann variables minimally couple with $GG$ gauge fields. The saddle point of $L_{\rm eff}$, still labeled by $\bar\rho,\bar a_0$, Higgses $GG$ down to $IGG$. Next, we demonstrate how to extract microscopic gauge dynamics via VMPI. 

\subsubsection{Gauge Ward identity in the $\chi$-language and long-wavelength gauge kernel}
In this section, we will start with the action $S_{\kappa,\chi,a_0}$ in Eq.(\ref{eq:S_kappa_chi_a0}). Integrating out the matter field $\kappa$ and other bosonic fields in $\chi$, our goal is to obtain an effective action $S^{IGG}$ for the acoustic $U(1)_I$ gauge field $\mathbf a_\mu$ \emph{only}. We assume that the matter field is fully gapped, and the practice outlined below provides a route to compute the microscopic dispersion of the gauge photon, as well as possible CS terms. 

On the other hand, when the matter field is gapless, as in the situation of a composite Fermi liquid, the low-energy effective theory should not be of $\mathbf a_\mu$ only. Instead, one should only integrate out the high-energy matter fields. Assuming all other bosonic fields are massive, the low-energy effective theory is qualitatively described by gapless fermions coupled with the dynamical acoustic gauge field $\mathbf a_\mu$. The discussion below will then be useful to compute the topological terms (i.e., CS-term in the present example) in the gauge dynamics, which come from integrating out high-energy matter fields only.

After integrating out the matter field $\kappa$ in $S_{\kappa,\chi,a_0}$, one obtains the action $S_{\chi,a_0}$:
\begin{align}
S_{\chi,a_0}=&\sum_{\mathbf q}\int\frac{d\omega}{4\pi}\Big[\llangle \tilde\chi(-\mathbf q,-\omega),(-i\omega \hat{\mathcal F}+\hat{\mathcal G})_T^{-1}\tilde\chi(\mathbf q,\omega)\rrangle  \notag\\
&+\llangle \chi(-\mathbf q,-\omega),\hat{\mathcal K}_{\rm HS}^{-1}\chi(\mathbf q,\omega)\rrangle\Big]+O((\chi,a_0)^3).
\end{align}

The equation of motion (eom) for $S_{\chi,a_0}$ giving the collective mode spectrum is:
\begin{align}
\hat{\mathcal P}_{\rm  HS}(-i\omega\hat{\mathcal F}+\hat{\mathcal G})_T^{-1}\big(\chi(\omega)+\hat{\mathcal B}\Lambda[a_0(\omega)]\big)+\hat{\mathcal K}^{-1}_{\rm HS}\chi(\omega)=0,\label{eq:chi_eigen_eqn_Hdet}
\end{align}
under the Gauss's law constraint on $(\chi,a_0)$ implemented by the $a_0$-eom: $\forall \delta a_0$,
\begin{align}
\llangle\hat{\mathcal B}\Lambda[\delta a_0(-\omega)],(-i\omega\hat{\mathcal F}+\hat{\mathcal G})_T^{-1}\big(\chi(\omega)+\hat{\mathcal B}\Lambda[a_0(\omega)]\big)\rrangle=0.\label{eq:chi_a0_eom}
\end{align}
This is the Hdet analog of the eom Eq.(\ref{eq:chi_eigen_eqn}), but with $a_0$ fluctuations. One can show that the $[\chi(\omega),a_0(\omega)]$ eigenmode satisfying Eq.(\ref{eq:chi_eigen_eqn_Hdet},\ref{eq:chi_a0_eom}) can be mapped to $[\kappa(\omega),a_0(\omega)]$ eigenmode satisfying Eq.(\ref{eq:kappa_eigen_eqn_Hdet},\ref{eq:kappa_a0_eom}) via:
\begin{align}
\chi\rightarrow\kappa:\kappa(\omega)=-(-i\omega\hat{\mathcal F}+\hat{\mathcal G})_T^{-1}\big(\chi(\omega)+\hat{\mathcal B}\Lambda[a_0(\omega)]\big),\label{eq:kappa_chi_eigenmode_map_Hdet}
\end{align}
which is the Hdet analog of Eq.(\ref{eq:kappa_chi_eigenmode_map})

The nonphysical pure gauge modes in the $\chi$-language, from Eq.(\ref{eq:kappa_chi_a0_gauge_transformation}), are:
\begin{align}
\text{pure gauge: } \chi^{\rm pg}(\omega)=\boldsymbol\delta_{\theta(\omega)}\chi,\;\;a^{\rm pg}_0(\omega)=i\omega\theta(\omega)+\delta_{\theta(\omega)}\bar a_0.\label{eq:chi_pure_gauge}
\end{align}
Parallel to the gauge Ward identity in the $\kappa$-language Eq.(\ref{eq:kappa_gauge_Ward}), due to gauge invariance in $S_{\chi,a_0}$,  they automatically satisfy the $\chi,a_0$-eom Eq.(\ref{eq:chi_eigen_eqn_Hdet},\ref{eq:chi_a0_eom}), i.e., the \emph{gauge Ward identity} in the $\chi$-language:
\begin{align}
&\hat{\mathcal P}_{\rm  HS}(-i\omega\hat{\mathcal F}+\hat{\mathcal G})_T^{-1}\big(\chi_{\theta(\omega)}+\hat{\mathcal B}\Lambda[i\omega\theta(\omega)]\big)+\hat{\mathcal K}^{-1}_{\rm HS}\boldsymbol\delta_{\theta(\omega)}\chi=0,\notag\\
&\llangle\hat{\mathcal B}\Lambda[\delta a_0(-\omega)],(-i\omega\hat{\mathcal F}+\hat{\mathcal G})_T^{-1}\big(\chi_{\theta(\omega)}+\hat{\mathcal B}\Lambda[i\omega\theta(\omega)]\big)\rrangle=0.
\label{eq:chi_gauge_Ward}
\end{align}
To see this, using Eq.(\ref{eq:B_F_relation},\ref{eq:gauge_kappa_Q_chi_Q_relations}), we have:
\begin{align}
&\hat{\mathcal P}_T\Big[\chi_{\theta(\omega)}+\hat{\mathcal B}\Lambda[i\omega\theta]\Big]=-\hat{\mathcal G}\kappa_{\theta(\omega)}+i\omega\hat{\mathcal P}_T\hat{\mathcal B}\Lambda[\theta(\omega)]\notag\\
=&-(-i\omega\hat{\mathcal F}+\hat {\mathcal G})\kappa_{\theta(\omega)}.
\end{align}
Using Eq.(\ref{eq:gauge_kappa_Q_chi_Q_relations}) again, the $\chi$-language identity Eq.(\ref{eq:chi_gauge_Ward}) is mapped back to the $\kappa$-language Eq.(\ref{eq:kappa_gauge_Ward}).

Next, we show the concrete consequence of this gauge invariance using the \textbf{bond tensor} basis. Using notations developed in Eq.(\ref{eq:u_A_m_def},\ref{eq:many_bond_tensors},\ref{eq:Lambda_A_m_def}), the pure gauge modes in Eq.(\ref{eq:chi_pure_gauge}) can be represented as:
\begin{align}
\boldsymbol\delta_{\theta}\chi(\mathbf q,\omega)=&\sum_{A,m}\theta^A_m(\mathbf q,\omega)\mathsf u^A_{m}(\mathbf q),\notag\\
\Lambda[a^{\rm pg}_0](\mathbf q,\omega)=&\sum_{A,m}\theta^A_m(\mathbf q,\omega)[i\omega\Lambda^A_m(\mathbf q)+\Lambda[\delta_m^A\bar a_0](\mathbf q)].
\end{align}

Different from the TDHF demo, due to the presence of $a_0$, the relevant fields live not only in $V_{\rm HS}$, but also in a temporal subspace $\subset M_{\rm h}$:
\begin{align}
V_{a_0}(\mathbf q)\equiv \text{span}\{\Lambda_m^A(\mathbf q)\}_{A=0,...,N_G-1;m=0,...,N_{\rm subl.}-1}.
\end{align}
Due to Eq.(\ref{eq:Lambda_delta_m_a_expansion}), $\Lambda[\delta_m^A\bar a_0](\mathbf q)\in V_{a_0}$. The full $\chi,a_0$ fields live in a subspace:
\begin{align}
V_{\chi,a_0}(\mathbf q)\equiv V_{\rm HS}(\mathbf q)\oplus V_{a_0}(\mathbf q),
\end{align}
and a general dynamical field is denoted as:
\begin{align}
\mathcal X(\mathbf q,\omega)\equiv\begin{pmatrix}\chi(\mathbf q,\omega)\\
\Lambda[a_0](\mathbf q,\omega)
\end{pmatrix}.
\end{align}

Now we define a temporal bond tensor $\mathsf w^{0}(\mathbf q)$ for the purpose of long-wavelength $IGG=U(1)_I$ analysis, whose tensor value is \emph{independent of $\mathbf q$} and is given by $\Lambda_{m=0}^{A=0,(p)}(\mathbf q\rightarrow\mathbf 0)$:
\begin{align}
\mathsf w^{0,(p)}_{\boldsymbol\upalpha\boldsymbol\upbeta;\boldsymbol\delta}(\mathbf q)\equiv {\rm q}^{(p)} \delta_{\boldsymbol\upalpha\boldsymbol\upbeta}\delta_{\boldsymbol\delta,\mathbf0}.\label{eq:Lambda_0_q_0_limit}
\end{align}
$\mathsf w^{0}(\mathbf q)\in V_{a_0}(\mathbf q)$. Together with $\mathsf w^x(\mathbf q),\mathsf w^y(\mathbf q)\in V_{\rm HS}(\mathbf q)$ from Eq.(\ref{eq:w_xy_tensors}), we introduce three vectors in $V_{\chi,a_0}(\mathbf q)$:
\begin{align}
\mathcal W^0(\mathbf q)\equiv
\begin{pmatrix}\mathbf 0\\ \mathsf w^0(\mathbf q)\end{pmatrix},\qquad
\mathcal W^u(\mathbf q)\equiv
\begin{pmatrix}\mathsf w^u(\mathbf q)\\ \mathbf 0\end{pmatrix},\quad u=x,y.
\label{eq:IGG_three_combined_vectors}
\end{align}
All three have $\mathbf q$-independent tensor values. These vectors define the long-wavelength acoustic $IGG$ fields $\mathbf a_\mu$, with $\mu=0,x,y$ (see below).

For every full-$GG$ gauge component $(A,m)$, define the combined pure gauge vector
\begin{align}
\mathcal U_m^A(\mathbf q,\omega)\equiv
\begin{pmatrix}
\mathsf u_m^A(\mathbf q)\\
i\omega\Lambda_m^A(\mathbf q)+\Lambda[\delta_m^A\bar a_0](\mathbf q)
\end{pmatrix}.
\end{align}
The pure gauge fluctuation written above is equivalently
\begin{align}
\mathcal X^{\rm pg}(\mathbf q,\omega)
=\sum_{A,m}\theta_m^A(\mathbf q,\omega)\mathcal U_m^A(\mathbf q,\omega).
\end{align}
For all $(A,m)\neq(0,0)$, we keep $\mathcal U_m^A(\mathbf q,\omega)$ with its exact $\mathbf q$-dependence. This is necessary for isolating their exact zero modes at finite $\mathbf q$. $IGG=U(1)_I$ is special because $\mathsf u^{A=0}_{m=0}(\mathbf q\rightarrow 0)=0$.

$\{\Lambda_m^A(\mathbf q)\}$ are linearly independent vectors for different values of $m$ and $A$ (at least in the vicinity of $\mathbf q=0$). We have
\begin{align}
\dim V_{\chi,a_0}(\mathbf q)=N_w+N_GN_{\rm subl.}.
\end{align}
We choose a set of $N_GN_{\rm subl.}+2$ (linearly independent) vectors $\in V_{\chi,a_0}(\mathbf q)$:
\begin{align}
\big\{\mathcal W^0,\mathcal W^x,\mathcal W^y\big\}\cup\big\{\mathcal U_m^A(\mathbf q,\omega)\,|\,(A,m)\neq(0,0)\big\}.
\end{align}
One can show that $\forall v(\mathbf q,\omega)\in $ the set above satisfies $v(\mathbf q,\omega)=v^\dagger(-\mathbf q,-\omega)$ (Hermicity condition) since they describe Hermitian matrices $\in M_{\rm h}$. We choose other $N_w-2$ (linearly independent) vectors $\{\mathcal E^B(\mathbf q)\}\subset V_{\chi,a_0}$, $(B=1,...,N_w-2)$, also satisfying $\mathcal E^B(\mathbf q)=\mathcal E^{B\dagger}(-\mathbf q)$, so that they form a complete basis of $V_{\chi,a_0}$ satisfying the Hermicity condition. $\{E^B(\mathbf q)\}$ describe the other bosonic fields that $\kappa$ couples to, which are neither nonphysical pure gauge modes nor the physical $IGG$ transverse mode. Note that, in this basis, $U_m^A$ has $\omega$-dependence, which has a well-defined $\omega\rightarrow 0$ limit.

A general field configuration can therefore be expanded as
\begin{align}
\mathcal X(\mathbf q,\omega)=&\sum_{\mu=0,x,y}\mathbf a_\mu(\mathbf q,\omega)\mathcal W^\mu(\mathbf q)
+\sum_{B=1}^{N_w-2}\eta_B(\mathbf q,\omega)\mathcal E^B(\mathbf q)\notag\\
&+\sum_{(A,m)\neq(0,0)}z_m^A(\mathbf q,\omega)\mathcal U_m^A(\mathbf q,\omega).
\label{eq:Hdet_full_basis}
\end{align}
We collect these expansion fields into
\begin{align}
\mathcal V\equiv(\mathbf a_0,\mathbf a_x,\mathbf a_y,\eta,z)^T.
\end{align}
Due to the Hermicity condition of the basis choice, $\mathcal V(-\mathbf q,-\omega)=\mathcal V^*(\mathbf q,\omega)$. The quadratic action takes the form
\begin{align}
S_{\chi,a_0}=\sum_{\mathbf q}\int\frac{d\omega}{4\pi}
\mathcal V^T(-\mathbf q,-\omega)D(\mathbf q,\omega)\mathcal V(\mathbf q,\omega)
+O(\mathcal V^3),
\end{align}
where
\begin{align}
D(-\mathbf q,-\omega)=D(\mathbf q,\omega)^*,\qquad
D(\mathbf q,\omega)=D(\mathbf q,\omega)^\dagger.
\end{align}

For each $(A,m)\neq(0,0)$, the corresponding linearized gauge transformation is simply
\begin{align}
z_m^A(\mathbf q,\omega)\rightarrow z_m^A(\mathbf q,\omega)+\theta_m^A(\mathbf q,\omega).
\end{align}
The gauge Ward identity Eq.(\ref{eq:chi_gauge_Ward}) therefore dictates that the action is independent of every $z_m^A$. Equivalently, the exact full-$GG$ kernel has the block form
\begin{align}
D(\mathbf q,\omega)=
\begin{pmatrix}
D_{\mathbf a\mathbf a}&D_{\mathbf a\eta}&\mathbf 0\\
D_{\eta\mathbf a}&D_{\eta\eta}&\mathbf 0\\
\mathbf 0&\mathbf 0&\mathbf 0
\end{pmatrix}.
\label{eq:Hdet_full_D}
\end{align}
Thus all optical $GG$ pure gauge modes and all non-$IGG$ acoustic pure gauge modes have been isolated as exact null vectors. Dropping the $z$ coordinates is only a quotient by these nonphysical redundancies, and we should focus on the other four nonzero blocks.

For example, the $3\times3$ block involving the acoustic $IGG$ fields follows directly from $S_{\chi,a_0}$:
\begin{align}
D_{\mathbf a_\mu\mathbf a_\nu}=&\llangle \mathsf W^\mu(-\mathbf q),(-i\omega\hat{\mathcal F}+\hat{\mathcal G})_T^{-1}(\mathbf q)\mathsf W^\nu(\mathbf q)\rrangle\notag\\
&+(1-\delta_{\mu0})(1-\delta_{\nu0})\llangle\mathsf w^\mu(-\mathbf q),\hat{\mathcal K}^{-1}_{\rm HS}(\mathbf q)\mathsf w^\nu(\mathbf q)\rrangle,\label{eq:D_aa_element}
\end{align}
where
\begin{align}
\mathsf W^0\equiv\hat{\mathcal B}\mathsf w^0,\;\;\mathsf W^x\equiv\mathsf w^x,\;\;\mathsf W^y\equiv\mathsf w^y.
\end{align}

We next isolate the remaining acoustic $IGG$ redundancy. Since $T^{A=0}$ is in the center of the Lie algebra Eq.(\ref{eq:gg_def}),
\begin{align}
\Lambda[\delta_0^0\bar a_0](\mathbf q)=0,
\end{align}
and the exact acoustic $IGG$ pure gauge vector is
\begin{align}
\mathcal U_0^0(\mathbf q,\omega)=
\begin{pmatrix}
\mathsf u_0^0(\mathbf q)\\
i\omega\Lambda_0^0(\mathbf q)
\end{pmatrix}.
\label{eq:exact_acoustic_IGG_vector}
\end{align}
$\{\mathcal E^B\}$ can be chosen so that this vector has no $z$ component, allowing us to write its exact decomposition in the retained $(\mathbf a,\eta)$ space as
\begin{align}
\mathcal U_0^0(\mathbf q,\omega)=&\sum_{\mu=0,x,y}\mathcal V_{0,\mathbf a_\mu}(\mathbf q,\omega)\mathcal W^\mu(\mathbf q)+\sum_B\mathcal V_{0,\eta_B}(\mathbf q,\omega)\mathcal E^B(\mathbf q).
\label{eq:exact_acoustic_IGG_decomposition}
\end{align}
From Equations (\ref{eq:u_0_q_0_limit},\ref{eq:Lambda_0_q_0_limit}), the $\mathbf q$-independent tensor values $\mathsf w^0,\mathsf w^x,\mathsf w^y$ lead to:
\begin{align}
\mathsf u_0^0(\mathbf q)=&-i\mathbf q_x\mathsf w^x(\mathbf q)-i\mathbf q_y\mathsf w^y(\mathbf q)+O(\mathbf q^2),\notag\\
\Lambda_0^0(\mathbf q)=&\mathsf w^0(\mathbf q)+O(|\mathbf q|).
\end{align}
Consequently,
\begin{align}
\mathcal V_{0,\mathbf a}(\mathbf q,\omega)=&
\begin{pmatrix}
i\omega\\-i\mathbf q_x\\-i\mathbf q_y
\end{pmatrix}
+O((|\mathbf q|,\omega)^2),\notag\\
\mathcal V_{0,\eta}(\mathbf q,\omega)=&O((|\mathbf q|,\omega)^2).
\label{eq:null_vec_limit}
\end{align}
The leading-order vector therefore has the standard form of $U(1)_I$ gauge transformation in the continuum limit Eq.(\ref{eq:standard_U1_gauge_transform}).

After the exact $z$-null directions have been removed, the remaining exact acoustic Ward identity is
\begin{align}
\begin{pmatrix}
D_{\mathbf a\mathbf a}&D_{\mathbf a\eta}\\
D_{\eta\mathbf a}&D_{\eta\eta}
\end{pmatrix}
\begin{pmatrix}
\mathcal V_{0,\mathbf a}(\mathbf q,\omega)\\
\mathcal V_{0,\eta}(\mathbf q,\omega)
\end{pmatrix}=\mathbf 0.
\label{eq:acoustic_gauge_Ward}
\end{align}
Only the long-wavelength form in Eq.(\ref{eq:null_vec_limit}) is a gradient expansion.

Apart from $\mathbf a_\mu$, we now assume that all physical $\eta$ fields are gapped, so that $D_{\eta\eta}$ is invertible in a neighborhood of $(\mathbf q,\omega)=(\mathbf 0,0)$. Integrating them out gives the acoustic $IGG$ action
\begin{align}
S^{IGG}=\sum_{\mathbf q}\int\frac{d\omega}{4\pi}
\sum_{\mu,\nu=0,x,y}
\mathbf a_\mu(-\mathbf q,-\omega)\Pi_{\mu\nu}(\mathbf q,\omega)
\mathbf a_\nu(\mathbf q,\omega),
\end{align}
where
\begin{align}
\Pi\equiv D_{\mathbf a\mathbf a}
-D_{\mathbf a\eta}D_{\eta\eta}^{-1}D_{\eta\mathbf a}.
\label{eq:Hdet_D_to_Pi}
\end{align}
The second row of Eq.(\ref{eq:acoustic_gauge_Ward}) gives
\begin{align}
\mathcal V_{0,\eta}=-D_{\eta\eta}^{-1}D_{\eta\mathbf a}\mathcal V_{0,\mathbf a},
\end{align}
and substitution into the first row yields the exact reduced Ward identity
\begin{align}
\Pi(\mathbf q,\omega)\mathcal V_{0,\mathbf a}(\mathbf q,\omega)=\mathbf 0.
\label{eq:exact_reduced_IGG_Ward}
\end{align}
By Hermiticity, $\mathcal V_{0,\mathbf a}^\dagger\Pi=0$ as well. Defining $q_\mu\equiv(-\omega,\mathbf q_x,\mathbf q_y)$ and keeping the leading gradient order, Eq.(\ref{eq:exact_reduced_IGG_Ward}) becomes the familiar continuum Ward identity
\begin{align}
\sum_\nu\Pi_{\mu\nu}(\mathbf q,\omega)q_\nu=0,
\qquad
\sum_\mu q_\mu\Pi_{\mu\nu}(\mathbf q,\omega)=0.
\end{align}
When the integrated-out $\kappa$ matter field and $\eta$ fields are all gapped, regularity constrains the leading long-wavelength kernel to the Chern--Simons plus Maxwell form in Eq.(\ref{eq:CS_Max_Pi}). In an approximate Hdet treatment this gives the $3\times3$ kernel $\Pi_{\mu\nu}$ containing $\mathbf a_0$. In an exact Hdet treatment, where Gauss's law is already implemented and $\mathbf a_0$ is absent, one obtains the spatial reduced kernel $\mathsf S_{uv}$ in Eq.(\ref{eq:reduced_CS_Max_kernel}).

The \emph{limitation} of the quadratic gauge action $S^{IGG}$ is that the underlying compactness of the $U(1)$-gauge field is missed, which can be traced back to the Gaussian order approximation in VMPI. But this is only a technical challenge. One could recover the compactness of the $U(1)$-gauge field by resorting to the original VMPI without expanding locally around a fixed $\bar\rho$. We will soon see such an example in Sec.\ref{sec:flat_connection}. For instance, a \emph{single} topological $U(1)$-gauge monopole event is not included in the Gaussian theory. In a compact $U(1)$ pure gauge theory without a CS term, it is well-known that the symmetry-allowed monopole events could lead to confinement of the gauge field\cite{Polyakov1977}. One can still use the original VMPI to compute the quantum numbers carried by monopoles, i.e., whether they are symmetry-allowed. 

It is well-known that if the monopole carries a nonzero $U(1)_Q$-symmetry charge (e.g., bosonic electron number), the single monopole event is forbidden in the presence of the $U(1)_Q$ symmetry: The photon phase is nothing but the dual description of the superfluid phase spontaneously breaking $U(1)_Q$ symmetry (denoted as $U(1)_Q$-SF phase), and the gapless photon is the $U(1)_Q$ Goldstone mode written in the gauge-field formulation\cite{Peskin1978, DasguptaHalperin1981, FisherLee1989}.

\subsubsection{Flat connections on a torus and Chern-Simons level}\label{sec:flat_connection}
The CS-term features the first time derivative. Parallel to the discussion for Type-I/Type-II Goldstone modes in our TDHF demo Eq.(\ref{eq:chi_kernel_expansion}--\ref{eq:B_ab_calculation}), here we show that the CS-level $k$ can be directly computed using the spatial gauge fields $\mathbf a_u(\mathbf q=0,t)$-- the flat connections. This point is particularly important for an \emph{exact} Hdet calculation, in which case one cannot read $k$ out from $\mathsf S(\mathbf q\neq 0,\omega)$ as in Eq.(\ref{eq:reduced_CS_Max_kernel}).

Define:
\begin{align}
\Pi_{uv}(\mathbf 0,\omega)=i\omega \mathsf B_{uv}+O(\omega^2)
\end{align}
The goal is to compute the $\mathsf B_{uv}$. First, we show that $i\omega\mathsf B_{uv}$ is just the $\omega$-linear term in the full kernel $D_{\mathbf a_u\mathbf a_v}$ in Eq.(\ref{eq:Hdet_full_D}). This is because in the Schur complement Eq.(\ref{eq:Hdet_D_to_Pi}), the acoustic gauge Ward identity Eq.(\ref{eq:acoustic_gauge_Ward}) at $\omega=0$ but arbitrary small $\mathbf q$ dictates that $D_{\mathbf a_u\eta}(\mathbf 0,0)=D_{\eta\mathbf a_u}(\mathbf 0,0)=0$. The second term in Eq.(\ref{eq:Hdet_D_to_Pi}) at most scales as $\omega^2$.

Consequently, we should compute $\mathsf B_{uv}$ directly from $D_{\mathbf a_u\mathbf a_v}$. Based on Eq.(\ref{eq:D_aa_element}), the small $\omega$-expansion $(-i\omega\hat{\mathcal F}+\hat{\mathcal G})_T^{-1}=\hat{\mathcal G}_T^{-1}+i\omega\hat{\mathcal G}_T^{-1}\hat{\mathcal F}\hat{\mathcal G}_T^{-1} +O(\omega^2)$ leads to:
\begin{align}
\mathsf B_{uv}=\frac{1}{\theta_u\theta_v}\llangle\kappa_u (\mathbf 0),\hat{\mathcal F}\kappa_v(\mathbf 0) \rrangle,
\end{align}
where we introduced small parameters $\theta_u\ll 1$ and defined:
\begin{align}
\kappa_u(\mathbf q=\mathbf  0)\equiv -\hat{\mathcal G}_T^{-1} \cdot [\theta_u\mathsf w^u(\mathbf q=\mathbf 0)].
\end{align}

Following the discussion below Eq.(\ref{eq:G_inv_superoperator}) and the definition of the $\hat{\mathcal F}$ superoperator Eq.(\ref{eq:gauge_Berry_F}), we arrive at:
\begin{align}
\mathsf B_{uv}=\frac{2}{\theta_u\theta_v}\left.{\rm Im}\big[\langle\Phi(\kappa_u)|\Phi(\kappa_v)\rangle\big]\right|_{\theta_u\theta_v-{\rm part}}.
\end{align}
If we parameterize $\mathbf a_u(\mathbf q=0,t)$ as the flux variable $\varphi_u$ as Eq.(\ref{eq:torus_flux_varphi}), $\mathsf B_{xy}$ is nothing but the Berry curvature of Hdet wavefunctions $|\Phi(\varphi_x,\varphi_y)\rangle$ that is obtained by adding flat connections on top of $\bar\rho$.

Note that due to the Gaussian order expansion, the prescription above is for the local Berry's curvature $\mathsf B_{uv}$ at a certain $\bar\rho$. To recover the compactness of the gauge flux, one simply computes $\mathsf B_{uv}(\varphi_x,\varphi_y)$ for a sequence of $\bar\rho(\varphi_x,\varphi_y)$ by varying $\varphi_x,\varphi_y$, and there is no fundamental principle forbidding the $\varphi_x,\varphi_y$-dependence in $\mathsf B_{uv}$:
\begin{align}
\mathsf B_{xy}(\varphi_x,\varphi_y)=2{\rm Im}[\langle\partial_{\varphi_x}\Phi(\varphi_x,\varphi_y)|\partial_{\varphi_y}\Phi(\varphi_x,\varphi_y)\rangle]\label{eq:B_xy_Hdet_formula}
\end{align}
In general, the Berry's phase part of the Lagrangian for $\varphi_x,\varphi_y$ is:
\begin{align}
&L_{\rm Berry}(\varphi_x,\varphi_y)=\mathsf A_x(\varphi_x,\varphi_y)\dot \varphi_x+\mathsf A_y(\varphi_x,\varphi_y)\dot \varphi_y,\notag\\
&\text{where }\partial_{\varphi_x}\mathsf A_y-\partial_{\varphi_y}\mathsf A_x=\mathsf B_{xy}(\varphi_x,\varphi_y).
\end{align}
The large gauge transformation leads to compactness $\varphi_u+2\pi\simeq\varphi_u$. When the matter fields are all gapped, $\varphi_u$ flux insertion does not cause many-body level crossing, and $\varphi_u+2\pi$ and $\varphi_u$ label the same physical Hdet state. Consequently, the CS-level is quantized to an integer $k$:
\begin{align}
\int_0^{2\pi} d\varphi_x \int _0^{2\pi}d\varphi_y \;\mathsf B_{xy}(\varphi_x,\varphi_y)=2\pi k.
\end{align}

It is worth noting that the ground state degeneracy $|k|$ $(k\neq0)$ is not modified when $\mathsf B_{xy}(\varphi_x,\varphi_y)$ is nonuniform, as long as $\mathsf B_{xy}(\varphi_x,\varphi_y)$ is sign-definite, possibly up to isolated zeros. Mathematically, this result can be shown via geometric quantization\cite{BatesWeinstein1997}. Physically, the simpler way to see the result is to perform a smooth change of variable for the $\varphi_x,\varphi_y$ space: $X(\varphi_x,\varphi_y),Y(\varphi_x,\varphi_y)$, with identification $X\simeq X+2\pi$,$Y\simeq Y+2\pi$ such that in the new coordinate $\mathsf B_{xy}(X,Y)=\frac{k}{2\pi}$ is uniform. Consequently, the Lagrangian is $L_{\rm Berry}(X,Y)=\frac{k}{2\pi}Y\dot X$, and canonical quantization leads to $[\hat Y,\frac{k}{2\pi}\hat X]=i$. The $2\pi$-periodicity of $\hat Y$ dictates that $\frac{k}{2\pi}\hat X$ can only take integer eigenvalues. The $2\pi$-periodicity of $\hat X$ means only $|k|$ states are physically distinct.

Nevertheless, if one computes $\mathsf B_{xy}(\varphi_x,\varphi_y)$ using Eq.(\ref{eq:B_xy_Hdet_formula}) by the mean-field wavefunction (i.e., at the 0th-order projective-expansion approximation), $\mathsf B^{\rm 0th}_{xy}(\varphi_x,\varphi_y)$ will be uniform (i.e., $\varphi_x,\varphi_y$-independent), and the CS-level $k$ is simply the summation of the Chern numbers carried by each parton mean-field state. $k=C^{(1)}+C^{(2)}$. It is very likely, although we cannot prove, that generally $k$ is not modified beyond the mean-field level even in an exact treatment. 

\begin{figure}
    \centering
    \includegraphics[width=0.95\linewidth]{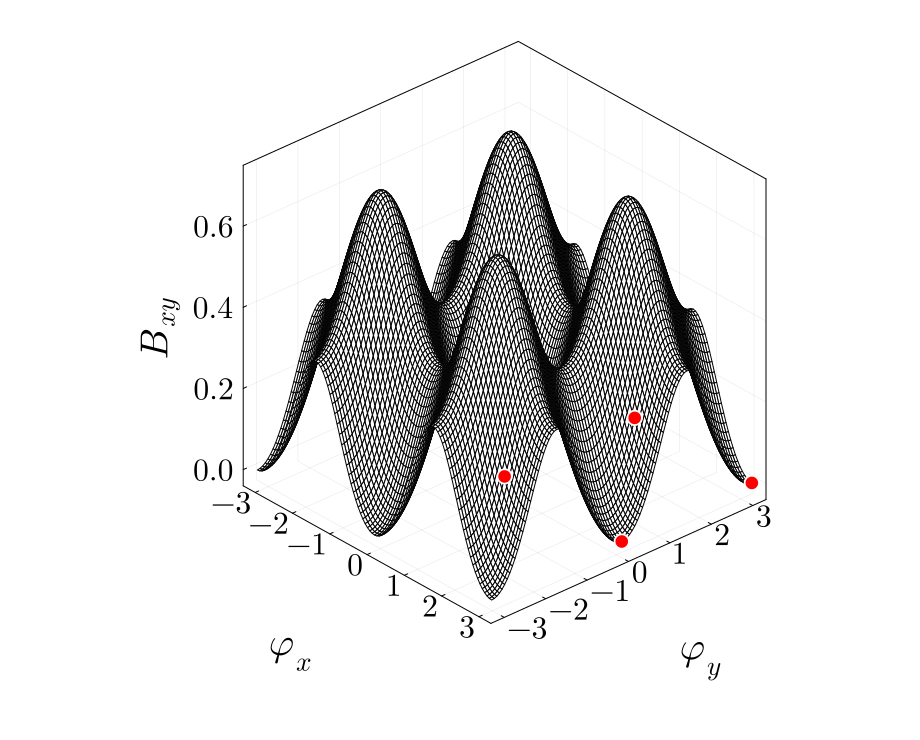}
    \caption{The exact Berry's curvature of flat-gauge-connections $\mathsf B_{xy}(\varphi_x,\varphi_y)$ for $\nu=\frac12$ Laughlin state on a square torus, independent of system size. Red dots mark the zeros at $(0,0)$,$(\pi,0)$,$(0,\pi)$,$(\pi,\pi)$.}
    \label{fig:B_xy_nu12}
\end{figure}

We have numerically checked that, however, for the exact bosonic Laughlin $\nu=1/2$ wavefunction (an Hdet wavefunction with the number of fusion channels $R=1$), $\mathsf B_{xy}(\varphi_x,\varphi_y)$ is non-uniform, does not change sign, and has isolated zeros, and the total integral remains quantized to $k=2$ as in the mean-field calculation. The simplest way to see this is to consider Haldane-Rezayi's construction for the Laughlin $\nu=\frac{1}{k}$ wavefunctions on a torus\cite{HR1985}, where the dependence of the boundary-conditions $(\varphi_x,\varphi_y)$ only appears in the center-of-mass (c.m) part of the wavefunction $F_{k}^{\rm c.m.}(Z;\varphi_x,\varphi_y)$ (Eq.12 in Ref\cite{HR1985}), where $Z\equiv \sum_{i=1}^{N_e} z_i$. Based on the parton construction, each parton is in a $k=1$ IQH state. Therefore, the c.m. part of the electronic wavefunction can be obtained as $F_{k=2}^{\rm c.m.(e)}(Z;\varphi_x,\varphi_y)=F_{k=1}^{\rm c.m.(1)}(Z;\varphi_x,\varphi_y)\cdot F_{k=1}^{\rm c.m.(2)}(Z;-\varphi_x,-\varphi_y)$, whose Berry's curvature $\mathsf B_{xy}(\varphi_x,\varphi_y)$ can be easily computed independent of system-size, as shown in Fig.\ref{fig:B_xy_nu12}.

To study the low-energy CS-term with gapless matter such as CFL, one should not integrate out all matter fields. Instead, only the high-energy gapped partons should be integrated out, and the gapless parton forming the Fermi surface should be kept. We leave a detailed Hdet analysis of gapless matter fields to a future project.

\subsubsection{Zero-point motion of gauge fields in VMPI wavefunctions}
In the previous subsection, we outlined how to use VMPI to compute the ground-state degeneracy in a topologically ordered Hdet state, using flat connections $\varphi_x$ and $\varphi_y$ of the gauge field. Although the computational scheme is well-anticipated, conceptually it clarifies why VMPI is fundamentally necessary for treating quantum many-body systems. If one treats variational wavefunctions classically, one would run into confusion: there are infinite possible $\varphi_x,\varphi_y$, each one labels a variational ground state. Does that mean there is an infinite number of ground states? 

Here we present another example in which VMPI helps solve a known puzzle in QSL and FCI. From the effective field theory of traditional Abrikosov fermion QSL construction (see Sec.\ref{sec:QSL} for connection with the general Hdet construction), it is known that the quantum spin state described by a gapped fermion band structure featuring Chern number $C^{(\uparrow)}=1$ for $f^{(\uparrow)}$-parton and $C^{(\downarrow)}=-1$ for $f^{(\downarrow)}$-parton actually describes a $XY$-ordered state spontaneously breaking the $\mathbf S_z$-conservation. But previous Monte Carlo simulation shows that after exact projection, the physical spin wavefunction for such a parton bandstructure only gives a power-law correlation function $\langle \Phi(\bar\rho) |\mathbf S^+_\mathbf r \mathbf S_0^-|\Phi(\bar\rho)\rangle\sim r^{-\eta}$\cite{RanVishwanathLee2008NeelDwaveTransitionProjectedTBI}, contradicting a true long-range order.

In the Hdet language, as explained in detail in Sec.\ref{sec:QSL}, we map the spin-1/2 system into a hard-core boson problem (i.e., $\mathbf S^+_\mathbf r\rightarrow b_\mathbf r^\dagger$), and perform a particle-hole transformation for $f^{(\downarrow)}$-parton only: $f^{(\downarrow)}\rightarrow h^{(\downarrow)\dagger}$. The local fusion rule is simple: $b^\dagger_{\mathbf r}=f^{(\uparrow)\dagger}_{\mathbf r}h^{(\downarrow)\dagger}_{\mathbf r}$ -- only one fusion channel per site. The spin state is now constructed by fusing the gapped mean-field state with $C^{(\uparrow)}=1$ for $f^{(\uparrow)}$ (carrying gauge charge $+1$) and $C^{(\downarrow)}=-1$ for $h^{(\downarrow)}$ (carrying gauge charge $-1$). The monopole operator inserts a $2\pi$ gauge flux, and consequently creates both a $f^{(\uparrow)}$-parton and $h^{(\downarrow)}$-parton -- a spin flip $\mathbf S^+$ operator.

This already indicates that one should \emph{not} expect to have a long-range spin correlation in a \emph{single} Hdet spin wavefunction $|\Phi(\bar\rho)\rangle$: It is the monopole-antimonopole correlation function that is supposed to have true long-range order, which requires that the spin wavefunction can capture gauge-field fluctuations so that, at least, the magnetic flux in a \emph{local} region can change by $\pm 2\pi$. But for a single Hdet state $|\Phi(\bar\rho)\rangle$, there is no gauge field fluctuation (zero-point motion).

Again, a correct method should include the gauge photon's zero-point motion, which is built into the VMPI treatment: both as a microscopic field theory and as microscopic wavefunctions. At the Gaussian order, although total magnetic flux through the torus sample cannot change, locally the flux can change $\pm 2\pi$. If one measures the spin-spin correlation function in the VMPI ground state $|\Psi\rangle_{\rm VMPI,GS}$ (see Eq.(\ref{eq:VMPI_GS_wavefunction})) -- a linear superposition of Hdet states that may be challenging to simulate exactly, the true long-range order will be reproduced. Here, since the spatial gauge field $\mathbf a_u$ is a part of $\chi$-field, Eq.(\ref{eq:kappa_chi_eigenmode_map_Hdet}) will be useful to translate its zero-point motion into the variational manifold labeled by $\kappa$.

\subsubsection{Special considerations for type-(B) models: Fine-Grid as the UV completion}\label{sec:special_type_B}
The VMPI and effective theories for Hdet wavefunctions that we have constructed so far use the full real-space variables $\rho_{ij},\chi_{ij}$, etc. A careful reader may have realized that this framework can be problematic for type-(B) models. To see the issue, let's focus on a type-(B) model within the LLL--as mentioned previously, an FCI model within a partially filled Chern number $\pm 1$ band can always be mapped to a model within the partially filled LLL (see Ref\cite{HuXiaoRan2024} for details). In such a type-(B) model, $i,j$ in $\rho^{(p)}_{ij}$ label the parton local orbitals $\{|z^{(p)}_{(m)}\rangle_{nLL}\}$ (i.e., generalized coherent states Eq.(\ref{eq:generalized_coherent_state_def}) but after the change of viewpoint). 

 If one allows $\rho_{ij}$ to fluctuate arbitrarily, as described by the Thouless parameterization Eq.(\ref{eq:Hdet_Thouless_param}), the electronic Hdet state $|\Phi(\rho)\rangle$ will \emph{not} be a legitimate many-body state inside the LLL. Based on the derivation of the local fusion tensor Eq.(\ref{eq:boson_general_fusion},\ref{eq:boson_general_CFL_fusion},\ref{eq:fermion_general_fusion},\ref{eq:fermion_general_CFL_fusion}), $|\Phi(\rho)\rangle$ is  an Hdet state within the LLL only if each parton's mean-field state \emph{before the change of viewpoint} is a Slater determinant in the parton's LL basis. 

Concretely, if we denote the Bloch-basis for parton-$p$'s $nLL$ as $|\phi^{(p)}_{\mathbf k}\rangle_{nLL}$ as in Sec.\ref{sec:type_B_model_FCI}, and the associated fermion operator as $\mathsf f^{(p)}_{n,\mathbf k}$, the parton RDM \emph{before the change of viewpoint} is:
\begin{align}
\rho^{(p)}_{m,n;\mathbf q}(\mathbf k)\equiv \langle \mathsf f^{(p)\dagger}_{n,\mathbf k+\mathbf q} \mathsf f^{(p)}_{m,\mathbf k}\rangle.
\end{align}
$\rho^{(p)}_{m,n;\mathbf q}(\mathbf k)$ fully determines the filled parton state before the change of viewpoint and $\rho^{(p)}_{ij}$ after the change of viewpoint: $\rho^{(p)}_{m,n,\mathbf q}(\mathbf k)$ strongly constrains the form of $\rho^{(p)}_{ij}$. 

For example, consider a Laughlin wavefunction where all partons only fill their lowest LL: $\rho^{(p)}_{m,n;\mathbf q}(\mathbf k)=1$ iff $\mathbf q=0$ and $m=n=0$, and otherwise $\rho^{(p)}_{m,n;\mathbf q}(\mathbf k)=0$. One can choose a single local orbital $|z^{(p)}_{(0)}\rangle_{0LL}$ per parton species-$p$ on each $z\in$Fine-Grid to describe the Laughlin state. Although there are $(N_s^{(e)})^2$ such orbitals per parton species, i.e., $\rho^{(p)}_{ij}$ is a $(N_s^{(e)})^2\times(N_s^{(e)})^2$ matrix, $\rho^{(p)}_{ij}$ is completely fixed.

Therefore, only the fluctuation described by $\rho^{(p)}_{m,n;\mathbf q}(\mathbf k)$ (\emph{not} a completely general $\rho_{ij}$) corresponds to fluctuations of electronic Hdet wavefunctions inside the LLL. The right way to view the Hdet wavefunctions described by a general $\rho_{ij}$ is that they are inside a UV-completed many-body Fock space -- \emph{the Fine-Grid Fock space}, larger than the original many-body Fock space inside the LLL.

Precisely speaking, given the type-(B) Hamiltonian $H_{LLL}$ within the LLL, one should consider a UV-completed lattice model defined on the Fine-Grid (FG) \emph{after the change of viewpoint}, where the electrons' discrete-coherent-state basis is considered as orthonormal (whose second-quantized operator is $c_z$):
\begin{align}
H_{\rm FG}\equiv H^{\rm FG}_{LLL}+\Delta \cdot\hat N_\perp,\label{eq:type_B_UV_completion}
\end{align}
where $H^{\rm FG}_{LLL}$ acting on the Fine-Grid Fock space is a Hamiltonian in terms of $c$-operators based on $H_{LLL}$, $\hat N_\perp$ is an operator measuring the electron number \emph{outside} the LLL, both of which will be defined shortly. $\Delta>0$ is an energy penalty. The point is that, in the $\Delta\rightarrow\infty$ limit, $H_{\rm FG}$ is mathematically equivalent to $H_{LLL}$ at finite energies. Consequently, in the VMPI and effective-theory treatment for type-(B) models, it is fine to allow general $\rho_{ij}$ fluctuations, but the Hamiltonian $H$ in the VMPI action Eq.(\ref{eq:Hdet_L_TDVP}) should be understood as $H_{\rm FG}$ (in the limit of $\Delta\rightarrow\infty$) instead of $H_{LLL}$. In this sense, all the previous discussions in the real-space basis, e.g., exact gauge Ward identities, directly apply to type-(B) models. 

Energetically, $\Delta\rightarrow\infty$ enforces the finite-energy fluctuations of $\rho^{(p)}_{ij}$ to be constrained by $\rho^{(p)}_{m,n;\mathbf q}(\mathbf k)$. In the VMPI discussion of Type-(A) model where the Hdet manifold is parameterized by $\rho^{(p)}_{ij}$, the gauge redundancies must be carefully dealt with so that, for example, nonphysical pure gauge modes are isolated and removed. However, if one chooses to use $\rho^{(p)}_{m,n;\mathbf q}(\mathbf k)$ to parameterize the Hdet manifold for the Type-(B) model (with a finite number of parton LLs), it appears to us that one effectively fixes the gauge and the pure-gauge modes are not present in our preliminary simulations. Consequently, the $a^{\rm full}_0$-insertion in VMPI may not be even needed in this case.

For example, in the $\rho^{(p)}_{m,n;\mathbf q}(\mathbf k)$ parameterization, the single-body parton Hamiltonian at $\mathbf q=0$ (i.e., respecting the commuting subgroup of the magnetic translation symmetry used to define the Bloch-basis $\mathbf k$) of a type-(B) FCI model is:
\begin{align}
\hat{\mathsf h}^{(p)}(\mathbf k)\equiv\sum_{n,m}\mathsf t^{(p)}_{nm}(\mathbf k)\mathsf f^{(p)\dagger}_{n,\mathbf k}\mathsf f^{(p)}_{m,\mathbf k},\;\mathsf t^{(p)}_{nm}(\mathbf k)\equiv\frac{\partial E(\rho)}{\partial\rho^{(p)}_{m,n;\mathbf q=0}(\mathbf k) }.\label{eq:type_B_eff_h}
\end{align}
Here, $\hat{\mathsf h}^{(p)}(\mathbf k)$ should be understood as $\hat{\mathsf h}^{(p)}(\mathbf k;\rho)$, since it is a Hamiltonian depending on $\rho^{(p')}_{m,n;\mathbf 0}(\mathbf k)$ for all parton species $p'=1,...,n_{\rm p}$. Finding the saddle point $\bar\rho$ boils down to the self-consistency condition: $\rho^{(p)}_{m,n;\mathbf q}(\mathbf k)$ is nonzero only for $\mathbf q=0$,  and the ground state of $\hat{\mathsf h}^{(p)}(\mathbf k)$ is also $\rho^{(p)}_{m,n;\mathbf 0}(\mathbf k)$. As advertised in Sec.\ref{sec:type_B_model_FCI}, the FCI physics causes the hybridization between the parton LLs, mathematically described by $\rho^{(p)}_{m,n;\mathbf 0}(\mathbf k)$. In addition, the finite-energy collective modes at $\mathbf q$ should be computed using VMPI via the Thouless parameters $\kappa^{(p)}_{m,n;\mathbf q}(\mathbf k)$ of $\rho^{(p)}_{m,n;\mathbf q}(\mathbf k)$ near the saddle.

Next, we sharply define $\hat N_\perp$ and $H^{\rm FG}_{LLL}$ in Eq.(\ref{eq:type_B_UV_completion}). $\hat N_\perp$ is defined as
\begin{align}
\hat N_\perp\equiv&\sum_{z,w\in {\rm FG}} c_z^\dagger\Big[\delta_{zw}-[p_{LLL}]_{zw}\Big]c_w,\;\;\text{ where:}\notag\\
[p_{LLL}]_{zw}\equiv& \frac{1}{N_s^{(e)}}\langle z^{(e)}|w^{(e)} \rangle_{0LL},
\end{align}
and $\langle z^{(e)}|w^{(e)} \rangle_{0LL}$ is the overlap between discrete-coherent states before the change of viewpoint in Eq.(\ref{eq:z_w_overlap}) (with $m=0$). Equivalently, if $\{|\phi^{(e)}_k\rangle\}$ is an orthonormal LLL basis before the change of viewpoint (e.g., the Bloch basis), then the matrix:
\begin{align}
[p_{LLL}]_{zw}=\frac{1}{N_s^{(e)}}\sum_{k}\langle z^{(e)}|\phi^{(e)}_k\rangle \langle \phi^{(e)}_k|w^{(e)}\rangle.
\end{align}
After the change of viewpoint, according to Eq.(\ref{eq:change_of_viewpoint}), one should interpret $|\phi^{(e)}_k\rangle$ as the linear superposition of an orthonormal basis $\{|z^{(e)}\rangle\}$:
\begin{align}
|\phi^{(e)}_k\rangle=\sum_{z\in {\rm FG}} \langle z^{(e)}|\phi^{(e)}_k\rangle \cdot |z^{(e)}\rangle,
\end{align}
and the factor $\frac{1}{N_s^{(e)}}$ in $[p_{LLL}]_{zw}$ is due to the fact that $\frac{1}{\sqrt{N_s^{(e)}}}|\phi^{(e)}_k\rangle$ is the normalized state after the change of viewpoint (see Eq.(\ref{eq:generalized_coherent_state_resolution_of_identity})): $\sum_{z,w\in {\rm FG}} c_z^\dagger[p_{LLL}]_{zw}c_w$ measures the electron number inside the LLL, defined as the subspace spanned by $\{\frac{1}{\sqrt{N_s^{(e)}}}|\phi^{(e)}_k\rangle\}$.

Now we define $H^{\rm FG}_{LLL}$ acting on the Fine-Grid Fock space. The physical requirement for $ H^{\rm FG}_{LLL}$ is that when restricted to the LLL Fock space, it is equivalent to the original $H_{LLL}$. Due to the overcompleteness of the discrete-coherent-state, $H^{\rm FG}_{LLL}$ satisfying this requirement is not unique. However, in Ref\cite{WangHaldane2019}, a canonical choice for $H^{\rm FG}_{LLL}$ is provided, together with a prescription to compute $H^{\rm FG}_{LLL}$ for an arbitrary finite-size sample.

When the torus size $L$ is much larger than the magnetic length: $L\gg l^{(e)}$, the form of $H^{\rm FG}_{LLL}$ is particularly simple. In this case\cite{WangHaldane2019}, up to exponentially small finite-size corrections, the prescription is to replace the LLL-projected density operator $\boldsymbol\rho^{(e)}(\mathbf q)=\bar{\boldsymbol\rho}^{(e)}(\mathbf q)e^{-\frac{\mathbf q^2(l^{(e)})^2}{4}}$ by the Fine-Grid density operator:
\begin{align}
\boldsymbol\rho^{(e)}(\mathbf q)\rightarrow \hat n^{(e)}_{\mathbf q,{\rm FG}}\equiv\sum_{z\in {\rm FG}}e^{i\mathbf q\cdot \mathbf r_z}\hat n_z^{(e)}, \;\;\hat n_z^{(e)}\equiv c_z^\dagger c_z.\label{eq:Wang_Haldane_prescription}
\end{align}
Here, $e^{-\frac{\mathbf q^2(l^{(e)})^2}{4}}$ is the form factor from the LLL projection, and $\bar{\boldsymbol\rho}^{(e)}(\mathbf q)$ are the operators satisfying the magnetic translation algebra (i.e., GMP-algebra\cite{GirvinMacDonaldPlatzman1986}). 

For example, let's consider a type-(B) FCI model inside the LLL involving a periodic potential breaking the Galilean invariance and a bare-Coulomb interaction:
\begin{align}
H_{LLL}&=\lambda\cdot\sum_{\mathbf Q\in Q_{\rm period}} \boldsymbol\rho^{(e)}(\mathbf Q)\notag\\
&+\frac{1}{2A}\sum^{q\in {\rm BZ_{FG}}}_{\mathbf q\neq 0} \frac{2\pi}{|\mathbf q|} :\boldsymbol\rho^{(e)}(\mathbf q)\boldsymbol\rho^{(e)}(-\mathbf q):,
\end{align}
where $A$ is the sample area, $Q_{\rm period}$ is a finite set of $\mathbf Q$ reciprocal vectors of the periodic potential while $\lambda$ is its strength, $\rm BZ_{FG}$ is the Brillouin Zone of the Fine-Grid. According to the prescription Eq.(\ref{eq:Wang_Haldane_prescription}), the corresponding Fine-Grid Hamiltonian is nothing but:
\begin{align}
H^{\rm FG}_{LLL}=\lambda\cdot\sum_{\mathbf Q\in Q_{\rm period}} \hat n^{(e)}_{\mathbf Q,{\rm FG}}+\frac{1}{2A}\sum^{\mathbf q\in {\rm BZ_{FG}}}_{\mathbf q\neq 0} \frac{2\pi}{|\mathbf q|} :\hat n^{(e)}_{\mathbf q,{\rm FG}}\hat n^{(e)}_{-\mathbf q,{\rm FG}}:.
\end{align}
In Sec.\ref{sec:FCI_saddle}, we will exactly follow this prescription to simulate FCI models.

\section{Simulating Hdet wavefunctions: Projective expansion}\label{sec:proj_expansion}
\subsection{Projective Expansion: General setup}\label{sec:PE_setup}
\subsubsection{PE as a bookkeeping device}
In this section, continuing the discussion of Hdet wavefunctions admitting a locality structure in Sec.\ref{sec:locality_def}, we introduce a general and practical framework to simulate these Hdet wavefunctions: the Projective-Expansion (PE). The Hdet wavefunction involves local fusion gates (or equivalently, projection) that drastically change the wavefunction from the mean-field state $|\Psi^{MF}\rangle$ (a direct product of Slater determinants of partons) to the Hdet state $|\Phi^{(e)}\rangle$ whose unnormalized version $|\Psi^{(e)}\rangle$ is given in Eq.(\ref{eq:general_Hdet_locality_state}). PE is a systematic way to approximately simulate the physical state $|\Phi^{(e)}\rangle$, improvable order by order via an expansion scheme. The first question that one may ask is: "What is PE expanding about?". 

Indeed, the full projection $\prod_i\hat{\mathbf C}_i$ has no intrinsically small parameter to expand about. As we will see, PE can be viewed as a bookkeeping device, with a formal expansion parameter $\epsilon$. For example, in PE for the expectation value $\langle\hat{O}_D^{(e)}\rangle$ of a physical operator $\hat{O}_D^{(e)}$ on an Hdet state ($D$ is the spatial region where the observable has nonzero support), we will define a prescription to introduce a real parameter $\epsilon$: $\langle\hat{ O}_D^{(e)}\rangle(\epsilon)$, such that when $\epsilon=0$, the mean-field result is obtained, while when $\epsilon=1$, the exactly projected Hdet result is obtained. The PE calculation boils down to computing the power series
\begin{align}
\langle\hat{ O}_D^{(e)}\rangle(\epsilon)=\langle\hat{O}_D^{(e)}\rangle_{(0)}+\epsilon\cdot \langle\hat{ O}_D^{(e)}\rangle_{(1)}+\epsilon^2\cdot \langle\hat{ O}_D^{(e)}\rangle_{(2)}+....
\end{align}
\emph{Throughout this paper, we use the subscript $_{(m)}$ to denote the $\epsilon$-expansion order $m$.} 

In practice, the PE will be truncated at a finite order, since the calculation becomes increasingly complicated as the order increases. Precisely, the calculation complexity follows a power-law w.r.t. both the system-size and the fusion-channel number $R$, whose exponents increase linearly as the PE order increases.  At the end of the calculation, one sets $\epsilon=1$ at the truncation order, yielding an approximate observable:
\begin{align}
\langle\hat{ O}_D^{(e)}\rangle_{[0]}=&\langle\hat{O}_D^{(e)}\rangle_{(0)},\notag\\
\langle\hat{ O}_D^{(e)}\rangle_{[1]}=&\langle\hat{O}_D^{(e)}\rangle_{(0)}+\langle\hat{O}_D^{(e)}\rangle_{(1)},\notag\\
\langle\hat{ O}_D^{(e)}\rangle_{[2]}=&\langle\hat{O}_D^{(e)}\rangle_{(0)}+\langle\hat{O}_D^{(e)}\rangle_{(1)}+\langle\hat{O}_D^{(e)}\rangle_{(2)},...
\end{align}
\emph{Throughout this section, we use the subscript $_{[m]}$ to denote the PE quantity after collecting terms up to order-$m$.} 

The 0th-order PE produces mean-field-level results. If one takes the infinite-order limit, the exactly projected Hdet observable is recovered. As we will see, even in the first few orders, PE already provides satisfactory improvement beyond mean-field. 

Conceptually, the PE is analogous to the $\epsilon$-expansion technique\cite{WilsonFisher1972} in field theory, which is very useful in computing critical exponents. In the $\epsilon$-expansion for the 3D classical Ising critical point, there is also no intrinsically small parameter. The physical spatial dimension is $3$, while the $\epsilon$-expansion considers a formal spatial dimension $4-\epsilon$, obtains a power series of $\epsilon$, and sets $\epsilon=1$ at a truncated order. The closest analogous technique to PE in literature, to our knowledge, is the diagrammatic-expansion technique for Gutzwiller wavefunctions\cite{Gebhard1990, Bunemann2012}, although it was developed for electronic d.o.f. in the absence of partons.

Following the VMPI formulation of the microscopic effective theories in Sec.\ref{sec:Hdet_eff}, one needs to be able to compute two kinds of quantities for \emph{normalized} Hdet states $|\Phi(\rho)\rangle$ to construct the microscopic VMPI action: 
\begin{enumerate}
\item the static observable $\langle \Phi(\rho)|\hat O^{(e)}_D|\Phi(\rho)\rangle$ (denoted as \emph{static-properties}).
\item the Berry's phase via the wavefunction overlap $\langle \Phi(\rho+\delta\rho)|\Phi(\rho)\rangle$ (denoted as \emph{dynamical-properties}).
\end{enumerate}
We will start with the general PE setup and the general diagrammatic technique for closed-form results. Then we apply PE in model systems for benchmarking tests. Before all this, we would like to emphasize that \emph{the PE approximation order can differ for static and dynamical properties.} For example, even if one uses the second-order PE to compute the static properties, one could still choose zeroth-order PE to compute Berry's phases: from VMPI, all we require is the approximation scheme to be statically and dynamically conserving; mixed orders of PE for the two properties are allowed choices.

\subsubsection{PE for static properties}
Now we describe the general PE prescription for static properties of Hdet states. First, we add a tunable real parameter into the local fusion gate Eq.(\ref{eq:onsite_fusion_gate}) for electronic orbital-$i$:
\begin{align}
&\hat {\mathbf C}_i(\epsilon)\equiv \exp(\gamma(\epsilon)^{\frac{1}{2}}\hat{\mathbf F}_i)\bullet(|0_i^{(e)}\rangle\langle 0_i^{(\text{all } p)}|)\notag\\
&=  \sum_{n=0}^{\infty} \gamma(\epsilon)^{\frac{n}{2}}\Big[\frac{\hat{\mathbf F}_i^n}{n!}\bullet|0_i^{(e)}\rangle\langle 0_i^{(\text{all } p)}|\Big]\equiv\sum_{n=0}^{\infty} \gamma(\epsilon)^{\frac{n}{2}}\hat{\mathbf C}_{i,n},\label{eq:onsite_fusion_gate_epsilon}
\end{align}
where, for later discussions, we defined $\hat{\mathbf C}_{i,n}$ based on electron number sectors. (Recall for fermionic electrons the summation of $n$ truncates to $n=1$. For bosonic electrons the summation truncates at a finite $n$ determined by the fusion channels. And with the change of viewpoint discussed in Sec.\ref{sec:change_of_viewpoint}, electron local orbitals are treated as orthonormal, even if the underlying electron basis is overcomplete.) 

Comparing with the original local fusion gate Eq.(\ref{eq:onsite_fusion_gate}), we multiplied a factor $\gamma(\epsilon)^\frac{1}{2}$ to the local fusion operator $\hat{\mathbf F}_i$, where $\gamma(\epsilon)$ is a positive real function that will be defined later. Clearly, the exact Hdet state obtained by the full projection $\prod_i\hat{\mathbf C}_i(\epsilon)$ only differs from that obtained by $\prod_i\hat{\mathbf C}_i$ by an overall factor $\gamma(\epsilon)^\frac{N_e}{2}$, and consequently it is perfectly fine to use $\hat{\mathbf C}_i(\epsilon)$ to construct the Hdet state $|\Phi(\rho)\rangle$.

The goal of the calculation is: 
\begin{align}
&\langle\hat O^{(e)}_D\rangle\equiv\langle\Phi(\rho)|\hat O^{(e)}_D|\Phi(\rho)\rangle\notag\\
=&\frac{\langle \Psi^{MF}(\rho)|\hat{\mathbf O}_D(\epsilon)\prod_{i\neq D}\hat{\mathbf P}_i(\epsilon)|\Psi^{MF}(\rho)\rangle}{\langle \Psi^{MF}(\rho)|\prod_i\hat{\mathbf P}_i(\epsilon)|\Psi^{MF}(\rho)\rangle},\label{eq:PE_goal}
\end{align}
where
\begin{align}
\hat{\mathbf P}_i(\epsilon)\equiv& \hat{\mathbf C}_i(\epsilon)^\dagger \hat{\mathbf C}_i(\epsilon),\notag\\
\hat {\mathbf O}_D(\epsilon)\equiv& \prod_{i\in D}\hat{\mathbf C}_i(\epsilon)^\dagger\cdot  \hat O_D^{(e)}\cdot\prod_{i\in D}\hat{\mathbf C}_i(\epsilon).\label{eq:PE_O_D_def}
\end{align}
At this point, although we formally introduced $\epsilon$-dependence in many operators, the goal $\langle\hat O^{(e)}_D\rangle$ is independent of $\epsilon$. Due to the contraction of the electronic d.o.f in $\hat{\mathbf C}_i(\epsilon)^\dagger \hat{\mathbf C}_i(\epsilon)$, $\hat{\mathbf P}_i(\epsilon)$ depends on $\epsilon$ through $\gamma(\epsilon)$, \emph{not} $\gamma(\epsilon)^\frac{1}{2}$. Similarly, $\hat {\mathbf O}_D(\epsilon)$ also only depends on $\gamma(\epsilon)$ for any particle-number conserving operator $\hat O^{(e)}_D$.

Precisely, because different electron-number sectors are orthogonal, we have:
\begin{align}
\hat{\mathbf C}^\dagger_{i,n}\hat{\mathbf C}_{i,m}=\hat{\mathbf C}^\dagger_{i,n}\hat{\mathbf C}_{i,n}\delta_{mn},
\end{align}
and consequently,
\begin{align}
\hat{\mathbf P}_i(\epsilon)=\sum_{n=0}^{\infty}\gamma(\epsilon)^n\Big[\hat{\mathbf C}^\dagger_{i,n}\hat{\mathbf C}_{i,n}\Big]\equiv \sum_{n=0}^{\infty}\gamma(\epsilon)^n\hat{\mathbf P}_{i,n},\label{eq:PE_P_i_as_P_i_n}
\end{align}
where the operator $\hat{\mathbf P}_{i,n}\equiv\hat{\mathbf C}^\dagger_{i,n}\hat{\mathbf C}_{i,n}$ acts on the local parton sub-Fock space with exactly $n$ partons per species. Therefore, $\hat{\mathbf P}_{i,n}\hat{\mathbf P}_{i,m}=0$, if $m\neq n.$, leading to
\begin{align}
[\hat{\mathbf P}_{i,n},\hat{\mathbf P}_{i,m}]=0,\;\;\forall n,m.\label{eq:P_i_n_commuting}
\end{align}
We also introduce the $\epsilon$-expansion of $\hat {\mathbf O}_D(\epsilon)$ operator:
\begin{align}
\hat {\mathbf O}_D(\epsilon)=\sum_{m\ge 0}\epsilon^m\hat {\mathbf O}_{D,(m)}
\end{align}

Next, we define an operator $\hat{\mathbf Q}_i(\epsilon)$ acting on the local \emph{parton Fock space} interpolating between the identity operator $\hat{\mathbf 1}$ and $\hat{\mathbf P}_i(\epsilon)$:
\begin{align}
\hat{\mathbf Q}_i(\epsilon)\equiv\hat{\mathbf 1}+\epsilon\cdot\Big(\hat{\mathbf P}_i(\epsilon)-\hat{\mathbf 1}\Big).\label{eq:PE_Q_def}
\end{align}
We then replace $\hat{\mathbf P}_i(\epsilon)$ in Eq.(\ref{eq:PE_goal}) by $\hat{\mathbf Q}_i(\epsilon)$, and define:
\begin{align}
&\langle\hat O^{(e)}_D\rangle(\epsilon)\equiv\frac{\langle \Psi^{MF}(\rho)|\hat{\mathbf O}_D(\epsilon)\prod_{i\notin D}\hat{\mathbf Q}_i(\epsilon)|\Psi^{MF}(\rho)\rangle}{\langle \Psi^{MF}(\rho)|\prod_i\hat{\mathbf Q}_i(\epsilon)|\Psi^{MF}(\rho)\rangle},\label{eq:PE_goal_epsilon}
\end{align}
$\langle\hat O^{(e)}_D\rangle(\epsilon)$ will depend on $\epsilon$. Since $\hat{\mathbf Q}_i(1)=\hat{\mathbf P}_i(1)$, we have:
\begin{align}
\langle\hat O^{(e)}_D\rangle(\epsilon=1)=\langle\hat O^{(e)}_D\rangle
\end{align}

If we know the function $\gamma(\epsilon)$, one can expand $\langle\hat O^{(e)}_D\rangle(\epsilon)$ in Eq.(\ref{eq:PE_goal_epsilon}) as a power series of $\epsilon$ by expanding both the numerator and denominator, as advertised. The technical details of the expansion will be given shortly in Sec.\ref{sec:PE_linked_cluster}. The condition to fix $\gamma(\epsilon)$ is the electron-number sum-rule:
\begin{align}
\sum_i\langle \hat n_i^{(e)}\rangle (\epsilon)=N_e, \;\;\;\forall\epsilon.\label{eq:PE_N_sum_rule}
\end{align}
Practically, one should consider the power series expansion of $\gamma(\epsilon)$:
\begin{align}
\gamma(\epsilon)=\gamma_{(0)}+\epsilon\cdot\gamma_{(1)}+\epsilon^2\cdot\gamma_{(2)}+...,
\end{align}
where expansion coefficients $\gamma_{(m)}$'s are determined by Eq.(\ref{eq:PE_N_sum_rule}) order by order.

For instance, for the 0th order $\langle \hat n_i^{(e)}\rangle_{(0)}$, all $\hat{\mathbf Q}_i(\epsilon)$ in Eq.(\ref{eq:PE_goal_epsilon}) will be identity since $\epsilon=0$ should be taken. $\hat {\mathbf n}_i(\epsilon=0)=\hat{\mathbf C}_i^\dagger(\epsilon=0) \hat n_i^{(e)}\hat{\mathbf C}_i(\epsilon=0)$ will be proportional to $\gamma_{(0)}$ for fermionic electrons. $\sum_i\langle \hat n_i^{(e)}\rangle_{(0)}=N_e$ can fix $\gamma_{(0)}$, which can be viewed as the mean-field fugacity for electrons. Due to Eq.(\ref{eq:PE_N_sum_rule}), we have $\sum_i\langle \hat n_i^{(e)}\rangle_{(m)}=0$ for $m\ge1$, which will fix higher order $\gamma_{(m)}$'s. Iteratively, PE for the expectation value of any electron operator $\hat O^{(e)}_D$ is well-defined. 

\emph{Crucially}, one only needs to compute operator expectation values on the mean-field state in PE (see Eq.(\ref{eq:PE_goal_epsilon})), which can be done straightforwardly using Wick's theorem. Naturally, \emph{PE is a statically-conserving approximation at any order} (see Eq.(\ref{eq:stat_conserving_def}) for definition).

\subsubsection{PE for dynamical properties: quantum geometric tensor}\label{sec:PE_dynamical}
We now move on to computing the Hdet wavefunction overlaps using PE, for the purpose of VMPI. Consider two parton RDMs: $\rho_f$ and $\rho_i$, both in the vicinity of the saddle point $\bar\rho$. We define the unnormalized PE overlap:
\begin{align}
\mathcal N_{fi}(\epsilon)\equiv \langle\Psi^{MF}(\rho_f)|\prod_i\hat{\mathbf Q}_i(\epsilon)|\Psi^{MF}(\rho_i)\rangle.
\end{align}
Here the $\gamma(\epsilon)$ function in $\hat{\mathbf Q}_i(\epsilon)$ can be determined by the particle-number sum rule Eq.(\ref{eq:PE_N_sum_rule}) on the saddle point mean-field state $|\Psi^{MF}(\bar\rho)\rangle$. In this prescription, $\gamma(\epsilon)$ is independent of $\rho_i,\rho_f$.

The normalized PE overlap is defined as:
\begin{align}
\mathcal S_{fi}(\epsilon)\equiv \frac{\mathcal N_{fi}(\epsilon)}{\sqrt{\mathcal N_{ff}(\epsilon)\cdot \mathcal N_{ii}(\epsilon)}}.
\end{align}
We have the exact limit: $\mathcal S_{fi}(\epsilon=1)=\langle \Phi(\rho_f)|\Phi(\rho_i)\rangle$.

For the purpose of VMPI, the central quantity of interest is Berry's curvature $F_{ab}$ defined in Eq.(\ref{eq:general_Berry_matrix}). For simplicity of presentation, we will instead compute the full quantum geometric tensor $\mathcal Q_{ab}$, defined as follows. Consider two real numbers $a,b$ that parameterize the two states $\rho_f(a)$ and $\rho_i(b)$, 
\begin{align}
\mathcal Q_{ab}(\epsilon)\equiv& \partial_a\partial_b \left.\log\mathcal S_{fi}(\epsilon)\right|_{\rho_f=\rho_i=\bar\rho}\notag\\
=&\partial_a\partial_b \left.\log\mathcal N_{fi}(\epsilon)\right|_{\rho_f=\rho_i=\bar\rho},\label{eq:PE_Q_tensor}
\end{align}
where the second equality follows the fact that $N_{ff}$ ($N_{ii}$) only depends on $a$ ($b$). The exact quantum metric and Berry's curvature introduced in Eq.(\ref{eq:general_quantum_metric},\ref{eq:general_Berry_matrix}) correspond to the real and imaginary parts of $\mathcal  Q_{ab}(\epsilon=1)$:
\begin{align}
\mathcal  Q_{ab}(\epsilon=1)=g_{ab}(\bar\rho)+\frac{i}{2}F_{ab}(\bar\rho).
\end{align}

Based on the Thouless parameterization for the tangent space $T_{\bar\rho}$, we know that any small deformation of $\rho$ near $\bar\rho$ is equivalent to applying a fermion bilinear operator on the mean-field state:
\begin{align}
\partial_a\left.|\Psi^{MF}(\rho_f(a))\rangle\right|_{\rho_f=\bar\rho}\equiv& \hat X_a|\Psi^{MF}(\bar\rho)\rangle,\notag\\
\partial_b\left.|\Psi^{MF}(\rho_i(b))\rangle\right|_{\rho_i=\bar\rho}\equiv& \hat X_b|\Psi^{MF}(\bar\rho)\rangle,\label{eq:PE_partial_a}
\end{align}
where $\hat X_a,\hat X_b$ are fermion bilinears of partons. Using Eq.(\ref{eq:PE_Q_tensor}), one finds that:
\begin{align}
&\mathcal Q_{ab}(\epsilon)=\frac{\langle\Psi^{MF}(\bar\rho)|\hat X_a^\dagger\hat{\mathbb Q}(\epsilon)\hat X_b|\Psi^{MF}(\bar\rho)\rangle}{\langle\Psi^{MF}(\bar\rho)|\hat{\mathbb Q}(\epsilon)|\Psi^{MF}(\bar\rho)\rangle}\notag\\
&-\frac{\langle\Psi^{MF}(\bar\rho)|\hat X_a^\dagger\hat{\mathbb Q}(\epsilon)|\Psi^{MF}(\bar\rho)\rangle}{\langle\Psi^{MF}(\bar\rho)|\hat{\mathbb Q}(\epsilon)|\Psi^{MF}(\bar\rho)\rangle}\frac{\langle\Psi^{MF}(\bar\rho)|\hat{\mathbb Q}(\epsilon)\hat X_b|\Psi^{MF}(\bar\rho)\rangle}{\langle\Psi^{MF}(\bar\rho)|\hat{\mathbb Q}(\epsilon)|\Psi^{MF}(\bar\rho)\rangle},\label{eq:PE_Q_tensor_X}
\end{align}
where we defined a shorthand notation:
\begin{align}
\hat{\mathbb Q}(\epsilon)\equiv\prod_i\hat{\mathbf Q}_i(\epsilon).\label{eq:PE_bulk_Q}
\end{align}
Obviously, $\hat{\mathbb Q}(\epsilon=0)=\hat{\mathbf 1}$, and the zeroth order $\mathcal Q_{ab}(\epsilon=0)$ reduces to the mean-field quantum geometric tensor. Iteratively, the finite-order PE for $\mathcal Q_{ab}(\epsilon)$ can be performed in the same fashion as the static observable $\langle \hat O^{(e)}_D\rangle(\epsilon)$. The technical details will be presented shortly.

Naturally, \emph{PE is a dynamically-conserving approximation at any order} (see Eq.(\ref{eq:dyna_conserving_def}) for definition).

\subsection{Technical details}\label{sec:PE_technical}
\subsubsection{Ordered-cumulants and the linked-cluster theorem}\label{sec:PE_linked_cluster}
We have formulated PE mathematically in Eq.(\ref{eq:PE_goal_epsilon}) and Eq.(\ref{eq:PE_Q_tensor_X}), for static and dynamical properties, respectively. Together with the sum rule Eq.(\ref{eq:PE_N_sum_rule}) that fixes $\gamma(\epsilon)$, PE is already well-defined. However, it is worth mentioning some useful technical details in practical calculations. It turns out that PE is organized via the linked-cluster theorem\cite{Kubo1962} as in many other contexts of quantum many-body theories, from which we can obtain the closed-form for an arbitrary order PE quantity.

To facilitate the discussion, we will start with introducing some notations and definitions. For any operator $\hat A$ acting on the parton Hilbert space, define:
\begin{align}
\langle \hat A \rangle_\rho\equiv \langle \Psi^{MF}(\rho)|\hat A|\Psi^{MF}(\rho)\rangle,
\end{align}
where $|\Psi^{MF}(\rho)\rangle$ is a \emph{normalized} direct product of parton's Slater determinants.
The local PE operator $\hat{\mathbf Q}_i(\epsilon)$ in Eq.(\ref{eq:PE_Q_def}) can be expanded as a power series:
\begin{align}
\hat{\mathbf Q}_i(\epsilon)\equiv \hat{\mathbf 1}+\sum_{m\geq 1} \epsilon^m \hat{\boldsymbol q}_{i,(m)}.\label{eq:PE_Q_expansion}
\end{align}
The detailed form of the operator $\hat{\boldsymbol q}_{i,(m)}$ will depend on the fusion gate $\hat {\mathbf C}_i$ and $\gamma(\epsilon)$, however, it is important to note that all $\hat{\boldsymbol q}_{i,(m)}$ commute with each other:
\begin{align}
[\hat{\boldsymbol q}_{i,(m)},\hat{\boldsymbol q}_{j,(n)}]=0,\;\;\forall i,j,m,n. \label{eq:q_i_m_commuting}
\end{align}
If $i\neq j$, the operators are acting on different electronic orbitals and commute. If $i=j$, the commuting relation follows from Eq.(\ref{eq:P_i_n_commuting}), since any $\hat{\boldsymbol q}_{i,(m)}$ will be a linear superposition of $\hat{\mathbf 1}$ and $\hat{\mathbf P}_{i,n}$'s. 

For later purposes, we also introduce the logarithm of $\hat{\mathbf Q}_i(\epsilon)$:
\begin{align}
\log\hat{\mathbf Q}_i(\epsilon)\equiv \hat {\mathbf L}_i(\epsilon)\equiv \sum_{m\ge 1}\epsilon ^m \hat{\boldsymbol l}_{i,(m)}.\label{eq:PE_L_l_def}
\end{align}
Obviously,
\begin{align}
[\hat{\boldsymbol l}_{i,(m)},\hat{\boldsymbol l}_{j,(n)}]=0,\;\;\forall i,j,m,n,\label{eq:l_i_m_commuting}
\end{align}
since $\hat{\boldsymbol l}_{i,(m)}$ can be represented in terms of $\hat{\boldsymbol q}_{i,(n)}$. For example:
\begin{align}
\hat{\boldsymbol l}_{i,(1)}=&\hat{\boldsymbol q}_{i,(1)},&\hat{\boldsymbol l}_{i,(2)}=&\hat{\boldsymbol q}_{i,(2)}-\frac12 \hat{\boldsymbol q}_{i,(1)}^2, &...\label{eq:PE_l_using_q}
\end{align}

\textbf{Ordered-cumulants:} For any order list of operators, $\hat A_1,\hat A_2,...,\hat A_n$, we now define the ordered-cumulants as:
\begin{align}
\mathcal C_\rho(\hat A_1,...,\hat A_n)\equiv\left.\partial_{s_1}...\partial_{s_n}G_{\{\hat A_l\}}(\{s_l\})\right|_{s_1=...=s_n=0},\label{eq:PE_cumulant_def}
\end{align}
where
\begin{align}
G_{\{\hat A_l\}}(\{s_l\})\equiv \log\left\langle e^{s_1\hat A_1}...e^{s_n\hat A_n}\right\rangle_{\rho}\label{eq:PE_cumulant_G_s_def}
\end{align}
If we denote 
\begin{align}
\langle \hat A_{a_1}\hat A_{a_2}...\hat A_{a_m}\rangle_\rho\equiv M_{a_1a_2...a_m},
\end{align}
the first few ordered-cumulants are:
\begin{align}
&\mathcal C_\rho(\hat A_1)=M_1,\notag\\
&\mathcal C_\rho(\hat A_1,\hat A_2)=M_{12}-M_1M_2,\notag\\
&\mathcal C_\rho(\hat A_1,\hat A_2,\hat A_3)=M_{123}-M_{1}M_{23}-M_{12}M_{3}\notag\\
&\;\;\;\;\;\;\;\;\;\;\;\;\;\;\;\;\;\;\;\;\;\;\;\;\;\;-M_{13}M_{2}+2M_{1}M_{2}M_3,\notag\\
&\mathcal C_\rho(\hat A_1,\hat A_2,\hat A_3,\hat A_4)=
M_{1234}-M_1M_{234}-M_{12}M_{34}\notag\\
&-M_{13}M_{24}-M_{14}M_{23}-M_{123}M_4-M_{124}M_3-M_{134}M_2\notag\\
&+2M_1M_2M_{34}+2M_1M_3M_{24}+2M_1M_4M_{23}+2M_{12}M_3M_4\notag\\
&+2M_{13}M_2M_4+2M_{14}M_2M_3-6M_1M_2M_3M_4 .\label{eq:PE_cumulants_first_few_orders}
\end{align}
Namely, disconnected expectation-value terms are all removed, and only linked clusters survive.

The ordered-cumulant features some useful properties, which we list below. 

\textbf{(1) Cumulant is multilinear:}
\begin{align}
\mathcal C_\rho(..,a\hat A+b\hat B,..)=a\mathcal C_\rho(..,\hat A,..)+b\mathcal C_\rho(..,\hat B,..).
\end{align}

\textbf{(2) Cumulant for commuting operators:}
\begin{align}
&\text{If }[\hat A_{l},\hat A_{l+1}]=0,\notag\\
&\text{then }\mathcal C_\rho(..,\hat A_l,\hat A_{l+1},..)=\mathcal C_\rho(..,\hat A_{l+1},\hat A_{l},..).
\end{align}

\textbf{(3) Cumulant for identity operator:} If $\hat A_l=\hat{\mathbf 1}$ for some $l$, all higher order 
\begin{align}
\mathcal C_\rho(..,\hat{\mathbf 1},..)=0
\end{align}
except for $\mathcal C_\rho(\hat{\mathbf 1})=1$. 

\textbf{(4) Cumulant for uncorrelated operators: } In our PE simulation of Hdet states, $\rho$ is a direct product of Slater determinants for each parton species. For example, for two-parton case, $\rho=\rho^{(1)}\otimes \rho^{(2)}$. Consider a sequence of operators: $\{\hat A_l\}$ ($l=1,..,n$), \emph{every $\hat A_l$ only acts on a single species parton-$p_l$'s Hilbert space}. It is easy to show that:
\begin{align}
\mathcal C_\rho(\hat A_1,...,\hat A_n)=0,\text{ unless all $p_l$ are identical.}\label{eq:PE_cumulant_uncorrelated}
\end{align}

\textbf{(5) Product-cumulant identities:} 
\begin{align}
&\mathcal C_\rho(\hat X_1,...,\hat X_p,\hat A_1\cdot \hat A_2,\hat Y_1,...,\hat Y_q)\notag\\
=&\mathcal C_\rho(\hat X_1,...,\hat X_p,\hat A_1, \hat A_2,\hat Y_1,...,\hat Y_q)\notag\\
+&\sum_{S\subseteq\{1,...,p\}}\sum_{T\subseteq\{1,...,q\}}\mathcal C_\rho(\hat {\mathbf X}_S,\hat A_1,\hat {\mathbf Y}_T)\cdot \mathcal C_\rho(\hat {\mathbf X}_{S^C},\hat A_2,\hat {\mathbf Y}_{T^C}),\label{eq:PE_product_cumulant}
\end{align}
where $\hat {\mathbf X}_S$ means the subsequence of $\hat X_1,...,\hat X_p$ selected by the subset $S$\cite{LeonovShiryaev1959}. Superscript-$^C$ means complement. For example:
\begin{align}
&\mathcal C_\rho(\hat A_1\cdot \hat A_2,\hat Y)\notag\\
=&\mathcal C_\rho(\hat A_1,\hat A_2,\hat Y)+\mathcal C_\rho(\hat A_1,\hat Y)\mathcal C_\rho(\hat A_2)+\mathcal C_\rho(\hat A_1)\mathcal C_\rho(\hat A_2,\hat Y).
\end{align}
As applications of Eq.(\ref{eq:PE_product_cumulant}), for $\hat A,\hat B$ satisfying $\hat A\hat B=\hat A$, we have identities:
\begin{align}
\mathcal C_\rho(\hat A,\hat B)=&(1-\langle\hat B\rangle_\rho)\langle \hat A\rangle_\rho\notag\\
\mathcal C_\rho(\hat A,\hat B,\hat X)=&(1-\langle\hat B\rangle_\rho)\mathcal C_\rho(\hat A,\hat X)-\langle\hat A\rangle_\rho \mathcal C_\rho(\hat B,\hat X)\notag\\
\mathcal C_\rho(\hat X,\hat A,\hat B)=&(1-\langle\hat B\rangle_\rho)\mathcal C_\rho(\hat X,\hat A)-\langle\hat A\rangle_\rho \mathcal C_\rho(\hat X,\hat B).\label{eq:PE_cumulant_AB_A}
\end{align}
For the special case $\hat A^2=\hat A$:
\begin{align}
\mathcal C_\rho(\hat A,\hat A,\hat X)=&(1-2\langle\hat A\rangle_\rho)\mathcal C_\rho(\hat A,\hat X)\notag\\
\mathcal C_\rho(\hat X,\hat A,\hat A)=&(1-2\langle\hat A\rangle_\rho)\mathcal C_\rho(\hat X,\hat A).\label{eq:PE_cumulant_idempotent}
\end{align}

\textbf{(6) Cumulant-bound:}
Physically, if $\rho$ describes a gapped short-range state with correlation length $\xi$, let $\hat A_1,\hat A_2,...,\hat A_n$ be bounded local operators with supports $D_1,...,D_n$, then the ordered-cumulant will decay exponentially for a bipartition of the operators that are far apart in the real-space\cite{HastingsKoma2006, AmpelogiannisDoyon2022}:
\begin{align}
|\mathcal C_\rho(\hat A_1,...,\hat A_n)|< C_n\prod_l\|\hat A_l\|\exp\Big[-\frac{{\rm dist}(D_I,D_{I^C})}{\xi}\Big],\label{eq:PE_cumulant_bound}
\end{align}
where, $\|\hat A\|\equiv {\rm sup}_{\|\psi\|=1}\|\hat A|\psi\rangle\|$ is the operator norm of $\hat A$, $C_n$ is some constant independent of $\{\hat A_l\}$. $I$ and $I^C$ denote a bipartition of operators: $I\cup I^C=\{1,2,..,n\}$, and 
\begin{align}
D_I\equiv& \cup_{l\in I} D_l,&D_{I^C}\equiv& \cup_{l\notin I} D_l.
\end{align}
\emph{As an intuitive interpretation, the linked clusters have a real-space meaning: any given operator must be adjacent to some other operator to have a sizable cumulant}. 

\textbf{(7) Cumulant factorization:} This is a consequence of the combination of the cumulant-bound Eq.(\ref{eq:PE_cumulant_bound}) and the product-cumulant identity Eq.(\ref{eq:PE_product_cumulant}). For a gapped short-range state, consider an operator $\hat A=\hat A_1\cdot \hat A_2$: if the supports of $\hat A_1$ and $\hat A_2$: $D_{A_1}$ and $D_{A_2}$ are far apart:
\begin{align}
{\rm dist}(D_{A_1},D_{A_2})=R\gg\xi, 
\end{align}
assuming $\hat X_i$ and $\hat Y_j$ each has a support whose length-scale $\lesssim \xi$, then a fixed finite-order cumulant factorizes:
\begin{align}
&\mathcal C_\rho(\hat X_1,...,\hat X_p,\hat A_1\cdot \hat A_2,\hat Y_1,...,\hat Y_q)\notag\\
=&\sum_{S\subseteq\{1,...,p\}}\sum_{T\subseteq\{1,...,q\}}\mathcal C_\rho(\hat {\mathbf X}_S,\hat A_1,\hat {\mathbf Y}_T)\notag\\
&\cdot \mathcal C_\rho(\hat {\mathbf X}_{S^C},\hat A_2,\hat {\mathbf Y}_{T^C})+O(e^{-\frac{R}{(p+q+1)\xi}}),\label{eq:PE_cumulant_factorize}
\end{align}
because $\mathcal C_\rho(\hat X_1,...,\hat X_p,\hat A_1, \hat A_2,\hat Y_1,...,\hat Y_q)= O(e^{-\frac{R}{(p+q+1)\xi}})$ (the denominator in the exponent is from the minimal bipartition distance achieved by equally spaced $X,Y$ operators between $A_1,A_2$).
 For example:
\begin{align}
&\mathcal C_\rho(\hat A_1\cdot \hat A_2,\hat Y)\notag\\
=&\mathcal C_\rho(\hat A_1,\hat Y)\mathcal C_\rho(\hat A_2)+\mathcal C_\rho(\hat A_1)\mathcal C_\rho(\hat A_2,\hat Y)+O(e^{-\frac{R}{2\xi}}).
\end{align}

\textbf{(8) Full Taylor expansion of $G_{\{\hat A_l\}}(\{s_l\})$:}Ordered-cumulants also encode the full Taylor expansion of $G_{\{\hat A_i\}}(\{s_i\})$ defined in Eq.(\ref{eq:PE_cumulant_G_s_def}):
\begin{align}
&G_{\{\hat A_l\}}(\{s_l\})\notag\\
=&\sum_{k_1,k_2,...\ge0}\Big[\prod_{l=1}^n\frac{s_l^{k_l}}{k_l!}\Big]\cdot\mathcal C_\rho(\hat A_1^{[k_1]},\hat A_2^{[k_2]},...,\hat A_n^{[k_n]}),\notag\\
\equiv&\sum_{\{k_l\ge0\}}\Big[\prod_{l=1}^n\frac{s_l^{k_l}}{k_l!}\Big]\cdot\mathcal C_\rho(\{\hat A_l^{[k_l]}\}_{l\in\{1,..,n\}})\label{eq:cumulant_full_Taylor}
\end{align}
where the notation $\hat A_l^{[k_l]}$ means a consecutive ordered block of $k_l$ copies of $\hat A_l$ with empty blocks being omitted, and we introduced shorthand notation in the last line. For instance:
\begin{align}
\mathcal C_\rho(\hat A_1^{[2]},\hat A_4^{[3]})\equiv \mathcal C_\rho(\hat A_1,\hat A_1,\hat A_4,\hat A_4,\hat A_4).
\end{align}
We also \emph{define} that when $k_l=0$, $\forall l$, the empty cumulant is zero: $\mathcal C_\rho(\text{empty})\equiv 0$.

To prove Eq.(\ref{eq:cumulant_full_Taylor}), one may first introduce more independent variables $e^{s_l\hat A_l}\rightarrow e^{s_{l,1}\hat A_l}...e^{s_{l,k_l}\hat A_l}$. Performing the first derivative for each variable $s_{l,\alpha}$ once is equivalent to performing $k_l$-th derivative on $s_l$, i.e.:
\begin{align}
\left.\partial_{s_1}^{k_1}...\partial_{s_n}^{k_n}G_{\{\hat A_i\}}(\{s_i\})\right|_{\{s_l=0\}}=\mathcal C_\rho(\hat A_1^{[k_1]},...,\hat A_n^{[k_n]}).
\end{align}

\subsubsection{Cumulant formulation of PE}
\textbf{Summation notations: } We introduce a few useful notations for the summations in the cumulant expansion. The primed summation means the summation of \emph{ordered sequences} of pairwise distinct electronic orbitals: 
\begin{align}
\sum'_{i_1,i_2,...i_k} \text{means } i_a\neq i_b,\forall a\neq b.
\end{align}
Here, ordered sequence means that the sequences $2,5,3$ and $3,2,5$ are different and both will appear in the sum.
For a subset $R$ of all electronic orbitals in the system, we denote its complement by $R^C$; $R\cup R^C$ contains all electronic orbitals. We define the primed summation with the superscript "$R$" as the summation of pairwise distinct electronic orbitals \emph{all} in $R$:
\begin{align}
\sum'^{R}_{i_1,i_2,...i_k}\equiv \sum'_{i_1,i_2,...i_k\in R}.
\end{align}
Finally, we define the primed summation with the superscript "$(R)$" as the summation of pairwise distinct electronic orbitals in which at least one orbital $\in R$:
\begin{align}
\sum'^{(R)}_{i_1,i_2,...i_k}\equiv \sum'_{i_1,i_2,...i_k}-\sum'^{R^C}_{i_1,i_2,...i_k}.\label{eq:PE_notation_(R)}
\end{align}

\textbf{Static PE:} We are now ready to perform the explicit PE calculation. Consider the static PE calculation Eq.(\ref{eq:PE_goal_epsilon}). To apply the cumulant expansion, we first consider a related quantity:
\begin{align}
\widetilde {O}_D(\epsilon)\equiv \frac{\langle\hat{\mathbf O}_D(\epsilon)\prod_{i\in D^C}\hat{\mathbf Q}_i(\epsilon)\rangle_\rho}{\langle \prod_{i\in D^C}\hat{\mathbf Q}_i(\epsilon) \rangle_\rho}
\end{align}
$\widetilde {O}_D(\epsilon)$ and our goal $\langle \hat O_D\rangle(\epsilon)$ in Eq.(\ref{eq:PE_goal_epsilon}) differ in their denominators. We then introduce another quantity for any subset $R$ of all electronic orbitals:
\begin{align}
F_R(\epsilon)\equiv\log\left\langle \prod_{i\in R}\hat{\mathbf Q}_i(\epsilon)\right\rangle_\rho.
\end{align}
Using $F_{D^C}(\epsilon)$ and  $F_{\rm all-orb.}(\epsilon)$, we have
\begin{align}
\langle \hat O_D\rangle(\epsilon)=\widetilde {O}_D(\epsilon)\exp(-\Delta_D(\epsilon)),\notag\\
\text{where } \Delta_D(\epsilon)\equiv F_{\rm all-orb.}(\epsilon)-F_{D^C}(\epsilon).\label{eq:PE_O_D_using_tilde_O}
\end{align}

This manipulation of expression is convenient since both $\widetilde {O}_D(\epsilon)$ and $F_R(\epsilon)$ can be computed via the cumulant expansion.

We now write $\widetilde {O}_D(\epsilon)$ in a cumulant-friendly form:
\begin{align}
&\widetilde {O}_D(\epsilon)=\left.\partial_u\log\left\langle e^{u\hat{\mathbf O}_{D}(\epsilon)}\prod_{i\in D^C}e^{t_i\hat{\mathbf L}_i(\epsilon)}\right\rangle_\rho\right|_{u=0,t_i=1}.\label{eq:PE_tilde_O_D_log_form}
\end{align}
Using Eq.(\ref{eq:cumulant_full_Taylor}), we have the full Taylor expansion in terms of $u,t_i$ for the logarithmic object. $\partial_u$ selects the $u$-linear term, we arrive at the closed-form,
\begin{align}
&\widetilde {O}_{D}(\epsilon)=\sum_{\{r_i\ge 0\}}\Big[\prod_i\frac{1}{r_i!}\Big]\mathcal C_\rho(\hat{\mathbf O}_D(\epsilon),\{\hat{\mathbf L}_i(\epsilon)^{[r_i]}\}_{i\in D^C}),\notag\\
\end{align}
where we followed the shorthand notation introduced in Eq.(\ref{eq:cumulant_full_Taylor}). The summation can be organized by the pairwise distinct electronic orbitals: $\{i_1,...,i_k\}\subset D^C$ whose $r_{i_a}\equiv \mathsf r_a\ge 1$, while other orbitals' $r=0$: 
\begin{align}
&\widetilde {O}_{D}(\epsilon)\notag\\
=&\sum_{k\ge 0}\frac{1}{k!}\sum_{i_1,...,i_k}^{'D^C}\sum_{\mathsf r_1,..\mathsf r_k\ge 1}\frac{\mathcal C_\rho(\hat{\mathbf O}_{D}(\epsilon),\hat{\mathbf L}_{i_1}(\epsilon)^{[\mathsf r_1]},..,\hat{\mathbf L}_{i_k}(\epsilon)^{[\mathsf r_k]})}{\mathsf r_1!\cdots\mathsf r_k!},\label{eq:PE_tilde_O_D_expansion}
\end{align}
where $1/k!$ factor is due to the summation convention of ordered sequences. Collecting terms $\propto\epsilon^m$ on the RHS, $\widetilde {O}_{D,(m)}$ can be found. The first few orders are:
\begin{align}
&\widetilde {O}_{D,(0)}=\mathcal C_\rho(\hat{\mathbf O}_{D,(0)}),\notag\\
&\widetilde {O}_{D,(1)}=\mathcal C_\rho(\hat{\mathbf O}_{D,(1)})+\sum'^{D^C}_i\mathcal C_\rho(\hat{\mathbf O}_{D,(0)},\hat{\boldsymbol q}_{i,(1)}),\notag\\
&\widetilde {O}_{D,(2)}=\mathcal C_\rho(\hat{\mathbf O}_{D,(2)})+\sum'^{D^C}_i\mathcal C_\rho(\hat{\mathbf O}_{D,(1)},\hat{\boldsymbol q}_{i,(1)})\notag\\
&+\sum'^{D^C}_i\mathcal C_\rho(\hat{\mathbf O}_{D,(0)},\hat{\boldsymbol q}_{i,(2)})+\frac{1}{2}\sum'^{D^C}_{ij}\mathcal C_\rho(\hat{\mathbf O}_{D,(0)},\hat{\boldsymbol q}_{i,(1)},\hat{\boldsymbol q}_{j,(1)})\notag\\
&-\sum^{'D^C}_{i}\mathcal C_\rho(\hat{\mathbf O}_{D,(0)},\hat{\boldsymbol q}_{i,(1)})\mathcal C_\rho(\hat{\boldsymbol q}_{i,(1)}).\label{eq:PE_O_D_tilde_first_few_orders}
\end{align}
Here, for the purpose of algorithms later, in the second-order expression, we write $-\frac12\mathcal C_\rho(\hat{\mathbf O}_{D,(0)},\hat{\boldsymbol q}_{i,(1)}^2)+\frac12\mathcal C_\rho(\hat{\mathbf O}_{D,(0)},\hat{\boldsymbol q}_{i,(1)},\hat{\boldsymbol q}_{i,(1)})$ as a product of cumulants $-\mathcal C_\rho(\hat{\mathbf O}_{D,(0)},\hat{\boldsymbol q}_{i,(1)})\mathcal C_\rho(\hat{\boldsymbol q}_{i,(1)})$, so that the expansion does \emph{not} involve any cumulant with repeated electronic orbital labels. It turns out that this is always possible at any order.

Next, we use the same trick to compute $F_{D^C}(\epsilon)$ by considering a similar logarithmic object as in Eq.(\ref{eq:PE_tilde_O_D_log_form}), but fixing $u=0$ and without $\partial_u$. We have:
\begin{align}
F_{D^C}(\epsilon)=\sum_{k\ge 1}\frac{1}{k!}\sum_{i_1,..,i_k}^{'D^C}\sum_{\mathsf r_1,..,\mathsf r_k\ge 1}\frac{\mathcal C_\rho(\hat{\mathbf L}_{i_1}(\epsilon)^{[\mathsf r_1]},..,\hat{\mathbf L}_{i_k}(\epsilon)^{[\mathsf r_k]})}{\mathsf r_1!\cdots\mathsf r_k!},
\end{align}
If we compute $F_{\rm all-orb.}(\epsilon)$, the only difference is to remove the superscript "$D^C$" of the summation constraint. Using the notation Eq.(\ref{eq:PE_notation_(R)}), we have the closed-form:
\begin{align}
\Delta_D (\epsilon)=\sum_{k\ge 1}\frac{1}{k!}\sum_{i_1,..,i_k}^{'(D)}\sum_{\mathsf r_1,..,\mathsf r_k\ge 1}\frac{\mathcal C_\rho(\hat{\mathbf L}_{i_1}(\epsilon)^{[\mathsf r_1]},..,\hat{\mathbf L}_{i_k}(\epsilon)^{[\mathsf r_k]})}{\mathsf r_1!\cdots\mathsf r_k!}.\label{eq:PE_Delta_expansion}
\end{align}
The first few orders are ($\Delta_{D,(0)}=0$):
\begin{align}
\Delta_{D,(1)}=&\sum'^{(D)}_i\mathcal C_\rho(\hat{\boldsymbol q}_{i,(1)})=\sum_{i\in D}\mathcal C_\rho(\hat{\boldsymbol q}_{i,(1)})\notag\\
\Delta_{D,(2)}=&\sum'^{(D)}_i\mathcal C_\rho(\hat{\boldsymbol q}_{i,(2)})+\frac{1}{2}\sum'^{(D)}_{i,j}\mathcal C_\rho(\hat{\boldsymbol q}_{i,(1)},\hat{\boldsymbol q}_{j,(1)})\notag\\
&-\frac{1}{2}\sum'^{(D)}_{i}\mathcal C_\rho(\hat{\boldsymbol q}_{i,(1)})^2.\label{eq:PE_Delta_first_few_orders}
\end{align}

Using Eq.(\ref{eq:PE_O_D_using_tilde_O}), Eq.(\ref{eq:PE_tilde_O_D_expansion},\ref{eq:PE_Delta_expansion}) fully determines the PE of $\langle\hat O_D^{(e)}\rangle(\epsilon)=\sum_{n\ge 0}\epsilon^n\langle\hat O_D^{(e)}\rangle_{(n)}$:
\begin{align}
&\langle\hat O_D^{(e)}\rangle_{(n)}=\widetilde O_{D,(n)}\notag\\
&+\sum_{s=0}^{n-1}\Big[\sum_{k=1}^{n-s}\frac{(-1)^k}{k!}\sum_{r_1,...,r_k\ge1}^{r_1+...+r_k=n-s}\Delta_{D,(r_1)}\cdots\Delta_{D,(r_k)}\Big]\widetilde O_{D,(s)}.\label{eq:PE_O_D_final}
\end{align}
To the first few orders:
\begin{align}
\langle\hat O_D^{(e)}\rangle_{(0)}=&\widetilde O_{D,(0)},\notag\\
\langle\hat O_D^{(e)}\rangle_{(1)}=&\widetilde O_{D,(1)}-\Delta_{D,(1)}\widetilde O_{D,(0)},\notag\\
\langle\hat O_D^{(e)}\rangle_{(2)}=&\widetilde O_{D,(2)}-\Delta_{D,(1)}\widetilde O_{D,(1)}\notag\\
+&\Big[\frac{1}{2}\Delta_{D,(1)}^2-\Delta_{D,(2)}\Big]\widetilde O_{D,(0)}.\label{eq:PE_O_D_final_first_few_orders}
\end{align}

\textbf{PE for quantum geometric tensor:} Next, consider the dynamical PE calculation Eq.(\ref{eq:PE_Q_tensor_X}). We begin by writing it in a cumulant-friendly form:
\begin{align}
&\mathcal Q_{ab}(\epsilon)=\frac{\langle\hat X_a^\dagger \hat{\mathbb Q}(\epsilon) \hat X_b\rangle_{\bar\rho}}{\langle\hat{\mathbb Q}(\epsilon)\rangle_{\bar\rho}}-\frac{\langle\hat X_a^\dagger\hat{\mathbb Q}(\epsilon)\rangle_{\bar\rho}}{\langle\hat{\mathbb Q}(\epsilon)\rangle_{\bar\rho}}\frac{\langle\hat{\mathbb Q}(\epsilon)\hat X_b\rangle_{\bar\rho}}{\langle\hat{\mathbb Q}(\epsilon)\rangle_{\bar\rho}}\notag\\
=&\left.\partial_u\partial_v\log\left\langle e^{u\hat X_a^\dagger}\prod_{i}e^{t_i\hat{\mathbf L}_i(\epsilon)} e^{v\hat X_b}\right\rangle_{\bar\rho}\right|_{u=v=0,t_i=1}.
\end{align}
Performing a similar analysis, we have the closed form:
\begin{align}
&\mathcal Q_{ab}(\epsilon)\notag\\
=&\sum_{k\ge 0}\frac{1}{k!}\sum_{i_1,..,i_k}^{'}\sum_{\mathsf r_1,..,\mathsf r_k\ge 1}\frac{\mathcal C_{\bar\rho}(\hat X_a^\dagger,\hat{\mathbf L}_{i_1}(\epsilon)^{[\mathsf r_1]},..,\hat{\mathbf L}_{i_k}(\epsilon)^{[\mathsf r_k]},\hat X_b)}{\mathsf r_1!\cdots\mathsf r_k!}
\label{eq:PE_Q_tensor_final_expansion}
\end{align}

The first few orders are:
\begin{align}
&\mathcal Q_{ab,(0)}=\mathcal C_{\bar\rho}(\hat X_a^\dagger, \hat X_b),\notag\\
&\mathcal Q_{ab,(1)}=\sum_i\mathcal C_{\bar\rho}(\hat X_a^\dagger, \hat{\boldsymbol q}_{i,(1)},\hat X_b),\notag\\
&\mathcal Q_{ab,(2)}=\sum_i\mathcal C_{\bar\rho}(\hat X_a^\dagger, \hat{\boldsymbol q}_{i,(2)},\hat X_b)\notag\\
&+\frac{1}{2}\sum'_{i,j}\mathcal C_{\bar\rho}(\hat X_a^\dagger, \hat{\boldsymbol q}_{i,(1)},\hat{\boldsymbol q}_{j,(1)},\hat X_b)\notag\\
&-\sum_i\mathcal C_{\bar\rho}(\hat{\boldsymbol q}_{i,(1)})\mathcal C_{\bar\rho}(\hat X_a^\dagger, \hat{\boldsymbol q}_{i,(1)},\hat X_b)\notag\\
&-\sum_i\mathcal C_{\bar\rho}(\hat X_a^\dagger,\hat{\boldsymbol q}_{i,(1)})\mathcal C_{\bar\rho}( \hat{\boldsymbol q}_{i,(1)},\hat X_b).\label{eq:PE_Q_tensor_final_first_few_orders}
\end{align}

\subsubsection{PE is a local expansion} \label{sec:PE_local}
Here we comment on the limitation of PE at finite orders. Due to the cumulant-bound Eq.(\ref{eq:PE_cumulant_bound}), the physical picture of the PE should be viewed as a local expansion. Therefore, for any short-range property, we expect finite-order PE to perform well. These include the energy density of a short-range Hamiltonian, Berry's phase calculation for quasiparticles and finite-momentum collective modes, etc. On the other hand, many long-range properties are not expected to be well-captured by finite-order PE. These include the long-range tail of correlation functions, the Berry's phase calculation in the momentum $\rightarrow 0$ limit, etc. These limitations will be shown in our demonstrations in the next subsection. Nevertheless, since PE is a conserving approximation, both statically and dynamically, leading to exact Ward identities in VMPI, the inaccuracy at small momentum is not expected to qualitatively change the physics.

\begin{figure}
\includegraphics[width=0.45\textwidth]{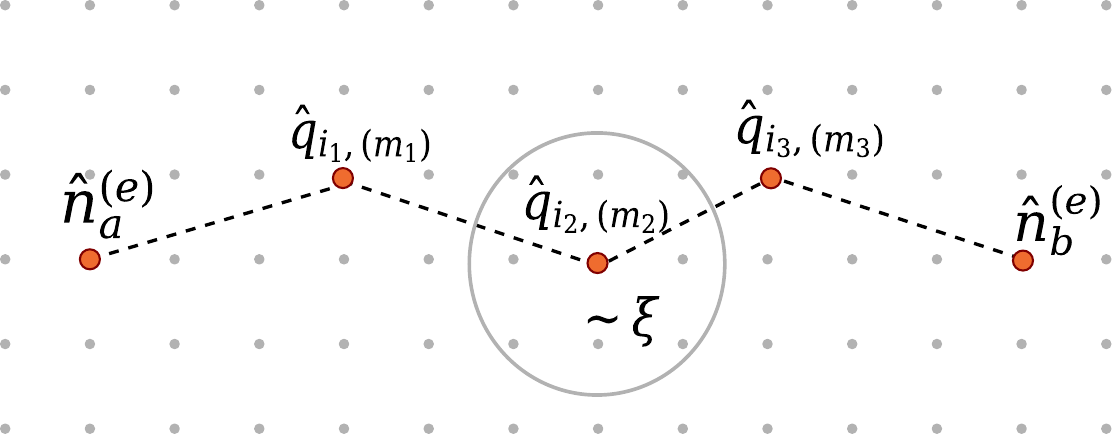}
\caption{Schematic illustration of PE corrections to long-range observables.}
\label{fig:cumulant_bridge}
\end{figure}

For example, suppose one wants to compute the PE for the static connected density-density correlation function between electronic orbitals: $a,b$: $\langle\hat n^{(e)}_a\hat n^{(e)}_b\rangle-\langle\hat n^{(e)}_a\rangle\langle\hat n^{(e)}_b\rangle$, according to  Eq.(\ref{eq:PE_tilde_O_D_expansion},\ref{eq:PE_Delta_expansion}), the PE corrections to the mean-field result (i.e. zeroth order PE result) come in via the insertion of the $\hat {\boldsymbol q}_{i,(m)}$ operators in the ordered-cumulants. Let's visualize each $\hat {\boldsymbol q}_{i,(m)}$ operator as a dot located at electronic orbital-$i$. When the $a$ and $b$ orbitals are far apart, there needs to be at least a sequence of dots forming a bridge-like line connecting $a$ and $b$ with the dots separating from their nearest neighbors by $\sim \xi$ (see Fig.\ref{fig:cumulant_bridge}). The longer-range correlation will then need a higher-order PE to capture.

\subsubsection{Fusion-correlation-matrix and fusion-channel-space algorithm}\label{sec:channel_space_algorithm}
In practice, since PE is defined in terms of operators acting on the parton mean-field state, the cumulants can be computed using Wick's theorem. In this subsection, we introduce a practical algorithm to compute the cumulants, which speeds up computation.

To present the algorithm, we consider a generic single-body electronic operator $c^{\dagger}_ic_j$ and a two-body electronic operator $c^{\dagger}_ic^{\dagger}_j c_kc_l$. Here, $c_i$ can be either fermionic or bosonic, depending on the parity of the number of parton species $n_{\rm p}$. The goal is to compute the static expectation values using PE:
\begin{align}
\textbf{Goal: }& &\langle c^{\dagger}_ic_j\rangle(\epsilon)&,&\langle  c^{\dagger}_i c^{\dagger}_j c_k c_l\rangle(\epsilon).\label{eq:PE_generic_1_2_body_terms}
\end{align}
To simplify the presentation, \emph{we focus on the case that $ c_i$ is either a fermion or a hard-core boson, and only briefly comment on the usual boson case at the end of the subsection}. Consequently, we assume in $c^{\dagger}_ic^{\dagger}_j c_kc_l$, $i\neq j$ and $k\neq l$ (but $c^\dagger$ and $c$ operators can be on the same site). In this case, the electron-number-sector decomposition in Eq.(\ref{eq:PE_P_i_as_P_i_n}) truncates to $n=0,1$:
\begin{align}
\hat{\mathbf P}_i(\epsilon)=\hat{\mathbf P}_{i,0}+\gamma(\epsilon )\hat{\mathbf P}_{i,1},\label{eq:PE_P_truncation_01}
\end{align}
where $\hat{\mathbf P}_{i,0}$ is the local parton vacuum projector:
\begin{align}
\hat{\mathbf P}_{i,0}=\prod_{p=1}^{n_{\rm p}}\hat{\mathbf P}^{(p)}_{i,0}, \;\;\hat{\mathbf P}^{(p)}_{i,0}\equiv \prod_{\mathbf m_i^{(p)}}(1-f^{(p)\dagger}_{i,\mathbf m^{(p)}_i} f^{(p)}_{i,\mathbf m^{(p)}_i}),
\end{align}
where we used the formulation introduced in Eq.(\ref{eq:onsite_fusion_operator}). Next, we introduce a few shorthand notations:
\begin{align}
\hat{\mathbf F}_i=&c_i^\dagger\sum_{\alpha=1}^R\lambda_\alpha \hat D_{i\alpha}, \text{ where }\hat D_{i\alpha}\equiv f^{(1)}_{i,\mathbf m^{(1)}_{\alpha,i}}...f^{(n_{\rm p})}_{i,\mathbf m^{(n_{\rm p})}_{\alpha,i}},\notag\\
\hat {\mathbf n}_i\equiv&  \hat{\mathbf P}_{i,1}=\sum_{\alpha,\beta=1}^{R}\lambda^*_\alpha\lambda_\beta \hat D^\dagger_{i,\alpha} \hat{\mathbf P}_{i,0}\hat D_{i\beta},\label{eq:PE_hat_D_def}
\end{align}
where, for simplicity of presentation, we assumed the number of fusion channels $R_i=R$ and the fusion amplitudes $\lambda_{i,\alpha}=\lambda_\alpha$ are independent of the electronic orbital-$i$. We have local PE operators defined in Eq.(\ref{eq:PE_Q_expansion}):
\begin{align}
\hat{\boldsymbol q}_{i,(1)}=&\hat{\mathbf P}_{i,0}-\mathbf 1+\gamma_{(0)}\hat{\mathbf n}_i,\notag\\
\hat{\boldsymbol q}_{i,(m)}=&\gamma_{(m-1)}\hat{\mathbf n}_i,\;\;\forall m\ge 2.
\end{align}

We now write down the parton operators corresponding to $c^{\dagger}_ic_j$ and $c^{\dagger}_ic^{\dagger}_j c_kc_l$ according to Eq.(\ref{eq:PE_O_D_def}). It is easy to show that:
\begin{align}
&[\hat {\mathbf c}^\dagger_i\hat {\mathbf c}_j](\epsilon)\equiv \prod_{a\in \{i,j\}}\hat{\mathbf C}_a(\epsilon)^\dagger\cdot  c_i^\dagger c_j\cdot\prod_{b\in \{i,j\}}\hat{\mathbf C}_b(\epsilon) \notag\\
&=\gamma(\epsilon)\sum_{\alpha,\beta=1}^R\lambda_\alpha^*\lambda_\beta \hat D_{i\alpha}^\dagger \hat{\mathbf P}_{i,0}\hat{\mathbf P}_{j,0}\hat D_{j\beta}\equiv\gamma(\epsilon)[\hat {\mathbf c}^\dagger_i\hat {\mathbf c}_j],\notag\\
&[\hat {\mathbf c}^\dagger_i\hat {\mathbf c}^\dagger_j\hat {\mathbf c}_k\hat {\mathbf c}_l](\epsilon)\equiv \prod_{a\in \{i,j,k,l\}}\hat{\mathbf C}_a(\epsilon)^\dagger\cdot  c_i^\dagger c_j^\dagger c_k c_l\cdot\prod_{b\in \{i,j,k,l\}}\hat{\mathbf C}_b(\epsilon)\notag\\
&=\gamma(\epsilon)^2\sum_{\alpha,\beta,\delta,\eta=1}^R\lambda_\alpha^*\lambda^*_\beta\lambda_\delta\lambda_\eta \hat D_{i\alpha}^\dagger \hat D_{j\beta}^\dagger\prod_{a\in\{i,j,k,l\}}\hat{\mathbf P}_{a,0}\hat D_{k\delta}\hat D_{l\eta}\notag\\
&\equiv\gamma(\epsilon)^2[\hat {\mathbf c}^\dagger_i\hat {\mathbf c}^\dagger_j\hat {\mathbf c}_k\hat {\mathbf c}_l],
\end{align}
where we separate out the $\epsilon$-independent operators in $[\hat {\mathbf c}^\dagger_i\hat {\mathbf c}_j]$ and $[\hat {\mathbf c}^\dagger_i\hat {\mathbf c}^\dagger_j\hat {\mathbf c}_k\hat {\mathbf c}_l]$. Note that when $i=j$, $[\hat {\mathbf c}^\dagger_i\hat {\mathbf c}_i]=\hat {\mathbf n}_i$. Using Eq.(\ref{eq:PE_cumulants_first_few_orders},\ref{eq:PE_O_D_tilde_first_few_orders},\ref{eq:PE_Delta_first_few_orders}), in order to compute the goal Eq.(\ref{eq:PE_generic_1_2_body_terms}), a typical component of a cumulant has a form:
\begin{align}
\mathsf A^{[q=1]}_{(m)}(\{s_n\})\equiv &\langle[\hat {\mathbf c}^\dagger_i\hat {\mathbf c}_j]\hat {\mathbf n}_{s_1}...\hat {\mathbf n}_{s_m}\rangle_{\rho}, \;\;s_n\in \{i,j\}^C\notag\\
\mathsf A^{[q=2]}_{(m)}(\{s_n\})\equiv&\langle[\hat {\mathbf c}^\dagger_i\hat {\mathbf c}^\dagger_j\hat {\mathbf c}_k\hat {\mathbf c}_l]\hat {\mathbf n}_{s_1}...\hat {\mathbf n}_{s_m}\rangle_{\rho},\;s_n\in \{i,j,k,l\}^C.\label{eq:PE_A_B}
\end{align}
where the PE electronic orbitals $\{s_n\}$ are \emph{pairwise} distinct. If one identifies $\hat {\mathbf n}_{s_n}$ as contributed from $\hat {\boldsymbol q}_{s_n,(1)}$, then the subscript $_{(m)}$ in $\mathsf A^{[q]}_{(m)}$ can be identified as the PE order. The superscript $^{[q]}$ in $\mathsf A^{[q]}_{(m)}$ denotes whether the goal electronic operator is single-body or two-body.

There are two complexities involved in computing these expectation values: 
\begin{enumerate}
    \item The presence of the onsite parton vacuum projectors $\hat {\mathbf P}_{a,0}$, which makes Wick's theorem expansion complicated.
    \item The summation of the fusion channel indices $\{\alpha,\beta,...\}$, which scales as $O(R^{2(q+m)})$ for $\mathsf  A^{[q]}_{(m)}$, making computation costly when $R$ is sizable. (Note that there is a double summation for fusion channels in each $\hat{\mathbf n}_{s_n}$)
\end{enumerate}
The purpose of this subsection is to introduce algorithms to tackle these two complexities:
\begin{enumerate}
    \item We will introduce the \emph{fusion-correlation-matrix}, such that in terms of Wick's theorem expansion, the onsite parton vacuum projectors $\hat {\mathbf P}_{a,0}$ can be effectively replaced by identity: for each set of fixed fusion channels $\{\alpha,\beta,...\}$, only one determinant need to be computed for each parton-$p$.
    \item We will introduce the \emph{fusion-channel-space algorithm}: the complexity in computing $\mathsf  A^{[q]}_{(m)}$ is reduced to $O(R^{q+m+1})$.
\end{enumerate}

\textbf{Fusion-correlation-matrix: }
Now let's pay attention to the operator expressions on the RHS of Eq.(\ref{eq:PE_A_B}). Each $[\hat {\mathbf c}^\dagger_i\hat {\mathbf c}_j]$, $[\hat {\mathbf c}^\dagger_i\hat {\mathbf c}^\dagger_j\hat {\mathbf c}_k\hat {\mathbf c}_l]$, and $\hat {\mathbf n}_{s_n}$ operator has a form that the parton creation operators $\hat D^\dagger$ on the left, annihilation operators $\hat D$ on the right, and the projectors $\hat{\mathbf P}_{a,0}$ in the middle. Notice that the projector $\hat{\mathbf P}_{a,0}$ is fermion parity even, they commute with each other and with any parton operator on a distinct electronic orbital. Consequently, in the operator expressions on the RHS of Eq.(\ref{eq:PE_A_B}), after including a fixed fermion sign: 
\begin{align}
\sigma_{(m)}^{[q]}\equiv (-1)^{n_{\rm p}\cdot[qm+\frac{m(m-1)}{2}]},
\end{align}
one can safely move all $\hat D^\dagger$ to the left (but keeping their order), all $\hat D$ operators to the right (but keeping their order), and all $\hat{\mathbf P}_{a,0}$ in the middle. Denoting the combined fusion channel index as the bold-font Greek letter: 
\begin{align}
\boldsymbol \alpha\equiv& (\alpha_1,...,\alpha_{q+m}),& W_{\boldsymbol\alpha}\equiv& \prod_{r=1}^{q+m}\lambda_{\alpha_r},
\end{align}
also denoting:
\begin{align}
\bar{\mathbf D}^{[q=1]}_{\boldsymbol\alpha}\equiv& \hat D_{i\alpha_1}^\dagger\hat D^\dagger_{s_1\alpha_2}...\hat D^\dagger_{s_m\alpha_{m+1}},\notag\\
\mathbf D^{[q=1]}_{\boldsymbol\beta}\equiv& \hat D_{j\beta_1}\hat D_{s_1\beta_2}...\hat D_{s_m\beta_{m+1}},\notag\\
\bar{\mathbf D}^{[q=2]}_{\boldsymbol\alpha}\equiv& \hat D_{i\alpha_1}^\dagger\hat D_{j\alpha_2}^\dagger\hat D_{s_1\alpha_3}^\dagger...\hat D_{s_m\alpha_{m+2}}^\dagger,\notag\\
\mathbf D^{[q=2]}_{\boldsymbol\beta}\equiv& \hat D_{k\beta_1}\hat D_{l\beta_2}\hat D_{s_1\beta_3}...\hat D_{s_m\beta_{m+2}},\notag\\
\mathbf P^{[q]}\equiv& \prod_{p=1}^{n_{\rm p}}\mathbf P^{(p),[q]},\notag\\
\mathbf P^{(p),[q=1]}\equiv& \hat{\mathbf P}^{(p)}_{i,0}\hat{\mathbf P}^{(p)}_{j,0}\hat{\mathbf P}^{(p)}_{s_1,0}...\hat{\mathbf P}^{(p)}_{s_m,0},\notag\\
\mathbf P^{(p),[q=2]}\equiv& \hat{\mathbf P}^{(p)}_{i,0}\hat{\mathbf P}^{(p)}_{j,0}\hat{\mathbf P}^{(p)}_{k,0}\hat{\mathbf P}^{(p)}_{l,0}\hat{\mathbf P}^{(p)}_{s_1,0}...\hat{\mathbf P}^{(p)}_{s_m,0},
\end{align}
we have:
\begin{align}
\mathsf A^{[q]}_{(m)}=\sigma_{(m)}^{[q]}\sum_{\boldsymbol\alpha,\boldsymbol\beta}W_{\boldsymbol\alpha}^*W_{\boldsymbol\beta}\langle \bar{\mathbf  D}_{\boldsymbol\alpha}\mathbf P^{[q]}\mathbf D_{\boldsymbol\beta}\rangle_\rho.
\end{align}

Defining vectors $\tilde{\boldsymbol \alpha}^{(p)},\tilde{\boldsymbol \beta}^{(p)}$ of the combined parton orbital indices $(i,\mathbf m^{(p)})$ of length $q+m$:
\begin{align}
&\tilde{\boldsymbol \alpha}^{(p),[q=1]} \equiv [(i,\mathbf m^{(p)}_{\alpha_1}),(s_1,\mathbf m^{(p)}_{\alpha_2}),...,(s_m,\mathbf m^{(p)}_{\alpha_{m+1}})],\notag\\
&\tilde{\boldsymbol \beta}^{(p),[q=1]} \equiv [(j,\mathbf m^{(p)}_{\beta_1}),(s_1,\mathbf m^{(p)}_{\beta_2}),...,(s_m,\mathbf m^{(p)}_{\beta_{m+1}})],\notag\\
&\tilde{\boldsymbol \alpha}^{(p),[q=2]} \notag\\
&\equiv [(i,\mathbf m^{(p)}_{\alpha_1}),(j,\mathbf m^{(p)}_{\alpha_2}),(s_1,\mathbf m^{(p)}_{\alpha_3}),...,(s_m,\mathbf m^{(p)}_{\alpha_{m+2}})],\notag\\
&\tilde{\boldsymbol \beta}^{(p),[q=2]} \notag\\
&\equiv [(k,\mathbf m^{(p)}_{\beta_1}),(l,\mathbf m^{(p)}_{\beta_2}),(s_1,\mathbf m^{(p)}_{\beta_3}),...,(s_m,\mathbf m^{(p)}_{\beta_{m+2}})],\label{eq:PE_algorithm_tilde_alpha_beta}
\end{align}
also defining
\begin{align}
\bar{\mathbf f}^{(p),[q]}_{\boldsymbol\alpha}\equiv& f^{(p)\dagger}_{\tilde{\boldsymbol \alpha}^{(p),[q]}_1}...f^{(p)\dagger}_{\tilde{\boldsymbol \alpha}^{(p),[q]}_{q+m}},\notag\\
\mathbf f^{(p),[q]}_{\boldsymbol\beta}\equiv &f^{(p)}_{\tilde{\boldsymbol \beta}^{(p),[q]}_1}...f^{(p)}_{\tilde{\boldsymbol \beta}^{(p),[q]}_{q+m}},
\end{align}
we have:
\begin{align}
\mathsf A^{[q]}_{(m)}=&\sigma_{(m)}^{[q]}\sum_{\boldsymbol\alpha,\boldsymbol\beta}W_{\boldsymbol\alpha}^*W_{\boldsymbol\beta}\prod_{p=1}^{n_{\rm p}} \mathsf B^{(p),[q]}_{\boldsymbol\alpha\boldsymbol\beta},\notag\\
\text{where: } \mathsf B^{(p),[q]}_{\boldsymbol\alpha\boldsymbol\beta}\equiv& \langle\bar{\mathbf f}^{(p),[q]}_{\boldsymbol\alpha} \mathbf P^{(p),[q]}\mathbf f^{(p),[q]}_{\boldsymbol\beta}\rangle_{\rho^{(p)}}.\label{eq:PE_algorithm_A_B}
\end{align}
Therefore, the calculation for a fixed set of fusion channels $\boldsymbol\alpha,\boldsymbol\beta$ boils down to computing $\mathsf B^{(p),[q]}_{\boldsymbol\alpha\boldsymbol\beta}$. If there were no projector $\mathbf P^{(p),[q]}$, this is straightforward: up to the fermion sign:
\begin{align}
\tilde{\sigma}^{(p),[q]}_{(m)}\equiv (-1)^{\frac{(q+m)(q+m-1)}{2}},
\end{align}
one only needs to compute $\det[\rho^{(p)}_{\tilde {\boldsymbol\beta}^{(p),[q]},\tilde {\boldsymbol\alpha}^{(p),[q]}}]$, where the $(q+m)\times(q+m)$ submatrix $\rho^{(p)}_{\tilde {\boldsymbol\beta}^{(p),[q]},\tilde {\boldsymbol\alpha}^{(p),[q]}}$ selects the $\tilde {\boldsymbol\beta}^{(p),[q]}$-indexed rows and $\tilde {\boldsymbol\alpha}^{(p),[q]}$-indexed columns of the parton RDM: $\rho^{(p)}_{ba}\equiv \langle f_a^\dagger f_b\rangle$. But with the projector $\mathbf P^{(p),[q]}$, naively, the computation is complicated.

To solve this complexity, we define the fusion-correlation-matrix as follows. First, we denote \emph{all} parton-$p$ orbitals associated with the electronic-orbital labels on which $\mathbf P^{(p),[q]}$ acts as $\mathcal S^{(p),[q]}$:
\begin{align}
\mathcal S^{(p),[q=1]}\equiv &\{ (a,\mathbf m^{(p)}_a): a\in \{i,j,s_1,...s_m\},\forall \mathbf m_a^{(p)}\},\notag\\
\mathcal S^{(p),[q=2]}\equiv &\{ (a,\mathbf m^{(p)}_a): a\in \{i,j,k,l,s_1,...s_m\},\forall \mathbf m_a^{(p)}\}.
\end{align}
Since there may be many parton-$p$ orbitals per electronic orbital, the number of elements in $\mathcal S^{(p),[q]}$: $|\mathcal S^{(p),[q]}|$ can be much larger than the number of electronic orbitals that $\mathbf P^{(p),[q]}$ acts on. The \emph{fusion-correlation-matrix} for parton-$p$ is defined as:
\begin{align}
\rho^{(p),[q]}_{\rm fus}\equiv \rho^{(p)}_{\mathcal S^{(p),[q]},\mathcal S^{(p),[q]}}\cdot\Big[\mathbf 1-\rho^{(p)}_{\mathcal S^{(p),[q]},\mathcal S^{(p),[q]}}\Big]^{-1},
\end{align}
which is a $|\mathcal S^{(p),[q]}|\times |\mathcal S^{(p),[q]}|$ matrix.

It turns out that a known property of Gaussian Grassmann state\cite{CheongHenley2004FreeFermionDensityMatrix} leads to:
\begin{align}
\mathsf B^{(p),[q]}_{\boldsymbol\alpha\boldsymbol\beta}=\tilde{\sigma}^{(p),[q]}_{(m)}\cdot Z^{(p),[q]}\cdot \det\Big\{\big[\rho^{(p),[q]}_{\rm fus}\big]_{\tilde {\boldsymbol\beta}^{(p),[q]},\tilde {\boldsymbol\alpha}^{(p),[q]}}\Big\},
\end{align}
where the $(q+m)\times(q+m)$ submatrix of $\rho^{(p),[q]}_{\rm fus}$ appears inside the determinant, and $Z^{(p),[q]}$ is a factor independent of fusion channels:
\begin{align}
Z^{(p),[q]}\equiv\det\Big[\mathbf 1-\rho^{(p)}_{\mathcal S^{(p),[q]},\mathcal S^{(p),[q]}}\Big].
\end{align}
Namely, using the fusion-correlation-matrix, even with the projector $\mathbf P^{(p),[q]}$, one still only needs to compute one determinant per fixed set of fusion channels $\boldsymbol\alpha,\boldsymbol\beta$. In particular, for low-order PE, $q+m$ is small (e.g., for a single-body term, 2nd-order PE needs $q+m=3$). 

Typically, the determinant is not computed numerically using the permutation-group-based definition, whose complexity scales exponentially with matrix size. Instead, it is usually computed using Gaussian elimination and is applicable to large matrices. However, for the small matrices relevant for PE, both the Gaussian-elimination-based algorithm and the definition-based algorithm give comparable speed for a single determinant. We will shortly see that using the definition-based algorithm will significantly reduce the computation cost for the summation of fusion channels $\boldsymbol\alpha,\boldsymbol\beta$.

\textbf{Fusion-channel-space algorithm: }
We still need to sum over the fusion channels $\boldsymbol\alpha,\boldsymbol\beta$ to compute $\mathsf A^{[q]}_{(m)}$, whose computation cost naively scales as $O(R^{2(q+m)})$. We will show that using the definition-based algorithm for the determinant reduces this cost to $O(R^{q+m+1})$. To facilitate the discussion, we introduce some notations. First, we collect all the fermion signs
\begin{align}
\sigma^{[q]}\equiv \sigma^{[q]}_{(m)}\prod_{p=1}^{n_{\rm p}}\tilde \sigma^{(p),[q]}_{(m)}=(-1)^{\frac{q(q-1)}{2}\cdot n_{\rm p}},
\end{align}
which is independent of $m$. Second, we collect all the factors $Z^{(p),[q]}$:
\begin{align}
\mathsf Z^{[q]}\equiv \prod_{p=1}^{n_{\rm p}} Z^{(p),[q]}.
\end{align}
Third, we introduce a shorthand notation for the matrix $\big[\rho^{(p),[q]}_{\rm fus}\big]_{\tilde {\boldsymbol\beta}^{(p),[q]},\tilde {\boldsymbol\alpha}^{(p),[q]}}$:
\begin{align}
[K^{(p),[q]}_{\boldsymbol\beta\boldsymbol\alpha}]_{b,a}\equiv \big[\rho^{(p),[q]}_{\rm fus}\big]_{\tilde {\boldsymbol\beta}_b^{(p),[q]},\tilde {\boldsymbol\alpha}_a^{(p),[q]}},\;\;a,b=1,...,q+m.
\end{align}
\emph{Crucially}, for fixed $a,b$, $[K^{(p),[q]}_{\boldsymbol\beta\boldsymbol\alpha}]_{ba}$ depends on $\boldsymbol\alpha,\boldsymbol\beta$ only via $\alpha_a$ and $\beta_b$ as indicated in Eq.(\ref{eq:PE_algorithm_tilde_alpha_beta}). This fact allows us to define, for each pair of $a,b\in 1,...,q+m$, an $R\times R$ \emph{fusion-channel-space matrix} $L^{(p),[q]}_{b,a}$:
\begin{align}
 [K^{(p),[q]}_{\boldsymbol\beta\boldsymbol\alpha}]_{b,a}\equiv\big[L^{(p),[q]}_{b,a}\big]_{\beta_b,\alpha_a}.\label{eq:PE_fusion_channel_space_matrix_def}
\end{align}

Altogether, based on Eq.(\ref{eq:PE_algorithm_A_B}), we have:
\begin{align}
\mathsf A^{[q]}_{(m)}=\sigma^{[q]}\mathsf Z^{[q]}\sum_{\boldsymbol\alpha\boldsymbol\beta}W^*_{\boldsymbol\alpha}W_{\boldsymbol\beta}\prod_{p=1}^{n_{\rm p}}\det K^{(p),[q]}_{\boldsymbol\beta\boldsymbol\alpha}\label{eq:PE_A_q_m_1}
\end{align}

Now use the definition of determinants, denoting:
\begin{align}
&\sum_{\pi_1\in S_{q+m}}\sum_{\pi_2\in S_{q+m}}...\sum_{\pi_{n_{\rm p}}\in S_{q+m}}\equiv \sum_{\boldsymbol\pi},\notag\\
&\boldsymbol\pi\equiv(\pi_1,...\pi_{n_{\rm p}}),\;\; (-1)^{\boldsymbol\pi}\equiv \prod_{p=1}^{n_{\rm p}}(-1)^{\pi_p},
\end{align}
we have
\begin{align}
&\prod_{p=1}^{n_{\rm p}}\det K^{(p),[q]}_{\boldsymbol\beta\boldsymbol\alpha}=\sum_{\boldsymbol\pi}(-1)^{\boldsymbol\pi}\prod_{b=1}^{q+m}\prod_{p=1}^{n_{\rm p}}[K^{(p),[q]}_{\boldsymbol\beta\boldsymbol\alpha}]_{b,\pi_p(b)}\notag\\
=&\sum_{\boldsymbol\pi}(-1)^{\boldsymbol\pi}\prod_{b=1}^{q+m}\prod_{p=1}^{n_{\rm p}}\big[L^{(p),[q]}_{b,\pi_p(b)}\big]_{\beta_{b}\alpha_{\pi_p(b)}},
\end{align}
where we used Eq.(\ref{eq:PE_fusion_channel_space_matrix_def}). The consequence is that, for fixed $\boldsymbol\alpha$ and $\boldsymbol\pi$, the $\boldsymbol\beta$-summation in Eq.(\ref{eq:PE_A_q_m_1}) factorizes:
\begin{align}
&\mathsf C^{[q]}_{(m)}(\boldsymbol\alpha;\boldsymbol\pi)\equiv \sum_{\boldsymbol\beta}W_{\boldsymbol\beta}\prod_{b=1}^{q+m}\prod_{p=1}^{n_{\rm p}}\big[L^{(p),[q]}_{b,\pi_p(b)}\big]_{\beta_{b}\alpha_{\pi_p(b)}}\notag\\
=&\prod_{b=1}^{q+m}\Big[\sum_{\beta=1}^R \lambda_\beta\prod_{p=1}^{n_{\rm p}}\big[L^{(p),[q]}_{b,\pi_p(b)}\big]_{\beta,\alpha_{\pi_p(b)}}\Big],
\end{align}
whose computational complexity is $O(R)$. And finally:
\begin{align}
\mathsf A^{[q]}_{(m)}=\sigma^{[q]}\mathsf Z^{[q]}\sum_{\boldsymbol\alpha}W^*_{\boldsymbol\alpha}\sum_{\boldsymbol\pi}(-1)^{\boldsymbol\pi}\mathsf C^{[q]}_{(m)}(\boldsymbol\alpha;\boldsymbol\pi),\label{eq:PE_A_q_m_2}
\end{align}
where the $\boldsymbol\alpha$ summation contributes another factor of $R^{q+m}$ in complexity. If one accounts for all other contributions to the complexity, this fusion-channel-space algorithm to compute $\mathsf A^{[q]}_{(m)}$ has complexity $O\Big(\big[(q+m)!\big]^{n_{\rm p}} (q+m) n_{\rm p}R^{q+m+1}\Big)$, while the naive algorithm using Gaussian-elimination-based determinant algorithm gives $O\big(n_{\rm p}(q+m)^3 R^{2(q+m)}\big)$. For practical low-order PE, when the number of fusion channels $R$ becomes sizable, the fusion-channel-space algorithm has an advantage.

Finally, we briefly remark on the situation that the electron is a usual boson instead of a hard-core boson, where the truncation to $n=0,1$ in Eq.(\ref{eq:PE_P_truncation_01}) must be generalized:
\begin{align}
\hat{\mathbf P}_i(\epsilon)=\hat{\mathbf P}_{i,0}+\gamma(\epsilon )\hat{\mathbf P}_{i,1}+...+\gamma(\epsilon )^{n_{\rm max}}\hat{\mathbf P}_{i,n_{\rm max}},
\end{align}
where $n_{\rm max}$ is the maximal boson number allowed by the fusion gate from fermionic partons. For example, consider the $n=2$ sector $\hat{\mathbf P}_{i,2}$, which involves $\hat{\mathbf C}_{i,2}$ according to Eq.(\ref{eq:onsite_fusion_gate_epsilon}). $\hat{\mathbf C}_{i,2}$ will involve two summation of fusion channels $\sum_{\alpha,\beta}$ and associated fusion operators: $\hat D_{i\alpha}\hat D_{i\beta}$ defined in Eq.(\ref{eq:PE_hat_D_def}). Roughly speaking, the $n=2$ sector $\hat{\mathbf P}_{i,2}$ can be treated as two separate $n=1$ sectors in terms of the current fusion-channel-space algorithm. Generally, the complexity w.r.t. the PE order $m$ will increase from $O(R^m)$ in the hard-core case to $O(R^{mn_{\rm max}})$.

\subsection{Benchmarking results}\label{sec:benchmark}
In this subsection, we will start with one of the simplest Hdet states, the bosonic Laughlin $\nu=\frac{1}{2}$ FQH state with only one local fusion channel $R=1$, and test the PE for both \emph{static properties and the quantum geometric tensor}. The advantage of this pedagogical example is that the exact results are available via Monte Carlo, since $R=1$. In addition, the PE calculation can be performed analytically, yielding simple closed-form results. We then move on to the static properties of the fermionic $\nu=\frac{1}{3}$ Laughlin state and Jain's $\nu=\frac{2}{5}$ FQH state. As mentioned before, Jain's $\nu=\frac{2}{5}$ FQH state already has 4 local fusion channels $R=4$, and cannot be exactly simulated by Monte Carlo for large systems. In these examples, we will test PE up to the second order.

In all the FQH examples mentioned above, the saddle points $\bar\rho$ for Hdet states are known: partons filling LLs. Finally, we study the FCI model systems, where the saddle points $\bar\rho$ must be found self-consistently. In principle, one can improve $\bar\rho$ order by order using PE. But as a benchmark test, we will only apply PE at the zeroth order (i.e., mean-field level) and find the zeroth-order-PE saddle point $\bar\rho_{(0)}$, as well as the effective parton Hamiltonian $\hat {\mathsf h}^{(p)}(\mathbf k;\bar\rho_{(0)})$ as in Eq.(\ref{eq:type_B_eff_h}). In particular, we aim to test PE in physical regimes far from the FQH regime. We will show that the static saddle-point results already give reasonable energetics. When limited to a small-sized system where exact diagonalization(ED) can be performed, the exact Hdet wavefunctions based on $\bar\rho_{(0)}$ can be directly compared with the ED wavefunctions, showing much improved overlap.

Although the examples presented in this subsection belong to the Type-(B) models defined in Sec.\ref{sec:two_model_types}, PE is generally applicable to both type-(A) and type-(B) models.

\subsubsection{Bosonic Laughlin's $\nu=\frac{1}{2}$ state}
We fix the unit so that the electron's magnetic length $l^{(e)}=1$ throughout this subsection. Since $q^{(1)}=q^{(2)}=\frac{1}{2}$:
\begin{align}
(l^{(1)})^2=(l^{(2)})^2=2(l^{(e)})^2=2.
\end{align}

To set up the PE calculation, we will use the discrete-coherent-state basis for both electrons and partons on the Fine-Grid on a torus sample with $N_s^{(e)}$ magnetic flux quanta, as introduced in Sec.\ref{sec:discrete_coherent_state_basis}. Namely, the Fine-Grid is a lattice of $N_s^{(e)}\times N_s^{(e)}$ sites: $\{z;z\in \text{Fine-Grid}\}$, whose real-space lattice basis vectors are the minimally allowed electronic magnetic translation vectors. On every Fine-Grid site-$z$, there is also a coherent state for each parton species in their LLL. The relevant local quantum states on site-$z$ are labeled as $|z^{(e)}\rangle,|z^{(1)}\rangle,|z^{(2)}\rangle$, whose corresponding second-quantized operators are denoted as $b_z,f_z^{(1)},f_z^{(2)}$. Recall the change of viewpoint in Sec.\ref{sec:change_of_viewpoint}, the coherent states on different sites are treated as orthonormal, for both electrons and partons.

The local fusion gate is:
\begin{align}
\hat{\mathbf C}_z(\epsilon)=|0^{(e)}\rangle\langle0^{(\text{all }p)}|+\gamma(\epsilon)^{\frac{1}{2}} b_z^\dagger|0^{(e)}\rangle\langle0^{(\text{all }p)}|f^{(1)}_{z}f^{(2)}_{z},
\end{align}
The local projection operator in Eq.(\ref{eq:PE_O_D_def}) is:
\begin{align}
\hat{\mathbf P}_z(\epsilon)=&|0^{(\text{all }p)}\rangle\langle0^{(\text{all }p)}|+\gamma(\epsilon) \hat n_z^{(1)}\hat n_z^{(2)}\notag\\
=&(\hat{\mathbf 1}-\hat n_z^{(1)})(\hat{\mathbf 1}-\hat n_z^{(2)})+\gamma(\epsilon) \hat n_z^{(1)}\hat n_z^{(2)}\notag\\
=&\hat{\mathbf 1}-\hat n_z^{(1)}-\hat n_z^{(2)}+\big(1+\gamma(\epsilon)\big) \hat n_z^{(1)}\hat n_z^{(2)}
\end{align}
where $\hat n_z^{(p)}\equiv f^{(p)\dagger}_{z}f^{(p)}_{z}$, leading to the local PE operator in Eq.(\ref{eq:PE_Q_def}):
\begin{align}
\hat{\mathbf Q}_z(\epsilon)=&\hat{\mathbf 1}+\epsilon\Big(-\hat n_z^{(1)}-\hat n_z^{(2)}+\big(1+\gamma(\epsilon)\big) \hat n_z^{(1)}\hat n_z^{(2)}\Big)\notag\\
=&\hat{\mathbf 1}+\epsilon \hat{\boldsymbol q}_{z,(1)}+\epsilon^2 \hat{\boldsymbol q}_{z,(2)}+O(\epsilon^3),\label{eq:PE_boson_laughlin_q}
\end{align}
where the expansion defined in Eq.(\ref{eq:PE_Q_expansion}) is
\begin{align}
\hat{\boldsymbol q}_{z,(1)}=&-\hat n_z^{(1)}-\hat n_z^{(2)}+\big(1+\gamma_{(0)}\big) \hat n_z^{(1)}\hat n_z^{(2)},\notag\\
\hat{\boldsymbol q}_{z,(2)}=&\gamma_{(1)} \hat n_z^{(1)}\hat n_z^{(2)}.\label{eq:PE_boson_laughlin_q_first_few_orders}
\end{align}
When computing PE, the parton number sum rule will be repeatedly used as an operator identity:
\begin{align}
\sum_{z} \hat n_z^{(p)}=N_e\cdot \hat{\mathbf 1}\label{eq:PE_exact_operator_sum_rule}
\end{align}
All higher-order cumulants satisfy:
\begin{align}
\sum_z \mathcal C_{\rho} (..., \hat n_z^{(p)},....)=N_e\mathcal C_{\rho} (..., \hat {\mathbf 1},....)=0,\label{eq:PE_exact_operator_sum_rule_cumulant}
\end{align}
except for $\sum_z \mathcal C_{\rho} (\hat n_z^{(p)})=N_e$.

\textbf{Static PE, Preparation:} Our goal here is to compute the pair-correlation function in the Hdet state $|\Phi(\bar\rho)\rangle$ in the thermodynamic limit using low order PE:
\begin{align}
\textbf{Goal: } g_{[m]}(|z-w|)\equiv \lim_{N_e\rightarrow\infty}\frac{1}{\bar n^2}\langle \Phi(\bar\rho)|\hat n_z^{(e)}\hat n_w^{(e)}|\Phi(\bar\rho)\rangle_{[m]},
\end{align}
Fortunately, due to $R=1$, the exact Hdet state can be efficiently simulated by Monte Carlo, so that the PE results $g_{[m]}(r)$ with $m=0,1,2$ can be directly compared with the exact pair correlation function $g(r)$. Here, the electron density per site is:
\begin{align}
\bar n\equiv \bar{n}^{(e)}=\langle \Phi(\bar\rho)|\hat n_z^{(e)}| \Phi(\bar\rho)\rangle=\frac{N_e}{(N_s^{(e)})^2}=\frac{1}{2N^{(e)}_{s}}.
\end{align}
Note $\nu=\frac{N_e}{N_s^{(e)}}=\frac{1}{2}$. Since the parton number of each species equals the electron number, we also have:
\begin{align}
\bar{n}^{(e)}=\bar{n}^{(1)}=\bar{n}^{(2)}=\bar n.
\end{align}

We will soon apply Wick's theorem to compute the expectation values $\langle ...\rangle_{\bar\rho}$. According to Eq.(\ref{eq:fdagf_FG}), the fundamental expectation value is $(z,w\in \text{Fine-Grid})$:
\begin{align}
\langle f^{(p)\dagger}_wf^{(p)}_z\rangle_{\bar\rho}=\bar n^{(p)}\langle z^{(p)}|w^{(p)}\rangle_{0LL}= \bar n\; e^{-\frac{|w|^2+|z|^2}{4(l^{(p)})^2}+\frac{\bar w z}{2(l^{(p)})^2}},
\end{align}
where we used the overlap between two coherent states for parton-$p$ ($p=1,2.$) before the change of viewpoint in Eq.(\ref{eq:z_w_overlap}). Although Eq.(\ref{eq:z_w_overlap}) is obtained for the disk geometry, in the thermodynamic limit, the result for the torus sample is the same, provided care is taken to choose the $z,w$ torus image positions to be the closest pair.

We define a dimensionless operator:
\begin{align}
\hat d_z\equiv \frac{\hat n_z^{(1)}\hat n_z^{(2)}}{\bar n^2}.
\end{align}
For any \emph{pairwise distinct} set of  sites: $Z\equiv\{z_1,...,z_N\}\in \text{Fine-Grid}$, define:
\begin{align}
\hat d_Z=\hat d_{\{z_1,...,z_N\}}\equiv\prod_{i=1}^N\hat d_{z_i},
\end{align}
Wick's theorem leads to:
\begin{align}
\left\langle \hat d_Z \right\rangle_{\bar\rho}=[\det \mathsf M_Z]^2,
\end{align}
where the $N\times N$ matrix $\mathsf M_Z$ is defined as:
\begin{align}
[\mathsf M_Z]_{ij}\equiv e^{-\frac{|z_i|^2+|z_j|^2}{4(l^{(p)})^2}+\frac{\bar z_i z_j}{2(l^{(p)})^2}}, \;\;(l^{(p)})^2=2.
\end{align}
The pair correlation function ($z\neq w$) of the parton mean-field state is nothing but:
\begin{align}
g^{(p)}(|z-w|)=\frac{1}{\bar n^2}\langle \hat n^{(p)}_z\hat n^{(p)}_w\rangle_{\bar\rho}= \det\mathsf M_{\{z,w\}},
\end{align}
giving the well-known result in the thermodynamic limit:
\begin{align}
g^{(p)}(r)=1-e^{-\frac{r^2}{2(l^{(p)})^2}}=1-e^{-\frac{r^2}{4}}.\label{eq:PE_boson_laughlin_parton_pair_corr}
\end{align}
For later purposes, we also define the connected parton mean-field pair-correlation function ($z\neq w$):
\begin{align}
h^{(p)}(|z-w|)=\frac{1}{\bar n^2}\mathcal C_{\bar\rho}(\hat n^{(p)}_z,\hat n^{(p)}_w),
\end{align}
in the thermodynamic limit:
\begin{align}
h^{(p)}(r)=g^{(p)}(r)-1=-e^{-\frac{r^2}{4}}.\label{eq:PE_boson_laughlin_parton_connected_pair_corr}
\end{align}

Assuming $z,w,x,y$ are pairwise distinct, we list a few useful ordered-cumulants involving $\hat d_z$:
\begin{align}
\mathcal C_{\bar\rho}(\hat d_z)=&1\notag\\
\mathcal C_{\bar\rho}(\hat d_z,\hat d_x)=&[\det\mathsf M_{\{z,x\}}]^2-1\notag\\
\mathcal C_{\bar\rho}(\hat d_z,\hat d_x,\hat d_y)=&[\det\mathsf M_{\{z,x,y\}}]^2-[\det\mathsf M_{\{z,x\}}]^2\notag\\
-&[\det\mathsf M_{\{z,y\}}]^2-[\det\mathsf M_{\{x,y\}}]^2+2,\label{eq:PE_boson_laughlin_dz_cumulants}
\end{align}
and a few ordered cumulants involving $\hat d_{\{z,w\}}$:
\begin{align}
\mathcal C_{\bar\rho}(\hat d_{\{z,w\}})=&[\det\mathsf M_{\{z,w\}}]^2\notag\\
\mathcal C_{\bar\rho}(\hat d_{\{z,w\}},\hat d_x)=&[\det\mathsf M_{\{z,w,x\}}]^2-[\det\mathsf M_{\{z,w\}}]^2\notag\\
\mathcal C_{\bar\rho}(\hat d_{\{z,w\}},\hat d_x,\hat d_y)=&[\det\mathsf M_{\{z,w,x,y\}}]^2-[\det\mathsf M_{\{z,w,x\}}]^2\notag\\
-&[\det\mathsf M_{\{z,w,y\}}]^2+2[\det\mathsf M_{\{z,w\}}]^2\notag\\
-&[\det\mathsf M_{\{z,w\}}]^2[\det\mathsf M_{\{x,y\}}]^2.\label{eq:PE_boson_laughlin_dzdw_cumulants}
\end{align}

For the purpose of computing $g(r)$, let's write down the relevant $\hat{\mathbf O}_D(\epsilon)$ operators for the single-site electron density $\hat O^{(e)}_D=\hat n_z^{(e)}$ and electron's pair-density $\hat O^{(e)}_D=\hat n_z^{(e)}\hat n_w^{(e)}$ ($z\neq w$). Following the definition Eq.(\ref{eq:PE_O_D_def}), we have:
\begin{align}
\hat{\mathbf n}_z(\epsilon)=&\gamma(\epsilon)\hat n_z^{(1)}\hat n_z^{(2)}=\bar n^2\gamma(\epsilon) \hat d_z,\notag\\
[\hat{\mathbf n}_z\hat{\mathbf n}_w](\epsilon)=&\hat{\mathbf n}_z(\epsilon)\hat{\mathbf n}_w(\epsilon)=\bar n^4\gamma(\epsilon)^2\hat d_{\{z,w\}}.
\end{align}
For reasons that will become clear soon, we write $\gamma(\epsilon)$ in terms of $\tilde\gamma(\epsilon)$, defined as:
\begin{align}
\tilde\gamma(\epsilon)\equiv \bar n\gamma(\epsilon).
\end{align}
To the first few orders:
\begin{align}
\hat{\mathbf n}_{z,(0)}=&\bar n \tilde\gamma_{(0)}\hat d_z,\notag\\
\hat{\mathbf n}_{z,(1)}=&\bar n \tilde\gamma_{(1)}\hat d_z,\notag\\
\hat{\mathbf n}_{z,(2)}=&\bar n \tilde\gamma_{(2)}\hat d_z,\notag\\
[\hat{\mathbf n}_z\hat{\mathbf n}_w]_{(0)}=&\bar n^2 \tilde\gamma_{(0)}^2\hat d_{\{z,w\}},\notag\\
[\hat{\mathbf n}_z\hat{\mathbf n}_w]_{(1)}=&\bar n^2\cdot 2 \tilde\gamma_{(0)}\tilde\gamma_{(1)}\hat d_{\{z,w\}},\notag\\
[\hat{\mathbf n}_z\hat{\mathbf n}_w]_{(2)}=&\bar n^2\cdot [2\tilde\gamma_{(0)}\tilde\gamma_{(2)}+\tilde\gamma_{(1)}^2]\hat d_{\{z,w\}}.\label{eq:PE_boson_laughlin_nznw}
\end{align}

\textbf{Computing $\tilde\gamma_{(m)}$:} Let's first compute $\gamma_{(0)}$ using the sum-rule Eq.(\ref{eq:PE_N_sum_rule}). Using magnetic translation symmetry, to the zeroth-order, Eq.(\ref{eq:PE_O_D_tilde_first_few_orders},\ref{eq:PE_Delta_first_few_orders},\ref{eq:PE_O_D_final_first_few_orders},\ref{eq:PE_boson_laughlin_nznw}) lead to:
\begin{align}
\bar n=\langle \hat n^{(e)}_z\rangle_{(0)}=\widetilde n_{z,(0)}=\bar n\gamma_{(0)}\mathcal C_{\bar\rho}(\hat d_z)=\tilde\gamma_{(0)}\bar n.
\end{align}
Namely:
\begin{align}
\tilde \gamma_{(0)}=1.
\end{align}
In fact, we will see soon that to all orders, $\gamma_{(m)}$ scales as $\gamma_{(m)}\sim \frac{1}{\bar n}$, justifying the normalized $\tilde\gamma(\epsilon)$.
This scaling also simplifies the calculation in the thermodynamic limit: one can safely replace the $(1+\gamma_{(0)})$ by $\gamma_{(0)}$ in Eq.(\ref{eq:PE_boson_laughlin_q_first_few_orders}):
\begin{align}
\hat{\boldsymbol q}_{z,(1)}=&-\hat n_z^{(1)}-\hat n_z^{(2)}+\bar n \tilde\gamma_{(0)}\hat d_z+O(\bar n^2),\notag\\
\hat{\boldsymbol q}_{z,(2)}=&\bar n\tilde\gamma_{(1)} \hat d_z.\label{eq:PE_boson_laughlin_q_first_few_orders_1}
\end{align}
\emph{Basically, all $\hat{\boldsymbol q}_{z,(m)}$ and $\hat{\mathbf n}_{z,(m)}$ scale as $\bar n$, and $[\hat{\mathbf n}_z\hat{\mathbf n}_w]_{(m)}$ scales as $\bar n^2$.} We will ignore the $O(\bar n^2)$ term in $\hat{\boldsymbol q}_{z,(1)}$ below since it does not contribute in the thermodynamic limit.

Next, we fix $\tilde\gamma_{(1)}$. The sum-rule Eq.(\ref{eq:PE_N_sum_rule}) requires $\langle\hat n^{(e)}_z\rangle_{(1)}=0$. Eq.(\ref{eq:PE_O_D_final_first_few_orders}) demands us to compute $\widetilde n_{z,(1)}$ and $\Delta_{z,(1)}$. According to Eq.(\ref{eq:PE_O_D_tilde_first_few_orders},\ref{eq:PE_boson_laughlin_nznw},\ref{eq:PE_boson_laughlin_q_first_few_orders_1}), we have:
\begin{align}
\widetilde n_{z,(1)}&=\mathcal C_{\bar\rho}(\bar n\tilde\gamma_{(1)}\hat d_z)\notag\\
&+\sum_{w\neq z}\mathcal C_{\bar\rho}(\bar n\tilde\gamma_{(0)}\hat d_z,-\hat n_w^{(1)}-\hat n_w^{(2)}+\bar n\tilde\gamma_{(0)}\hat d_w)\notag\\
=&\bar n\tilde\gamma_{(1)}+\bar n\mathcal C_{\bar\rho}(\hat d_z,\hat n_z^{(1)}+\hat n_z^{(2)})+\bar n^2\sum_{w\neq z}\mathcal C_{\bar\rho}(\hat d_z,\hat d_w),
\end{align}
where the exact sum rule Eq.(\ref{eq:PE_exact_operator_sum_rule_cumulant}) is used.

Because $\hat d_z\hat n_z^{(p)}=\hat d_z$, using Eq.(\ref{eq:PE_cumulant_AB_A}):
\begin{align}
\mathcal C_{\bar\rho}(\hat d_z,\hat n_z^{(p)})=(1-\langle \hat n^{(p)}_z\rangle_{\bar\rho})\langle \hat d_z\rangle_{\bar\rho}=1-\bar n=1+O(\bar n).\label{eq:PE_boson_laughlin_dz_nz}
\end{align}
In the thermodynamic limit, it is safe to ignore the $O(\bar n)$-term above, and $\mathsf G_{(1)}\equiv \bar n\sum_{w\neq z}\mathcal C_{\bar\rho}(\hat d_z,\hat d_w)$ can be converted into a continuum integral: $\bar n\sum_{w\neq z}\rightarrow \frac{1}{2}\int\frac{d^2w}{2\pi}$, where the factor $\frac{1}{2}$ is from the filling fraction $\nu$. Using Eq.(\ref{eq:PE_boson_laughlin_dz_cumulants}), this integral can be computed analytically:
\begin{align}
\mathsf G_{(1)}=\int\frac{d^2w}{4\pi} \Big[[\det\mathsf M_{\{z,w\}}]^2-1\Big]=-\frac{3}{2}.
\end{align}
We arrive at:
\begin{align}
\widetilde n_{z,(1)}&=\bar n[\tilde\gamma_{(1)}+2]+\bar n \mathsf G_{(1)}\notag\\
&=\bar n[\tilde\gamma_{(1)}+\frac{1}{2}].
\end{align}

Next we compute $\Delta_{z,(1)}$, based on Eq.(\ref{eq:PE_Delta_first_few_orders}):
\begin{align}
\Delta_{z,(1)}=\mathcal C_{\bar\rho}(\hat{\boldsymbol q}_{z,(1)})=O(\bar n).
\end{align}
In fact, it is easy to see that to any order $\Delta_{D,(m)}=O(\bar n)$ for any finite operator support $D$. But only the $O(\bar n^0)$ order of $\Delta_D$ can contribute in the thermodynamic limit due to Eq.(\ref{eq:PE_O_D_final_first_few_orders}). Therefore, in the thermodynamic limit, in the present Fine-Grid calculation, \emph{we can safely set $\Delta_D(\epsilon)=0$}, and consequently:
\begin{align}
\langle \hat n^{(e)}_z\rangle_{(m)}=&\widetilde n_{z,(m)},&\langle \hat n^{(e)}_z\hat n^{(e)}_w\rangle_{(m)}=&\widetilde {n_{z}n_{w}}_{(m)},\;\;\forall m.
\end{align}
The sum-rule Eq.(\ref{eq:PE_N_sum_rule}) then dictates:
\begin{align}
\tilde\gamma_{(1)}=-\frac{1}{2}.
\end{align}

We move on to compute $\tilde\gamma_{(2)}$. We only need to compute $\widetilde n_{z,(2)}=0$.  According to Eq.(\ref{eq:PE_O_D_tilde_first_few_orders},\ref{eq:PE_boson_laughlin_nznw},\ref{eq:PE_boson_laughlin_q_first_few_orders_1}), we have:
\begin{align}
&\widetilde n_{z,(2)}=\bar n \tilde\gamma_{(2)}+\sum_{w\neq z}\mathcal C_{\bar\rho}(\bar n \tilde\gamma_{(1)}\hat d_z,-\hat n_w^{(1)}-\hat n_w^{(2)}+\bar n\hat d_w)\notag\\
&+\sum_{w\neq z}\mathcal C_{\bar\rho}(\bar n \hat d_z,\bar n\tilde\gamma_{(1)}\hat d_w)\notag\\
&+\frac{1}{2}\sum'^{z^C}_{w,x}\mathcal C_{\bar\rho}(\bar n \hat d_z,-\hat n_w^{(1)}-\hat n_w^{(2)}+\bar n\hat d_w,-\hat n_x^{(1)}-\hat n_x^{(2)}+\bar n\hat d_x)\notag\\
&-\sum_{w\neq z} \mathcal C_{\bar\rho}(\bar n\hat d_z,-\hat n_w^{(1)}-\hat n_w^{(2)}+\bar n\hat d_w)\mathcal C_{\bar\rho}(-\hat n_w^{(1)}-\hat n_w^{(2)}+\bar n\hat d_w)\label{eq:PE_boson_laughlin_nz_2}
\end{align}
The second and third terms have the same form as the term computed before in $\widetilde n_{z,(1)}$, we do not repeat the calculation. The last term $\propto \bar n^2$ can be safely ignored in the thermodynamic limit. Let's focus on the fourth term:
\begin{align}
&\frac{1}{2}\sum'^{z^C}_{w,x}\mathcal C_{\bar\rho}(\bar n \hat d_z,-\hat n_w^{(1)}-\hat n_w^{(2)}+\bar n\hat d_w,-\hat n_x^{(1)}-\hat n_x^{(2)}+\bar n\hat d_x)\notag\\
=&\sum'^{z^C}_{w,x}\Big[\bar n \mathcal C_{\bar\rho}(\hat d_z,\hat n_w^{(1)},\hat n_x^{(1)})+\bar n \mathcal C_{\bar\rho}(\hat d_z,\hat n_w^{(1)},\hat n_x^{(2)})\notag\\
&-2\bar n^2 \mathcal C_{\bar\rho}(\hat d_z,\hat d_w,\hat n_x^{(1)})+\frac{1}{2}\bar n^3\mathcal C_{\bar\rho}( \hat d_z,\hat d_w,\hat d_x)\Big]\notag\\
\equiv &A_z+B_z+C_z+D_z,
\end{align}
where we used the parton-species permutation symmetry for the Laughlin state and the invariance of cumulants under the rearrangement of adjacent commuting operators. 

Focusing on the last term $D_z$ first. Define the thermodynamic-limit cumulant summation: $\mathsf G_{(2)}\equiv \bar n^2\sum^{' z^C}_{w,x}\mathcal C_{\bar\rho}( \hat d_z,\hat d_w,\hat d_x)$, it can be converted into a continuum integral $\bar n^2\sum^{' z^C}_{w,x}\rightarrow (\frac{1}{2})^2\int\frac{d^2w}{2\pi}\frac{d^2x}{2\pi}$. Using Eq.(\ref{eq:PE_boson_laughlin_dz_cumulants}):
\begin{align}
\mathsf G_{(2)}=&\int\frac{d^2w}{4\pi}\frac{d^2x}{4\pi}\Big[[\det\mathsf M_{\{z,w,x\}}]^2-[\det\mathsf M_{\{z,w\}}]^2\notag\\
&-[\det\mathsf M_{\{z,x\}}]^2-[\det\mathsf M_{\{w,x\}}]^2+2\Big].
\end{align}
$\mathsf G_{(2)}$ can be computed analytically: 
\begin{align}
\mathsf G_{(2)}=&\frac{31}{6},&D_z=&\bar n\frac{\mathsf G_{(2)}}{2}.
\end{align}

To compute the first three terms, one needs to apply the exact sum rule Eq.(\ref{eq:PE_exact_operator_sum_rule_cumulant}) twice, once for $w$ and another for $x$. Noting $\sum_{x\neq z}\sum_w=\sum^{'z^C}_{w,x}+\sum_{w=z,x\neq z}+\sum_{w=x,x\neq z}$:
\begin{align}
A_z=&-\bar n\sum_{x\neq z}\mathcal C_{\bar\rho}(\hat d_z,\hat n^{(1)}_z,\hat n^{(1)}_x)-\bar n\sum_{x\neq z}\mathcal C_{\bar\rho}(\hat d_z,\hat n^{(1)}_x,\hat n^{(1)}_x)\notag\\
=&-\bar n\sum_{x\neq z}\mathcal C_{\bar\rho}(\hat d_z,\hat n^{(1)}_z,\hat n^{(1)}_x)-\bar n\sum_{x\neq z}(1-2\bar n)\mathcal C_{\bar\rho}(\hat d_z,\hat n^{(1)}_x)\notag\\
=&\bar n\mathcal C_{\bar\rho}(\hat d_z,\hat n^{(1)}_z,\hat n^{(1)}_z)+\bar n(1-2\bar n)\mathcal C_{\bar\rho}(\hat d_z,\hat n^{(1)}_z).\notag\\
=&\bar n(1-2\bar n)\mathcal C_{\bar\rho}(\hat d_z,\hat n^{(1)}_z)+\bar n(1-2\bar n)\mathcal C_{\bar\rho}(\hat d_z,\hat n^{(1)}_z).
\end{align}
where, in the second and fourth lines, we used the idempotent identity Eq.(\ref{eq:PE_cumulant_idempotent}), and in the third line, we used the exact sum rule for $\hat n^{(1)}_x$. Eq.(\ref{eq:PE_boson_laughlin_dz_nz}) leads to the thermodynamic-limit result after ignoring $O(\bar n^2)$ terms:
\begin{align}
A_z=2\bar n.
\end{align}
Consider $B_z$:
\begin{align}
B_z=&-\bar n\sum_{x\neq z}\mathcal C_{\bar\rho}(\hat d_z,\hat n^{(1)}_z,\hat n^{(2)}_x)-\bar n\sum_{x\neq z}\mathcal C_{\bar\rho}(\hat d_z,\hat n^{(1)}_x,\hat n^{(2)}_x).
\end{align}
Here, since $\hat n^{(1)}_x$ and $\hat n^{(2)}_x$ in the second term are for different parton species, this term is genuinely $\propto \bar n^2$ and can be safely ignored. Using Eq.(\ref{eq:PE_cumulant_AB_A}) for $\hat d_z\hat n^{(1)}_z=\hat d_z$
\begin{align}
&B_z=-\bar n\sum_{x\neq z}\mathcal C_{\bar\rho}(\hat d_z,\hat n^{(1)}_z,\hat n^{(2)}_x)\notag\\
&=-\bar n\sum_{x\neq z}\Big[(1-\langle \hat n^{(1)}_z\rangle_{\bar\rho})\mathcal C_{\bar\rho}(\hat d_z,\hat n^{(2)}_x)-\langle \hat d_z\rangle_{\bar\rho}\mathcal C_{\bar\rho}(\hat n^{(1)}_z,\hat n^{(2)}_x)\Big]\notag\\
&=\bar n\Big[\mathcal C_{\bar\rho}(\hat d_z,\hat n^{(2)}_z)-\mathcal C_{\bar\rho}(\hat n^{(1)}_z,\hat n^{(2)}_z)\Big]=\bar n,
\end{align}
where we implicitly ignored all higher order $O(\bar n^2)$ terms. \emph{The same will be done implicitly in later calculations}.

Finally, consider $C_z$:
\begin{align}
&C_z=2\bar n^2\sum_{w\neq z}\mathcal C_{\bar\rho}(\hat d_z,\hat d_w,\hat n_z^{(1)})+2\bar n^2\sum_{w\neq z}\mathcal C_{\bar\rho}(\hat d_z,\hat d_w,\hat n_w^{(1)})\notag\\
=&2\bar n^2\sum_{w\neq z}\Big[\big[\mathcal C_{\bar\rho}(\hat d_z,\hat d_w)-\mathcal C_{\bar\rho}(\hat n^{(1)}_z,\hat d_w)\big]\notag\\
&+\big[\mathcal C_{\bar\rho}(\hat d_z,\hat d_w)-\mathcal C_{\bar\rho}(\hat d_z,\hat n^{(1)}_w)\big]\Big]\notag\\
=&4\bar n^2\sum_{w\neq z}\mathcal C_{\bar\rho}(\hat d_z,\hat d_w)=4\bar n\cdot(-\frac{3}{2})=-6\bar n.
\end{align}

Collecting all terms in Eq.(\ref{eq:PE_boson_laughlin_nz_2}), we have:
\begin{align}
0=\widetilde n_{z,(2)}=\bar n\big[\tilde\gamma_{(2)}-\tilde\gamma_{(1)}-\frac{5}{12}],
\end{align}
leading to:
\begin{align}
\tilde\gamma_{(2)}=-\frac{1}{12}.
\end{align}

\textbf{Computing $g_{(m)}(r)$:} With all $\tilde\gamma$ determined up to the second order, we can compute the pair-correlation function up to the same order. The zeroth order $g_{(0)}(r)$ is straightforward, from Eq.(\ref{eq:PE_O_D_tilde_first_few_orders},\ref{eq:PE_boson_laughlin_parton_pair_corr},\ref{eq:PE_boson_laughlin_dzdw_cumulants},\ref{eq:PE_boson_laughlin_nznw}), we have:
\begin{align}
&g_{(0)}(|z-w|)\equiv \frac{1}{\bar n^2}\langle \hat n_z^{(e)}\hat n_w^{(e)}\rangle_{(0)}=\frac{1}{\bar n^2} \widetilde{n_zn_w}_{(0)}=\langle \hat d_{\{z,w\}}\rangle_{\bar\rho},\notag\\
&\Rightarrow g_{(0)}(r)=[g^{(p)}(r)]^2=\big(1-e^{-\frac{r^2}{4}}\big)^2.
\end{align}

To compute the first order $g_{(1)}(|z-w|)=\frac{1}{\bar n^2} \widetilde{n_zn_w}_{(1)}$, from Eq.(\ref{eq:PE_O_D_tilde_first_few_orders}), we need:
\begin{align}
&\frac{\widetilde{n_zn_w}_{(1)}}{\bar n^2}=\mathcal C_{\bar\rho}([\hat{\mathbf n}_z\hat{\mathbf n}_w]_{(1)})+\sum_{x\in\{z,w\}^C}\mathcal C_{\bar\rho}([\hat{\mathbf n}_z\hat{\mathbf n}_w]_{(0)},\hat{\boldsymbol q}_{x,(1)})\notag\\
=&2\tilde\gamma_{(1)}\langle \hat d_{\{z,w\}}\rangle_{\bar\rho}+\sum_{x\in\{z,w\}^C}\mathcal C_{\bar\rho}(\hat d_{\{z,w\}},-\hat n_x^{(1)}-\hat n_x^{(2)}+\bar n\hat d_x )\notag\\
=&-\langle \hat d_{\{z,w\}}\rangle_{\bar\rho}+4\mathcal C_{\bar\rho}(\hat d_{\{z,w\}},\hat n_z^{(1)})+\bar n\sum_{x\in\{z,w\}^C}\mathcal C_{\bar\rho}(\hat d_{\{z,w\}},\hat d_x )\notag\\
=&3\langle \hat d_{\{z,w\}}\rangle_{\bar\rho}+\bar n\sum_{x\in\{z,w\}^C}\mathcal C_{\bar\rho}(\hat d_{\{z,w\}},\hat d_x )\notag\\
\equiv&3\langle \hat d_{\{z,w\}}\rangle_{\bar\rho}+G_{(1)}(|z-w|).
\end{align}
where we used Eq.(\ref{eq:PE_boson_laughlin_nznw},\ref{eq:PE_boson_laughlin_q_first_few_orders_1}), $\tilde\gamma_{(0)}=1,\tilde\gamma_{(1)}=-1/2$, exact sum rule Eq.(\ref{eq:PE_exact_operator_sum_rule_cumulant}), and Eq.(\ref{eq:PE_cumulant_AB_A}) for $\hat d_{\{z,w\}}\hat n_z^{(1)}=\hat d_{\{z,w\}}$. The last term can be converted into a continuum integral: $\bar n\sum_{x\in\{z,w\}^C}\rightarrow \frac{1}{2}\int \frac{d^2x}{2\pi}$ in the thermodynamic limit:
\begin{align}
G_{(1)}(|z-w|)=\int \frac{d^2x}{4\pi} \Big([\det \mathsf M_{\{z,w,x\}}]^2-[\det \mathsf M_{\{z,w\}}]^2\Big),
\end{align}
where we used Eq.(\ref{eq:PE_boson_laughlin_dzdw_cumulants}). The integral can be carried out analytically, giving:
\begin{align}
G_{(1)}(r)=e^{-\frac{r^2}{8}}(1-e^{-\frac{r^2}{8}})^2-3g_{(0)}(r).
\end{align}
The second term in $G_{(1)}(r)$ cancels the term $3\langle \hat d_{\{z,w\}}\rangle_{\bar\rho}$, leading to:
\begin{align}
g_{(1)}(r)=e^{-\frac{r^2}{8}}(1-e^{-\frac{r^2}{8}})^2.
\end{align}

Finally, one can compute $g_{(2)}(r)$. The procedure is tedious but conceptually similar to the calculation of $\tilde\gamma_{(2)}$. Define the thermodynamic-limit cumulant summation $G_{(2)}(|z-w|)\equiv\bar n^2\sum^{'\{z,w\}^C}_{x,y}\mathcal C_{\bar\rho}(\hat d_{\{z,w\}},\hat d_x,\hat d_y)$, it can be converted to a double integral using Eq.(\ref{eq:PE_boson_laughlin_dzdw_cumulants}):
\begin{align}
&G_{(2)}(|z-w|)=\int \frac{d^2x}{4\pi}\frac{d^2y}{4\pi} \Big([\det \mathsf M_{\{z,w,x,y\}}]^2\notag\\
&-[\det \mathsf M_{\{z,w\}}]^2[\det \mathsf M_{\{x,y\}}]^2-[\det \mathsf M_{\{z,w,x\}}]^2\notag\\
&-[\det \mathsf M_{\{z,w,y\}}]^2+2[\det \mathsf M_{\{z,w\}}]^2\Big),
\end{align}
which we unfortunately cannot carry out analytically (although the numerical values of $G_{(2)}(r)$ are straightforward to compute). The final result is:
\begin{align}
g_{(2)}(r)=\frac{73}{12}g_{(0)}(r)+\frac{9}{2}G_{(1)}(r)+\frac{1}{2}G_{(2)}(r).
\end{align}

Using the factorization of cumulanants Eq.(\ref{eq:PE_cumulant_factorize}), it can be shown that $G_{(m)}(r\rightarrow\infty)$ satisfies:
\begin{align}
G_{(1)}(r\rightarrow\infty)=&2\mathsf G_{(1)},\notag\\
G_{(2)}(r\rightarrow\infty)=&2\mathsf G_{(2)}+2\mathsf G_{(1)}^2.
\end{align}
With the values of $\mathsf G_{(1)},\mathsf G_{(2)}$, we know that $g_{(1)}(r\rightarrow\infty)=g_{(2)}(r\rightarrow\infty)=0$. In addition, the determinant form of the cumulants immediately indicates $g_{(m)}(0)=0$, $\forall m$. These together lead to the expected behavior:
\begin{align}
g_{[0]}(r)=&g_{(0)}(r),\notag\\
g_{[m]}(r)=&g_{[m-1]}(r)+g_{(m)}(r),\;\; m\ge 1,
\end{align}
satisfy
\begin{align}
g_{[m]}(0)=&0,& g_{[m]}(r\rightarrow\infty)=&1.
\end{align}

\begin{figure}
    \centering
    \includegraphics[width=1.0\linewidth]{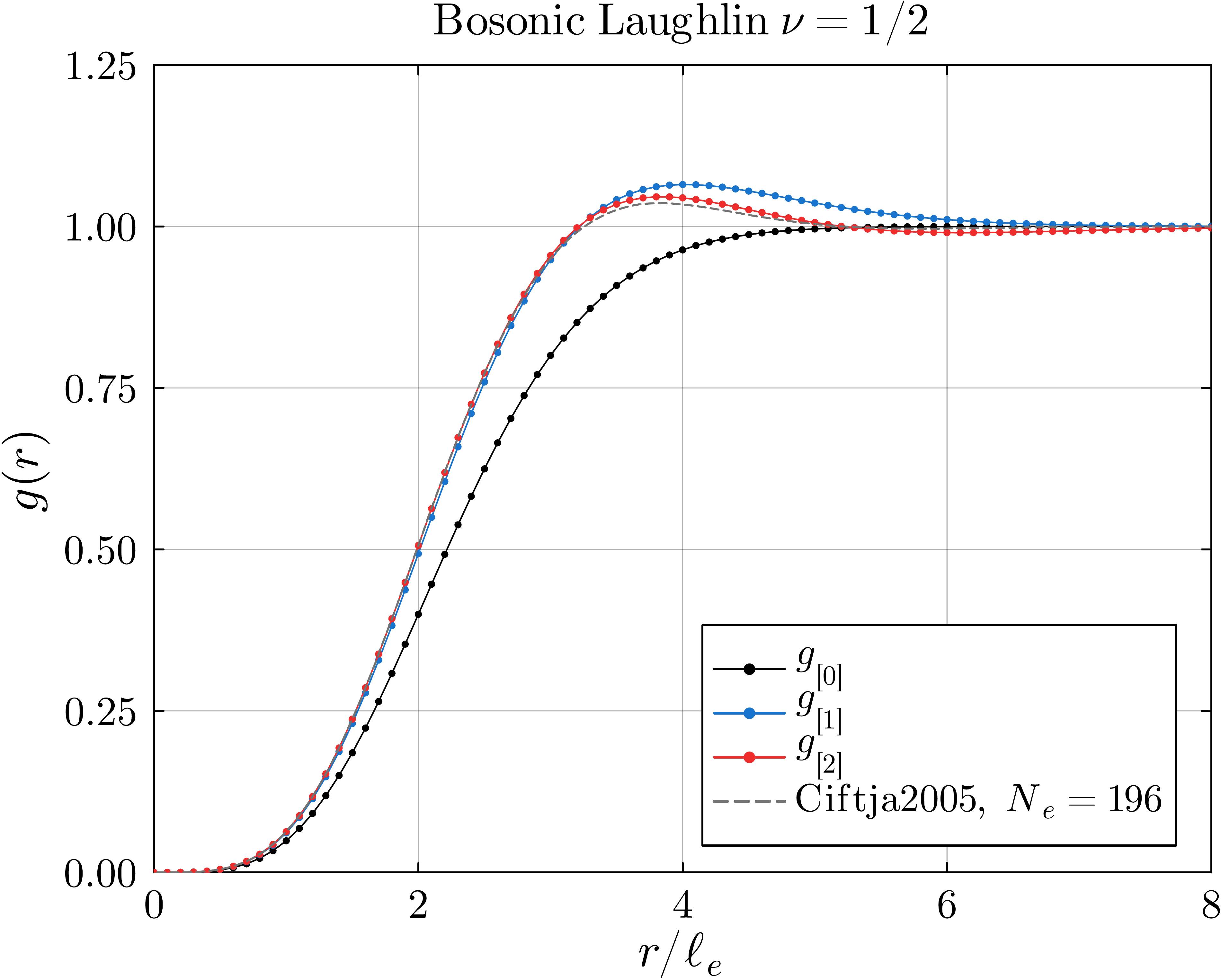}
    \caption{Pair-correlation function for $\nu=\frac12$ Laughlin state: comparison between the present thermodynamic-limit low-order PE and published $N_e=196$  Monte Carlo\cite{Ciftja2005} results.}
    \label{fig:nu12_pair_correlation}
\end{figure}

We plot these PE pair correlation functions, together with the previously published  Monte Carlo(MC) result ($N_e=196$)\cite{Ciftja2005} in Fig.\ref{fig:nu12_pair_correlation}. Clearly, as the PE order increases, the pair correlation becomes very accurate for the short-range part, while the longer-range part is still not fully captured. This is expected since PE is a local expansion. In fact, the integral of $g(r)$ should satisfy the exact electron-number sum rule:
\begin{align}
S\equiv\frac{1}{2}\int \frac{d^2z}{2\pi} [g(|z|)-1]=-1.
\end{align}
This sum rule is \emph{violated} by a finite-order PE:
\begin{align}
S_{[m]}\equiv\frac{1}{2}\int \frac{d^2z}{2\pi} [g_{[m]}(|z|)-1]\neq -1.
\end{align}
One may directly compute $S_{[m]}$ via a useful identity: 
\begin{align}
\frac{1}{2}\int \frac{d^2z}{2\pi} [G_{(m)}(|z|)-G_{(m)}(\infty)]=\mathsf G_{(m+1)},
\end{align}
which follows from Eq.(\ref{eq:PE_product_cumulant}). The results are:
\begin{align}
S_{[0]}=&-\frac{3}{2}, &S_{[1]}=&-\frac{5}{6},& S_{[2]}=&-\frac{31}{30}.
\end{align}

As the order increases, the sum-rule is better and better approximated. This violation of the electron-number sum rule suggests that the long-range Coulomb interaction may be challenging to simulate well using PE, which we will explore shortly in Sec.\ref{sec:FCI_saddle}.

\textbf{PE for quantum geometric tensors, Preparation:} When computing the quantum geometric tensor $\mathcal Q_{ab}$ for general tangent vectors $\hat X_a,\hat X_b$, one would go outside the LLL of partons, and $R>1$ fusion channels will be present. In this simple benchmark test, we avoid this technical complication by focusing on specific tangent vectors on the mean-field saddle point: $\hat X_a$ and $\hat X_b$ are generators of the pure gauge transformation. 

Since the $U(1)$ gauge charges of the two partons are ${\rm q}^{(1)}=1$, ${\rm q}^{(2)}=-1$, we define the gauge charge density:
\begin{align}
\hat n^g_{z}\equiv \hat n^{(1)}_{z}-\hat n^{(2)}_{z}.
\end{align}
The pure gauge generator at momentum $\mathbf p$ is:
\begin{align}
\hat G_{\mathbf p}\equiv \frac{1}{\sqrt{N^{(e)}_s}}\sum_z e^{i\mathbf p\cdot \mathbf r_z}\hat n^g_{z},
\end{align}
where $\mathbf r_z\equiv (\text{Re}[z],\text{Im}[z])$. $\hat G_{\mathbf p}=\hat G^\dagger_{-\mathbf p}$ generates a pure gauge transformation as a unitary rotation:
\begin{align}
|\Psi^{MF}(\bar\rho)\rangle \rightarrow e^{-i\sum_{\mathbf p}\theta_{\mathbf p}\hat G_{\mathbf p}}|\Psi^{MF}(\bar\rho)\rangle,
\end{align}
where Hermiticity requires $\theta_{\mathbf p}=\theta^*_{-\mathbf p}$. If we choose parameters $a={\rm Re}[\theta_{\mathbf p}]$, $b={\rm Im}[\theta_{\mathbf p}]$, the parton bilinear operators defined in Eq.(\ref{eq:PE_partial_a}) are $\hat X_a(\mathbf p)= (-i)[\hat G_{\mathbf p}+\hat G_{-\mathbf p}], \hat X_b(\mathbf p)= \hat G_{\mathbf p}-\hat G_{-\mathbf p}$. Due to magnetic translation symmetry, apart from isolated $\mathbf p$ points satisfying $-\mathbf p=\mathbf p$, it is easy to show that $\mathcal Q_{ab}(\epsilon)$ has the form:
\begin{align}
\mathcal Q_{ab}(\epsilon)= 2\mathcal Q_{\mathbf p}(\epsilon) \delta_{ab}, 
\end{align}
where the real quantity $\mathcal Q_{\mathbf p}(\epsilon)$ can be computed by considering
\begin{align}
\hat {\mathsf  X}_a=\hat {\mathsf  X}_b= \hat G_{\mathbf p}.
\end{align}
Namely, $\mathcal Q_{\mathbf p}(\epsilon)$ determines the quantum metric of the Hdet tangent states of pure gauge transformations. 

The advantage of this benchmark test is that, since the exact Hdet states are not changed by a pure gauge transformation, we know the exact result at $\epsilon=1$:
\begin{align}
\text{Exact result: }\mathcal Q_{\mathbf p}=0,\;\;\forall \mathbf p
\end{align}
The goal of this calculation is the low-order PE results in the \emph{thermodynamic-limit}:
\begin{align}
\textbf{Goal: }\lim_{N_e\rightarrow \infty}\mathcal Q_{\mathbf p,[m]}.
\end{align}

Using the general PE formula Eq.(\ref{eq:PE_Q_tensor_X},\ref{eq:PE_Q_tensor_final_first_few_orders}), we need to compute:
\begin{align}
\mathcal Q_{\mathbf p,(0)}=& \mathcal C_{\bar\rho}(\hat G_{-\mathbf p},\hat G_{\mathbf p}),\notag\\
\mathcal Q_{\mathbf p,(1)}=& \sum_x \mathcal C_{\bar\rho}(\hat G_{-\mathbf p},\hat{\boldsymbol q}_{x,(1)},\hat G_{\mathbf p}),\notag\\
\mathcal Q_{\mathbf p,(2)}=& \sum_x \mathcal C_{\bar\rho}(\hat G_{-\mathbf p},\hat{\boldsymbol q}_{x,(2)},\hat G_{\mathbf p})\notag\\
&+\frac{1}{2}\sum_{x\neq y}\mathcal C_{\bar\rho}(\hat G_{-\mathbf p},\hat{\boldsymbol q}_{x,(1)},\hat{\boldsymbol q}_{y,(1)},\hat G_{\mathbf p})\notag\\
&-\sum_x\mathcal C_{\bar\rho}(\hat{\boldsymbol q}_{x,(1)})\mathcal C_{\bar\rho}(\hat G_{-\mathbf p},\hat{\boldsymbol q}_{x,(1)},\hat G_{\mathbf p})\notag\\
&-\sum_x\mathcal C_{\bar\rho}(\hat G_{-\mathbf p},\hat{\boldsymbol q}_{x,(1)})\mathcal C_{\bar\rho}(\hat{\boldsymbol q}_{x,(1)},\hat G_{\mathbf p})\label{eq:PE_boson_laughlin_Q_p_first_few_orders}
\end{align}
These calculations will be performed in the real space: we write:
\begin{align}
\hat G_{\mathbf p}= &\frac{1}{\sqrt{N^{(e)}_s}}\sum_z e^{i\mathbf p\cdot \mathbf r_z}(\hat n^{(1)}_{z}-\hat n^{(2)}_{z})\notag\\
\hat G_{-\mathbf p}= &\frac{1}{\sqrt{N^{(e)}_s}}\sum_w e^{-i\mathbf p\cdot \mathbf r_w}(\hat n^{(1)}_{w}-\hat n^{(2)}_{w}).
\end{align}
Note that, unlike the static PE calculation, we now have contact terms at $z=w$ or $x=z$, etc. Only the PE operator sites are pairwise distinct, e.g., $x\neq y$.

\textbf{Computing $\mathcal Q_{\mathbf p,(m)}$: } At the zeroth order, noticing $C_{\bar\rho}(\hat n^{(1)}_{z},\hat n^{(2)}_{w})=0$ ($\forall z,w$) based on Eq.(\ref{eq:PE_cumulant_uncorrelated}), and 
\begin{align}
\mathcal C_{\bar\rho}(\hat n_z^{(1)},\hat n_w^{(1)})=\bar n\delta_{z,w}+\bar n^2 h^{(p)}(|z-w|),\label{eq:PE_boson_laughlin_C_nz1_nw1}
\end{align}
we have
\begin{align}
\mathcal Q_{\mathbf p,(0)}=& \frac{2}{N^{(e)}_s}\sum_{z,w}\mathcal C_{\bar\rho}\Big(e^{-i\mathbf p\cdot \mathbf r_w}\hat n^{(1)}_{w}, e^{i\mathbf p\cdot \mathbf r_z}\hat n^{(1)}_{z}\Big)\notag\\
=&\frac{1}{N_e}\left[\sum_z \bar n+\bar n^2\sum_{z,w}e^{i\mathbf p\cdot (\mathbf r_z-\mathbf r_w)}h^{(p)}(|z-w|)\right]\notag\\
=&\frac{1}{N_e}\left[N_e+\bar n N_e\sum_{z}e^{i\mathbf p\cdot \mathbf r_z} h^{(p)}(|z|)\right]\notag\\
=&1+\int \frac{d^2 z}{4\pi}e^{i\mathbf p\cdot \mathbf r_z} h^{(p)}(|z|)\equiv 1+h^{(p)}_{\mathbf p}=1-e^{-\mathbf p^2},\notag\\
\end{align}
where we used Eq.(\ref{eq:PE_boson_laughlin_parton_connected_pair_corr}), and replaced the summation by a continuum integral: $\bar n\sum_{z}\rightarrow \frac{1}{2}\int \frac{d^2z}{2\pi}$. $h^{(p)}_\mathbf p=-e^{-\mathbf p^2}$ is the Fourier integral. When $\mathbf p\rightarrow 0$, $\mathcal Q_{\mathbf p,(0)}\rightarrow 0$, which is expected since the $\mathbf p=0$ gauge transformation only multiplies overall phase factors for the parton Slater determinants. On the other hand, at large $\mathbf p$, the mean-field level $\mathcal Q_{\mathbf p,(0)}\rightarrow 1$, qualitatively different from the exact result. As we will see soon, the next order $\mathcal Q_{\mathbf p,[1]}=\mathcal Q_{\mathbf p,(0)}+\mathcal Q_{\mathbf p,(1)}$ will correct the large-$\mathbf p$ behavior drastically.

To compute the first order $\mathcal Q_{\mathbf p,(1)}$, due to the exact sum rule Eq.(\ref{eq:PE_exact_operator_sum_rule_cumulant}), the $\hat n_x^{(p)}$ operators in $\hat{\boldsymbol q}_{x,(1)}$ can be ignored, and one may replace $\hat{\boldsymbol q}_{x,(1)}$ by $\bar n \hat d_x=\frac{\hat n^{(1)}_x\hat n^{(2)}_x}{\bar n}$:
\begin{align}
&\mathcal Q_{\mathbf p,(1)}\notag\\
=&\frac{\bar n}{N^{(e)}_s}\sum_{z,w,x}e^{i\mathbf p\cdot (\mathbf r_z-\mathbf r_w)}\mathcal C_{\bar\rho}(\hat n_w^{(1)}-\hat n_w^{(2)},\hat d_x,\hat n_z^{(1)}-\hat n_z^{(2)})\notag\\
=&\frac{2\bar n}{N^{(e)}_s}\sum_{z,w,x}e^{i\mathbf p\cdot (\mathbf r_z-\mathbf r_w)}\Big[\mathcal C_{\bar\rho}(\hat n_w^{(1)},\hat d_x,\hat n_z^{(1)})\notag\\&
\qquad\qquad\qquad-\mathcal C_{\bar\rho}(\hat n_w^{(1)},\hat d_x,\hat n_z^{(2)})\Big]\notag\\
=&\frac{2}{\bar nN^{(e)}_s}\sum_{z,w,x}e^{i\mathbf p\cdot (\mathbf r_z-\mathbf r_w)}\Big[\bar n\mathcal C_{\bar\rho}(\hat n_w^{(1)},\hat n_x^{(1)},\hat n_z^{(1)})\notag\\&
\qquad\qquad\qquad-\mathcal C_{\bar\rho}(\hat n_w^{(1)},\hat n_x^{(1)})\mathcal C_{\bar\rho}(\hat n_x^{(2)},\hat n_z^{(2)})\Big],\notag\\
\end{align}
where we used Eq.(\ref{eq:PE_cumulant_uncorrelated},\ref{eq:PE_product_cumulant}). The first term vanishes due to the exact sum rule of $x$.

For the second term, first separate the case $z=w$ from the case $z\neq w$:
\begin{align}
&\mathcal Q_{\mathbf p,(1)}=\frac{-1}{\bar nN_e}\sum_{z,x}\mathcal C^2_{\bar\rho}(\hat n_z^{(1)},\hat n_x^{(1)})\notag\\
&+\frac{-1}{\bar n N_e}\sum^{z\neq w}_{z,w,x}e^{i\mathbf p\cdot (\mathbf r_z-\mathbf r_w)}\mathcal C_{\bar\rho}(\hat n_w^{(1)},\hat n_x^{(1)})\mathcal C_{\bar\rho}(\hat n_x^{(2)},\hat n_z^{(2)}).
\end{align}
Second, using Eq.(\ref{eq:PE_boson_laughlin_C_nz1_nw1}), omitting $O(\bar n)$ terms, we arrive at:
\begin{align}
&\mathcal Q_{\mathbf p,(1)}=-1-2\bar n\sum_z e^{i\mathbf p\cdot \mathbf r_z}h^{(p)}(|z|)\notag\\
&+\frac{-\bar n^3}{N_e}\sum'_{z,w,x}e^{i\mathbf p\cdot (\mathbf r_z-\mathbf r_w)}h^{(p)}(|w-x|)h^{(p)}(|x-z|)\notag\\
=&-1-2 h^{(p)}_{\mathbf p}-(h^{(p)}_{\mathbf p})^2=-(1+h^{(p)}_{\mathbf p})^2=-(1-e^{-\mathbf p^2})^2.
\end{align}
Importantly, the large-$\mathbf p$ limit of $\mathcal Q_{\mathbf p,(1)}$ is $-1$, exactly canceling that of $\mathcal Q_{\mathbf p,(0)}$. We have:
\begin{align}
\mathcal Q_{\mathbf p,[1]}\equiv \mathcal Q_{\mathbf p,(0)}+\mathcal Q_{\mathbf p,(1)}=e^{-\mathbf p^2}(1-e^{-\mathbf p^2}),
\end{align}
whose large-$\mathbf p$ behavior is exponentially close to the exact result.

\begin{figure}
    \centering
    \includegraphics[width=1.0\linewidth]{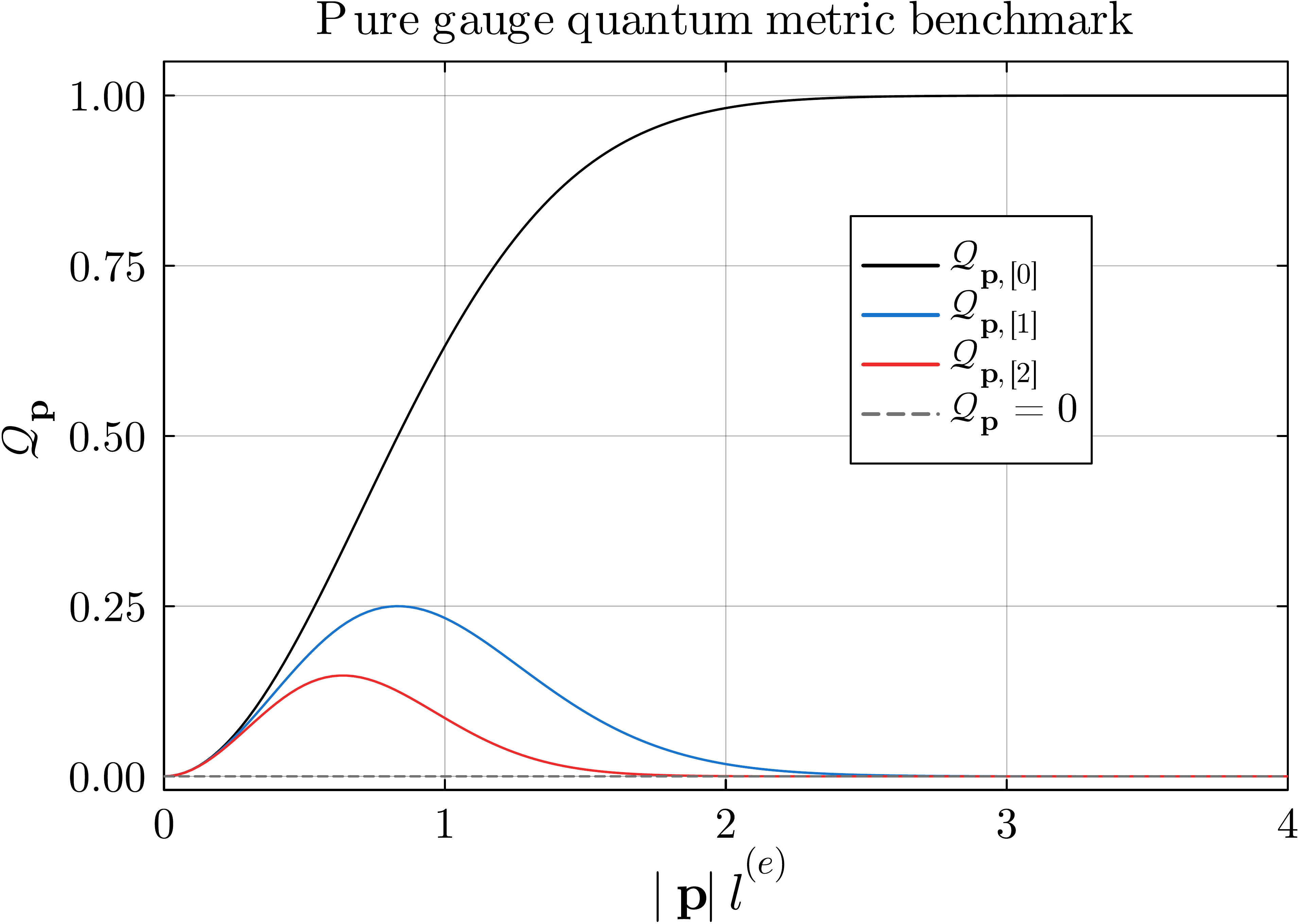}
    \caption{Pure gauge quantum metric $\mathcal Q_{\mathbf p}$ in bosonic Laughlin $\nu=\frac12$ state: comparison between low-order PE and exact results.}
    \label{fig:gauge_metric}
\end{figure}

At the second order, using Eq.(\ref{eq:PE_boson_laughlin_Q_p_first_few_orders}), there are totally four terms. The third term $\propto \bar n$ can be safely ignored in the thermodynamic limit. The fourth term vanishes since $\hat G_{\mathbf p}$ is odd under parton species permutation $1\leftrightarrow 2$, while $\hat {\boldsymbol q}_{x,(1)}$ is even. The first term repeats the $\mathcal Q_{\mathbf p,(1)}$ calculation, but replacing $\tilde\gamma_{(0)}=1$ by $\tilde\gamma_{(1)}=-\frac{1}{2}$, giving $\frac{1}{2}(1-e^{-\mathbf p^2})^2$. For the second term, one must carefully apply the exact sum rule Eq.(\ref{eq:PE_exact_operator_sum_rule_cumulant}) repeatedly, and the calculation is tedious. We only provide the final result:
\begin{align}
&\frac{1}{2}\sum_{x\neq y}\mathcal C_{\bar\rho}(\hat G_{-\mathbf p},\hat{\boldsymbol q}_{x,(1)},\hat{\boldsymbol q}_{y,(1)},\hat G_{\mathbf p})\notag\\
=&-\frac{1}{2}(1-e^{-\mathbf p^2})^2(1+2e^{-\mathbf p^2}).
\end{align}
Together with the first term:
\begin{align}
\mathcal Q_{\mathbf p,(2)}=&-e^{-\mathbf p^2}(1-e^{-\mathbf p^2})^2.\notag\\
\mathcal Q_{\mathbf p,[2]}\equiv&\mathcal Q_{\mathbf p,[1]}+\mathcal Q_{\mathbf p,(2)}=e^{-2\mathbf p^2}(1-e^{-\mathbf p^2}).
\end{align}
Comparing with $\mathcal Q_{\mathbf p,[1]}$, $\mathcal Q_{\mathbf p,[2]}$ has one additional suppression factor $e^{-\mathbf p^2}$, leading to a further improved approximation to the exact $\mathcal Q_{\mathbf p}=0$. We plot these PE results together with the exact result as a function of $|\mathbf p|$ in Fig.\ref{fig:gauge_metric}. The smaller $\mathbf{p}$ is, the higher PE order one has to go to in order to accurately approximate $\mathcal Q_{\mathbf p}$ -- as mentioned, PE is a local expansion.

\subsubsection{Fermionic Laughlin's $\nu=\frac{1}{3}$ state and Jain's $\nu=\frac{2}{5}$ state}
\textbf{Laughlin's $\nu=\frac{1}{3}$ state: pair correlation function. } For a general Laughlin state at filling fraction $\nu=\frac{1}{k}$, analytical calculations similar to the previous $k=2$ case can be carried out. For simplicity of presentation, we focus on $\nu=\frac{1}{3}$. We fix the unit so that the electron's magnetic length $l_e=1$, the electron mean density per Fine-Grid site is $\bar n=\frac{N_e}{(N_s^{(e)})^2}=\frac{1}{3N_s^{(e)}}$.
The local PE gate in Eq.(\ref{eq:PE_Q_def}) is:
\begin{align}
\hat{\mathbf Q}_z(\epsilon)=1+\epsilon\Big[\prod_{p=1}^3(1-\hat n^{(p)}_z)-1\Big]+\epsilon\gamma(\epsilon)\prod_{p=1}^3\hat n_z^{(p)},
\end{align}
and we define the $\hat d_z$ operator:
\begin{align}
\hat d_z=\frac{1}{\bar n^3}\prod_{p=1}^3 \hat n_z^{(p)},
\end{align}
together with $\hat d_Z=\prod_{z\in Z}\hat d_z$ for a set of pairwise distinct sites $Z=\{z_1,..,z_l\}$. Using Wick's theorem, we have:
\begin{align}
\langle \hat d_Z\rangle_{\bar\rho}=[\det \mathsf M_Z]^3,
\end{align}
where the correlation matrix:
\begin{align}
 [\mathsf M_Z]_{ij}\equiv e^{-\frac{|z_i|^2+|z_j|^2}{12}+\frac{\bar z_iz_j}{6}} =\frac{1}{\bar n}\langle f^{(p)\dagger}_{z_i}f^{(p)}_{z_j}\rangle_{\bar\rho}.
\end{align}

The relevant cumulant summations are:
\begin{align}
\mathsf G_{(1)}\equiv& \bar n\sum_{w\in z^C}\mathcal C_{\bar\rho}(\hat d_z,\hat d_w),\notag\\
\mathsf G_{(2)}\equiv &\bar n^2\sum'^{z^C}_{w,x}\mathcal C_{\bar\rho}(\hat d_z,\hat d_w,\hat d_x),\notag\\
G_{(1)}(|z-w|)\equiv &\bar n\sum_{x\in \{z,w\}^C}\mathcal C_{\bar\rho}(\hat d_{\{z,w\}},\hat d_x),\notag\\
G_{(2)}(|z-w|)\equiv &\bar n^2\sum'^{\{z,w\}^C}_{x,y}\mathcal C_{\bar\rho}(\hat d_{\{z,w\}},\hat d_x,\hat d_y).\notag\\
\end{align}
In the thermodynamic limit, these summations can be converted to continuum integrals via:
\begin{align}
\bar n\sum_x\rightarrow \frac{1}{3}\int\frac{d^2x}{2\pi }.
\end{align}
We have analytical results:
\begin{align}
\mathsf G_{(1)}=&-\frac{11}{6},\;\;\;\mathsf G_{(2)}=\frac{5027}{630};\;\;\;s\equiv e^{-\frac{r^2}{6}}:\notag\\
G_{(1)}(r)=&-\frac{11}3+3s^\frac12-2s^\frac23+7s+4s^\frac43\notag\\
&-4s^\frac53-7s^2+2s^\frac73-3s^\frac52+\frac{11}3s^3,
\end{align}
and we leave $G_{(2)}(r)$ as a numerical integral. 

Defining the normalized $\tilde\gamma(\epsilon)$:
\begin{align}
\tilde\gamma(\epsilon)\equiv \bar n^{2}\gamma(\epsilon)=\tilde\gamma_{(0)}+\epsilon\tilde\gamma_{(1)}+\epsilon^2\tilde\gamma_{(2)}+...,
\end{align}
$\tilde\gamma_{(m)}$ can be computed:
\begin{align}
\tilde\gamma_{(0)}=&1,\qquad\tilde\gamma_{(1)}=-3-\mathsf G_{(1)}=-\frac{7}{6},\notag\\
\tilde\gamma_{(2)}=&3+3\mathsf G_{(1)}+2\mathsf G_{(1)}^2-\frac{1}{2}\mathsf G_{(2)}=\frac{293}{1260},
\end{align}
leading to the first few orders:
\begin{align}
g_{(0)}(r)=&(1-s)^3,\notag\\
g_{(1)}(r)=&-2\mathsf G_{(1)}g_{(0)}(r)+G_{(1)}(r),\notag\\
g_{(2)}(r)=&[5\mathsf G_{(1)}^2-\mathsf G_{(2)}]g_{(0)}(r)-3\mathsf G_{(1)}G_{(1)}(r)+\frac12 G_{(2)}(r).
\end{align}

\begin{figure}
    \centering
    \includegraphics[width=1.0\linewidth]{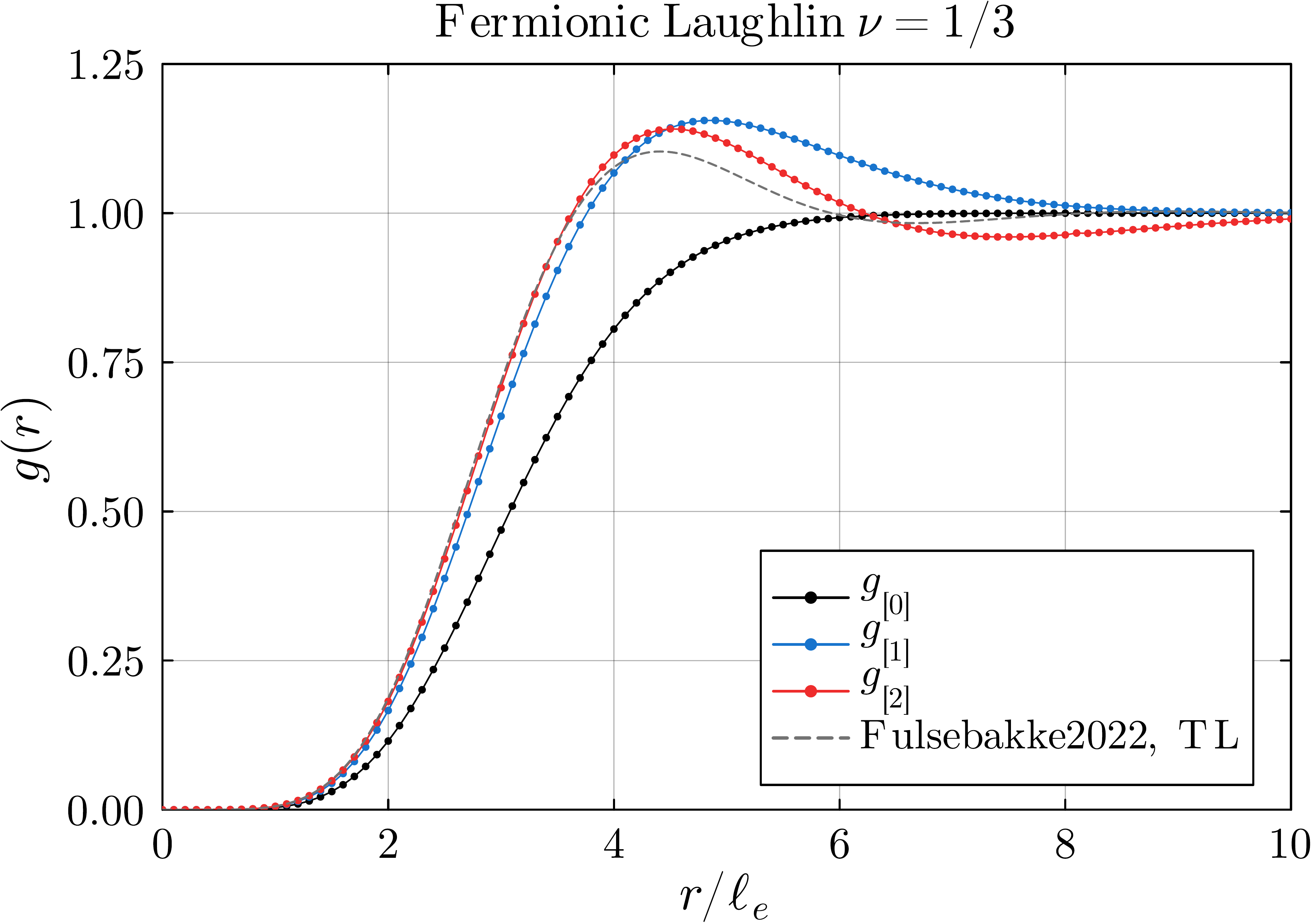}
    \caption{Pair-correlation function for $\nu=\frac13$ Laughlin state: comparison between the present low-order PE and published Monte Carlo (extrapolated to thermodynamic-limit (TL)) \cite{Fulsebakke2022} results. }
    \label{fig:nu13_pair_correlation}
\end{figure}

We plot the PE correlation functions $g_{[m]}(r)$ together with the previously published Monte Carlo\cite{Fulsebakke2022} $g(r)$ in Fig.\ref{fig:nu13_pair_correlation}. Again, the short-range part is well approximated, but the longer range tail would require higher order PE to capture. For example, $g(r)$ satisfies the exact sum rule:
\begin{align}
S\equiv \frac{1}{3}\int\frac{d^2z}{2\pi} [g(|z|)-1]=-1.
\end{align}
When replacing $g(|z|)$ by $g_{[m]}(|z|)$, the low-order PE violates this sum rule:
\begin{align}
S_{[0]}=&-\frac{11}{6},&S_{[1]}=&-\frac{121}{210},&S_{[2]}\approx&-1.132.
\end{align}

\textbf{Jain's $\nu=\frac{2}{5}$ state: pair correlation function. } As the first benchmark example with $R>1$, here we have the number of fusion channels $R=4$, whose Hdet locality structure is given in Sec.\ref{sec:locality_example_nu25}. The scaling $\gamma_{(m)}\propto \frac{1}{\bar n^2}$ remains true, where $\bar n\equiv \bar n^{(e)}=\frac{N_e}{(N_s^{(e)})^2}=\frac{2}{5 N_s^{(e)}}$. So, we define $\tilde\gamma_{(m)}=\bar n^2\gamma_{(m)}$ as before, as well as the operator:
\begin{align}
\hat d_z\equiv \frac{1}{\bar n^3}\sum_{\alpha,\beta=1}^4\lambda^*_\alpha\lambda_\beta \hat D_{z\alpha}^\dagger \hat D_{z\beta},\;\;\hat D_{z\alpha}\equiv\prod_{p=1}^3 f^{(p)}_{z,\mathbf m_\alpha^{(p)}}.
\end{align}
Since there are more than one parton-orbitals $\mathbf m^{(p)}$ per site, we define the parton total density:
\begin{align}
\hat n^{(p)}_z\equiv \sum_{\mathbf m^{(p)}} f^{(p)\dagger}_{z,\mathbf m^{(p)}} f^{(p)}_{z,\mathbf m^{(p)}},\;\; \hat s_z\equiv\sum_{p=1}^{3} \hat n^{(p)}_z.
\end{align}
The exact parton number sum-rule is: 
\begin{align}
\sum_{z\in {\rm Fine-Grid}}\hat n^{(p)}_{z}=N_e, \;\;\forall p\label{eq:nu25_parton_sum_rule}
\end{align}
which is used repeatedly to simplify the cumulants. In the thermodynamic limit, the relevant PE operators are:
\begin{align}
\hat{\boldsymbol q}_{z,(1)}=-\hat s_z+\bar n \tilde\gamma_0\hat d_z, \;\;\hat{\boldsymbol q}_{z,(m)}=\bar n\tilde\gamma_{(m-1)}\hat d_z,(m\ge2).
\end{align}

At this point, one could proceed with the PE process in the same fashion as the previous $R=1$ examples, given the fusion channels/amplitudes in Eq.(\ref{eq:nu25_example_local_fusion}) and the parton RDM in Eq.(\ref{eq:fdagf_FG}). The only challenge is that the most cumulant integrals are complicated to evaluate analytically, but the numerical evaluation is still straightforward. 

\textbf{Uniform-parton-orbitals prescription:} However, one PE-related technical issue needs to be addressed. In Eq.(\ref{eq:fdagf_FG}), where we defined the parton RDM after the change of viewpoint, one finds that there is a normalization factor $\frac{1}{N^{(p)}_{\rm coh},aLL}$--the number of parton orbitals per site per LL. If one literally follows the parton orbital construction in Eq.(\ref{eq:nu25_example_local_fusion}), one would find for parton $p=3$, per Fine-Grid site, the $0LL$ has only one orbital: $N^{(p)}_{{\rm coh},0LL}=1$, while two orbitals are present for $1LL$: $N^{(p)}_{{\rm coh},1LL}=2$; Namely, the RDM for the coherent states in the $1LL$ is effectively smaller by a factor of 2 compared with the $0LL$. A more natural way to define the parton RDM is to introduce parton orbitals so that $\frac{1}{\tilde N^{(p)}_{{\rm coh},aLL}}=\frac{1}{\tilde N^{(p)}_{\rm coh}}$, independent of the LL index-$a$. Practically, it means that we can choose:
\begin{align}
\tilde N^{(p)}_{\rm coh}=\max_{a\in \mathsf L^{(p)}}[N^{(p)}_{{\rm coh},aLL}],
\end{align}
where we introduced $\mathsf L^{(p)}$ for the collection of LL indices for parton-$p$: $\mathsf L^{(3)}=\{0,1\}$. In the present example, $\tilde N^{(3)}_{{\rm coh},0LL}=\tilde N^{(3)}_{{\rm coh},1LL}=2$, and the parton-RDM for the $0LL$ and $1LL$ are on the same footing. Despite letting $0LL$ have two orbitals per site: $\{|z_{(0)}^{(3)}\rangle_{0LL},|z_{(1)}^{(3)}\rangle_{0LL}\}$, only $|z_{(0)}^{(3)}\rangle_{0LL}$ participates in the fusion as Eq.(\ref{eq:nu25_example_local_fusion}). Note that, although $|z_{(1)}^{(3)}\rangle_{0LL}$ is unused in fusion, it still participates in the exact parton-number sum-rule Eq.(\ref{eq:nu25_parton_sum_rule}). We call this prescription \emph{uniform-parton-orbitals}.

This uniform-parton-orbitals prescription is immaterial if one performs a full projection, since it only gives an overall constant factor for the exact Hdet wavefunction. But, for the low-order PE calculation, this prescription matters. Both calculations, with or without the prescription, lead to the same $\epsilon\rightarrow 1$ exact result. But for PE, they give different expansion coefficients.

\textbf{Normalized-fusion-amplitude prescription:} A related issue is about the fusion amplitudes. We have computed these fusion amplitudes in Eq.(\ref{eq:nu25_example_local_fusion}), based on the real-space prescription result Eq.(\ref{eq:fermion_general_fusion},\ref{eq:fermion_special_fusion}). For fixed $(a,b,c)$ LL indices of parton-$(1,2,3)$, we have argued that the general form of the fusion amplitude Eq.(\ref{eq:fermion_general_fusion}) can be fixed by the Galilean invariance (see discussions near Eq.(\ref{eq:z_w_overlap})),
up to an overall factor, which we denote as $M_{(abc)}$ here. If we write down the two relevant fusion operators for $(a,b,c)=(0,0,0)$ and $(a,b,c)=(0,0,1)$, using Eq.(\ref{eq:nu25_example_local_F}), we have $\hat {\mathbf F}_z=\hat {\mathbf F}_{z,(0,0,0)}+\hat {\mathbf F}_{z,(0,0,1)}$, where
\begin{align}
\hat{\mathbf F}_{z,(0,0,0)}=&c_z^\dagger f^{(1)}_{z,1}f^{(2)}_{z,1}f^{(3)}_{z,1}\equiv c_z^\dagger \hat D_{z,(0,0,0)},\notag\\
\hat{\mathbf F}_{z,(0,0,1)}=&c_z^\dagger\Big[\frac{\sqrt 2}{5}f^{(1)}_{z,2}f^{(2)}_{z,1}f^{(3)}_{z,2}+\frac{\sqrt 2}{5}f^{(1)}_{z,1}f^{(2)}_{z,2}f^{(3)}_{z,2}\notag\\
&-\frac{4}{5}f^{(1)}_{z,1}f^{(2)}_{z,1}f^{(3)}_{z,3}\Big]\equiv c_z^\dagger \hat D_{z,(0,0,1)}.
\end{align}

As mentioned before, one way to understand the locality structure is to view the fusion as the map from the electronic Hilbert space to the parton Hilbert space (see discussion below Eq.(\ref{eq:parton_image_discussion})). The parton operators $\hat D_{z,(a,b,c)}$ introduced above can be viewed as the "parton-image-state" of the electron in the parton's $(a,b,c)$-LL sector. Note that, as parton many-body states, $\hat D_{z,(0,0,0)}$ is normalized, but $\hat D_{z,(0,0,1)}$ is \emph{not} (norm$=\frac{2}{\sqrt 5}$). Namely, the norm of the parton-image-state depends on the LL-indices $(a,b,c)$. 

Again, a more natural way to define the parton-image-state is to tune the overall factor $M_{(abc)}$ so that, the parton-image-state is \emph{always} normalized independent of $(a,b,c)$. After this is done, we have the following fusion operators: $\hat {\widetilde{\mathbf F}}_z=\hat {\widetilde{\mathbf F}}_{z,(0,0,0)}+\hat {\widetilde{\mathbf F}}_{z,(0,0,1)}$, where:
\begin{align}
{\widetilde{\mathbf F}}_{z,(0,0,0)}=&c_z^\dagger f^{(1)}_{z,1}f^{(2)}_{z,1}f^{(3)}_{z,1}\equiv c_z^\dagger \hat {\widetilde D}_{z,(0,0,0)},\notag\\
{\widetilde{\mathbf F}}_{z,(0,0,1)}=&c_z^\dagger\Big[\frac{1}{\sqrt {10}}f^{(1)}_{z,2}f^{(2)}_{z,1}f^{(3)}_{z,2}+\frac{1}{\sqrt {10}}f^{(1)}_{z,1}f^{(2)}_{z,2}f^{(3)}_{z,2}\notag\\
&-\frac{2}{\sqrt 5}f^{(1)}_{z,1}f^{(2)}_{z,1}f^{(3)}_{z,3}\Big]\equiv c_z^\dagger \hat{\widetilde D}_{z,(0,0,1)}.
\end{align}
$\hat{\widetilde D}_{z,(0,0,1)}=\frac{\sqrt 5}{2}\hat D_{z,(0,0,1)}$ is now normalized. We call this procedure \emph{normalized-fusion-amplitude} prescription.

Similarly, the normalized-fusion-amplitude prescription is immaterial for Jain's sequence if one performs the full projection, since it only gives an overall constant factor for the exact Hdet state. But for the low-order PE analysis, this prescription matters.

\begin{figure}
    \centering
    \includegraphics[width=1.0\linewidth]{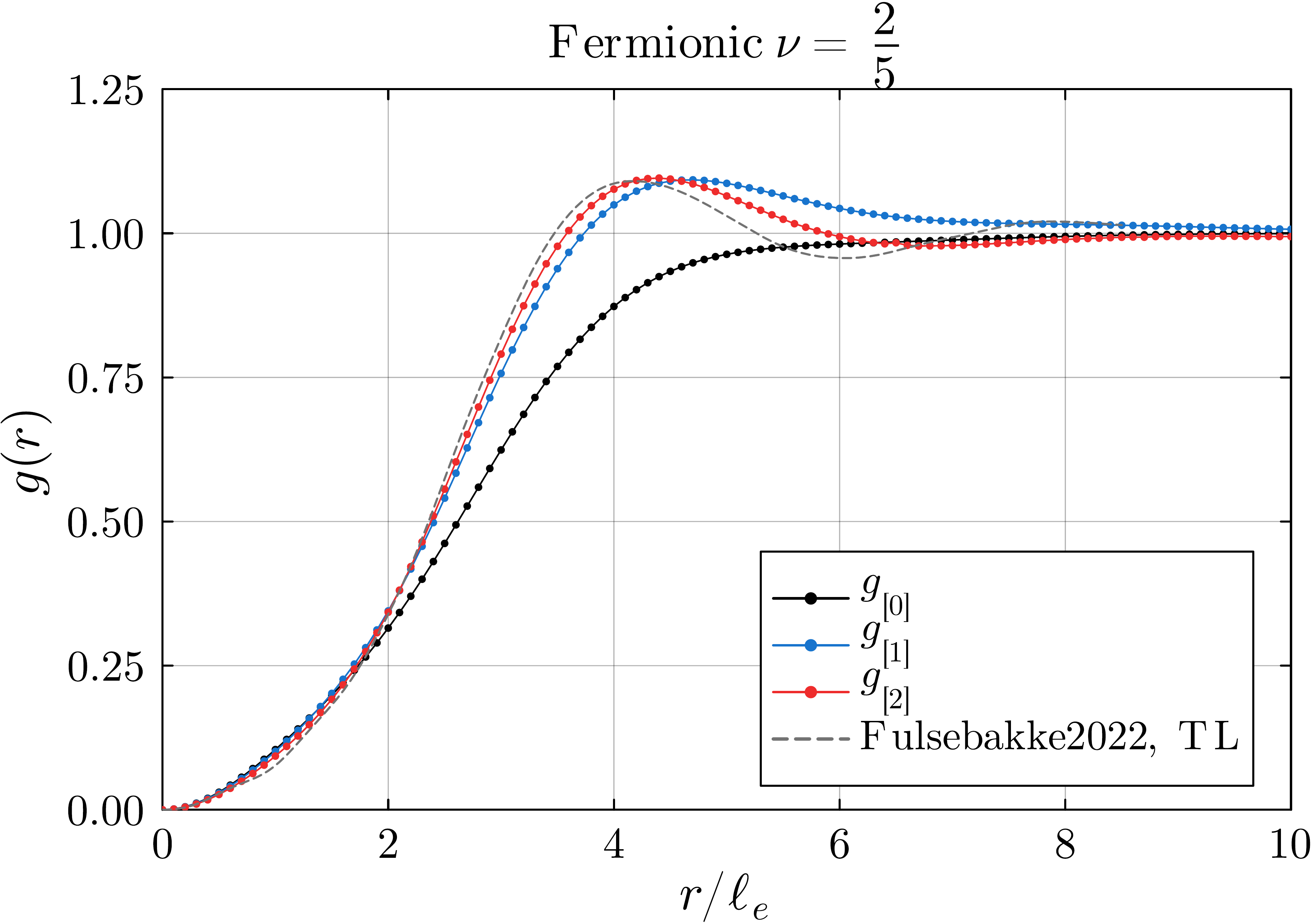}
    \caption{Pair-correlation function for $\nu=\frac25$: the present low-order PE results for Jain state and published Monte Carlo (extrapolated to thermodynamic-limit (TL))\cite{Fulsebakke2022} result for Jain-Kamilla state.}
    \label{fig:nu25_pair_correlation}
\end{figure}

\textbf{Pair correlation results: } We have applied both uniform-parton-orbitals and normalized-fusion-amplitude prescriptions. Performing the numerical integral for the cumulant expansion, we find:
\begin{align}
\tilde\gamma_{(0)}=8,\qquad\tilde\gamma_{(1)}\approx -9.837,\qquad \tilde\gamma_{(2)}\approx 2.118,
\end{align}
and we plot the thermodynamic-limit PE pair correlation function up to the second order in Fig.\ref{fig:nu25_pair_correlation}. Note that due to $R=4$, a large-size exact Monte Carlo (MC) simulation result is not available for comparison. On the other hand, Jain-Kamilla\cite{JainKamilla1997} developed a method to simulate the $\nu=\frac25$ FQH state, by writing it as a determinant with many-body coordinate-dependent entries, enabling a large-scale MC. Precisely speaking, the Jain-Kamilla $\nu=\frac25$ state is \emph{not} the original Jain's state, which we have shown to be an Hdet wavefunction with $R=4$. Therefore, our PE simulation for the original Jain's state should \emph{not} be directly compared with the Jain-Kamilla state. Nevertheless, since the MC (extrapolated to the thermodynamic limit) result for the pair correlation of the Jain-Kamilla state is available in the literature\cite{Fulsebakke2022}, we also present it in the same figure, as a reference.

Noticeably, the pair correlation of the 2nd order PE is slightly higher than the Jain-Kamilla state at short-range $r\lesssim 2$ (it is worth mentioning that Ref\cite{Fulsebakke2022} pointed out that their thermodynamic-limit extrapolation is not smooth in the small-$r$ regime), where the PE is supposed to be most accurate; while in the mid-range $2\lesssim r\lesssim 4$, the 2nd order PE curve is slightly lower. Overall, the trend of the PE results and the Jain-Kamilla state's published result broadly agree in the short- to mid-range. In the longer range, low-order PE is not expected to be accurate since it is a local expansion.

\subsubsection{Fermionic FCI models at $\nu=\frac{1}{3}$: zeroth-order PE saddle-point solutions and parton band structures}\label{sec:FCI_saddle}
\textbf{Models: }We start by introducing the two type-(B) FCI models inside the LLL in the present benchmark test: \emph{the Haldane-$V_1$ model and the bare-Coulomb model}. The many-body electron Hamiltonian of both models can be written in the following form:
\begin{align}
\hat H=\lambda\cdot\hat K+\hat U,
\end{align}
where $\lambda$ is a real parameter, $\hat K$ is a single-body kinetic energy term, and $\hat U$ is a normal-ordered two-body interaction term. 

The difference between the two models only lies in the interaction term: for the Haldane-$V_1$ model, $\hat U_{V_1}$ describes the Haldane-pseudopotential interactions inside the LLL with $V_1=1$ and all other pseudopotential components being zero, a short-range interaction model; while for the bare-Coulomb model, $\hat U_{\rm Coul.}$ describes the long-range bare Coulomb interaction in the LLL. Namely, when $\lambda=0$, these FCI systems go back to the well-studied FQH systems. We focus on $\nu=\frac{1}{3}$-filling: $\hat U_{V_1}$ has Laughlin's state as the exact zero energy ground state.

The FCI physics is turned on when $\lambda\neq 0$. The kinetic energy term $\hat K$, which is the same for both models, describes a square periodic potential projected into the electron's LLL:
\begin{align}
\hat K\equiv \sum_{u=x,y} [\bar{\boldsymbol\rho}(\mathbf q_u)+\bar{\boldsymbol\rho}(-\mathbf q_u)] e^{-\frac{|\mathbf q_u|^2 l_e^2}{4}},
\end{align}
where 
\begin{align}
\mathbf q_x\equiv(\frac{2\pi}{a},0),\;\;\mathbf q_y\equiv(0,\frac{2\pi}{a}),\;\; a\equiv \sqrt{2\pi} \cdot l_e.
\end{align}
Here $e^{-\frac{|\mathbf q_u|^2 l_e^2}{4}}$ is the LLL form factor, and $\bar{\boldsymbol\rho}(\mathbf q)$ is the Girvin-MacDonald-Platzman (GMP) density operator satisfying the magnetic translation algebra\cite{GirvinMacDonaldPlatzman1986}. $a$ is the unit cell linear size of the square potential, satisfying one flux per unit cell. 

At $\lambda=0$, the partons fill their LLL. $\lambda\neq 0$ breaks the electron's magnetic translation group in a FQH system down to the commuting magnetic translations of multiples of $a$ (denoted as square-potential-translation group). Consequently, the parton LLs will hybridize, and the saddle-point parton state (within the FCI phase) will fill a Chern band, which is a linear superposition of the parton LLs. \emph{One goal of the calculation is to find those superposition coefficients $A^{(p)}_{n,\mathbf k}$ in Eq.(\ref{eq:FCI_LL_superposition})}, where $n$ is the parton LL index. Here, as mentioned in Sec.\ref{sec:type_B_model_FCI}, the detailed analysis of the fractionalized translation and rotation symmetry analysis -- the projective symmetry group (PSG)\cite{Wen-Symmetric-QSL} of partons becomes important: only PSG-allowed parton LL hybridization is present. We have implemented the full square-potential-translation and $C_4$ rotation PSG into the present PE numerical study. 

\textbf{Simulation Details:} At low energy, we expect only the lowest few parton LLs to contribute significantly to the saddle point Chern band. In the present benchmark test, we have kept parton-LL index from $0$ up to $4$. Namely, there are 5 parton LLs per parton species in our simulation. If we label the parton LL indices as $(a,b,c)$ for the three parton species: $a,b,c=0,1,..4.$. The choice of $0$ up to $4$ LL indices is motivated by the $C_4$ symmetry in the present models: on a $C_4$-symmetric sample, PSG analysis shows that the high-symmetry parton-$\mathbf {k} $ points (e.g., the $\Gamma$-point) will only hybridize LLs whose indices are the same mod $4$. $0$ up to $4$ LLs allows nontrivial LL hybridization for all parton-$\mathbf k$ points for the filled band.

Note that, if one keeps all the fusion channels according to Eq.(\ref{eq:fermion_general_fusion}), based on Eq.(\ref{eq:fermion_general_R_abc}), the total number of fusion channels $R=3875$ would be too large. In our calculation, we keep only the fusion channels with $(a,b,c)$ of the form $(3,0,0), (0,2,0), (0,0,4)$, etc. Namely, among $(a,b,c)$, two are zeros. This reduces $R=103$. This $R$-reduction practice is motivated by two facts. First, note that this effectively means we set the factor $M_{(abc)}$ to zero for any $(a,b,c)$ that contains more than one nonzero value -- deviating from the literal real-space fusion prescription Eq.(\ref{eq:fermion_general_fusion}). However, this exercise is perfectly sound mathematically, because as long as the fusion amplitude is given by Eq.(\ref{eq:fermion_general_fusion}) up to an overall constant $M_{(abc)}$, the fusion for the overcomplete electron's discrete-coherent-states on the Fine-Grid is self-consistent. Despite setting many $M_{(abc)}=0$, the Hdet's full fusion tensor $T_{ijkl}$ is perfectly well-defined, giving a well-defined physical wavefunction inside the LLL. Second, as confirmed in our simulation, the dominant superposition weight in the filled parton Chern band is in the LLL: throughout the regimes in our simulation, the saddle point $|A^{(p)}_{n=0,\mathbf k}|^{2}>0.95$. Therefore, the fusion involving two or more nonzero parton LLs is a small effect anyway. 

With this reduced $R=103$, our PE calculation is performed with both uniform-parton-orbitals and normalized-fusion-amplitude prescriptions, and we utilized the fusion-channel-space algorithm in Sec.\ref{sec:channel_space_algorithm}. Before proceeding, we emphasize again that the rather large $R$ is a technical complexity induced by insisting on working within a Chern band (i.e., as a Type-(B) model introduced in Sec.\ref{sec:two_model_types}), which may not be physically necessary. If one works outside the topological band as in a Type-(A) model, the physically necessary $R$ describing the parton entanglement may not be as large.

On a finite torus, we apply the zeroth-order PE to approximately compute the Fine-Grid expectation values $\langle \Phi(\rho)|\hat n_z^{(e)}|\Phi(\rho)\rangle$ and $\langle \Phi(\rho)|\hat n_z^{(e)}\hat n_w^{(e)}|\Phi(\rho)\rangle$ ($z,w\in {\rm Fine-Grid}$). Following the prescription of Wang et.al.\cite{WangHaldane2019} Eq.(\ref{eq:Wang_Haldane_prescription}), both $\langle \Phi(\rho)|\hat K|\Phi(\rho)\rangle$ and $\langle \Phi(\rho)|\hat U|\Phi(\rho)\rangle$ can be computed, so is the variational energy $E(\rho)$. 

To find the saddle $|\Phi(\bar\rho)\rangle$, in terms of the general algorithm, one starts with some initial parton RDM $\rho_1$. By computing the derivative of $E(\rho_1)$, the effective single-body Hamiltonian $\hat h(\rho_1)$ can be found as in Eq.(\ref{eq:parton_eff_ham}). The RDM of the ground state of $\hat h(\rho_1)$ will be denoted as $\rho_2$, and the process continues until the self-consistent condition is achieved -- the ground state of $\hat h(\bar\rho)$ is also $\bar\rho$. For the stability of the algorithm, we have applied a standard linear mixing parameter $\beta\in (0,1)$: instead of $\rho_n$, we use $\tilde\rho_n$ defined as the McWeeny purification\cite{McWeeny1960} of $\rho_{n-1}+\beta(\rho_{n}-\rho_{n-1})$. Here, since we are dealing with a type-(B) model, following Eq.(\ref{eq:type_B_eff_h}), $\rho_i$ in this iteration should be understood as $\rho_{m,n;\mathbf 0}(\mathbf k)$ in the parton's LL basis, and the effective single-body Hamiltonian should be understood as $\hat{\mathsf h}(\mathbf k)$ in Eq.(\ref{eq:type_B_eff_h}). 

Varying $\lambda$, we find the zeroth-order PE optimal states on the 12x12 unit-cell sample with $48$ electrons, utilizing the full square-potential-translation and $C_4$ symmetry (i.e., PSG for partons). Each optimal state is labeled by $\{\bar \rho_{m,n;\mathbf 0}(\mathbf k)\}$ at $48$ parton-$\mathbf k$ points (Recall that due to the fractionalization of translation, the unit cell of partons is three times as large as that of electrons). They are PSG-related, forming 6 PSG-independent groups of $\mathbf k$ points. 

Within the Hdet construction, we expect the saddle point satisfies $IGG=U(1)_a\times U(1)_b$, corresponding to gauge changes: ${\rm q}_a^{(1)}=1,{\rm q}_a^{(2)}=-1,{\rm q}_a^{(3)}=0$, and ${\rm q}_b^{(1)}=1,{\rm q}_b^{(2)}=0,{\rm q}_b^{(3)}=-1$. However, throughout the regimes studied, \emph{We find that, the saddle states satisfy $\bar\rho^{(1)}=\bar\rho^{(2)}=\bar\rho^{(3)}$}, implying a larger $IGG$ in a generalized construction (see discussion below).

\begin{figure}
\includegraphics[width=0.4\textwidth]{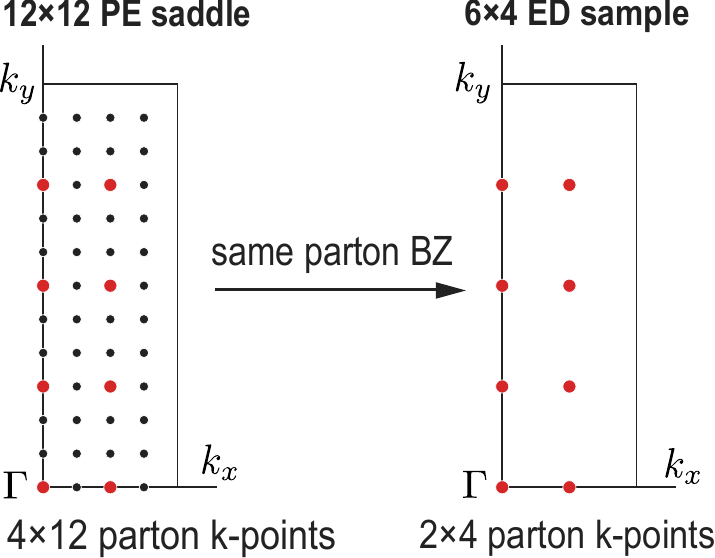}
\caption{Building Hdet state on the 6x4 ED sample based on the 12x12 PE saddle. Due to the symmetry fractionalization, we choose the parton unit cell three times enlarged along the $x$-direction (compared with the $a\times a$ electron's unit cell), and consequently the parton Brillouin Zone (BZ) is three times shrunk along $k_x$.}
\label{fig:kpoint_embedding}
\end{figure}

We then compare the Hdet wavefunction constructed from these (zeroth-order PE approximated) saddle points with wavefunctions obtained from exact diagonalization (ED) on the 6x4 sample with 8 electrons. In order to perform this comparison, we choose the 8 parton-$\mathbf k$ points from the 12x12 sample matching the 6x4 system(see Fig.\ref{fig:kpoint_embedding} for the illustration), and use the corresponding saddle-point $\{\bar \rho_{m,n;\mathbf 0}(\mathbf k)\}$ to construct the full fusion tensor $T_{ijkl}$ as in Eq.(\ref{eq:full_fusion_orig}) based on the local fusion gate $\hat{\mathbf C}_{z}$. The overlaps between the Hdet saddle-point wavefunctions $|\Phi(\bar\rho)\rangle$ and the ED ground states are then computed exactly by fully computing the many-body wavefunction $|\Phi(\bar\rho)\rangle$ in the electrons' Bloch basis in the LLL.

\begin{figure}
  \centering
\includegraphics[width=0.95\linewidth]{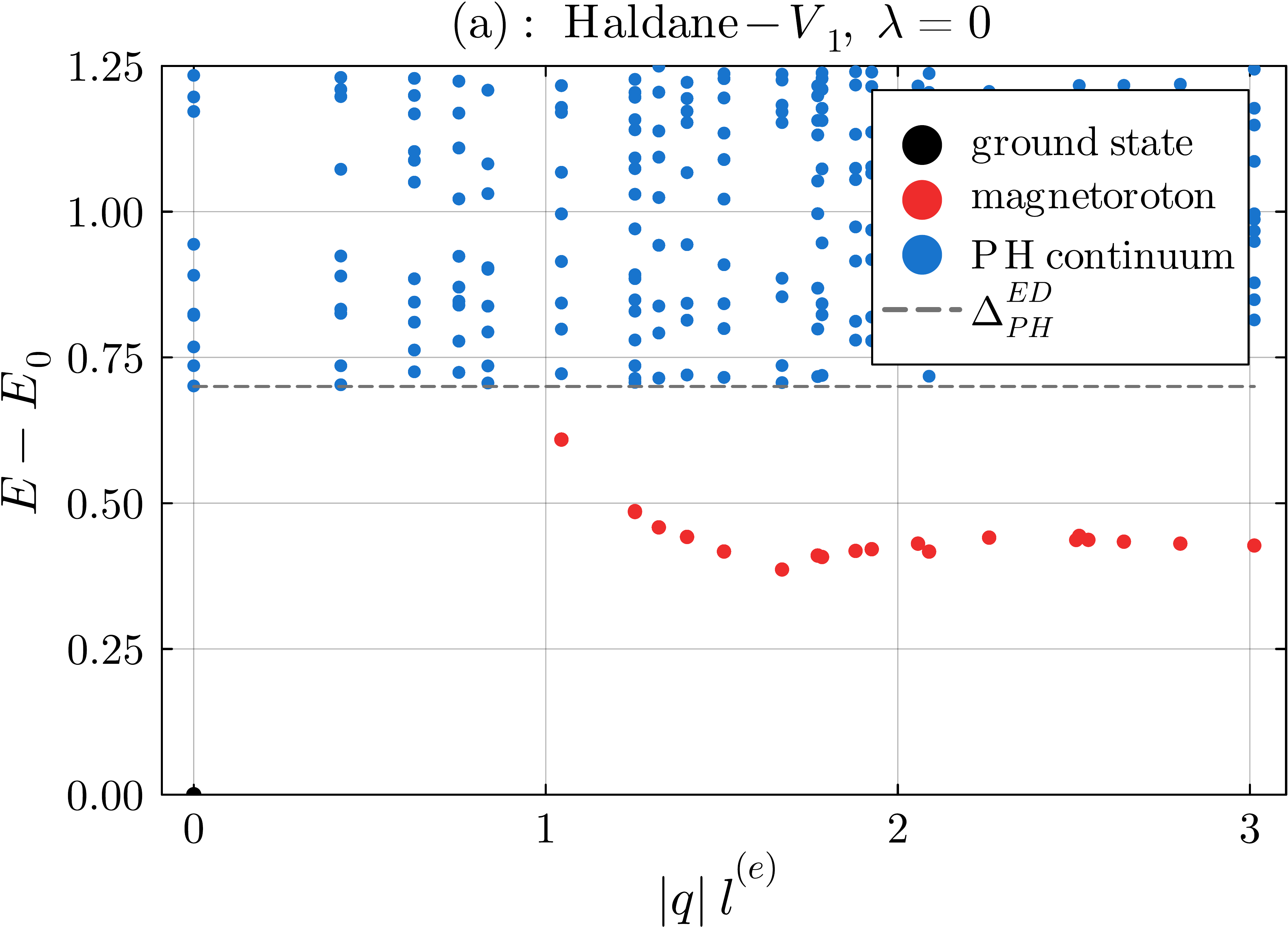}\\
\vspace{0.4cm}
\includegraphics[width=0.95\linewidth]{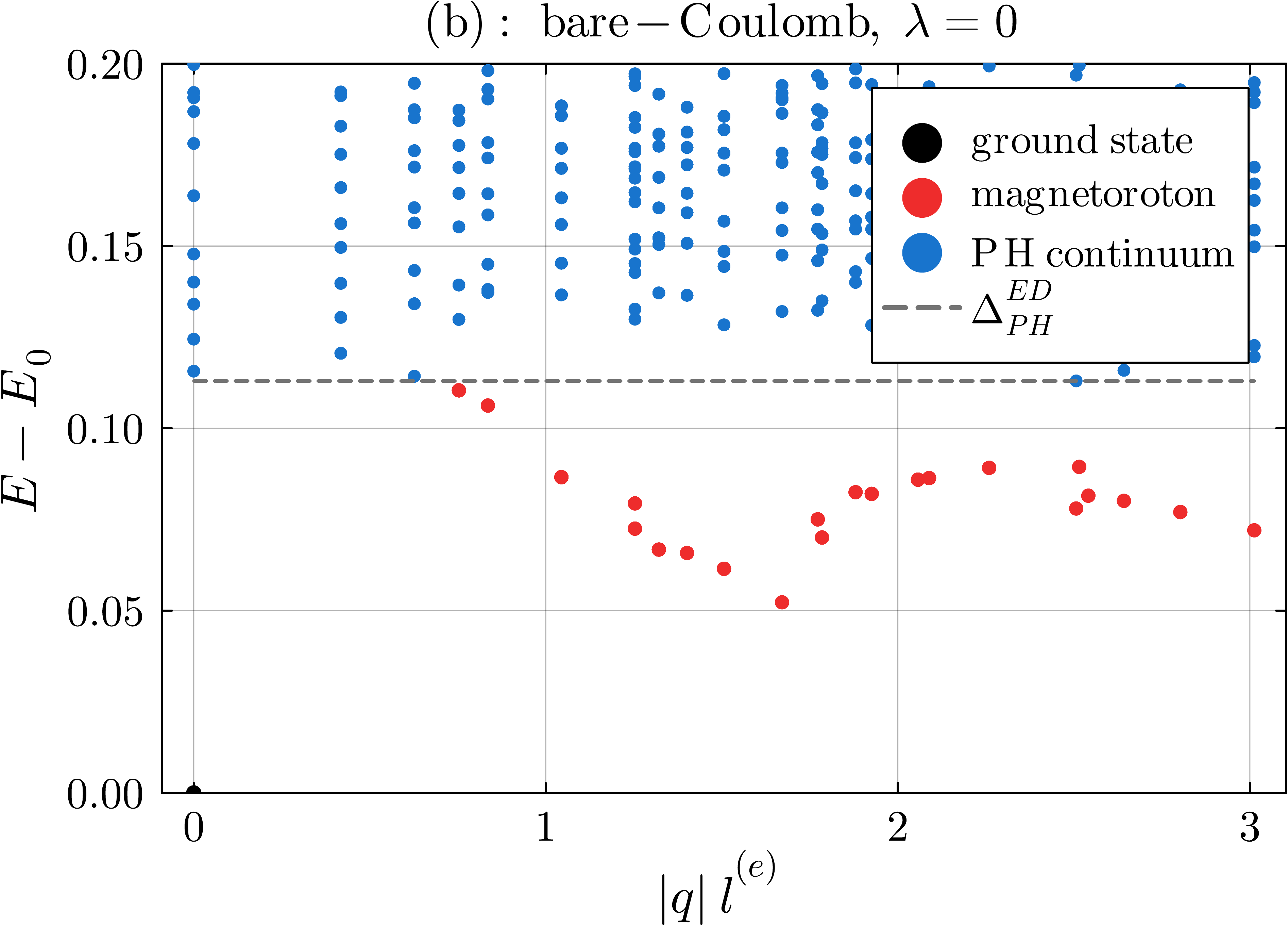}
\caption{Due to the high symmetry realized for $\lambda=0$, based on Haldane's prescription\cite{PhysRevLett.55.2095}, the magnetoroton modes can be cleanly separated from the quasiparticle-quasihole(PH) continuum in the many-body spectra in the present 6x4 ED.}
\label{fig:Haldane_prescription}
\end{figure}

\begin{figure}
    \centering
    \includegraphics[width=0.9\linewidth]{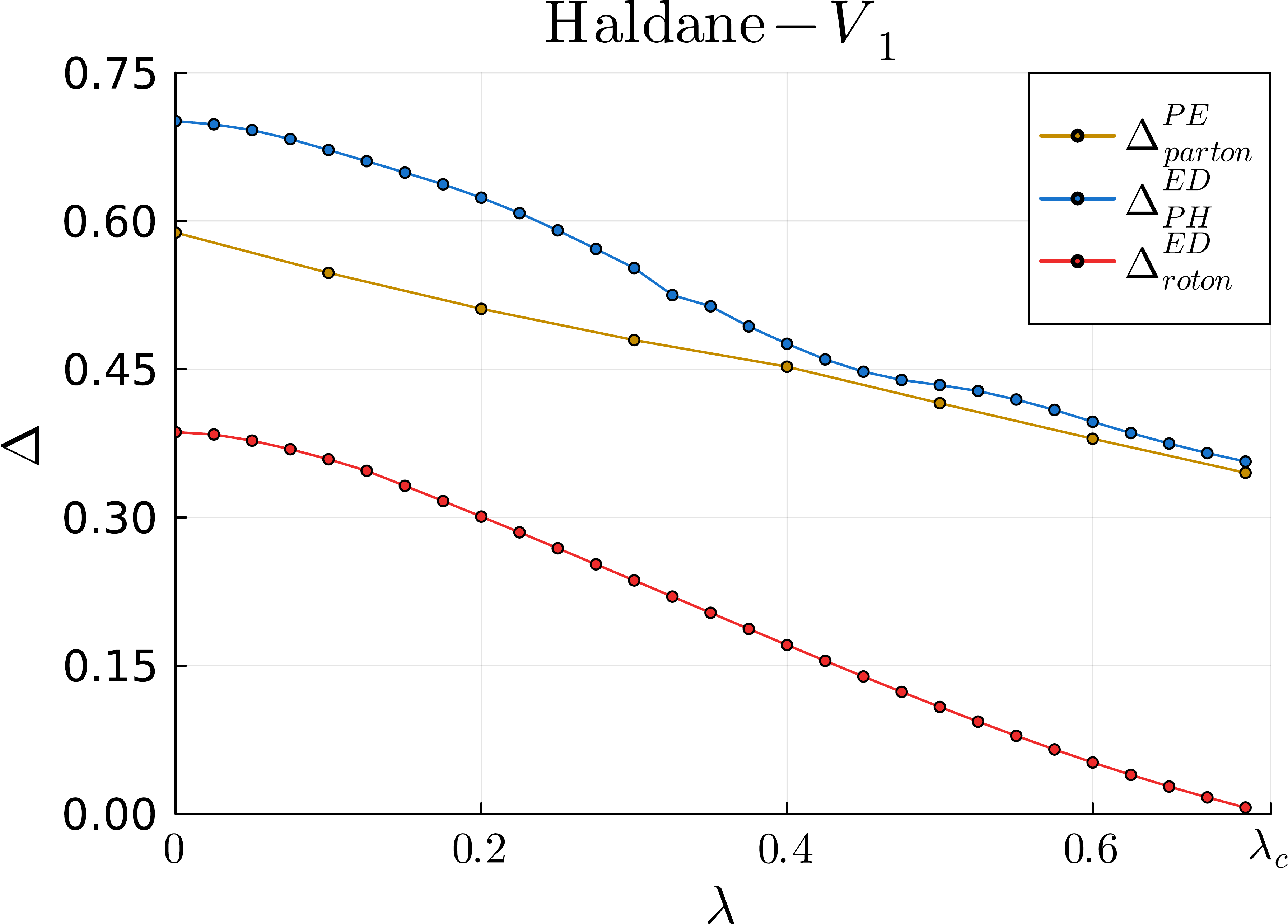}\\
    \vspace{0.5cm}
    \includegraphics[width=0.9\linewidth]{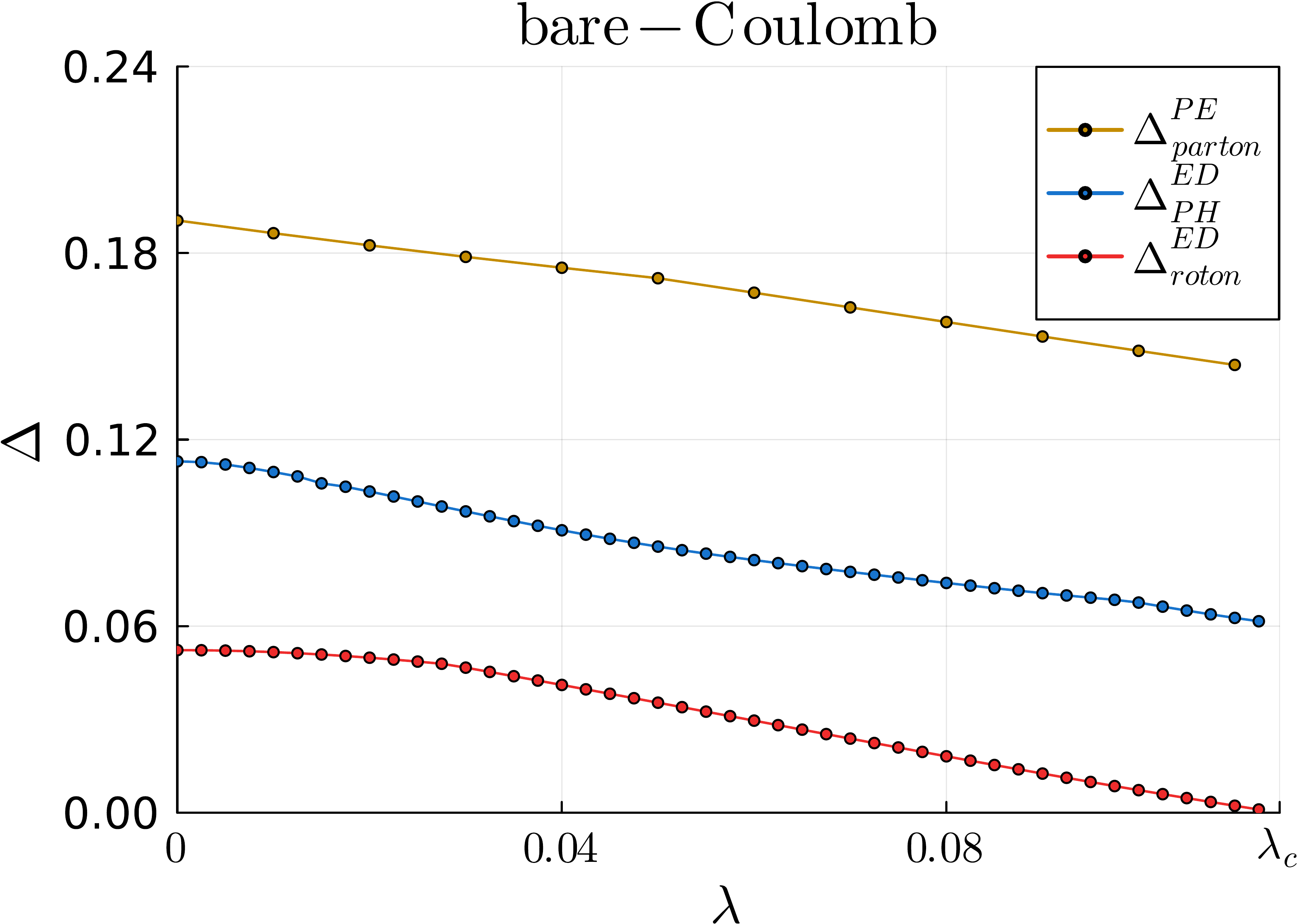}
    \caption{Comparison between $\Delta^{\rm PE}_{\rm  parton}$ and $\Delta^{\rm ED}_{\rm PH}$ in both models. For reference, $\Delta^{\rm ED}_{\rm roton}$ is also plotted, whose closing point is $\lambda_c$.}
    \label{fig:gap_comparison}
\end{figure}

\textbf{Results from exact diagonalization: }We first present the 6x4 ED results for both models as a preparation for the comparison with PE. Throughout the FCI phase, the 3-fold ground states in both models have center-of-mass (COM) momenta $K_{COM}=(0,0), (2,0), (4,0)$ due to sample-size quantization. When $\lambda=0$, the many-body gaps from the 3-fold ground state sector to the first excited state are $0.39$ (Haldane-$V_1$) and $0.052$ (bare-Coulomb) (in unit $l^{(e)}=1$), consistent with previously reported values\cite{Jolicoeur2017MagnetorotonED,RegnaultMaciejkoKivelsonSondhi2017FQHNematic}.  This many-body gap is due to the charge-neutral collective mode: it is the magnetoroton gap, instead of the quasiparticle-quasihole continuum gap (denoted as PH gap). The nice feature of the FQH $\lambda=0$ limit is that there is an existing prescription by Haldane\cite{PhysRevLett.55.2095} to clearly isolate the magnetoroton excitation branch from the PH continuum, which we show in Fig.\ref{fig:Haldane_prescription}. From this prescription, we find the PH gap to be $\Delta^{\rm ED}_{\rm PH}=0.70$ (Haldane-$V_1$) and $\Delta^{\rm ED}_{\rm PH}=0.113$(bare-Coulomb) at $\lambda=0$. As $\lambda$ increases, the many-body gap decreases and closes at $\lambda_c\approx0.72$ (Haldane-$V_1$) and $\lambda_c\approx0.115$ (bare-Coulomb) for the two models. (See Fig.\ref{fig:gap_comparison} for details). As a benchmark test, we do not attempt to investigate the nature of the phase transition and the nature of the phase realized at $\lambda>\lambda_c$. Instead, we focus on the FCI phase $0\leq\lambda<\lambda_c$.

\begin{figure}
    \centering
    \includegraphics[width=0.9\linewidth]{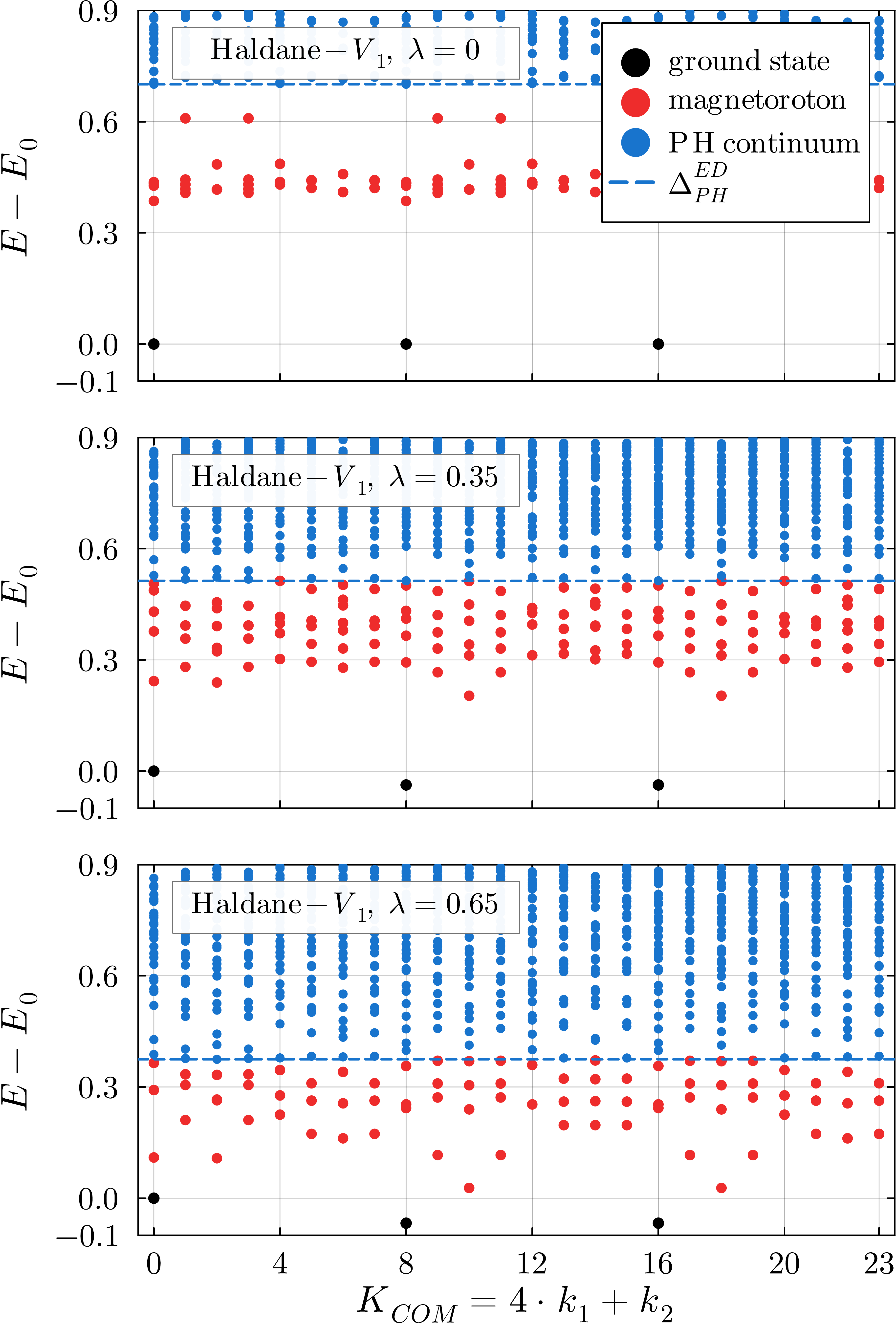}
    \caption{Representative ED spectra for Haldane-$V_1$ model. Note that the energy levels for $\lambda=0$ are identical to those in Fig.\ref{fig:Haldane_prescription}(a), but plotted differently.}
    \label{fig:representative_ED_V1}
\end{figure}

\begin{figure}
    \centering
    \includegraphics[width=0.9\linewidth]{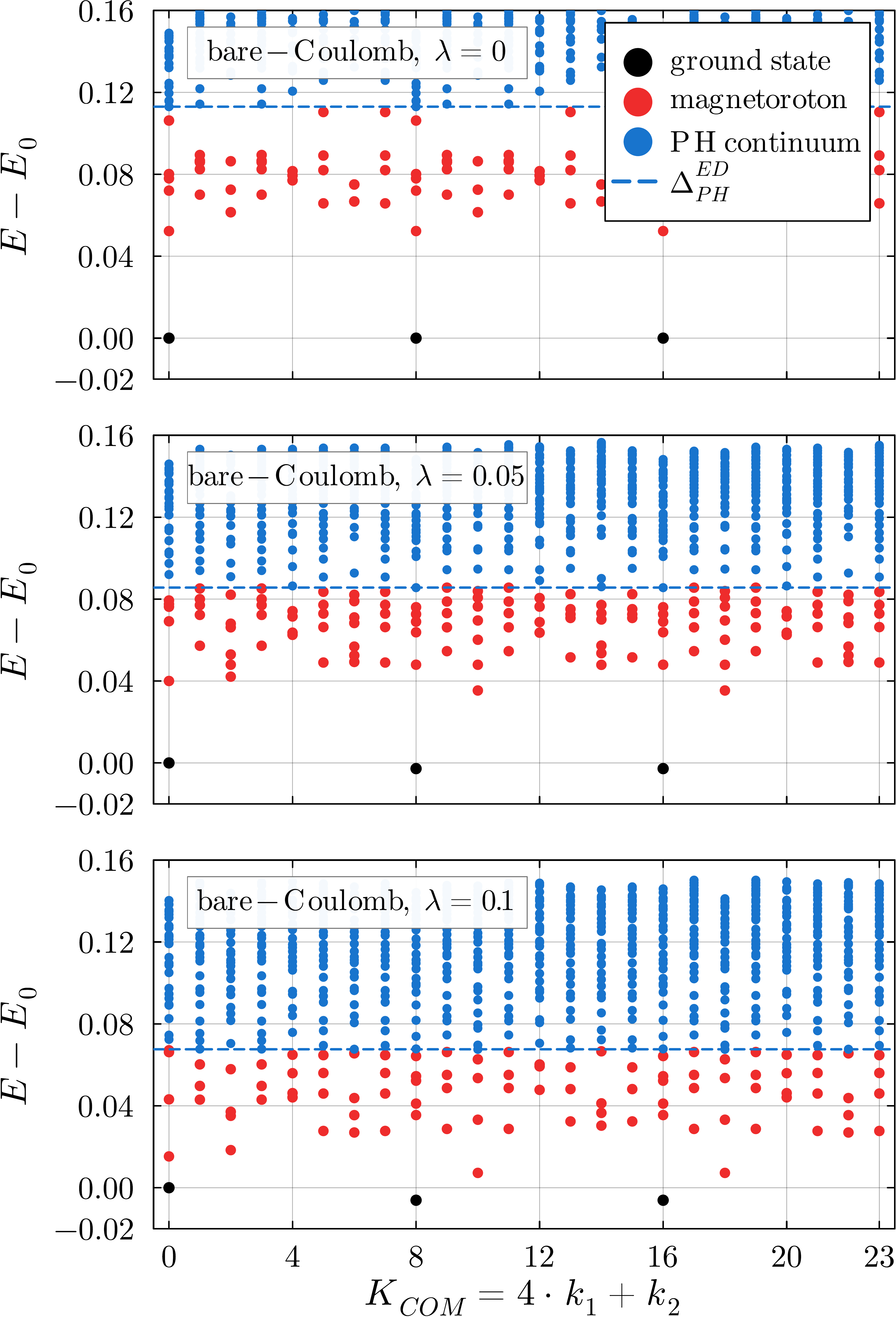}
    \caption{Representative ED spectra for bare-Coulomb model. Note that the energy levels for $\lambda=0$ are identical to those in Fig.\ref{fig:Haldane_prescription}(b), but plotted differently.}
    \label{fig:representative_ED_Coulomb}
\end{figure}

When $\lambda\neq0$, we are not aware of direct methods to sharply isolate the PH gap from the magnetoroton spectrum. To this end, we perform a calculation to adiabatically track the magnetoroton modes: we slowly increase $\lambda$ step by step from $0$. For each $\lambda$, we compute the overlap of the many-body eigenstates with the previous $\lambda$ eigenstates. Precisely, we choose the step $\Delta\lambda=0.025$ ($\Delta\lambda=0.0025$) for the Haldane-$V_1$ model (bare-Coulomb model). First, the eigenstate at step-$(n+1)$ with the maximal overlap with a magnetoroton mode at step-$n$ is marked as a "would-be magnetoroton" mode for the moment, unless that maximal overlap is below a threshold $O_{min}=0.9$. Second, the remaining modes are labeled as nonroton modes, whose lowest excitation energy from the top of the 3-fold ground-state sector is defined as $\Delta^{\rm ED}_{\rm PH}$. Finally, any "would-be magnetoroton" modes below $\Delta^{\rm ED}_{\rm PH}$ are marked as true magnetoroton modes for step-$(n+1)$, while those above $\Delta^{\rm ED}_{\rm PH}$ are marked as nonroton modes, since they are already inside the PH continuum. And the process continues iteratively. In this fashion, we obtain the magnetoroton modes and $\Delta^{\rm ED}_{\rm PH}$  in the range $0\leq\lambda<\lambda_c$. \emph{This calculation clearly shows that the magnetoroton gap $\Delta^{\rm ED}_{\rm roton}$ closes at $\lambda_c$ in both models, while the PH gap $\Delta^{\rm ED}_{\rm PH}$ remains finite (see Fig.\ref{fig:gap_comparison} for details). We plot the representative many-body spectra in Fig.\ref{fig:representative_ED_V1},\ref{fig:representative_ED_Coulomb}, in which $E_0$ is defined to be the top of the three-fold ground state sector. }

\begin{figure}
    \centering
    \includegraphics[width=0.49\linewidth]{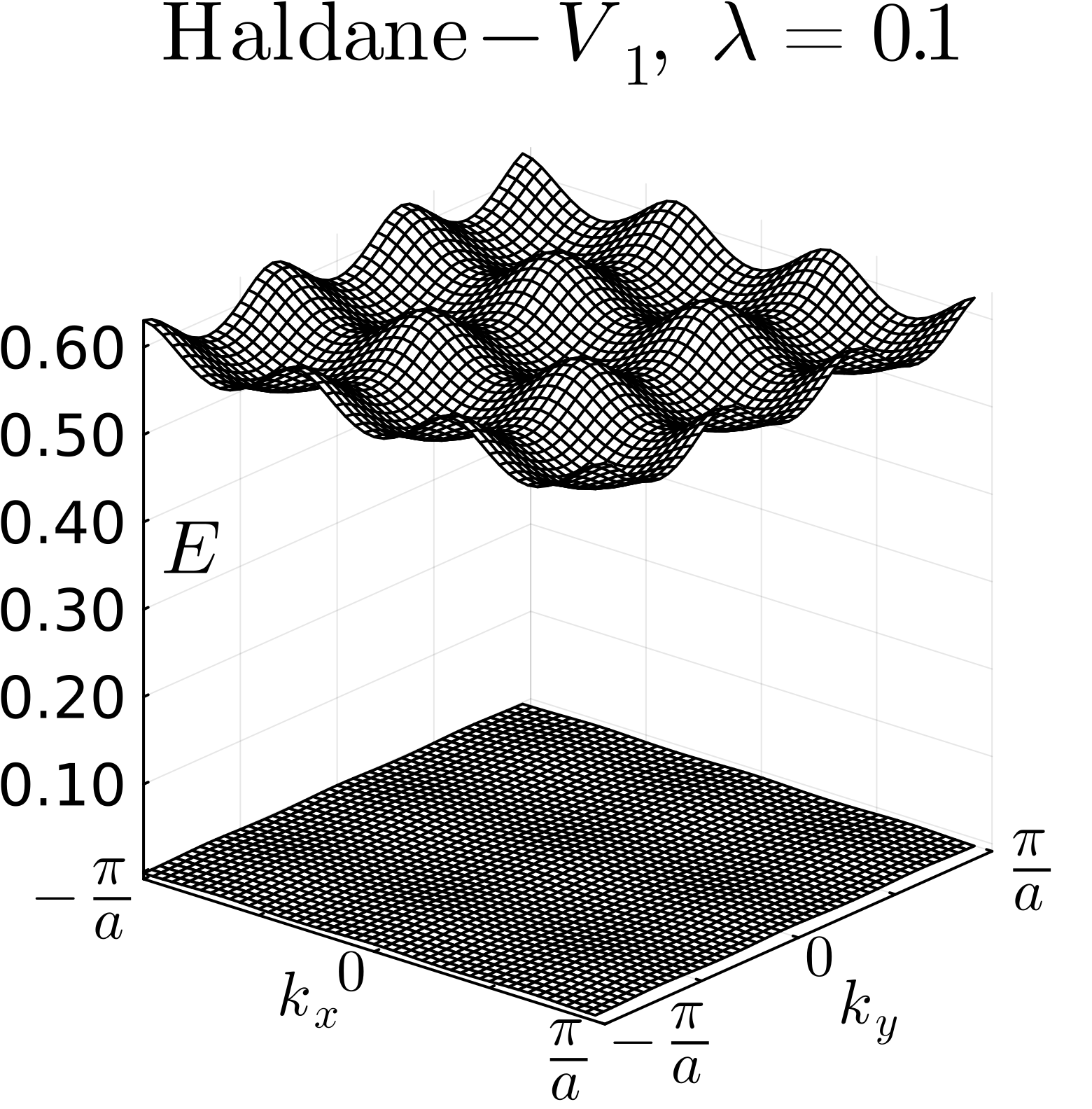}
    \includegraphics[width=0.49\linewidth]{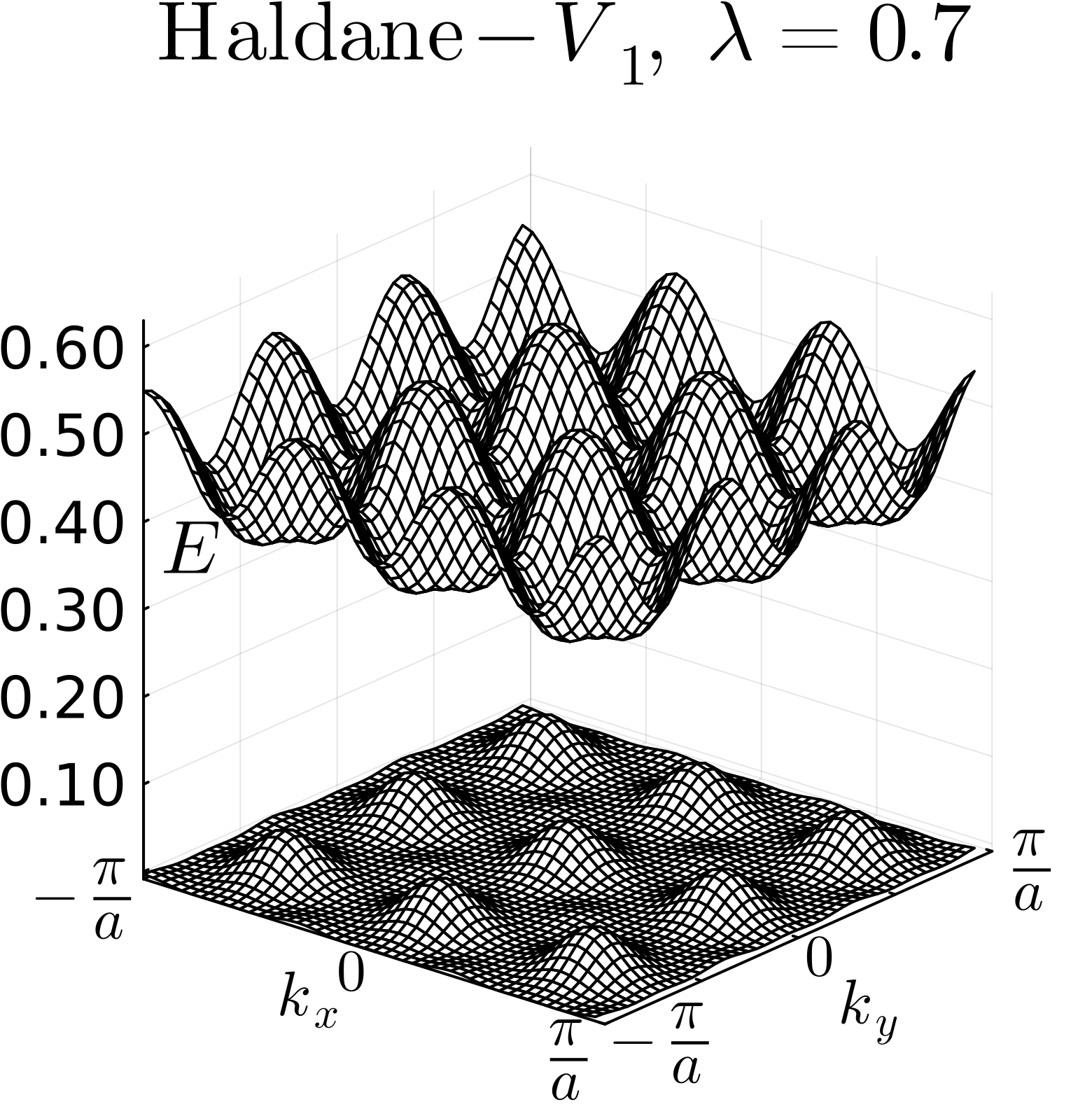}\\
    \vspace{0.5cm}
    \includegraphics[width=0.49\linewidth]{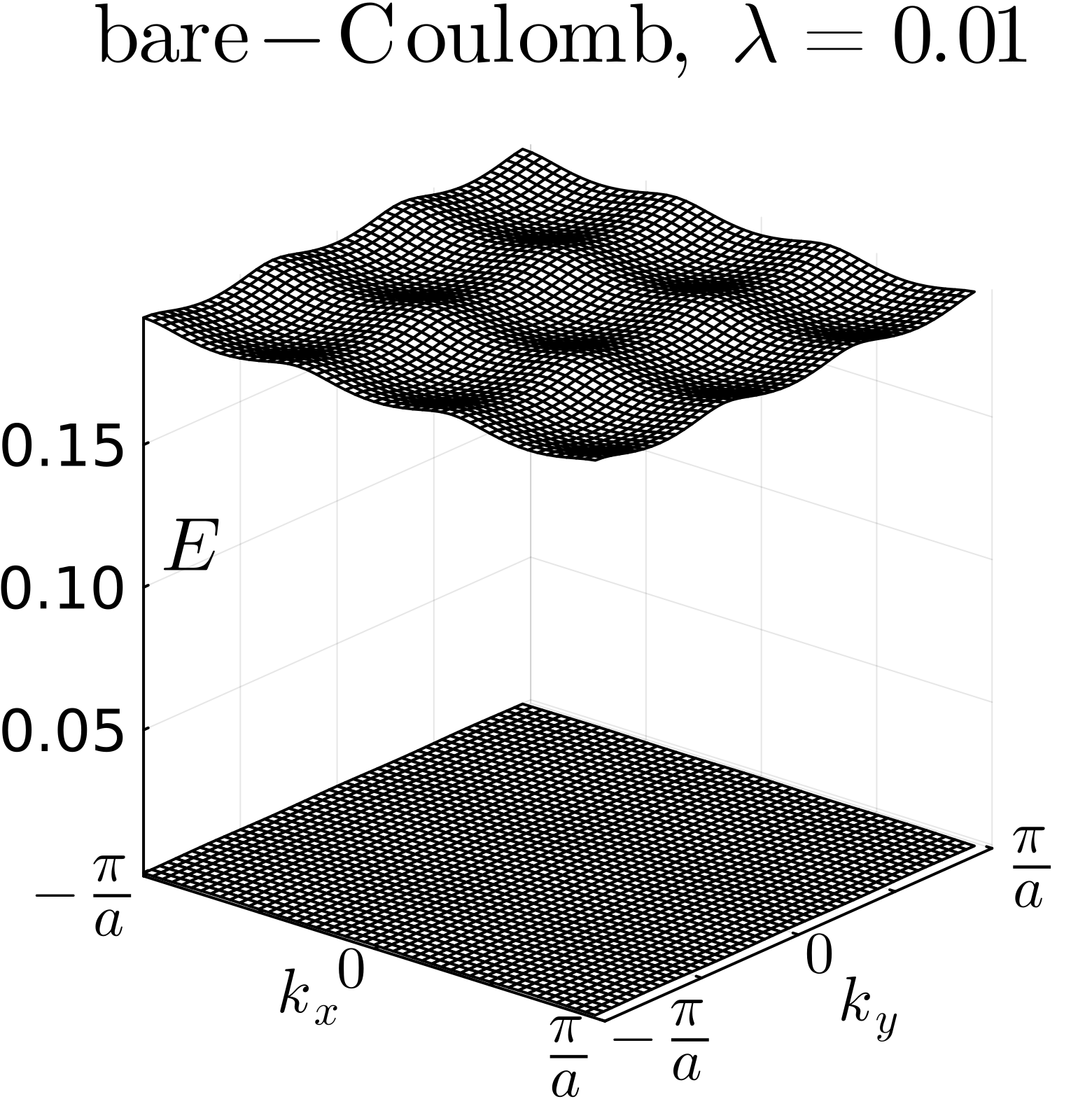}
    \includegraphics[width=0.49\linewidth]{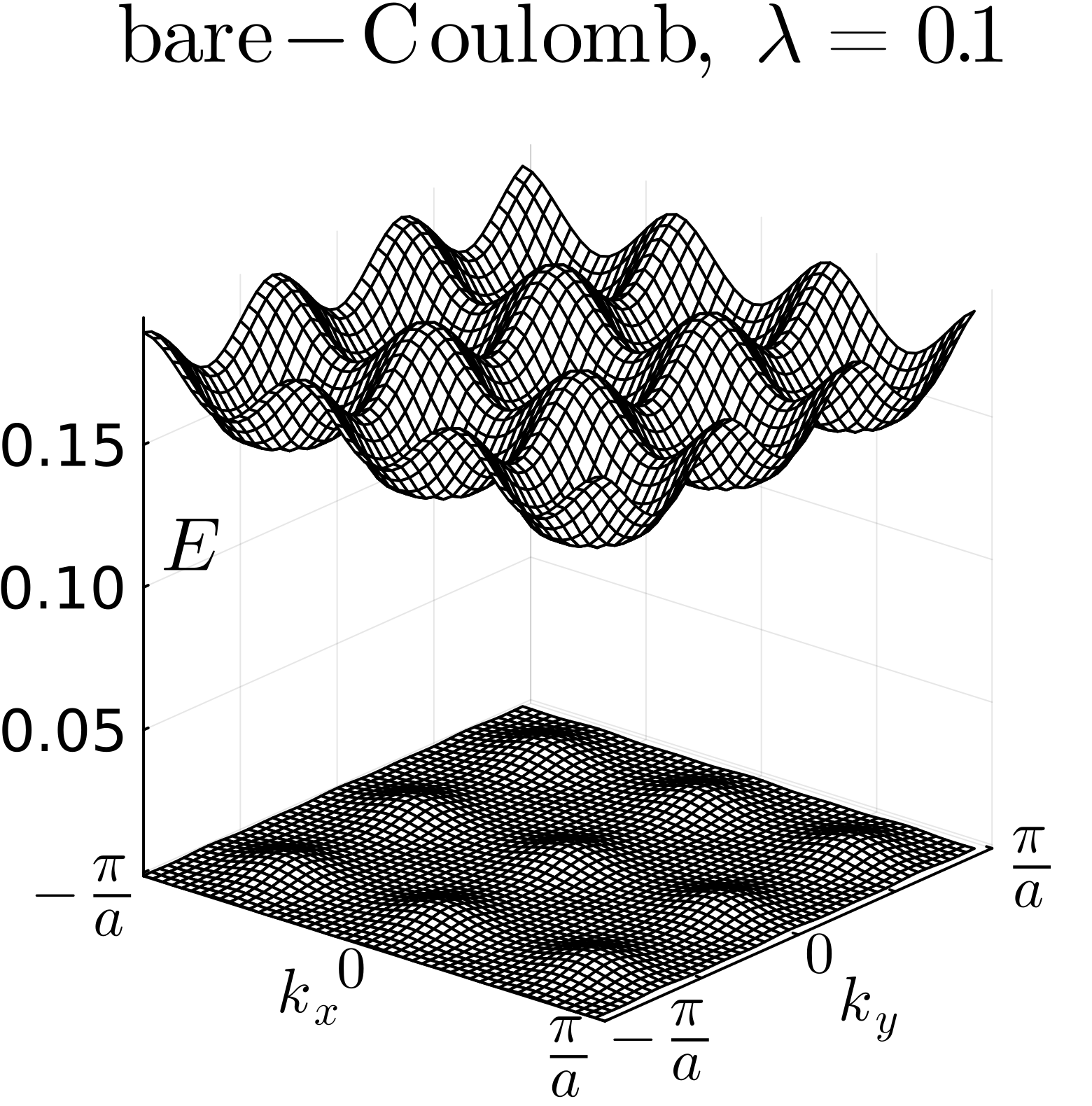}\\
    \caption{Representative parton band structures: the lowest two bands obtained by diagonalizing $\hat {\mathsf h}^{(p)}(\mathbf k)$ for the 12x12-sample saddle points are shown. These plots use the electron's Brillouin Zone (BZ) for better visualization, which contains three parton BZs. The 3x3 periodicity is a signature of the translation symmetry fractionalization\cite{Wen-Symmetric-QSL}.}
    \label{fig:parton_band}
\end{figure}

\textbf{Results from PE: Parton band structures:} Consider the band structure obtained from diagonalizing the $5\times5$  $\hat {\mathsf h}^{(p)}(\mathbf k)$ at the saddle point. \emph{This is the direct generalization of the static Hartree-Fock band structure to partons of Hdet states.} When $\lambda=0$, partons fill their LLL, with higher LLs all empty. Throughout the regime $0\leq\lambda<\lambda_c$, despite hybridization of the parton LLs leading to dispersion of the parton bands, we find that the band gap between the filled $C=1$ parton-band and higher energy bands does not close, for both models. We have plotted the representative parton band structures for the two models in Fig.\ref{fig:parton_band}. Note that, since $\bar\rho^{(1)}=\bar\rho^{(2)}=\bar\rho^{(3)}$ throughout the range $0\leq\lambda<\lambda_c$, the parton band structure is species-independent.

This parton band structure is not expected to capture the collective modes like the magnetorotons, similar to the fact that static Hartree-Fock cannot capture the collective modes. To capture magnetorotons, we should perform a dynamical PE calculation following VMPI as outlined in Sec.\ref{sec:PE_dynamical}. Namely, one should compute the full dynamical kernel as in Eq.(\ref{eq:kappa_eigen_eqn_Hdet}) by computing the matrices $\hat{\mathcal F},\hat{\mathcal G},\hat{\mathcal K}$ in the tangent space $T_{\bar\rho}$ (in the parton LL basis $\rho^{(p)}_{m,n;\mathbf q}$, see Sec.\ref{sec:special_type_B}), whose eigenspectrum contains the magnetoroton collective modes. Although this dynamical calculation is already well-defined,  due to the scope of this paper, we leave it as a future project. On the other hand, we do expect that this parton band structure can capture the PH continuum in the many-body excitation spectrum (at least when the PE order increases, and assuming the Berry curvature superoperator $\hat{\mathcal F}$ for these PH excitations can be well approximated by its mean-field value $\hat{\mathcal F}_{[0]}$.). This motivates us to plot the PH gap $\Delta^{\rm ED}_{\rm PH}$ from the ED together with the \emph{parton indirect band gap} $\Delta^{\rm PE}_{\rm  parton}$ from the parton band structure as a function of $\lambda$ in Fig.\ref{fig:gap_comparison} (We have checked that the parton band gap is independent of whether one uses the 4x12-$\mathbf k$-points from the 12x12 sample or the 2x4-$\mathbf k$-points from the 6x4 sample). 

We find that the trend of $\Delta^{\rm PE}_{\rm  parton}$ follows $\Delta^{\rm ED}_{\rm  PH}$ in both models: as $\lambda$ increases $\Delta$ decreases. In addition, in Haldane-$V_1$ model, $\Delta^{\rm PE}_{\rm  parton}$ and $\Delta^{\rm ED}_{\rm  PH}$ are quantitatively close to each other. In bare-Coulomb model, $\Delta^{\rm PE}_{\rm  parton}$ is shifted to higher energy by about $\sim 0.1$ from $\Delta^{\rm ED}_{\rm  PH}$. Since we are performing the zeroth-order PE, this inaccuracy is expected: as a local expansion, the long-range Coulomb interaction requires a higher-order PE.

\begin{figure}
    \centering
    \includegraphics[width=0.85\linewidth]{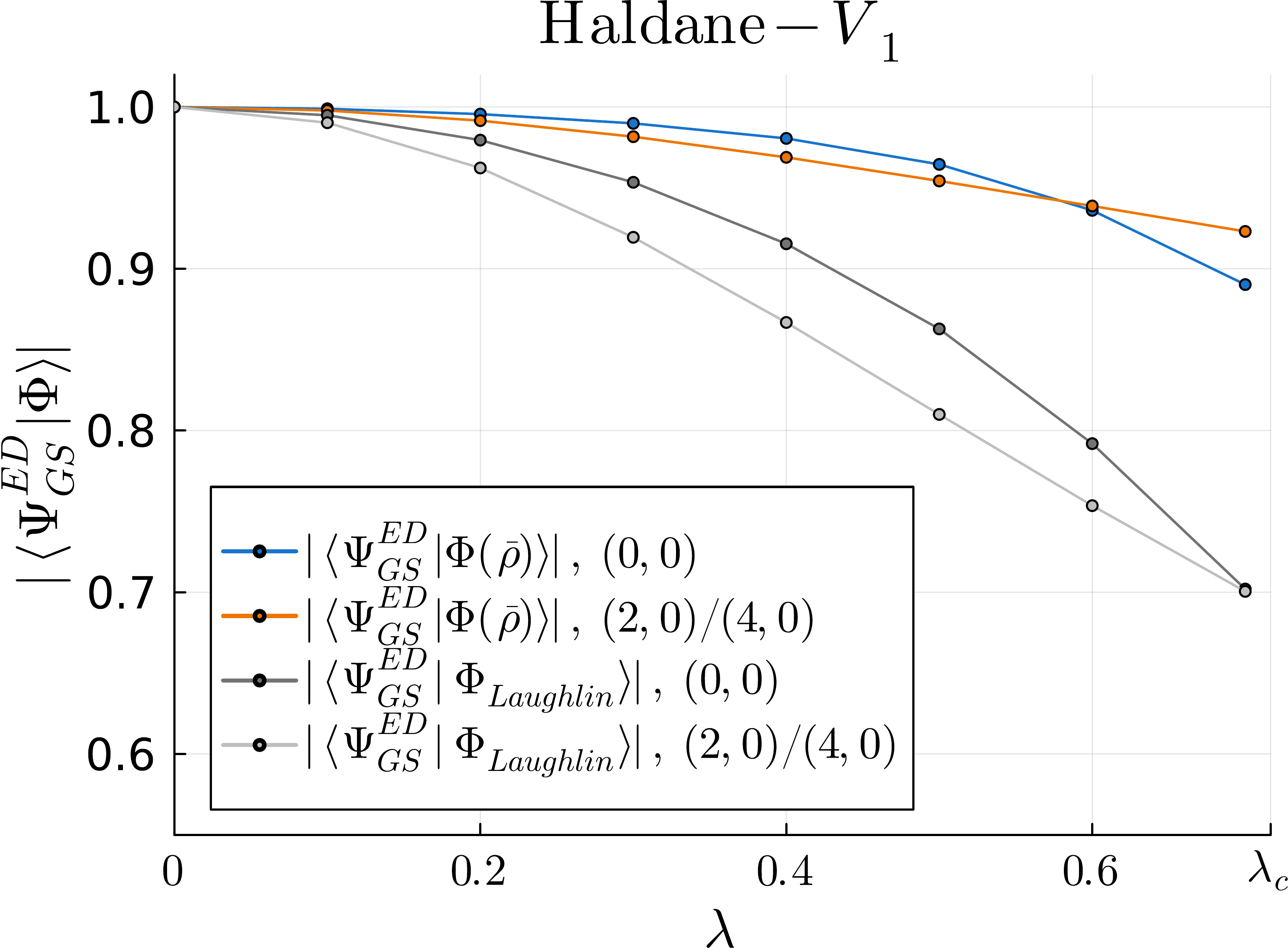}\\
    \vspace{0.5cm}
    \includegraphics[width=0.85\linewidth]{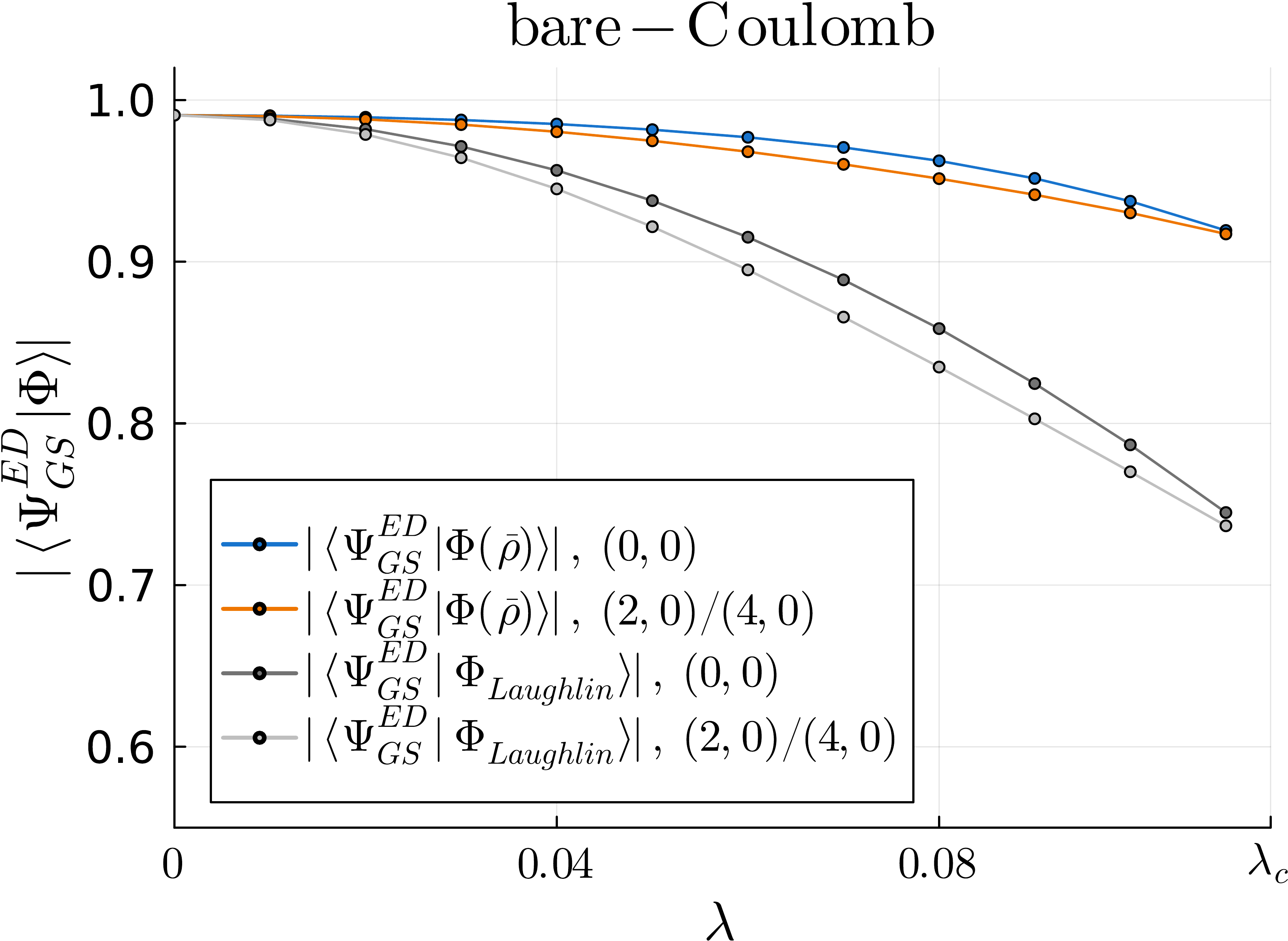}
    \caption{The overlaps between the ED ground states $|\Psi^{\rm ED}_{\rm GS}\rangle$ and the zeroth-order PE optimized Hdet states $|\Phi(\bar\rho)\rangle$. For comparison, the overlaps between $|\Psi^{\rm ED}_{\rm GS}\rangle$ and the Laughlin states $|\Phi_{\rm Laughlin}\rangle$ are also shown. The COM sectors $K_{COM}=(2,0),(4,0)$ are $x\rightarrow -x$ symmetry related so their curves are identical.}
    \label{fig:Hdet_ED_overlap}
\end{figure}

\textbf{Overlaps between the ED ground states and the saddle Hdet states: } One of the goals for this benchmark section is to test the performance of PE and the associated Hdet saddle-point states \emph{far from the FQH regime}. For this purpose, we compute the overlap between the Hdet state $|\Phi(\bar\rho)\rangle$ with the ED ground states $|\Psi^{\rm ED}_{\rm GS}\rangle$ in the three COM sectors as a function of $\lambda$, as shown in Fig.\ref{fig:Hdet_ED_overlap}. One important detail is that, based on the 12x12 sample PE saddle-point, given $\lambda$, we only construct one $|\Phi(\bar\rho)\rangle$ on the 6x4 sample from one fusion tensor $T_{ijkl}$. Due to the 6x4 sample size, the parton PSG for square-potential-translation is partially broken. Consequently, based on detailed PSG analysis, we know that $|\Phi(\bar\rho)\rangle$ is a simultaneous eigenstate of $T_{3\mathbf a_1}$ and $T_{\mathbf a_2}$: the translation operators of the three-time enlarged unit cell.  $|\Phi(\bar\rho)\rangle$ is really a linear superposition of the $K_{COM}=(0,0)$, $(2,0)$ and $(4,0)$ COM sectors. When we present the overlap $|\langle \Psi^{\rm ED}_{\rm GS}|\Phi(\bar\rho)\rangle|$ in each COM sector, we first project the Hdet state $|\Phi(\bar\rho)\rangle$ into that COM sector and normalize the projection image. As a comparison, we also present the overlap between the ED ground states and the Laughlin states $|\langle \Psi^{\rm ED}_{\rm GS} |\Phi_{\rm Laughlin}\rangle|$ in the three COM sectors. The Laughlin state contains no tunable variational parameter.

Not surprisingly, the overlap $|\langle \Psi^{\rm ED}_{\rm GS} |\Phi_{\rm Laughlin}\rangle|$ becomes quite poor when $\lambda$ is larger and near $\lambda_c$ ($\sim 70\%$ for both models), in both models -- a signature of strong deviation from the FQH regime. However, the Hdet saddle states optimized by the zeroth-order PE show much improved overlap: $|\langle \Psi^{\rm ED}_{\rm GS}|\Phi(\bar\rho)\rangle|$ $\sim 90\%$ even near $\lambda_c$, in both models. This result indicates that the Hdet wavefunctions together with PE can capture microscopic physics in FCI systems, even far from the FQH regime.

\textbf{Results from PE: $IGG=SU(3)$ vs. $IGG=U(1)^2$} As mentioned, we find throughout the regime $0\leq\lambda<\lambda_c$, in both models, the saddle-point satisfies $\bar\rho^{(1)}=\bar\rho^{(2)}=\bar\rho^{(3)}$: the parton mean-field states at the saddle satisfy the parton-permutation symmetry. One can further show that the $R=103$ fusion gate $\hat {\mathbf C}_{z}$ in the present study features the $SU(3)$ invariance in the parton species space. (In fact, as long as one keeps the parton LLs $(a,b,c)$ in an $S_3$ permutation symmetric fashion in $\hat {\mathbf C}_{z}$ for $q^{(1)}=q^{(2)}=q^{(3)}=\frac13$, $\hat {\mathbf C}_{z}$ is $SU(3)$ invariant.) 

As discussed near Eq.(\ref{eq:dyna_conserving_def}), this indicates that the $IGG=SU(3)$ for the saddle-point in the generalized Hpfaffian construction, whose low-energy effective theory is described by a fluctuating $SU(3)$ gauge field coupled to fermionic partons filling $C=1$ bands. After integrating out the matter fields, the $SU(3)$ level-1 Chern-Simons theory is obtained. The $IGG=U(1)^2$ within the Hdet construction may be viewed as a limitation of the formulation, which cannot capture the hybridization between parton species.

\section{Discussions and fused Gaussian States}\label{sec:discussion}
We have introduced a framework to microscopically investigate correlated quantum systems based on Hdet wavefunctions. After the original proposal for the Hdet wavefunction in Ref\cite{HuXiaoRan2024} involving some of us, many puzzles have been resolved (at least to a large extent) in the present paper: the locality structure, the effective theories (variational manifold path integral, VMPI), and practical simulation methods (projective-expansion, PE). In fact, both VMPI and PE are developed as general techniques, whose applications are not limited to Hdet states.

There are many future directions one may explore immediately. For example, we have primarily tested the projective expansion in benchmarking examples of type-(B) models, partially because there are many well-documented results in the FQH context. It would be interesting to also apply PE in examples of type-(A) models to test its performance. In addition, we have not yet demonstrated VMPI's full capabilities in concrete Hdet simulations, which requires full dynamical PE calculations and can capture collective-mode excitations. The algorithms for these calculations are already well-defined. For instance, we look forward to testing dynamical PE in future calculations, demonstrating the magnetorotons in the context of FQH and FCI or the gauge photons in the context of QSL. In addition, we plan to test the performance of the present framework in exotic states with gapless matter, such as the composite Fermi liquid\cite{HalperinLeeRead1993}, or the Dirac spin liquid\cite{AffleckMarston1988, Wen-Symmetric-QSL, Ran2007,PhysRevB.77.224413}.

The Hdet states generalize the previously known fermionic parton constructions\cite{Baskaran1987, AffleckMarston1988, Wen1991, Wen2004, LeeNagaosaWen2006}. In terms of conceptual progress, the main new physics captured by Hdet states with a locality structure is the entanglement between the partons. In terms of the language used in this paper, this is described by the number of local fusion channels $R$. Instead of $R=1$ in many traditional parton constructions, we have shown that there are important quantum states that necessarily require $R>1$ (e.g., $R=4$ in $\nu=\frac{2}{5}$ Jain's state). Intuitively, compared with $R=1$, the $R>1$ Hdet state can capture more sophisticated local entanglement in the physical electronic wavefunction. For generic microscopic models, we believe that allowing $R>1$ is crucial to improve energetics and microscopics.

In terms of practical simulations, although we believe PE is a useful method for approximately simulating Hdet states and provides reliable energetics, at least for short-range models, it is interesting to ask whether there are other numerical methods to systematically investigate Hdet wavefunctions. As long as those methods are statically and dynamically conserving as PE (see Eq.(\ref{eq:stat_conserving_def},\ref{eq:dyna_conserving_def}) for definitions), they can be combined with VMPI to produce microscopic effective theories. For example, at least when the partons are not filling topological bands (avoiding the known obstructions in the tensor-network formalism\cite{DubailRead2015, Read2017}), the Gaussian Grassmann tensor-network\cite{GuVerstraeteWen2010, Kraus2010} describing the parton mean-field state together with local fusion gates (see Fig.\ref{fig:fusion_gate_tn}) may be a natural way to simulate Hdet states, possibly accurately. Compared with generic Grassmann tensor-network methods\cite{Corboz2010, GuVerstraeteWen2010, Orus2014}, the advantage of using a Gaussian tensor-network with local fusion gates is that, due to VMPI, it immediately provides intuitive, mean-field-like pictures and direct access to fractionalized d.o.f. (i.e., partons). We look forward to combining tensor-network simulation techniques with the present framework to investigate correlated quantum states.

Finally, we remark that the Hdet states with a locality structure, the central topic of this paper, belong to a larger class of variational wavefunctions, which we coin the \emph{fused Gaussian states} (FGS). Similar to Hdet in Eq.(\ref{eq:onsite_fusion_operator},\ref{eq:general_Hdet_locality_state},\ref{eq:onsite_fusion_gate}), FGS is defined by applying the local fusion operator $\hat{\mathbf F}_{i}$, or equivalently the corresponding local fusion gate $\hat{\mathbf C}_i$ on a Gaussian mean-field parton state $|\Psi^{MF}\rangle$:
\begin{align}
|\Psi^{(e)}\rangle=\prod_i\hat{\mathbf C}_i|\Psi^{MF}\rangle.
\end{align}
Here, the fusion gate $\hat{\mathbf C}_i$ in FGS may not have the form Eq.(\ref{eq:onsite_fusion_gate}) as in an Hdet state. Instead, $\hat{\mathbf C}_i$ should be understood as a general linear map between the local Fock space of electrons and that of the partons, preserving fermion parity. In FGS, one also does not require all partons to be fermionic. For example, the Schwinger-boson construction for $\mathbf S=\frac{1}{2}$ QSL\cite{ArovasAuerbach1988, Sachdev1992, ReadSachdev1991,PhysRevB.84.020404}, and the slave-boson\cite{KotliarLiu1988, LeeNagaosaWen2006} or the slave-fermion\cite{KaneLeeRead1989SingleHoleAFM,AuerbachLarson1991DopedAFMInstability} constructions for unconventional superconductivity\cite{Anderson1987}, involving bosonic partons, can also be viewed as FGS-constructions when the parton mean-field states are Gaussian. Finally, $|\Psi^{MF}\rangle$ does not need to be a tensor product of decoupled parton species. Instead, any Gaussian state which may or may not involve hybridization and/or pairing of partons is allowed for $|\Psi^{MF}\rangle$. 

The key new ingredient emphasized here is about the fusion gate $\hat{\mathbf C}_i$: as a linear map, it may map one local electronic state to an \emph{entangled} many-body state (the parton-image state) in the local parton Fock space, resulting in the number of fusion channels $R>1$. This generalizes most traditional parton constructions where $R=1$. Again, we believe that allowing $R>1$ is important for capturing the microscopics in correlated electronic systems. Many of the techniques developed in this paper, particularly VMPI and PE, can be applied to general FGS after minor revisions, leading to computable, microscopic effective theories. 

We expect the FGS-based framework to find even broader applications, including nonabelian states in FCI (e.g., the FQAH analog of the $\nu=\frac{5}{2}$ FQH state \cite{MooreRead1991,GreiterWenWilczek1991}) and unconventional superconductivity. For instance, we already mentioned in Sec.\ref{sec:FCI_saddle}, the saddle-points for the FCI benchmarking models are found to be described by the $SU(3)_1$ effective theory in an Hpfaffian construction (see discussion near Eq.(\ref{eq:dyna_conserving_def}) for definition), where different parton species are allowed to hybridize and can be viewed as a FGS-construction. Furthermore, in Sec.\ref{sec:QSL}, we explained that the Zhang-Sachdev ancilla construction\cite{ZhangSachdev2020} in the undoped case is secretly a Hdet construction. In the presence of doping, it is easy to show that Zhang-Sachdev ancilla construction can be viewed as an FGS-construction. Precisely, in the formalism developed in Sec.\ref{sec:QSL}, introducing three fermionic partons $u,v,w$ per site, the fusion gate is:
\begin{align}
\hat {\mathbf C}^{\rm ZS}_{\mathbf r}=\sum_{n\in \{0,\uparrow,\downarrow,\uparrow\downarrow\}}|n_{\mathbf r}^{(e)}\rangle\langle \big(\uparrow_{\mathbf r,u}\downarrow_{\mathbf r,v}-\downarrow_{\mathbf r,u}\uparrow_{\mathbf r,v}\big)\cdot n_{\mathbf r}^{(w)}|,\label{eq:ZS_general_fusion}
\end{align}
which generalizes Eq.(\ref{eq:ZS_undoped_fusion}) to the doped case. Here $R=2$ due to the spin-singlet of $u,v$ partons in the fusion gate.

YR thanks Yuan-Ming Lu for helpful discussions. DX acknowledges support from the Department of Energy.

\appendix

\section{VMPI dictionary: half-filled antiferromagnet}
\label{sec:AFM_example}
In the application of VMPI to Slater determinants in Sec.\ref{sec:VMPI_TDHF} and Hdet states in Sec.\ref{sec:Hdet_eff}, to keep the discussion general, we have introduced a collection of abstract mathematical objects: e.g., the superoperator $\hat{\mathcal K}$, the bilinear form $\llangle.,.\rrangle$, Hubbard-Stratonovich subspace $V_{\rm HS}(\mathbf q)$, the bond tensor basis $\{w^A(\mathbf q)\}$, etc.  \emph{The sole purpose of this appendix is to concretely demonstrate these abstract objects in a textbook time-dependent Hartree-Fock (TDHF) calculation}, which may help readers to follow those general discussions.

Consider the one-band Hubbard model on the square lattice at half-filling:
\begin{align}
H=
-t\sum_{\langle \mathbf r,\mathbf r'\rangle,\sigma}
\left(c_{\mathbf r\sigma}^{\dagger}c_{\mathbf r'\sigma}+c_{\mathbf r'\sigma}^{\dagger}c_{\mathbf r\sigma}\right)
+U\sum_{\mathbf r} n_{\mathbf r\uparrow}n_{\mathbf r\downarrow},
\label{eq:app_Hubbard_model}
\end{align}
where $\mathbf r,\mathbf r'$ label lattice sites, $\sigma=\uparrow,\downarrow$, $n_{\mathbf r\sigma}=c_{\mathbf r\sigma}^{\dagger}c_{\mathbf r\sigma}$, and $\langle \mathbf r,\mathbf r'\rangle$ denotes nearest-neighbor pairs.

For a Slater determinant, define the one-body reduced density matrix by
\begin{align}
\rho_{\mathbf r\alpha,\mathbf r'\beta}\equiv\langle c_{\mathbf r'\beta}^{\dagger}c_{\mathbf r\alpha}\rangle,\qquad\alpha,\beta\in\{\uparrow,\downarrow\}.
\label{eq:app_global_rho_def}
\end{align}
The local spin-density matrix at site $\mathbf r$ is the $2\times2$ matrix
\begin{align}
(\rho_{\mathbf r})_{\alpha\beta}\equiv \rho_{\mathbf r\alpha,\mathbf r\beta}=\langle c_{\mathbf r\beta}^{\dagger}c_{\mathbf r\alpha}\rangle.
\label{eq:app_local_rho_def}
\end{align}
Define the local charge density and spin-density vector by
\begin{align}
n_{\mathbf r}\equiv \operatorname{tr}_{\sigma}\rho_{\mathbf r},\qquad m_{\mathbf r}^a\equiv \operatorname{tr}_{\sigma}(\rho_{\mathbf r}\sigma^a),\qquad a=x,y,z,
\label{eq:app_n_m_def}
\end{align}
where $\operatorname{tr}_{\sigma}$ is the trace over the two spin indices and $\sigma^a$ are Pauli matrices.  Then
\begin{align}
\rho_{\mathbf r}=\frac{1}{2}\left(n_{\mathbf r}\mathbf 1+\mathbf m_{\mathbf r}\cdot\boldsymbol\sigma\right).
\label{eq:app_rho_spin_decomp}
\end{align}

Using Wick's theorem in a Slater determinant, it is easy to show that the Hubbard interaction energy is
\begin{align}
E_U(\rho)=U\sum_{\mathbf r}\langle n_{\mathbf r\uparrow}n_{\mathbf r\downarrow}\rangle
=\frac{U}{4}\sum_{\mathbf r}\left(n_{\mathbf r}^2-|\mathbf m_{\mathbf r}|^2\right).
\label{eq:app_EU_spin_form}
\end{align}

\subsection{The saddle point analysis}
The static Hartree--Fock one-body kernel is the derivative of the energy functional, as in Eq.~(\ref{eq:TDHF_eff_hopping}).  The interaction contribution at site $\mathbf r$ is
\begin{align}
t_{U,\mathbf r}(\rho)=\frac{U}{2}\left(n_{\mathbf r}\mathbf 1-\mathbf m_{\mathbf r}\cdot\boldsymbol\sigma\right).
\label{eq:app_tU_general}
\end{align}
We choose the N\'eel saddle
\begin{align}
\bar n_{\mathbf r}=1,\qquad\bar{\mathbf m}_{\mathbf r}=m\eta_{\mathbf r}\hat z,\qquad\eta_{\mathbf r}=(-1)^{r_x+r_y},
\label{eq:app_AFM_saddle}
\end{align}
and define
\begin{align}
\Delta\equiv \frac{Um}{2}.
\label{eq:app_AFM_gap}
\end{align}
The self-consistent real-space kernel is then
\begin{align}
t(\bar\rho)_{\mathbf r\alpha,\mathbf r'\beta}=&-t\,\delta_{\langle \mathbf r,\mathbf r'\rangle}\delta_{\alpha\beta}+\delta_{\mathbf r\mathbf r'}\left[\frac{U}{2}\delta_{\alpha\beta}
-\Delta\eta_{\mathbf r}(\sigma^z)_{\alpha\beta}\right].
\label{eq:app_AFM_tbar_realspace}
\end{align}
The term $U/2$ is a uniform chemical-potential shift at half filling and will be omitted in the band Hamiltonian below.

Using the doubled magnetic unit cell with sublattices $s=A,B$, introducing Pauli matrices $\tau^a$ that act on the sublattice index,  in the magnetic Brillouin zone (BZ),  the saddle Hamiltonian is
\begin{align}
\bar h_{\mathbf k}=\epsilon_{\mathbf k}\tau^x-\Delta\tau^z\sigma^z,\qquad\epsilon_{\mathbf k}=-2t(\cos k_x+\cos k_y),
\label{eq:app_AFM_hk}
\end{align}
with single-particle energies
\begin{align}
\varepsilon_{\mathbf k,\pm}=\pm E_{\mathbf k},\qquad E_{\mathbf k}=\sqrt{\epsilon_{\mathbf k}^2+\Delta^2}.
\label{eq:app_AFM_Ek}
\end{align}
At half filling, the two negative-energy bands are filled.  The self-consistency condition may be written as
\begin{align}
1=\frac{U}{2N_{\rm uc}}\sum_{\mathbf k\in {\rm BZ}}\frac{1}{E_{\mathbf k}},
\label{eq:app_AFM_gap_equation}
\end{align}
where $N_{\rm uc}$ is the number of magnetic unit cells.

Now define the Hessian superoperator as in Eq.~(\ref{eq:real_space_Hessian}).  Let $M_{\rm h}$ be the real vector space of Hermitian one-body matrices, equipped with the bilinear form
\begin{align}
\llangle X,Y\rrangle\equiv\sum_{\mathbf r\alpha,\mathbf r'\beta}X_{\mathbf r\alpha,\mathbf r'\beta}Y_{\mathbf r'\beta,\mathbf r\alpha}.
\label{eq:app_Mh_pairing}
\end{align}
For an onsite fluctuation $X_{\mathbf r}$ at site $\mathbf r$, define
\begin{align}
\delta n_{\mathbf r}[X]\equiv\operatorname{tr}_{\sigma}X_{\mathbf r},\qquad\delta m_{\mathbf r}^a[X]\equiv\operatorname{tr}_{\sigma}(X_{\mathbf r}\sigma^a).
\label{eq:app_delta_n_m_X}
\end{align}
Expanding Eq.~(\ref{eq:app_EU_spin_form}) to second order gives
\begin{align}
\delta^2 E_U[X]=\frac{U}{4}\sum_{\mathbf r}\left[\delta n_{\mathbf r}[X]^2-\sum_{a=x,y,z}\delta m_{\mathbf r}^a[X]^2\right].
\label{eq:app_delta2_EU}
\end{align}
By definition,
\begin{align}
\delta^2E_U[X]=\frac{1}{2}\llangle X,\hat{\mathcal K}X\rrangle.
\label{eq:app_K_from_delta2}
\end{align}
The unique onsite Hermitian matrix $[\hat{\mathcal K}X]_{\mathbf r}$ that reproduces Eq.~(\ref{eq:app_delta2_EU}) for all $X_{\mathbf r}$ is
\begin{align}
[\hat{\mathcal K}X]_{\mathbf r}=\frac{U}{2}\left[\operatorname{tr}_{\sigma}(X_{\mathbf r})\mathbf 1-\sum_{a=x,y,z}\operatorname{tr}_{\sigma}(X_{\mathbf r}\sigma^a)\sigma^a\right].
\label{eq:app_Hubbard_K_local}
\end{align}

If we denote
\begin{align}
X_{\mathbf r}=x_{\mathbf r}^0\mathbf 1+\mathbf x_{\mathbf r}\cdot\boldsymbol\sigma
\label{eq:app_X_spin_decomp}
\end{align}
then Eq.~(\ref{eq:app_Hubbard_K_local}) becomes
\begin{align}
[\hat{\mathcal K}X]_{\mathbf r}=Ux_{\mathbf r}^0\mathbf 1-U\sum_a x_{\mathbf r}^a\sigma^a.
\label{eq:app_Hubbard_K_eigenchannels}
\end{align}
Consequently, as matrices in $M_{\rm h}$, the onsite charge component $x_{\mathbf r}^0\mathbf 1 $ has eigenvalue $+U$, while each onsite spin component $x_{\mathbf r}^a\sigma^a $ has eigenvalue $-U$. Now the full HS space $V_{\rm HS}\subset M_{\rm h}$ in Eq.~(\ref{eq:V_HS_def}) is simple: it is spanned by those $x_{\mathbf r}^0\mathbf 1 $, $x_{\mathbf r}^a\sigma^a$, four matrices per site-$\mathbf r$. The superoperator $\hat{\mathcal K}_{\rm HS}^{-1}$ is simply
\begin{align}
\hat{\mathcal K}_{\rm HS}^{-1}[x_{\mathbf r}^0\mathbf 1 ]=\frac{1}{U}[x_{\mathbf r}^0\mathbf 1 ]\qquad \hat{\mathcal K}_{\rm HS}^{-1}[x_{\mathbf r}^a\sigma^a]=-\frac{1}{U}[x_{\mathbf r}^a\sigma^a]
\label{eq:app_AFM_KHS_inverse_full}
\end{align}

\subsection{Goldstone subspace}
The broken spin generators are
\begin{align}
\hat Q_x=\frac12\sum_{\mathbf r} c_{\mathbf r}^{\dagger}\sigma^x c_{\mathbf r},
\qquad
\hat Q_y=\frac12\sum_{\mathbf r} c_{\mathbf r}^{\dagger}\sigma^y c_{\mathbf r}.
\label{eq:app_AFM_generators_full}
\end{align}
Their single-body matrices are $Q_x=\sigma^x/2$ and $Q_y=\sigma^y/2$.  Following Eq.~(\ref{eq:delta_rho_Q_realspace}), introduce the infinitesimal angles
$\theta_x,\theta_y$ and define
\begin{align}
\delta\rho_{Q_a}\equiv \mathrm{i}\theta_a[Q_a,\bar\rho],\qquad a=x,y.
\label{eq:app_AFM_delta_rho_Q_full}
\end{align}
The onsite saddle-point RDM is
\begin{align}
\bar\rho_{\mathbf r}=\frac12\left(\mathbf 1+m\eta_{\mathbf r}\sigma^z\right).
\label{eq:app_AFM_rhobar_i_full}
\end{align}
Therefore
\begin{align}
(\delta\rho_{Q_x})_{\mathbf r}&=\theta_x\frac{m\eta_{\mathbf r}}{2}\sigma^y,&(\delta\rho_{Q_y})_{\mathbf r}&=-\theta_y\frac{m\eta_{\mathbf r}}{2}\sigma^x.
\label{eq:app_AFM_delta_Q_full}
\end{align}
Similarly, following Eq.~(\ref{eq:chi_Q_realspace}) and Eq.~(\ref{eq:realspace_Ward}),
\begin{align}
\chi_{Q_a}=\mathrm{i}\theta_a[Q_a,t(\bar\rho)]=\hat{\mathcal K}\delta\rho_{Q_a}.
\label{eq:app_AFM_chi_Q_def_full}
\end{align}
Using Eq.~(\ref{eq:app_AFM_tbar_realspace}),
\begin{align}
\chi_{Q_x}&=-\theta_x\Delta\tau^z\sigma^y,&\chi_{Q_y}&=+\theta_y\Delta\tau^z\sigma^x,
\label{eq:app_AFM_chi_Q_full}
\end{align}
where we use the sublattice space $\tau^z$ to replace the staggered sign $\eta_{\mathbf r}$.

Using the magnetic unit cell and the corresponding Brillouin Zone (BZ), we now choose the bond tensor $\{w^A(\mathbf q)\}$ basis of $V_{\rm HS}(\mathbf q)$ (see Eq.(\ref{eq:T_VHS_q_def})) so that the first two vectors span the Goldstone subspace. Note that, since there are two sites per magnetic unit cell, $V_{\rm HS}(\mathbf q)$ is 8-dimensional.  A convenient choice is
\begin{align}
w^1&=-\Delta\tau^z\sigma^y,&w^2&=+\Delta\tau^z\sigma^x,\notag\\
w^3&=-\Delta\sigma^y,&w^4&=+\Delta\sigma^x,\notag\\
w^5&=\mathbf 1,&w^6&=\tau^z,\notag\\
w^7&=\sigma^z,&w^8&=\tau^z\sigma^z.
\label{eq:app_AFM_w_basis_full}
\end{align}
Note these are the tensor values for $\{w^A(\mathbf q)\}$, and are $\mathbf q$-independent. The physical meaning of $\mathbf q$ in $\{w^A(\mathbf q)\}$ is carried in Eq.(\ref{eq:lattice_Fourier}). In the language of $(\mathbf R,\boldsymbol\upalpha)$, these $w^A_{\boldsymbol\upalpha\boldsymbol\upbeta;\boldsymbol\delta}$ correspond to $\boldsymbol\delta=\mathbf 0$, and $\boldsymbol\upalpha$($\boldsymbol\upbeta$) labels the spin and the sublattice within one magnetic unit cell (a 4-dimensional index). The $4\times 4$ matrix $w^A_{\boldsymbol\upalpha\boldsymbol\upbeta;\boldsymbol\delta=\mathbf 0}$ is written as the tensor product of $\tau$ and $\sigma$ matrices above. We then have, as in Eq.(\ref{eq:chi_a_u_a}):
\begin{align}
\chi_{Q_x}=\theta_x w^1(\mathbf q=0),\qquad \chi_{Q_y}=\theta_y w^2(\mathbf q=0).
\label{eq:app_AFM_chiQ_wbasis_full}
\end{align}
Following the discussion below Eq.~(\ref{eq:chi_a_u_a}),
\begin{align}
V_{\rm G}(\mathbf q)=\operatorname{span}\{w^1(\mathbf q),w^2(\mathbf q)\},
\label{eq:app_AFM_VG_full}
\end{align}
and the complement subspace is
\begin{align}
V_{\perp}(\mathbf q)=\operatorname{span}\{w^3(\mathbf q),w^4(\mathbf q),w^5(\mathbf q),w^6(\mathbf q),w^7(\mathbf q),w^8(\mathbf q)\},
\label{eq:app_AFM_Vperp_full}
\end{align}
satisfying $V_{\rm HS}(\mathbf q)=V_{\rm G}(\mathbf q)\oplus V_{\perp}(\mathbf q)$

\subsection{Full kernel $D$, reduced kernel $\Pi$, and long-wavelength analysis}
Following Eq.(\ref{eq:chi_Goldstone_decomp}), we expand the $\chi$-field as
\begin{align}
\chi(\mathbf q,\omega)=\sum_{a=1}^{2}\phi_a(\mathbf q,\omega)w^a(\mathbf q)+\sum_{B=3}^{8}\eta_B(\mathbf q,\omega)w^B(\mathbf q).
\label{eq:app_AFM_chi_full_expansion_full}
\end{align}
The action has the form of Eq.~(\ref{eq:phi_eta_action}).

Now separate the six-dimensional $V_{\perp}$ into the transverse uniform spin sector
\begin{align}
V_{\eta}(\mathbf q)=\text{span}\{w^3(\mathbf q),w^4(\mathbf q)\},
\label{eq:app_AFM_Veta_full}
\end{align}
and the charge/longitudinal sector
\begin{align}
V_L(\mathbf q)=\text{span}\{w^5(\mathbf q),w^6(\mathbf q),w^7(\mathbf q),w^8(\mathbf q)\}.
\label{eq:app_AFM_VL_full}
\end{align}
Since the saddle preserves the $U(1)$ spin rotation generated by $S^z$, the quadratic kernel can only mix $V_{\rm G}$ with $V_{\eta}$, but $V_{L}$ cannot mix with either $V_{\rm G}$ or $V_{\eta}$. Consequently the kernel $D$ in Eq.(\ref{eq:kernel_D_TDHF}) has the form
\begin{align}
D(\mathbf q,\omega)=\begin{pmatrix}
D_{\rm GG}&D_{{\rm G}\eta}&0\\
D_{\eta {\rm G}}&D_{\eta\eta}&0\\
0&0&D_{LL}
\end{pmatrix},
\label{eq:app_AFM_D_block_full}
\end{align}
with respect to
$V_{\rm G}\oplus V_{\eta}\oplus V_L$.  The Schur complement for the reduced kernel $\Pi$ in Eq.~(\ref{eq:D_red}) is
\begin{align}
\Pi&=D_{\rm GG}-\begin{pmatrix}D_{{\rm G}\eta}&0\end{pmatrix}\begin{pmatrix}D_{\eta\eta}&0\\0&D_{LL}\end{pmatrix}^{-1}
\begin{pmatrix}D_{\eta {\rm G}}\\0\end{pmatrix}\notag\\
&=D_{\rm GG}-D_{{\rm G}\eta}D_{\eta\eta}^{-1}D_{\eta {\rm G}}.
\label{eq:app_AFM_full_schur_reduction_full}
\end{align}

For small $\mathbf q$ and $\omega$, direct evaluation of Eq.(\ref{eq:kernel_D_TDHF}) gives the Goldstone block
\begin{align}
D_{\rm GG}(\mathbf q,\omega)&=C_{\rm AFM}\omega^2\mathbf 1_2+\sum_{u,v}\mathbf q_u\mathbf q_vG^{uv}_{\rm AFM}\mathbf 1_2+O_3,
\label{eq:app_AFM_DGG_full}
\end{align}
the Goldstone--transverse block
\begin{align}
D_{{\rm G}\eta}(\mathbf q,\omega)&=\mathrm{i}\omega M_{\rm AFM}\boldsymbol\epsilon+O_2,
\label{eq:app_AFM_DGeta_full}
\end{align}
and the transverse non-Goldstone block
\begin{align}
D_{\eta\eta}(\mathbf q,\omega)&=-\Omega_\eta\cdot\mathbf 1_2+O_1.
\label{eq:app_AFM_Detaeta_full}
\end{align}
Here $O_n$ means up to a $n$-th order polynomial of $|\mathbf q|,\omega$, and $\boldsymbol\epsilon_{12}=-\boldsymbol\epsilon_{21}=1,\boldsymbol\epsilon_{11}=\boldsymbol\epsilon_{22}=0$ is the fully antisymmetric tensor.

The coefficients appearing in Eqs.~(\ref{eq:app_AFM_DGG_full})--
(\ref{eq:app_AFM_Detaeta_full}) are
\begin{align}
&\Omega_\eta=\frac{2\Delta^2}{N_{\rm uc}}\sum_{\mathbf k\in {\rm BZ}}\frac{\epsilon_{\mathbf k}^2}{E_{\mathbf k}^3},\;\;\;\;\;M_{\rm AFM}=\frac{\Delta^3}{N_{\rm uc}}\sum_{\mathbf k\in {\rm BZ}}\frac{1}{E_{\mathbf k}^3},\notag\\
&C_{\rm AFM}=\frac{\Delta^2}{2N_{\rm uc}}\sum_{\mathbf k\in {\rm BZ}}\frac{1}{E_{\mathbf k}^3},\notag\\
&G^{uv}_{\rm AFM}=\frac{-\Delta^2}{2N_{\rm uc}}\sum_{\mathbf k\in {\rm BZ}}\Bigg[
\frac{\epsilon_{\mathbf k}\,\partial_{k_u}\partial_{k_v}\epsilon_{\mathbf k}}{E_{\mathbf k}^3}\notag\\
&\quad+\frac{(2\Delta^2-\epsilon_{\mathbf k}^2)(\partial_{k_u}\epsilon_{\mathbf k})(\partial_{k_v}\epsilon_{\mathbf k})}{E_{\mathbf k}^5}\Bigg].
\end{align}

Substituting Eqs.~(\ref{eq:app_AFM_DGG_full})--(\ref{eq:app_AFM_Detaeta_full}) into the full Schur complement
Eq.~(\ref{eq:app_AFM_full_schur_reduction_full}) gives
\begin{align}
\Pi_{ab}(\mathbf q,\omega)=\omega^2 \mathsf C^{\rm AFM}_{ab}+\sum_{u,v}\mathbf q_u\mathbf q_v\mathsf G^{{uv},{\rm AFM}}_{ab}+O_3,
\label{eq:app_AFM_Dred_expansion_full}
\end{align}
where
\begin{align}
\mathsf C^{\rm AFM}_{ab}=\Big(C_{\rm AFM}+\frac{M_{\rm AFM}^2}{\Omega_\eta}\Big)\delta_{ab},\;\;\mathsf G^{{uv},{\rm AFM}}_{ab}=G^{uv}_{\rm AFM}\delta_{ab}.\label{eq:C_G_AFM}
\end{align}

In the strong-coupling limit $\Delta\gg t$, one can show:
\begin{align}
\Omega_\eta=&\frac{8 t^2}{\Delta}+O(\frac{t^4}{\Delta^3}),&M_{\rm AFM}&=1+O(\frac{t^2}{\Delta^2}),\notag\\
C_{\rm AFM}=&\frac{1}{2\Delta}+O(\frac{t^2}{\Delta^3})&G^{uv}_{\rm AFM}=&-\delta_{uv}\frac{t^2}{\Delta}+O(\frac{t^4}{\Delta^3}).
\end{align}
Putting together, Eqs.~(\ref{eq:C_G_AFM}) leads to
\begin{align}
\mathsf C^{\rm AFM}_{ab}=\delta_{ab}\frac{\Delta}{8t^2}+O(\frac{1}{\Delta}).
\label{eq:app_AFM_Cstrong_full}
\end{align}
The Goldstone modes from kernel-$\Pi$ is then described by the velocity $c_{\rm AFM}$ at long wavelength, to the leading order of $\frac{t}{\Delta}$-expansion:
\begin{align}
\omega_{\mathbf q}^{\rm AFM}=c_{\rm AFM}|\mathbf q|+O(q^2),\;\;c_{\rm AFM}^2=-\frac{ G^{xx}_{\rm AFM}}{\mathsf C^{\rm AFM}_{11}}\simeq\frac{8t^4}{\Delta^2}\simeq2J^2,
\label{eq:app_AFM_dispersion_full}
\end{align}
with $J\equiv 4t^2/U$ after using $\Delta\simeq U/2$ at strong coupling. $c_{\rm AFM}=\sqrt 2J$ is consistent with the usual linear spin-wave analysis for the Heisenberg model.

\bibliography{Hdet_references}
\end{document}